\RequirePackage{fix-cm}
\RequirePackage{fixltx2e}
\documentclass[twoside,english,aip, jcp,floatfix]{revtex4-1}
\usepackage{mathpazo}

\usepackage[T1]{fontenc}
\usepackage[latin9]{inputenc}
\usepackage[a4paper]{geometry}
\usepackage{xcolor}
\usepackage{babel}
\usepackage{amsmath}
\usepackage{amssymb}
\usepackage{graphicx}
\usepackage[unicode=true,pdfusetitle,
 bookmarks=false,
 breaklinks=true,pdfborder={0 0 0},pdfborderstyle={},backref=false,colorlinks=true]
 {hyperref}
\hypersetup{
 citecolor = blue, linkcolor = blue,  urlcolor  = blue}

\makeatletter

\providecommand{\tabularnewline}{\\}

\renewcommand{\textendash}{--}

\usepackage{orcidlink}

\newcommand{\beginappendix}{%
        \setcounter{table}{0}
        \renewcommand{\thetable}{A\arabic{table}}%
        \setcounter{figure}{0}
        \renewcommand{\thefigure}{A\arabic{figure}}%
     }

\makeatother

\begin{document}

\title{{\Large{}Validation of uncertainty quantification metrics: a primer
based on the consistency and adaptivity concepts}}

\author{Pascal PERNOT \orcidlink{0000-0001-8586-6222}}

\affiliation{Institut de Chimie Physique, UMR8000 CNRS,~\\
Université Paris-Saclay, 91405 Orsay, France}
\email{pascal.pernot@cnrs.fr}

\begin{abstract}
\noindent The practice of uncertainty quantification (UQ) validation,
notably in machine learning for the physico-chemical sciences, rests
on several graphical methods (scattering plots, calibration curves,
reliability diagrams and confidence curves) which explore complementary
aspects of calibration, without covering all the desirable ones. For
instance, none of these methods deals with the reliability of UQ metrics
across the range of input features (\emph{adaptivity}). Based on the
complementary concepts of \emph{consistency} and \emph{adaptivity},
the toolbox of common validation methods for variance- and intervals-
based UQ metrics is revisited with the aim to provide a better grasp
on their capabilities. This study is conceived as an introduction
to UQ validation, and all methods are derived from a few basic rules.
The methods are illustrated and tested on synthetic datasets and representative
examples extracted from the recent physico-chemical machine learning
UQ literature. 
\end{abstract}
\maketitle

\section{Introduction}

\noindent The quest for confidence in the predictions of data-based
algorithms\citep{Vishwakarma2021,Gruich2023} has led to a profusion
of uncertainty quantification (UQ) methods in machine learning (ML).\citep{Pearce2018,Musil2019,Hirschfeld2020,Tran2020,Abdar2021,Gawlikowski2021,Tynes2021,Zelikman2021,Hu2022,Varivoda2022,He2023}
Not all of these UQ methods provide uncertainties that can be relied
on,\citep{Liu2021,Pernot2022b} notably if one expects uncertainty
to inform us on a range of plausible values for a predicted property.\citep{GUM,Irikura2004}
In pre-ML computational chemistry, UQ metrics consisted essentially
in \emph{standard uncertainty}, i.e. the standard deviation of the
distribution of plausible values (a \emph{variance-based} metric),
or \emph{expanded uncertainty}, i.e. the half-range of a prediction
interval, typically at the 95\,\% level (an \emph{interval-based}
metric).\citep{Ruscic2004,Irikura2004} The advent of ML methods provided
UQ metrics beyond this standard setup, for instance distances in feature
or latent space\citep{Janet2019,Hullermeier2021,Hu2022} or the recent
$\Delta$-metric,\citep{Korolev2022} which have no direct statistical
or probabilistic meaning. These metrics might however be converted
to variance- or interval-based metrics by calibration methods, such
as conformal inference.\citep{Angelopoulos2021,Vovk2012} Still, all
UQ metrics need to be validated to ensure that they are adapted to
their intended use.

The validation of UQ metrics is designed around the concept of \emph{calibration}.
The use of uncertainty as a proxy to identify unreliable predictions,
as done in \emph{active learning}, does not require the same level
of calibration as its use for the prediction of properties at the
molecule-specific level.\citep{Reiher2022} A handful of validation
methods exist that explore more or less complementary aspects of calibration.
A trio of methods seems to have recently taken the center-stage: the
\emph{reliability diagram} (or RMSE vs RMV plot), the \emph{calibration
curve} and the \emph{confidence curve}. They implement three different
approaches to calibration which are not necessarily independent, but
they do not cover the full spectrum of calibration requirements. In
particular, none of these methods addresses the essential reliability
of predicted uncertainties with respect to the input features. As
will be shown below, a UQ metric validated by this trio of methods
might still be unreliable for individual predictions. Moreover, it
appears that some methods are not always used in an appropriate context,
such as the \emph{oracle} confidence curve for variance-based UQ metrics.\citep{Pernot2022c}

The aim of this article is to design a principled validation framework
based on complementary calibration concepts and review the relevant
methods within this framework. The choice of methods is based on two
main criteria: (1) the nature of the UQ metric to be validated (distribution,
interval, variance, other...) and (2) the calibration target (average
calibration, \emph{consistency} or \emph{adaptivity}, to be defined
below).

\subsection{Related\emph{ }studies}

\noindent Most of the validation methods presented in this study derive
from the seminal work of Gneiting \emph{et al.}\citep{Gneiting2007a,Gneiting2014}
on the calibration of \emph{probabilistic forecasters.} Probabilistic
forecasters are models that provide for each prediction a distribution
of plausible values. A major lesson from Gneiting's work is that calibration
metrics estimated over a validation dataset (\emph{average calibration})
do not necessarily lead to useful\emph{ }UQ statements and that additional
properties are necessary to design reliable UQ methods. An example
is \emph{sharpness} which quantifies the concentration of the prediction
uncertainties: among a set of average-calibrated methods, one should
prefer the sharpest one. Pernot\citep{Pernot2022a} noted that, in
the absence of a target value, sharpness might not be useful for the
validation of individual UQ metrics,\emph{ }so that alternative statistics
are required for UQ validation.

The definition of calibration provided by Gneiting \emph{et al.} was
extended by Kuleshov\citep{Kuleshov2018}. The formulation of Kuleshov,
based on coverage probabilities, led to use \emph{calibration curves}
as a validation tool. But Levi\emph{ et al.\citep{Levi2020}} demonstrated
that calibration curves are not reliable and proposed instead to use
\emph{individual calibration\citep{Chung2020}, }based on the \emph{conditional
variance} of the errors with respect to the prediction uncertainty.
Implementation of this \emph{conditional calibration} equation led
to \emph{reliability diagrams} or \emph{RMSE vs RMV} plots, also called
\emph{calibration diagrams} by Laves \emph{et al.}\citep{Laves2020}.
In practice, individual calibration is generally unreachable, and
an alternative is to consider local calibration\citep{Luo2021}, which
is reflected in the implementation of conditional statistics through
binning schemes, as done for instance in reliability diagrams.

Pernot\citep{Pernot2022b} proposed the concept of \emph{tightness}
to differentiate local calibration from average calibration, and implemented
it in a \emph{Local Z-Variance (LZV) analysis} in which calibration
is estimated in subsets of the validation dataset. These subsets can
be designed according to any relevant property (predicted value, uncertainty
or input feature). He also established the link between LZV analysis
in uncertainty space and reliability diagrams. Both reliability diagrams
and LZV analysis derive from the concept of \emph{conditional calibration},
but, for reliability diagrams, the conditioning variable is prediction
uncertainty, while the LZV analysis opens a larger palette of conditioning
variables.\emph{ Conditional coverage} with respect to input features
was proposed by Vovk\citep{Vovk2012} to asses the \emph{adaptivity}\citep{Angelopoulos2021}
of conformal predictors. The adaptivity concept can therefore be linked
to conditional calibration in input feature space, a form of tightness. 

\emph{Confidence curves} derive from another validation approach,
known as \emph{sparsification error curves} mainly used in computer
vision.\citep{Ilg2018,Scalia2020} They were not intended to validate
calibration, but to estimate the correlation between absolute errors
and UQ metrics.\citep{Tynes2021} Pernot\citep{Pernot2022c} showed
that the standard reference curve (the so-called \emph{oracle}) used
for the evaluation of confidence curves is irrelevant for variance-based
UQ metrics, and he introduced a new \emph{probabilistic reference
}that enables to test conditional calibration in uncertainty space.

Tran \emph{et al.}\citep{Tran2020} and Scalia \emph{et} \emph{al}.\citep{Scalia2020}
published motivating overviews of some of the methods presented here
(reliability diagrams, calibration and confidence curves) applied
to ML-UQ in materials sciences.

\subsection{Contributions\emph{ }}

\noindent Building on the the works of Levi \emph{et al}.\emph{\citep{Levi2020},}
Pernot\citep{Pernot2022b} and Angelopoulos \emph{et al.}\citep{Angelopoulos2021}
about conditional calibration, I propose to distinguish two calibration
targets besides average calibration,\textcolor{orange}{{} }namely 
\begin{itemize}
\item \emph{consistency} as the conditional calibration with respect to
prediction uncertainty, and 
\item \emph{adaptivity} as the conditional calibration with respect to input
features. 
\end{itemize}
\noindent Consistency derives from the metrological consistency of
measurements,\citep{Kacker2010} while adaptivity is borrowed from
the conformal inference literature\citep{Angelopoulos2021}. \emph{Tightness},
as introduced by Pernot\citep{Pernot2022b}, should then be understood
as covering both consistency and adaptivity. Unless there is a monotonous
transformation between input features and prediction uncertainty,
consistency and adaptivity are distinct calibration targets. 

The distinction between consistency and adaptivity is important to
better define the objective(s) of each validation method. In fact,
adaptivity has not been considered in recent overviews of ML-UQ validation
metrics\citep{Tran2020,Scalia2020}, but was practically implemented
\textendash{} albeit unnamed \textendash{} through the LZV and Local
Coverage Probability (LCP) analyses (see for instance Fig.\,4 in
Pernot\citep{Pernot2022a}). Still, most ML-UQ validation studies
focus on calibration and consistency, and I will show that this is
not sufficient to ensure the reliability of an UQ method across the
input features space. Adaptivity is essential to achieve reliable
UQ at the molecule-specific level advocated by Reiher\citep{Reiher2022}.

\subsection{Structure of the article}

\noindent The theoretical bases of the calibration-consistency-adaptivity
framework are presented in the next section (Sect.\,\ref{sec:Validation-concepts-and})
for variance- and interval-based UQ metrics. A comprehensive set of
validation methods is reviewed next (Sect.\,\ref{sec:Validation-methods})
according to their application range and calibration target, in order
to appreciate their merits and limitations. Examples from the recent
materials and chemistry ML-UQ literature are treated as case studies
in Sect.\,\ref{sec:Application}. Recommendations are presented as
conclusions (Sect.\,\ref{sec:Conclusion}).

\section{Validation concepts and models\label{sec:Validation-concepts-and}}

\noindent A major distinction will be made below between \emph{variance-based}
UQ metrics and \emph{interval-based} UQ metrics. These occur in ML
as statistical summaries of empirical distributions or ensembles,
or as parameters of theoretical distributions (typically normal).
Other UQ metrics, such as distances in feature space or latent space,\citep{Janet2019,Hu2022}
or the $\Delta$-metric,\citep{Korolev2022} which have not the correct
dimension to be comparable to errors, need first to be converted to
one of those two metrics to be usable for validation.\textcolor{violet}{\citep{Vovk2012,Angelopoulos2021,Cauchois2021,Feldman2021}}

\subsection{Validation model for variance-based UQ metrics}

\subsubsection{Validation datasets\label{subsec:Validation-datasets}}

\noindent In order to validate variance-based UQ results, one needs\textcolor{orange}{{}
}a set of predicted values $V=\{V_{i}\}_{i=1}^{M}$, the corresponding
uncertainties $u_{V}=\{u{}_{V}\}_{i=1}^{M}$, reference data to compare
with $R=\{R_{i}\}_{i=1}^{M}$, and possibly their uncertainties $u_{R}=\{u_{R_{i}}\}_{i=1}^{M}$.
Most of the validation methods considered below require to transform
these to $E=R-V$ and $u_{E}=(u_{R}^{2}+u_{V}^{2})^{1/2}$. 

The minimal validation dataset, $C=\left\{ E,u_{E}\right\} $, enables
to test calibration and consistency but not adaptivity. It is therefore
better to include a relevant input feature $X$ in the dataset $C=\left\{ E,u_{E},X\right\} $,
extending the validation targets to adaptivity in $X$ space. Alternatively,
if validation is done \emph{a posteriori} and no relevant input feature
is available, adaptivity can be tested on $C=\left\{ E,u_{E},V\right\} $,
with some caveats presented in Sect.\,\ref{subsec:Using-latent-distances}.

\subsubsection{Generative model\label{subsec:Generative-model}}

\noindent For variance-based UQ metrics, validation methods are based
on a probabilistic model linking errors to uncertainties
\begin{equation}
E_{i}\sim D(0,u_{E_{i}})\label{eq:probmod}
\end{equation}
where $D(\mu,\sigma)$ is an unspecified probability density function
with mean $\mu$ and standard deviation $\sigma$. This model states
that errors should be unbiased ($\mu=0$) and that uncertainty describes
their dispersion as a standard deviation, following the metrological
standard.\citep{GUM} 

It is essential to note that the relation between $E$ and $u_{E}$
is asymmetric: small uncertainties should be associated with small
errors, but small errors might be associated with small or large uncertainties,
while large errors should be associated with large uncertainties.
This considerably diminishes the interest of validation tests based
on the ranking of $|E|$ vs $u_{E}$ (e.g. correlation coefficients).

\subsubsection{Validation model\label{subsec:Validation-model-var}}

\paragraph{Average calibration.\label{par:Average-calibration.}}

\noindent According to the generative model, the validation of $u_{E}$
should be based on testing that it correctly describes the dispersion
of $E$.\citep{Pernot2022a,Pernot2022b} One can for instance check
that
\begin{equation}
\mathrm{Var}(E)\simeq<u_{E}^{2}>\label{eq:varEVal}
\end{equation}
where the average is taken over the validation dataset, and which
is valid only if $<E>\simeq0$.\citep{Pernot2022b} However, this
formula ignores the pairing between errors and uncertainties, and
a more stringent test is based on \emph{z}-scores ($Z=E/u_{E}$)\citep{Pernot2022b},
i.e.
\begin{equation}
\mathrm{Var}(Z)\simeq1\label{eq:varZval}
\end{equation}
Note that if the elements of $E$ and $u_{E}$ are obtained as the
means and standard deviations of small ensembles (say less than 30
elements) these formulas have to be transformed in order to account
for the uncertainty on the statistics.\citep{Pernot2022b} Unfortunately,
hypotheses need then to be made on the distribution of these small
ensembles. For a normal generative distribution of errors, the distribution
of the mean of $n$ values (ensemble size) is a Student's-\emph{t}
distribution with $\nu=n-1$ degrees of freedom, one should have $\mathrm{Var}(Z)\simeq(n-1)/(n-3)$.\citep{Pernot2022b} 

The satisfaction of one or both of these tests (Eqns.\,\ref{eq:varEVal}-\ref{eq:varZval})
validates \emph{average} \emph{calibration}, which is a minimal requirement,
but does not guarantee the usefulness of individual uncertainties,
as average calibration can be satisfied by a compensation of under-
and over-estimated values of $u_{E}$.

\paragraph{Conditional calibration: consistency and adaptivity.}

\noindent Based on Eq.\,\ref{eq:varZval}, one can evaluate the reliability
of individual uncertainties by \emph{conditional calibration}\citep{Vovk2012,Angelopoulos2021},
i.e., 
\begin{equation}
\mathrm{Var}(Z|A=a)\simeq1,\,\forall a\in\mathcal{A}\label{eq:varZvalCond-1-1}
\end{equation}
where $A$ is a variable with values $a$ in $\mathcal{A}$. As detailed
below, $A$ can be any relevant quantity, such as the uncertainty
$u_{E}$ or a feature $X$. The choice of the conditioning variable
depends on the question to be answered. Choosing the uncertainty $u_{E}$
will assess the calibration across the range of uncertainty values,
i.e. the \emph{consistency} between $E$ and $u_{E}$, while choosing
$X$ will assess \emph{adaptivity}.\textcolor{violet}{{} }

Using this terminology, one can see that the \emph{individual calibration}
proposed by Levi\emph{ et al.}\citep{Levi2020} 
\begin{equation}
\mathrm{Var}(E|u_{E}=\sigma)\simeq\sigma^{2},\,\forall\sigma>0\label{eq:varEvalCond}
\end{equation}
deriving from Eq.\,\ref{eq:varEVal} validates consistency. Although
it is a step forward from average calibration, it does not test the
adaptivity of the UQ metric under scrutiny. Nor does the \emph{local
Z-Variance }(LZV)\emph{ analysis}\citep{Pernot2022a,Pernot2022b}
in $u_{E}$ space, which can be derived from Eq.\,\ref{eq:varZval}
\begin{equation}
\mathrm{Var}(Z|u_{E}=\sigma)\simeq1,\,\forall\sigma>0\label{eq:varZvalCond}
\end{equation}

Practical implementation of conditional calibration tests to variance-
and interval- based UQ metrics requires to split the validation set
into subsets, generally based on the binning of the conditioning variable.
In these conditions, one is more testing \emph{local} than \emph{individual}
calibration\citep{Luo2021}, and Eq.\,\ref{eq:varEvalCond} leads
to \emph{reliability} \emph{diagrams}\citep{Levi2020}, while Eq.\,\ref{eq:varZvalCond}
leads to the \emph{local Z-Variance }(LZV)\emph{ analysis} in $u_{E}$
space. Note that the bin size in these methods should be small enough
to get as close as possible to individual calibration, but also large
enough to ensure a reasonable power for statistical testing. The LZV
analysis is easily applicable to any conditioning variable (see examples
in Pernot\citep{Pernot2022a}), which is not the case of the reliability
diagram which would need to superimpose reliability curves for each
subset and be difficult to analyze. 

An ideal variance-based UQ metric, i.e., one which provides reliable
individual uncertainties, should satisfy calibration, consistency
\emph{and} adaptivity. Consistency and adaptivity are therefore two
complementary aspects of \emph{tightness}, a term introduced in a
previous study\citep{Pernot2022b} to characterize local calibration. 

Calibration is a necessary condition to reach consistency or adaptivity.
In fact, consistency/adaptivity expressed as conditional calibration
should imply average calibration, but the splitting of the data into
subsets makes that the power of individual consistency/adaptivity
tests is smaller than for the full validation set. It is therefore
better to test average calibration separately, notably for small validation
datasets. Note that for homoscedastic datasets ($u_{E}=c^{te}$) consistency
is implied by calibration, but not adaptivity. As already mentioned,
as $X$, $V$ and $u_{E}$ are not necessarily related by monotonous
transformations, one should not expect these complementary conditional
calibration tests to provide identical results.

\subsection{Validation model for interval-based UQ metrics}

\noindent Prediction intervals can be extracted from predictive distributions,\citep{Gneiting2014}
generated by quantile regression or conformal predictors,\citep{Pearce2018,Hu2022}
or estimated from a variance-based UQ metric and an hypothetical generative
distribution\citep{Pernot2022a}.

\subsubsection{Validation datasets}

\noindent In order to validate interval-based UQ results, one needs
a set of reference data $R=\{R_{i}\}_{i=1}^{M}$, and possibly their
uncertainties $u_{R}=\{u_{R_{i}}\}_{i=1}^{M}$, and a series of prediction
intervals with prescribed coverage $P$, $I_{V}=\{I_{V,P}=\{I_{V_{i},P}\}_{i=1}^{M}\}_{P\in\mathcal{P}}$,
where $I_{V_{i},P}=[I_{V_{i},P}^{-},I_{V_{i},P}^{+}]$ and $\mathcal{P}$
is a set of coverage values, typically expressed as percentages in
$]0,\thinspace100[$. 

For convenience and consistency with the variance-based approach,
one might transform these data as errors and error intervals $C=\{V,I_{E}\}=\{V_{i},\{I_{E_{i},P}\}_{P\in\mathcal{P}}\}_{i=1}^{M}$
where $I_{E_{i},P}=[R_{i}-I_{V_{i},P}^{-},\thinspace R_{i}-I_{V_{i},P}^{+}]$
is the $P$\,\% interval for the prediction error at point $i$.\footnote{Note that this definition differs from the one in my previous studies,
which were based on the existence of a predicted value and an expanded
uncertainty, leading to zero-centered intervals that should contain
the error. In the present case the intervals are error-centered, but
should contain 0.} As for variance-based UQ metrics, the uncertainty on the reference
values, if any, should be accounted for, and the inclusion in the
dataset of pertinent input features $X=\{X_{i}\}_{i=1}^{M}$ is necessary
to test adaptivity.

\subsubsection{Validation model}

\paragraph{Average coverage.}

\noindent Within this setup, a method is considered to be calibrated
if prediction intervals have the correct empirical coverage. One defines
the \emph{prediction interval coverage probability} (PICP) as
\begin{equation}
\nu_{p}=\mathbb{P}\left(0\in I_{E,P}\right)\label{eq:validInt}
\end{equation}
where $\mathbb{P}$ is the probability function, and $I_{E,P}$ is
a $P=100p$\,\% prediction error interval. Using PICPs, a method
is calibrated in average, or marginally, if\emph{
\begin{equation}
\nu_{p}=p,\,\forall100p\in\mathcal{P}\label{eq:calibration-PICP}
\end{equation}
}which is equivalent to Kuleshov's definition\citep{Kuleshov2018}
when all probability levels can be estimated (for instance if one
has access to the full predictive distribution). For a limited set
of probabilities, one gets a milder calibration constraint. 

Application of Eq.\,\ref{eq:calibration-PICP} for a series of percentages
provides a\emph{ calibration curve}, where the values of $\nu_{p}$
are plotted against the target probabilities $p$.\citep{Kuleshov2018} 

\paragraph{Conditional coverage.}

Average calibration based on PICPs can be met by validation sets with
unsuitable properties\citep{Levi2020}, and, as for variance-based
UQ metrics, it is possible to build more stringent calibration tests
based on \emph{conditional coverage},\citep{Angelopoulos2021} i.e.,
\begin{equation}
\nu_{p}=\mathbb{P}\left(0\in I_{E,P}|A=a\right),\thinspace\forall a\in\mathcal{A}\label{eq:condCov}
\end{equation}
where $A$ is a property with values $a$ in $\mathcal{A}$. To get
the analog of consistency, one might consider to use $U_{P}$, the
half-range of $I_{E_{i},P}$, as a conditioning variable in order
to check that errors are consistent at all uncertainty scales. For
instance, Pernot used the expanded uncertainty $U_{95}$ to this aim.\citep{Pernot2022a}
For adaptivity testing, the conditioning variable can be any relevant
feature $X$.\citep{Pernot2022a}

In practice, the PICPs are estimated as frequencies over the validation
set
\begin{equation}
\nu_{p,M}=\frac{1}{M}\sum_{i=1}^{M}\boldsymbol{1}\left(0\in I_{E_{i},P}\right)\label{eq:PICP-freq}
\end{equation}
where $\boldsymbol{1}(x)$ is the \emph{indicator function} for proposition
$x$, taking values 1 when $x$ is true and 0 when $x$ is false.
Estimating a PICP amounts to count the number of times prediction
error intervals contain zero. 

Implementation of the conditional coverage tests requites binning
according to the conditioning variable, with PICP testing within each
bin, leading to the \emph{Local Coverage Probability} (LCP) analysis.\citep{Pernot2022a}

Note that, as for reliability diagrams it is also possible to consider
\emph{conditional calibration curves}, with the same problem of readability
and interpretability of overlapping curves.

\section{Validation methods\label{sec:Validation-methods}}

\noindent The principles exposed in the previous section are now developed
into practical methods grouped by validation target (average calibration,
consistency and adaptivity). All methods are illustrated on synthetic
datasets designed to reveal their potential limitations.

\subsection{Synthetic datasets\label{subsec:Synthetic-datasets}}

\noindent To illustrate the methods presented in this study, six datasets
of size $M=5000$ were designed to illustrate common validation problems.
They are summarized in Table\,\ref{tab:Summary-of-synthetic}. 
\begin{table*}[!t]
\begin{centering}
\begin{tabular}{cccc}
\hline 
Case & Calibration & Consistency & Adaptivity\tabularnewline
\hline 
A & yes & yes & yes\tabularnewline
B & yes & no & no\tabularnewline
C & yes/no & no & no\tabularnewline
D & no & no & no\tabularnewline
E & yes & yes & yes\tabularnewline
F & yes & - & no\tabularnewline
\hline 
\end{tabular}
\par\end{centering}
\caption{\label{tab:Summary-of-synthetic}Summary of the validation properties
expected from the synthetic datasets. }
\end{table*}

Case A features a quadratic model with additive \emph{heteroscedastic}
noise
\begin{align}
R_{i} & =a+X_{i}^{2}\\
u_{E_{i}} & =b\times|R_{i}|\\
V_{i} & =R_{i}+u_{E_{i}}\times N(0,1)
\end{align}
where $N(0,1)$ is the standard normal distribution. The errors are
thus in agreement with the generative model (Eq.\,\ref{eq:probmod}),
and, by construction, the set $C=\{X,E,u_{E}\}$ is \emph{calibrated},
\emph{consistent} and \emph{adaptive}. It should pass all the corresponding
tests.\textcolor{violet}{{} }

To generate a calibrated, non-consistent and non-adaptive dataset
(Case B), one preserves all the data of Case A, except the uncertainties
which are derived as perturbations of the mean uncertainty of Case
A
\begin{equation}
u_{E_{i}}=\sqrt{<u_{E}>_{A}}\times N(1,0.1)
\end{equation}
This transformation preserves the calibration, but the errors are
now inconsistent with $u_{E}$ and adaptivity is also lost. 

Case C is issued from Case A with a random ordering of $u_{E}$. This
preserves average calibration by Eq.\,\ref{eq:varEVal}, but breaks
calibration by Eq.\,\ref{eq:varZval}. Consistency and adaptivity
are also broken.

Case D is derived from Case A by a uniform scaling of $u_{E}$ by
a factor two. This set should fail all the tests for calibration,
consistency and adaptivity. Note that there is not much sense to test
consistency and adaptivity when average calibration is not satisfied.
This will be considered as illustrative of the expected diagnostics. 

Case E is similar to Case A, but the errors have now a Student's-$t_{\nu=4}$
distribution with four degrees of freedom
\begin{equation}
V_{i}=R_{i}+u_{E_{i}}\times t_{4}(0,1)
\end{equation}
where $t_{4}(0,1)$ is the $t$ distribution scaled to have a unit
variance. This dataset is calibrated, consistent and adaptive. However,
tests involving a normality hypothesis for the generative distribution
should fail.

Finally, Case F is an \emph{homoscedastic} dataset deriving from Case
A, where the uncertainties are replaced by a constant value (the root
mean square of Case A uncertainties). It is calibrated, and not adaptive.
For a homoscedastic dataset, consistency is identical to calibration.

\subsection{Testing average calibration\label{subsec:Average-calibration}}

\noindent The simplest way to assess average calibration for variance-based
UQ metrics is to test the equality in Eq.\,\ref{eq:varEVal} or in
Eq.\,\ref{eq:varZval}. Various statistical procedures have been
proposed, and it is best to avoid any assumption of normality for
the distribution of errors or \emph{z}-scores. See Pernot\citep{Pernot2022a}
for details.

Application to the synthetic datasets is presented in Table\,\ref{tab:Validation-of-average}.
For Cases A, B, E and F, both tests are satisfied. Case D is obviously
non calibrated. However, datasets with consistency problems (Case
C) might still check one of both metrics. 
\begin{table}[t]
\begin{centering}
\begin{tabular}{lr@{\extracolsep{0pt}.}lr@{\extracolsep{0pt}.}l}
\hline 
 & \multicolumn{2}{c}{$\frac{\mathrm{Var}(E)}{<u_{E}^{2}>}$} & \multicolumn{2}{c}{$\mathrm{Var}(Z)$}\tabularnewline
\hline 
Case A  & 0&98  & 1&02 \tabularnewline
Case B  & 0&97 & 1&02\tabularnewline
Case C  & 0&98  & 53&8{*}\tabularnewline
Case D  & 0&25{*}  & 0&26{*} \tabularnewline
Case E  & 1&09  & 0&96 \tabularnewline
Case F & 1&00 & 1&00\tabularnewline
\hline 
\end{tabular}
\par\end{centering}
\caption{\label{tab:Validation-of-average}Validation of average calibration
for cases A-F. An asterisk ({*}) means that a statistic exceeds its
admissible range.}
\end{table}

For interval-based UQ metrics, validation is done by testing Eq.\,\ref{eq:calibration-PICP}
for the probability levels that are available. The statistical procedure
comes with the same caveats as for variance-based UQ metrics and has
been detailed by Pernot\citep{Pernot2022a}. When a large set of probability
levels is available, one can build a calibration curve, as shown next.
\begin{figure*}[t]
\noindent \begin{centering}
\begin{tabular}{ccc}
\includegraphics[width=0.33\textwidth]{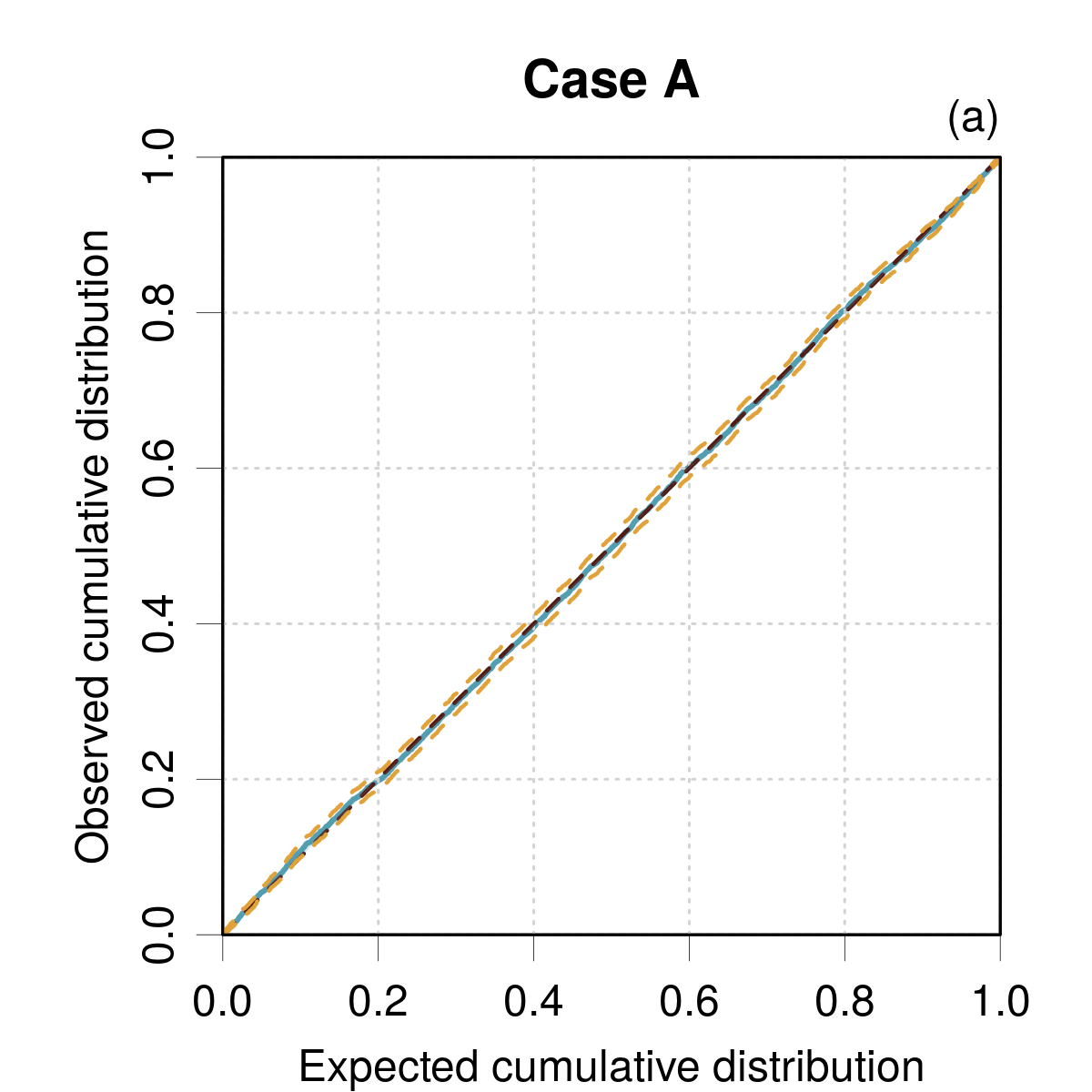} & \includegraphics[width=0.33\textwidth]{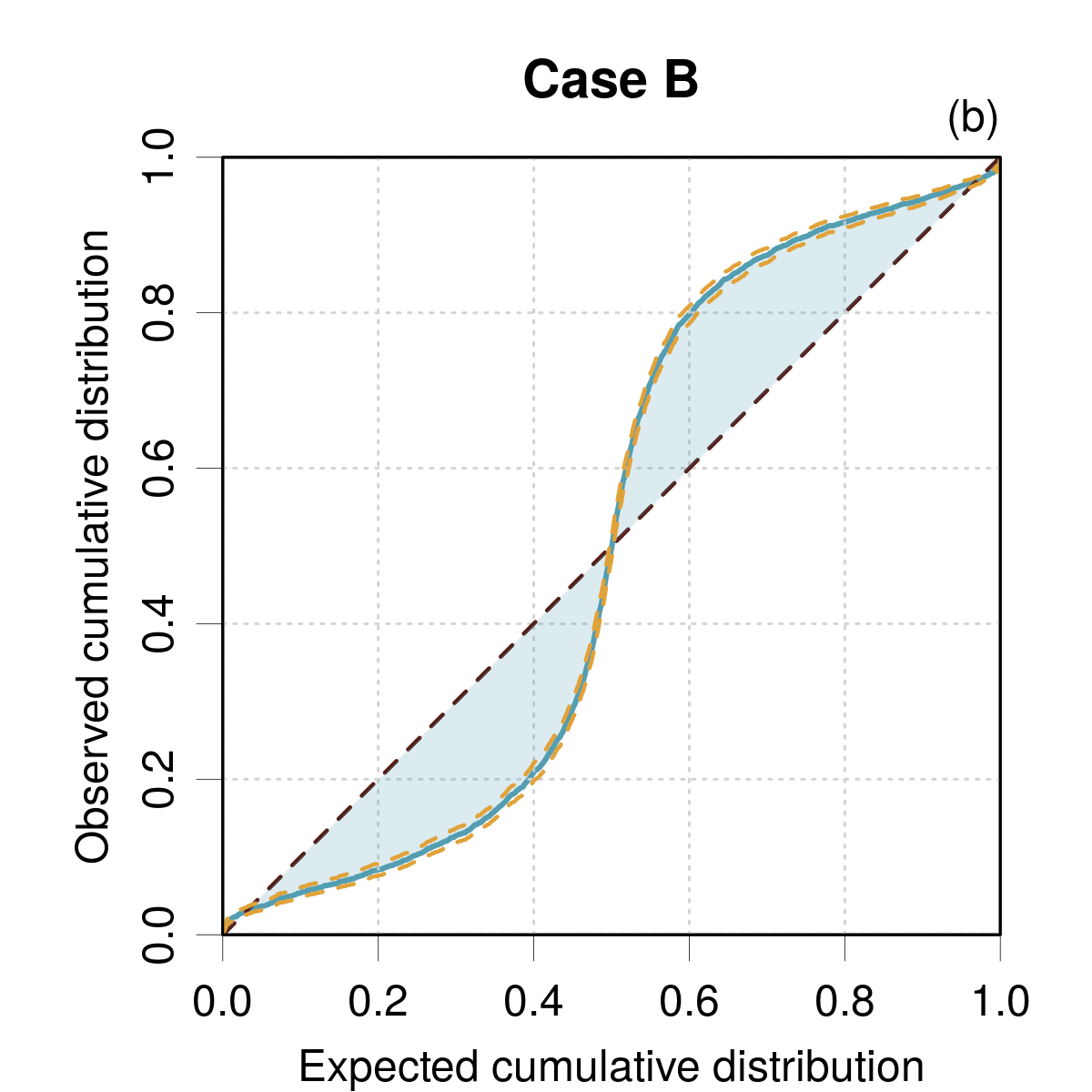} & \includegraphics[width=0.33\textwidth]{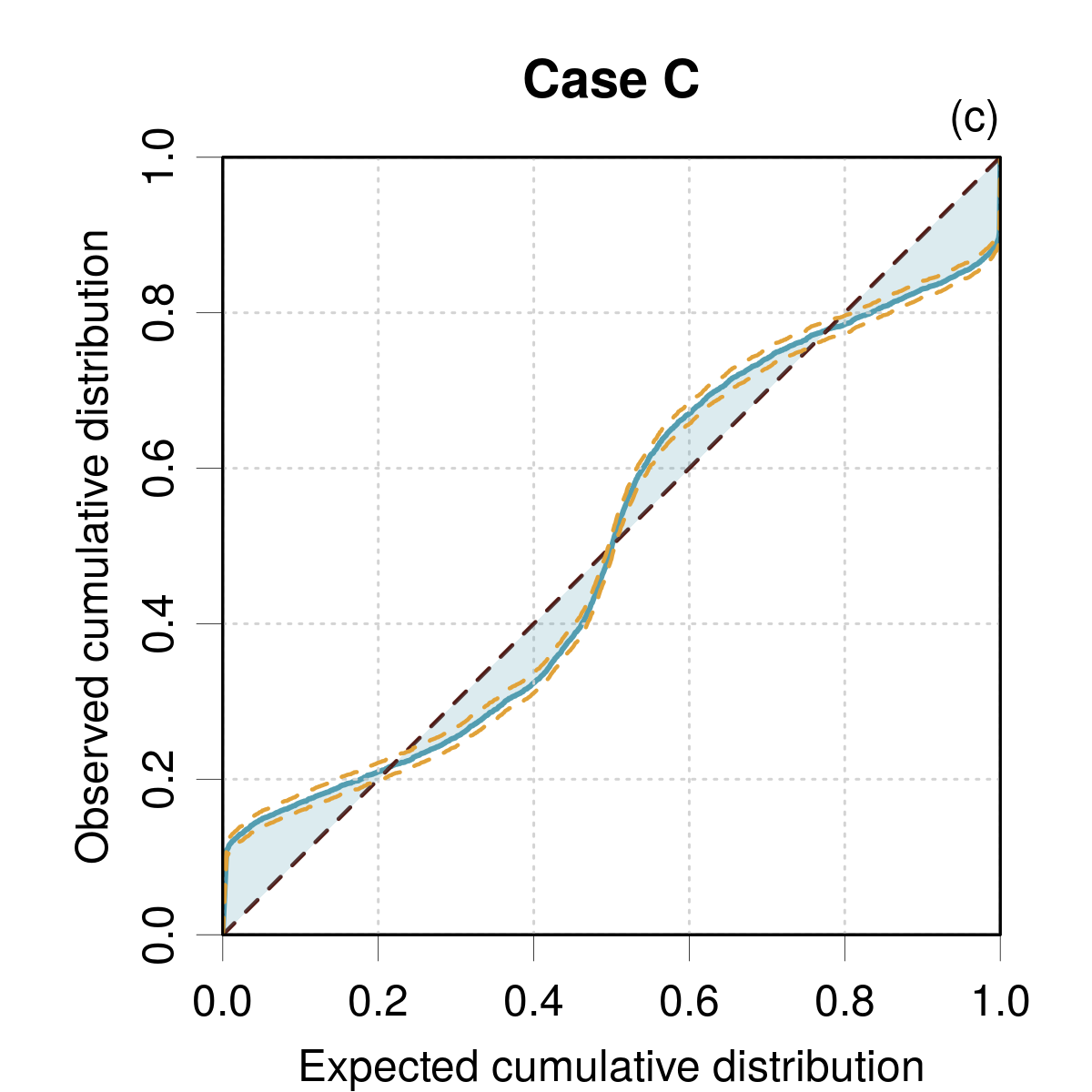}\tabularnewline
\includegraphics[width=0.33\textwidth]{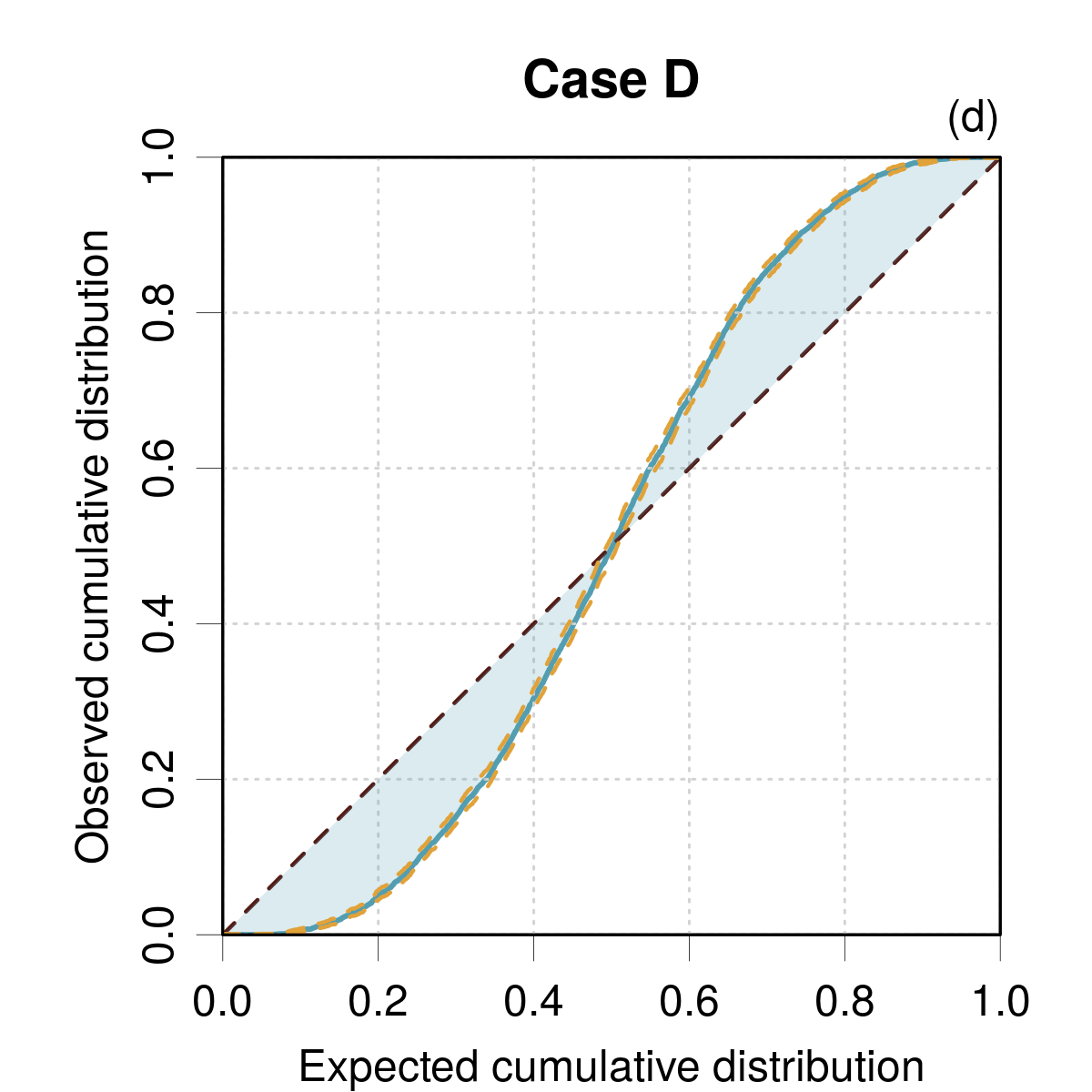} & \includegraphics[height=6cm]{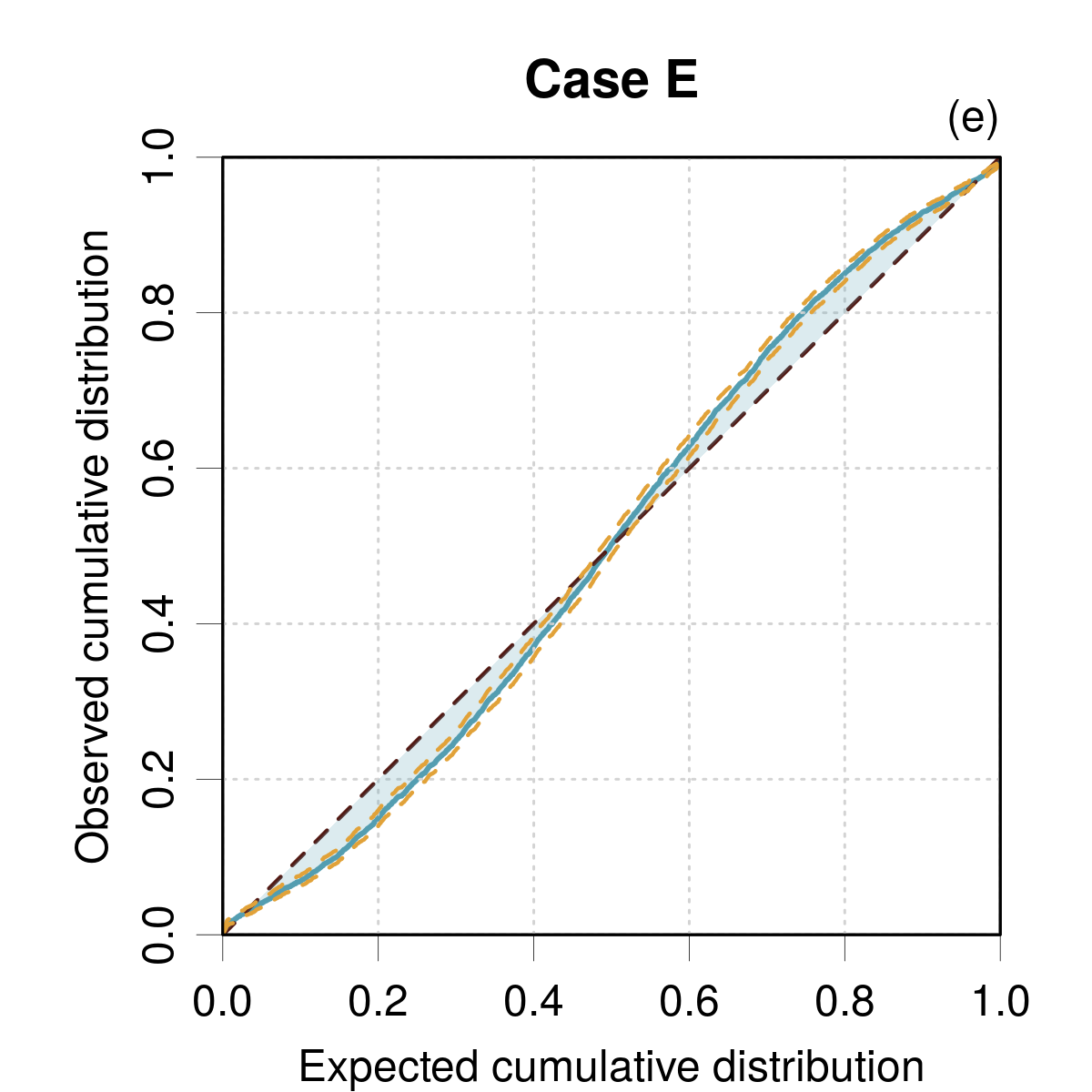} & \includegraphics[width=0.33\textwidth]{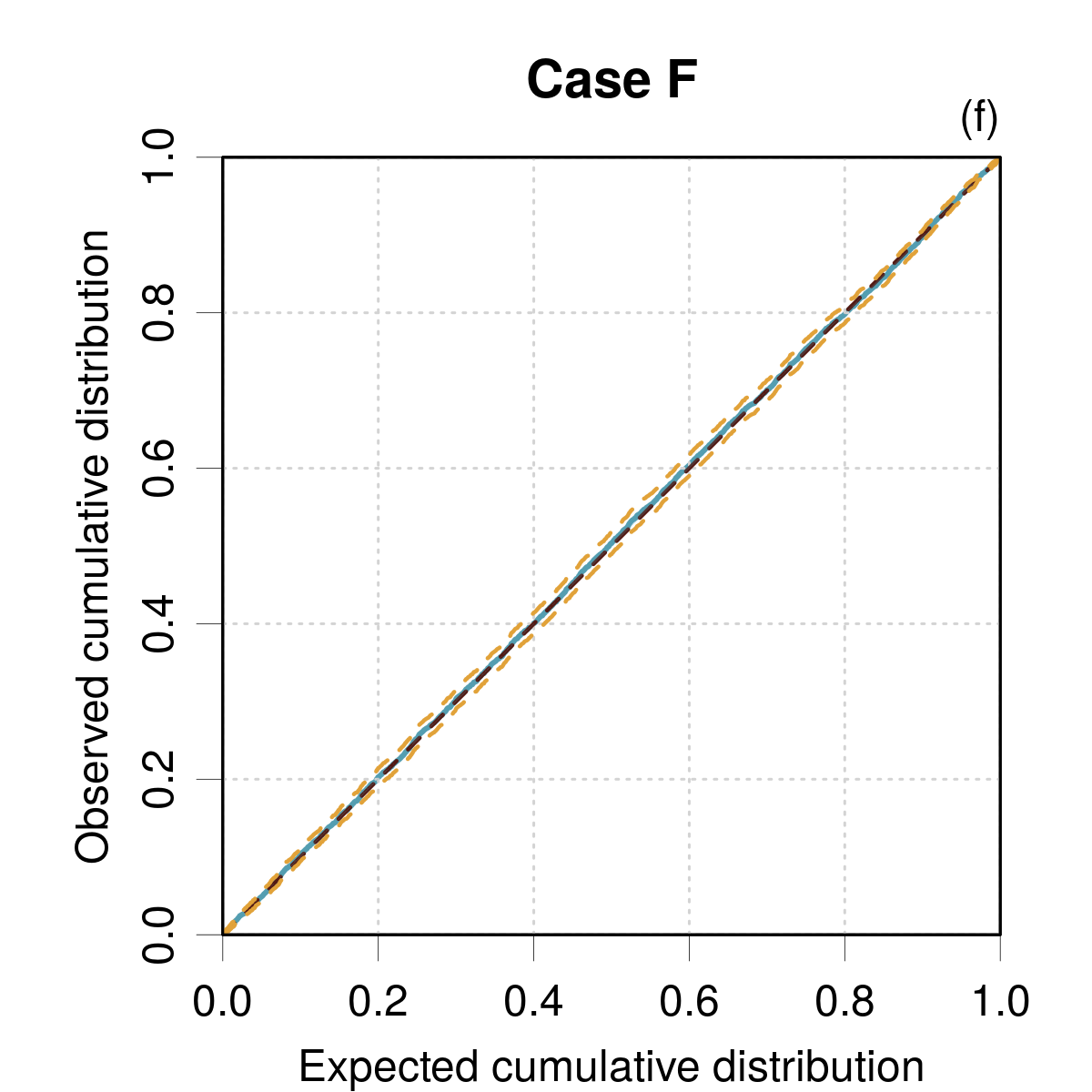}\tabularnewline
\end{tabular}
\par\end{centering}
\caption{\label{fig:Calibration-curves}Calibration curves for cases A-F. The
calibration curve is a solid blue line. The reference line and its
95\,\% confidence interval are shown as dashed lines. The shaded
area represents the calibration error (miscalibration area).}
\end{figure*}

\subsubsection{Calibration curve\label{subsec:Calibration-curve}}

\noindent In a calibration curve\citep{Kuleshov2018}, one estimates
the probability of errors $E$ to fall within a series of intervals
with coverage probability varying in $[0,1]$, and one plots the resulting
probability against the target one (Eq.\,\ref{eq:validInt}). Calibration
curves are applicable to homoscedastic datasets. 

The method is also referred to as \emph{confidence}- or \emph{intervals}-based
calibration\citep{Scalia2020,Pernot2022a,Pernot2022b} and formalized
as a variant of Eqns.\,\ref{eq:calibration-PICP}-\ref{eq:PICP-freq}

\emph{
\begin{equation}
\lim_{M\rightarrow\infty}\eta_{p,M}=p,\,\forall p\in[0,1]\label{eq:calibration-1}
\end{equation}
}where
\begin{equation}
\eta_{p,M}=\frac{1}{M}\sum_{i=1}^{M}\boldsymbol{1}\left(E_{i}<q_{p,i}\right)\label{eq:PICP-quantile}
\end{equation}
and $q_{p,i}$ is the quantile for probability $p$ at point $i$.

The ideal calibration curve is the identity line (Fig.\,\ref{fig:Calibration-curves},
Cases A and F). For validation, a 95\,\% confidence interval is plotted\citep{Pernot2022a},
either around the empirical curve or the identity line (the latter
is preferred when multiple curves are drawn for the conditional case
presented later). 

As the synthetic datasets provide only a variance-based UQ metric
($u_{E})$, the quantiles in Eq.\,\ref{eq:PICP-quantile} are calculated
by assuming a normal generative distribution, as usually done in the
literature. In such cases, deviations from the identity line are not
straightforward to interpret, as they might have their origin in the
non-consistency of uncertainties (Cases B and C), their non-calibration
(Case D) or a bad choice of reference distribution (Case E). 

It is interesting to contrast these results with the statistics for
average calibration obtained above (Table\,\ref{tab:Validation-of-average}).
Case E has satisfying calibration statistics, but presents a slightly
problematic calibration curve. This is due to the inadequate choice
of a normal generative distribution for this dataset based on a Student's
distribution. 

Also, Cases B and C present notably deviant calibration curves despite
the fact that they have at least one correct calibration metric. The
calibration curve obviously contains more information than the average
calibration metrics, but it does not provide direct information on
consistency or adaptivity. For instance, without additional information,
it is difficult to differentiate the diagnostics for Cases B and D.
For variance-based metrics, its use requires also an hypothesis on
the generative distribution to estimate the theoretical quantiles,
which complicates the interpretation of deviant curves. 

\subsection{Testing consistency\label{subsec:Consistency-testing}}

\noindent The validation literature for variance- or interval-based
UQ metrics has mostly focused on consistency tests, and several methods
are available, which are not fully equivalent. The aim of this section
is to highlight the main features of these methods. 
\begin{figure*}[t]
\noindent \begin{centering}
\begin{tabular}{ccc}
\includegraphics[width=0.33\textwidth]{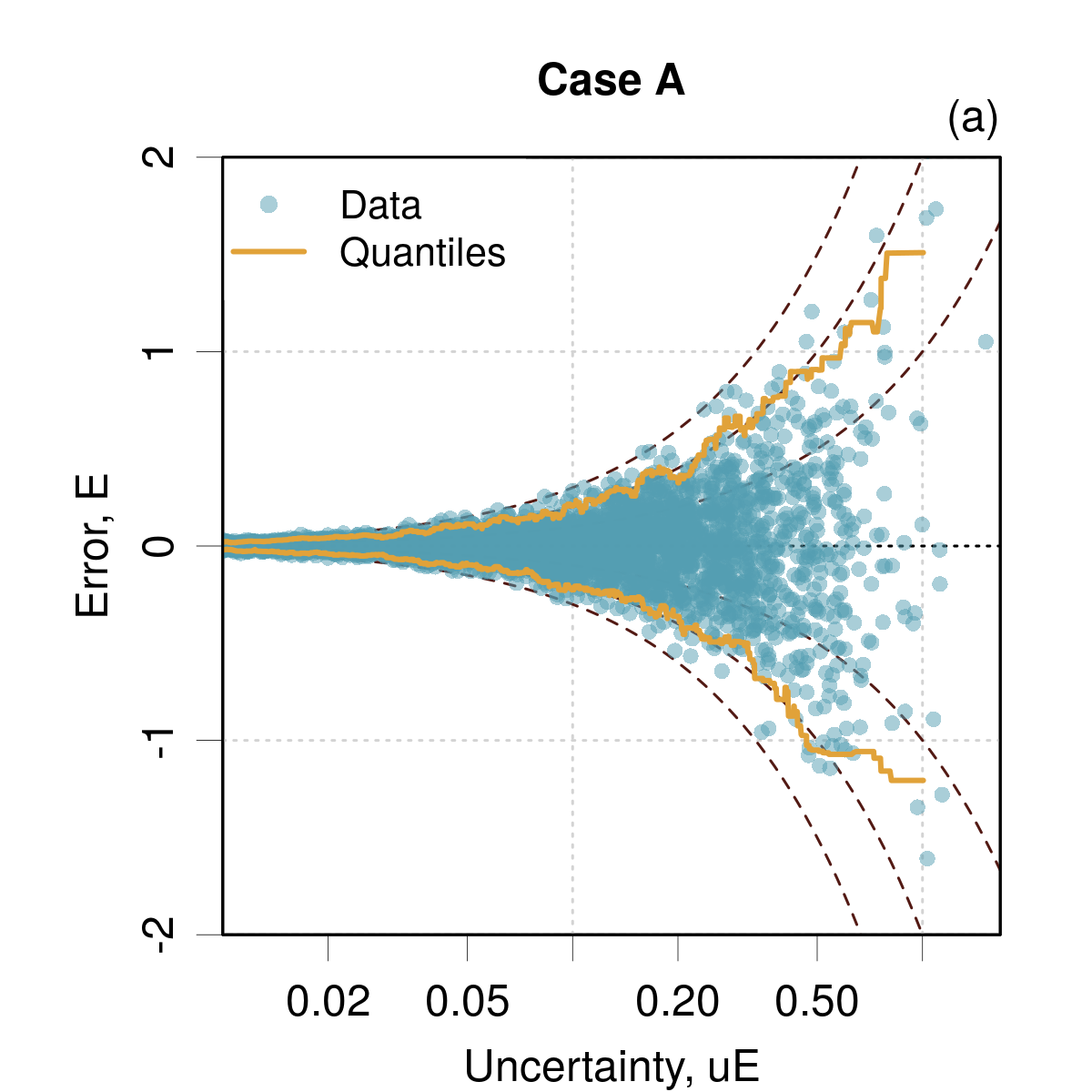} & \includegraphics[width=0.33\textwidth]{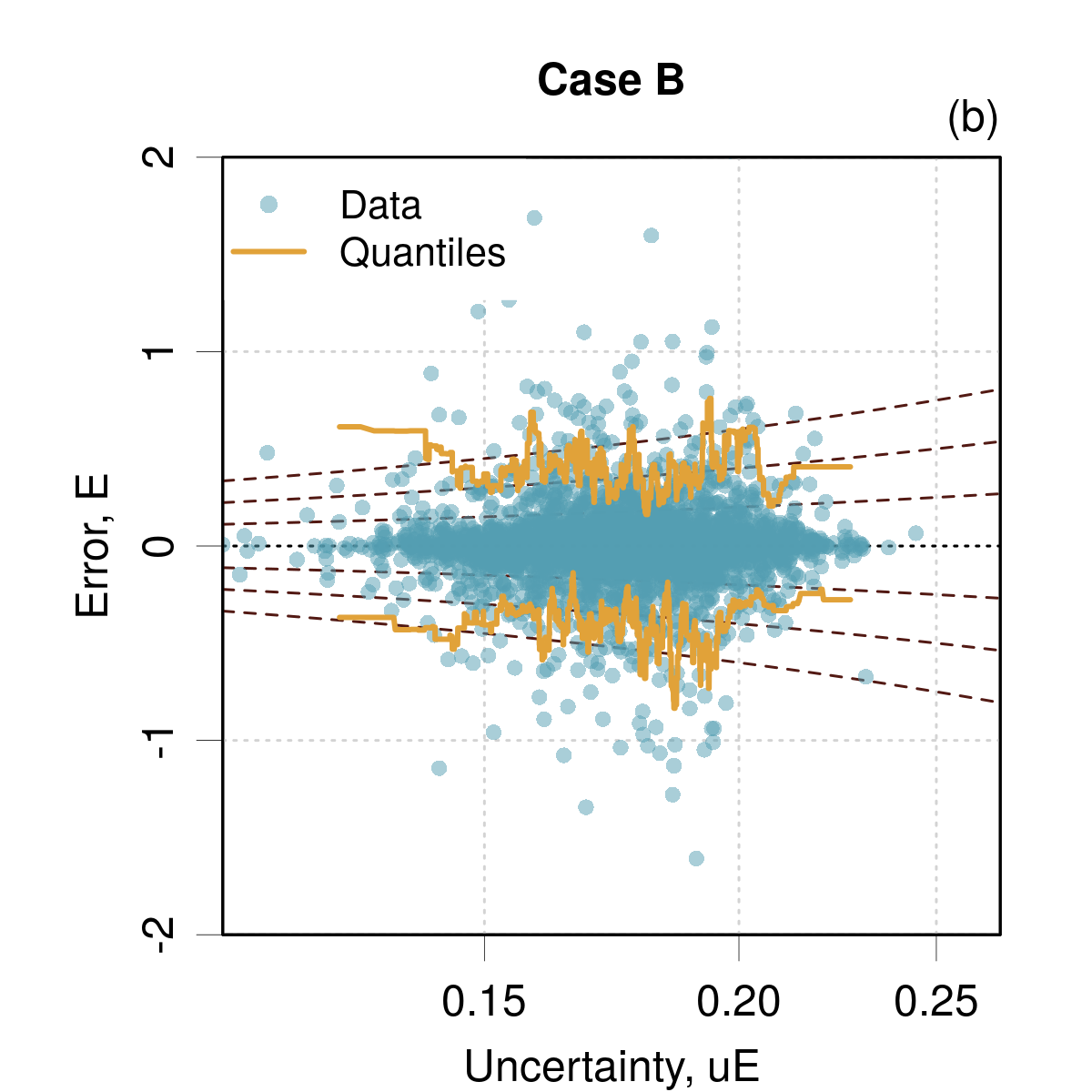} & \includegraphics[width=0.33\textwidth]{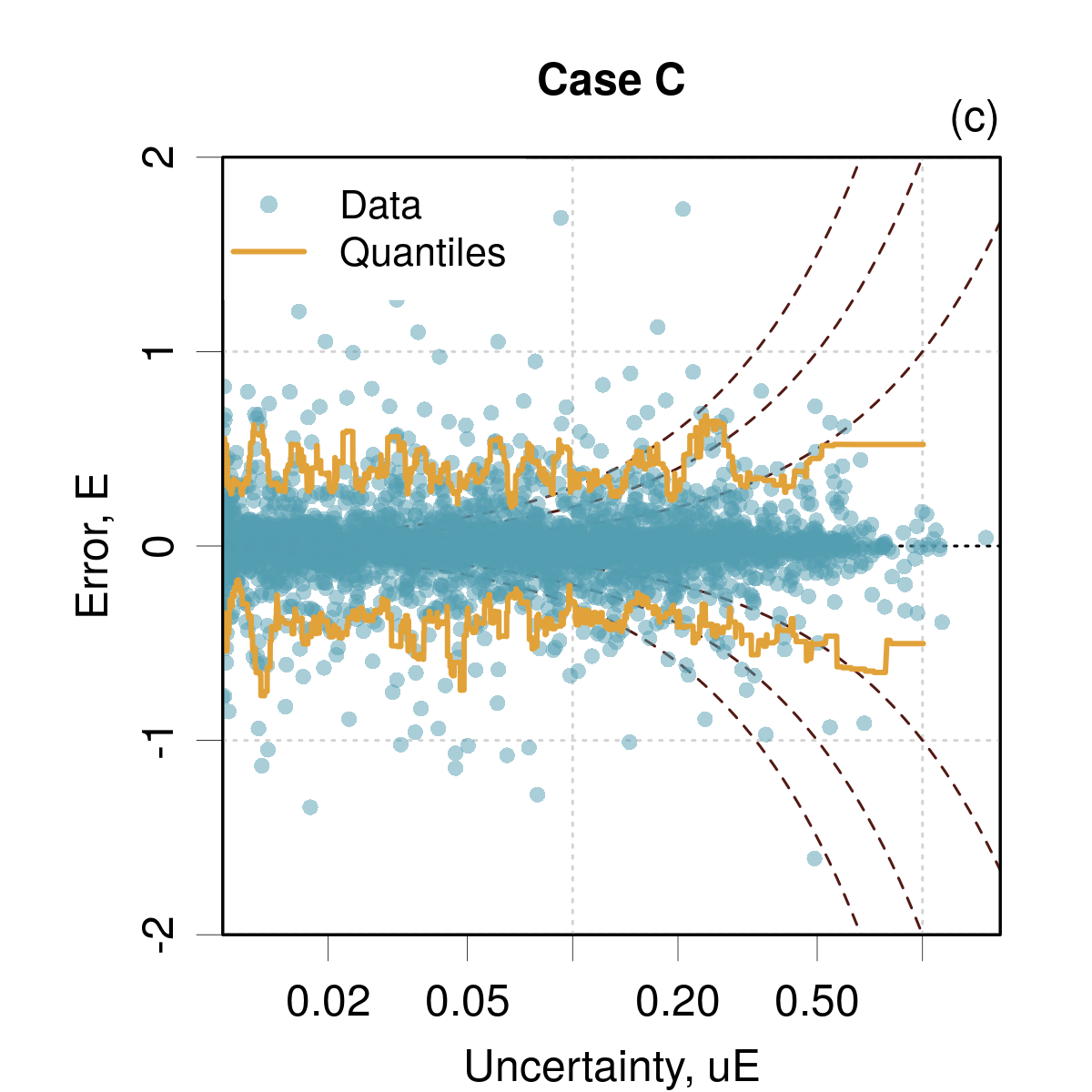}\tabularnewline
\includegraphics[width=0.33\textwidth]{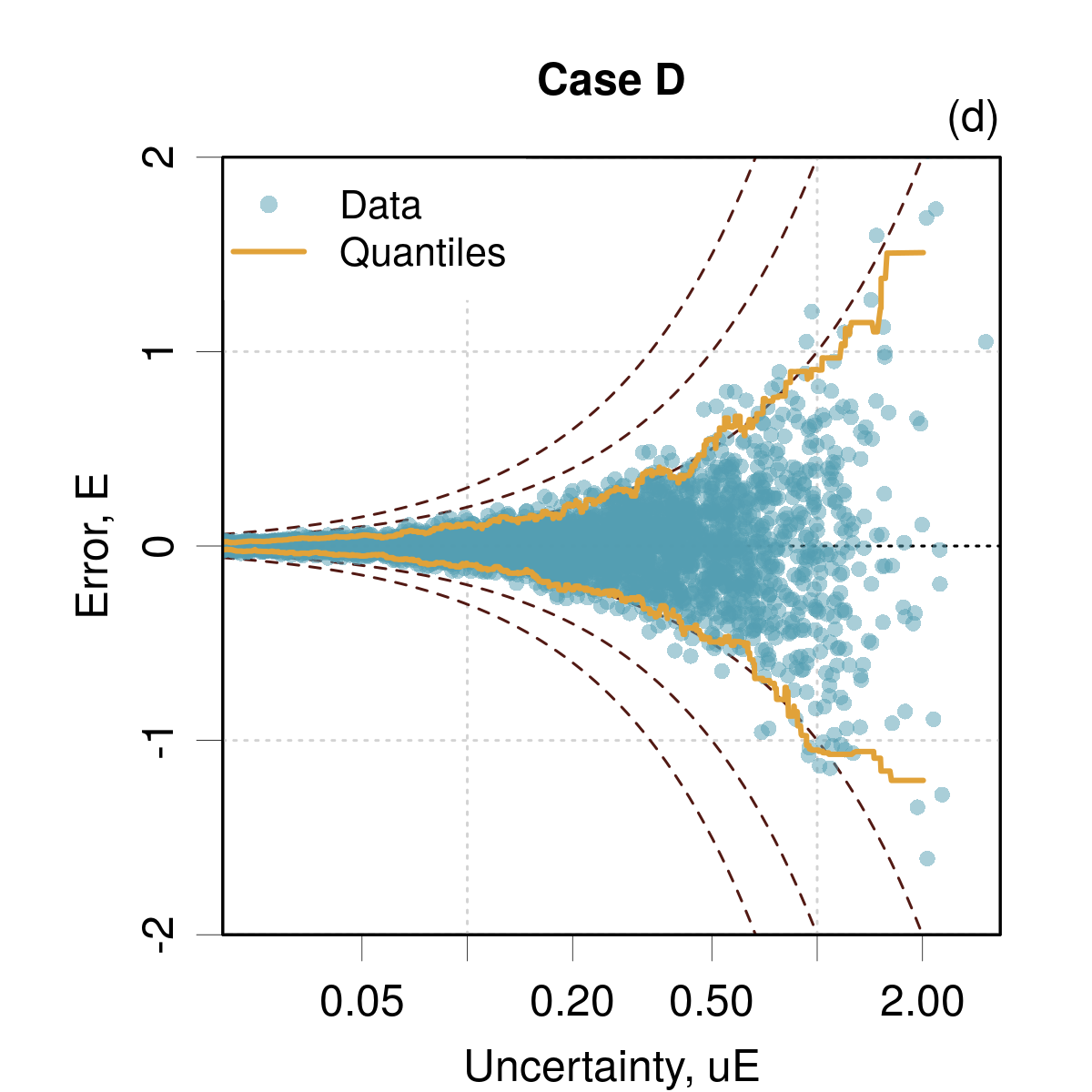} & \includegraphics[width=0.33\textwidth]{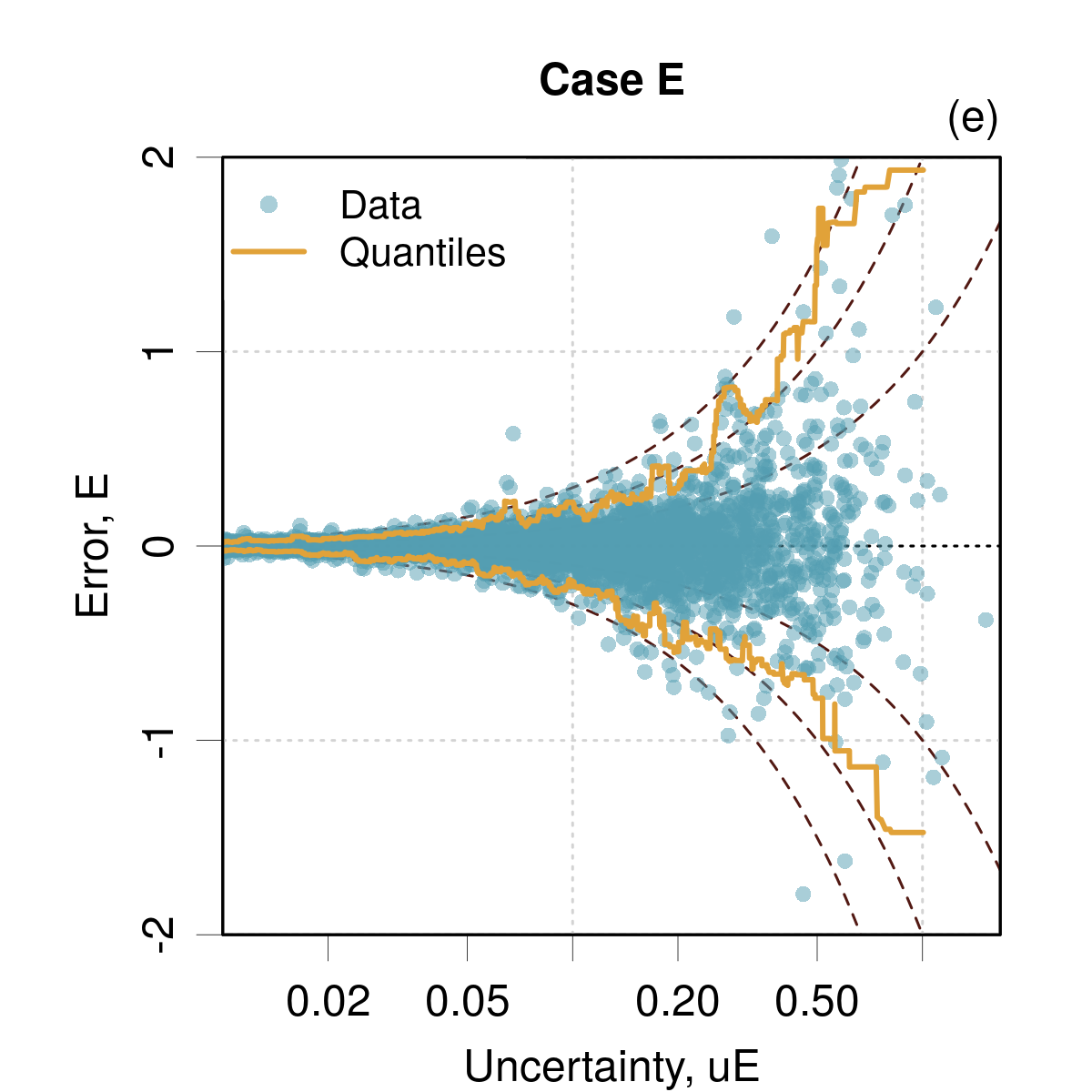} & \tabularnewline
\end{tabular}
\par\end{centering}
\caption{\label{fig:validSynth}''$E$ vs $u_{E}$'' validation plots for
cases A-E. The dashed lines correspond to the $y=\pm kx$ guidelines
with $k=1,2,3$. The orange lines represent moving quantiles for a
symmetric 95\,\% interval of the data. They are expected to be parallel
and close to the $y=\pm2x$ lines.}
\end{figure*}

\subsubsection{``Error vs. uncertainty'' plots}

\noindent Plotting the errors vs. the uncertainties can be very informative,
notably for datasets missing consistency \citep{Janet2019,Pernot2022b,Hu2022}.
Additional plotting of guiding lines and running quantiles is a welcomed
complement to facilitate the diagnostic.\citep{Pernot2022b} This
is obviously not applicable to homoscedastic datasets. 

Examples for the synthetic datasets are shown in Fig.\,\ref{fig:validSynth}.
The expected ``correct behavior'' is to have the running quantile
lines for a 95\% confidence interval to lie in the vicinity of and
parallel to the $E=\pm2\times u_{E}$ lines. Although the generative
distribution is not necessarily normal, one does not expect very large
deviations from $k=2$, as seen for instance for the Student's distribution
of Case E. Moreover, the cloud of points should be symmetrical around
$E=0$ (absence of bias).

The plots enable to sort out Cases B, C and D without ambiguity. For
Cases B and C, the running quantiles are more or less horizontal,
indicating an absence of link between the dispersion of $E$ and $u_{E}$.
For Case D, the running quantiles follow closely the $k=1$ lines,
pointing to a probable overestimation of uncertainties. 

For cases where consistency cannot be frankly rejected on the basis
of the shape or scale of the data cloud, it is imperative to perform
more quantitative tests as presented below. One should not conclude
on good consistency simply based on this kind of plot.

\subsubsection{Conditional calibration curves in uncertainty space\label{subsec:Conditional-calibration-curves}}

\noindent It is formally possible to construct \emph{conditional calibration
curves}, but, to my knowledge, they have not been proposed in the
ML-UQ literature. 

\begin{figure*}[t]
\noindent \begin{centering}
\begin{tabular}{ccc}
\includegraphics[width=0.33\textwidth]{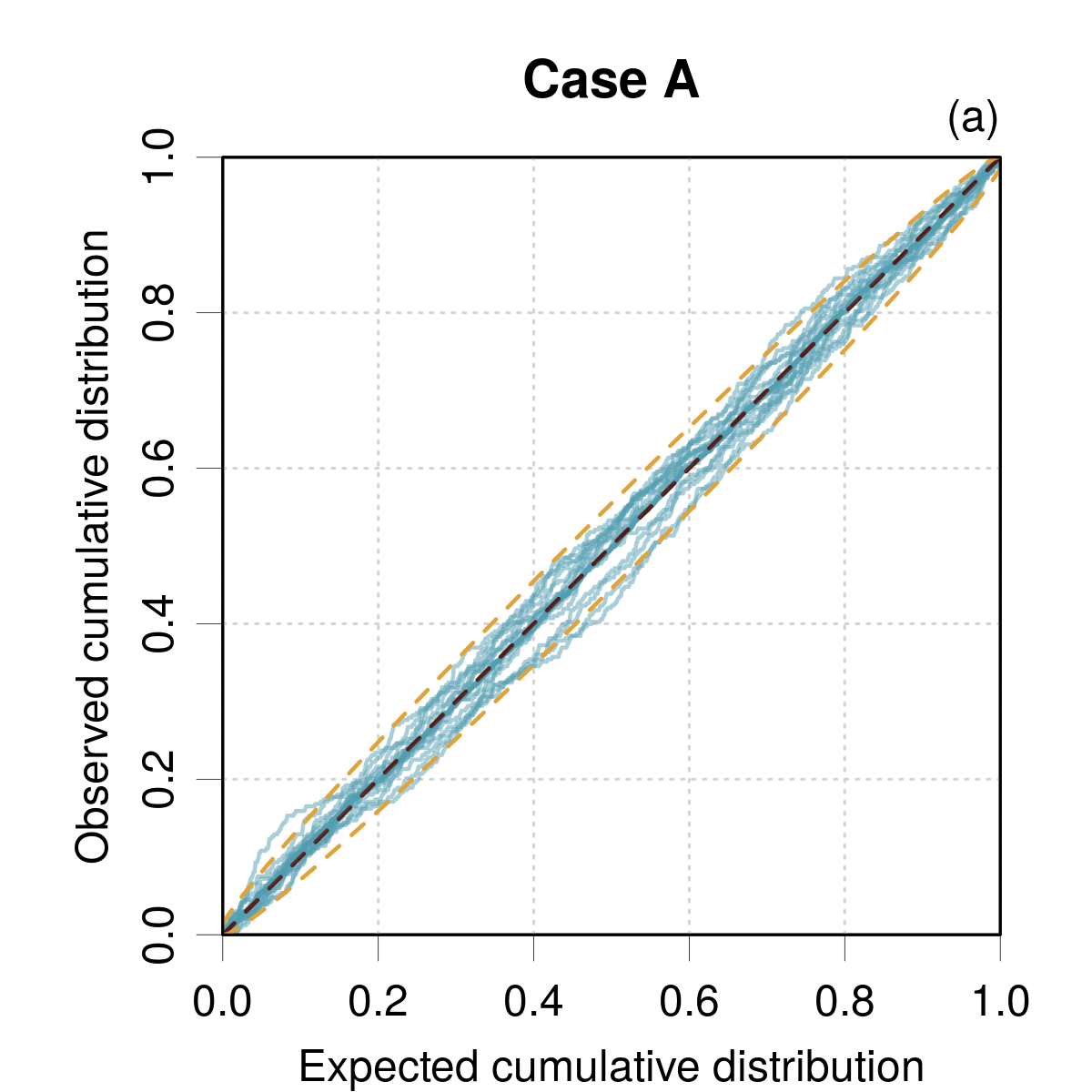} & \includegraphics[width=0.33\textwidth]{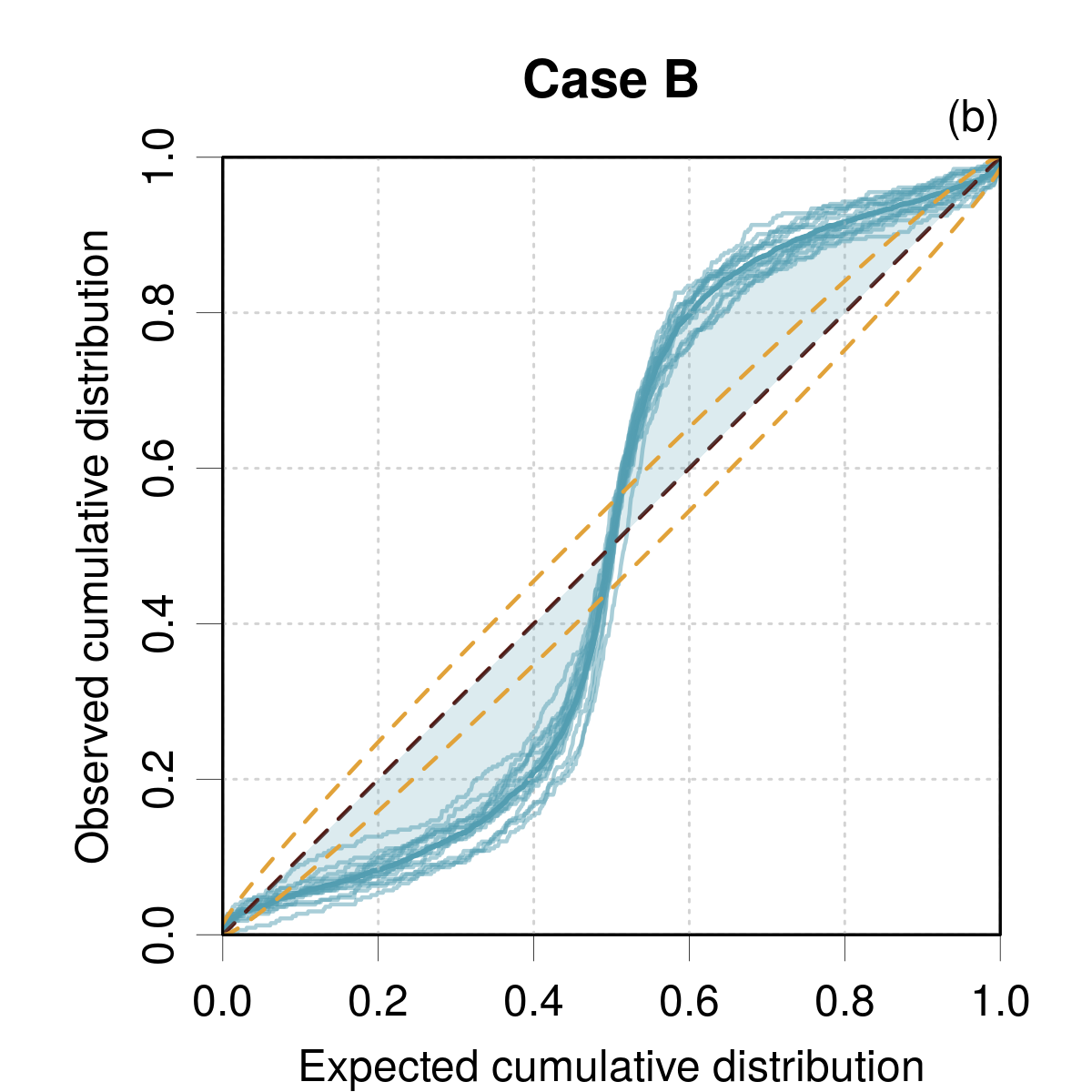} & \includegraphics[width=0.33\textwidth]{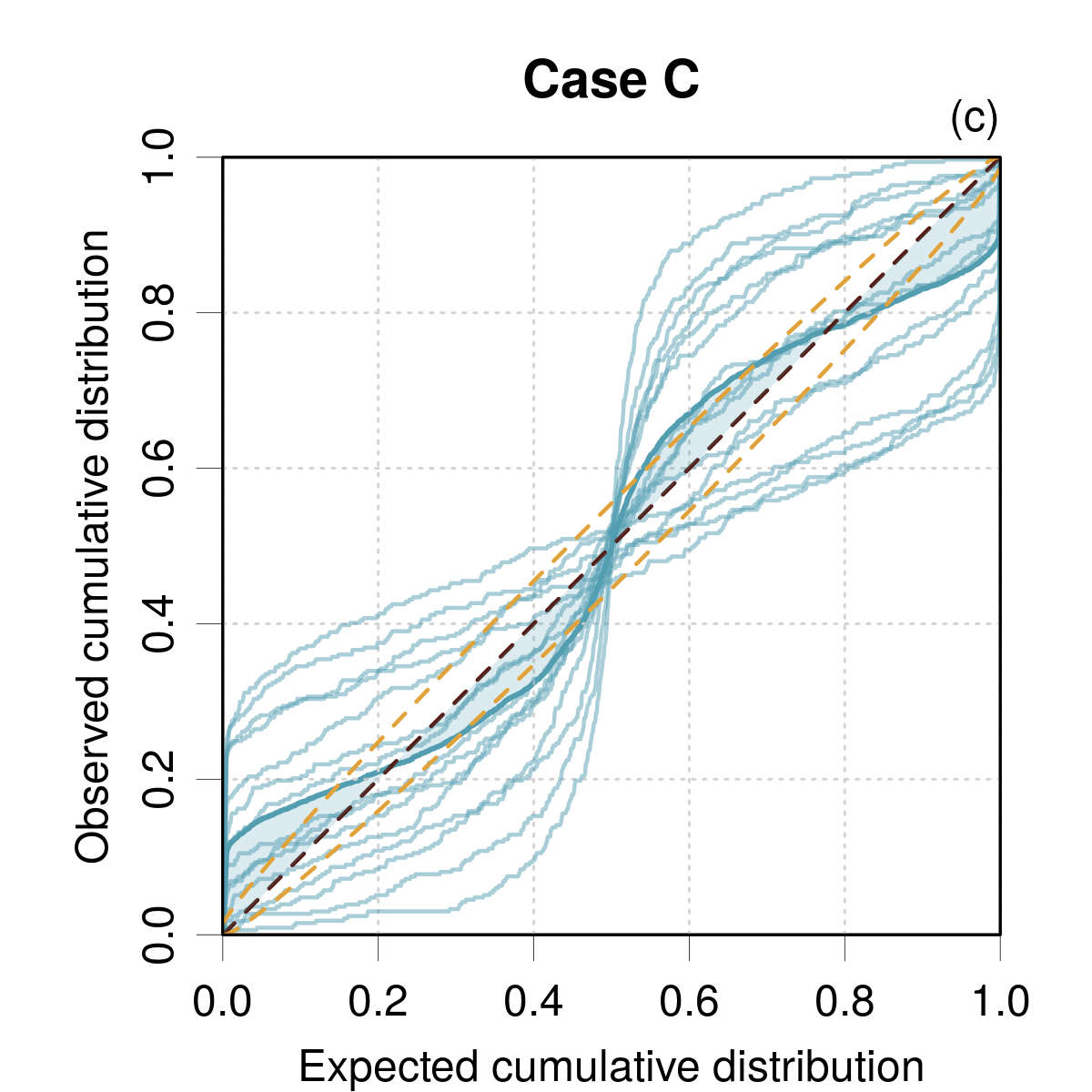}\tabularnewline
\includegraphics[width=0.33\textwidth]{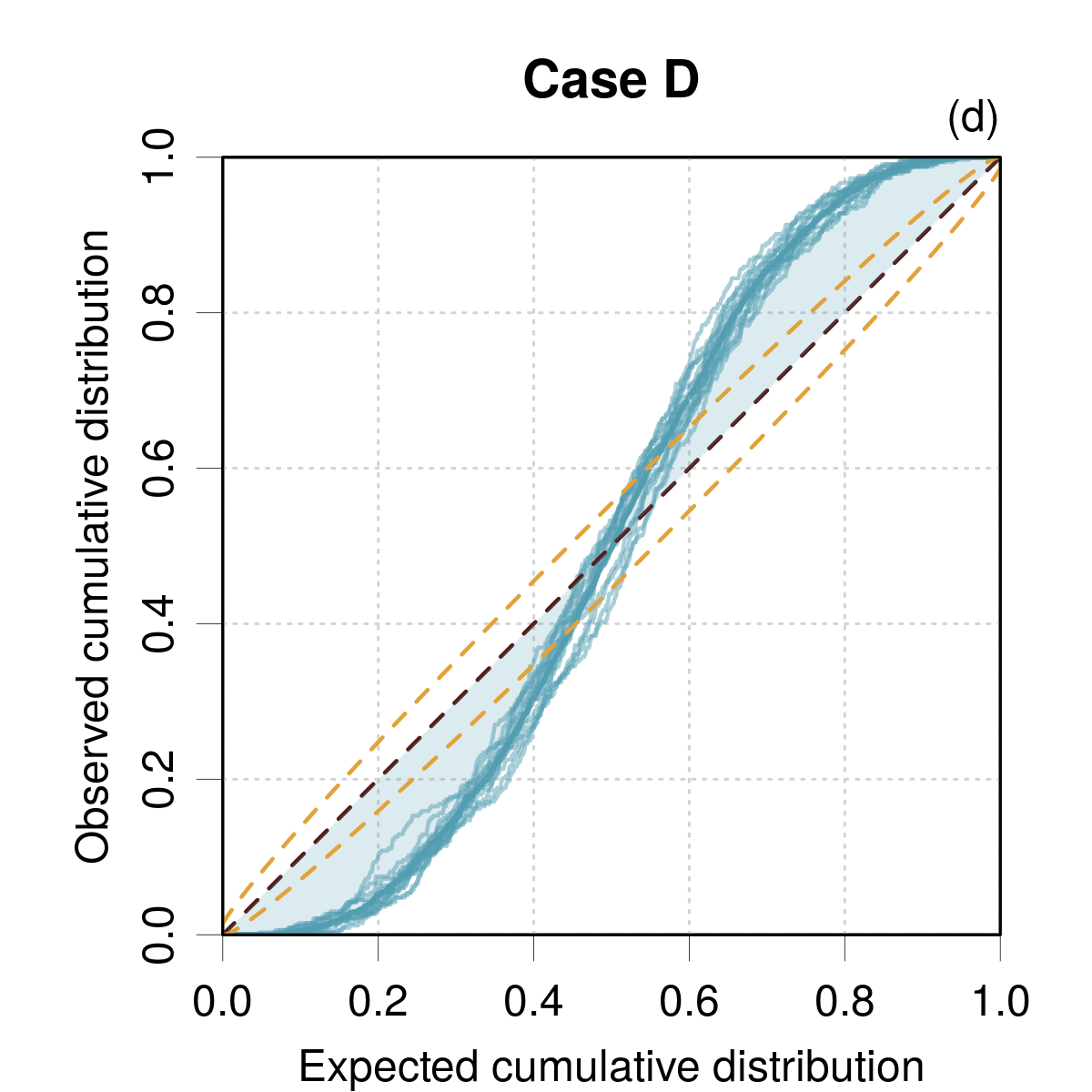} & \includegraphics[width=0.33\textwidth]{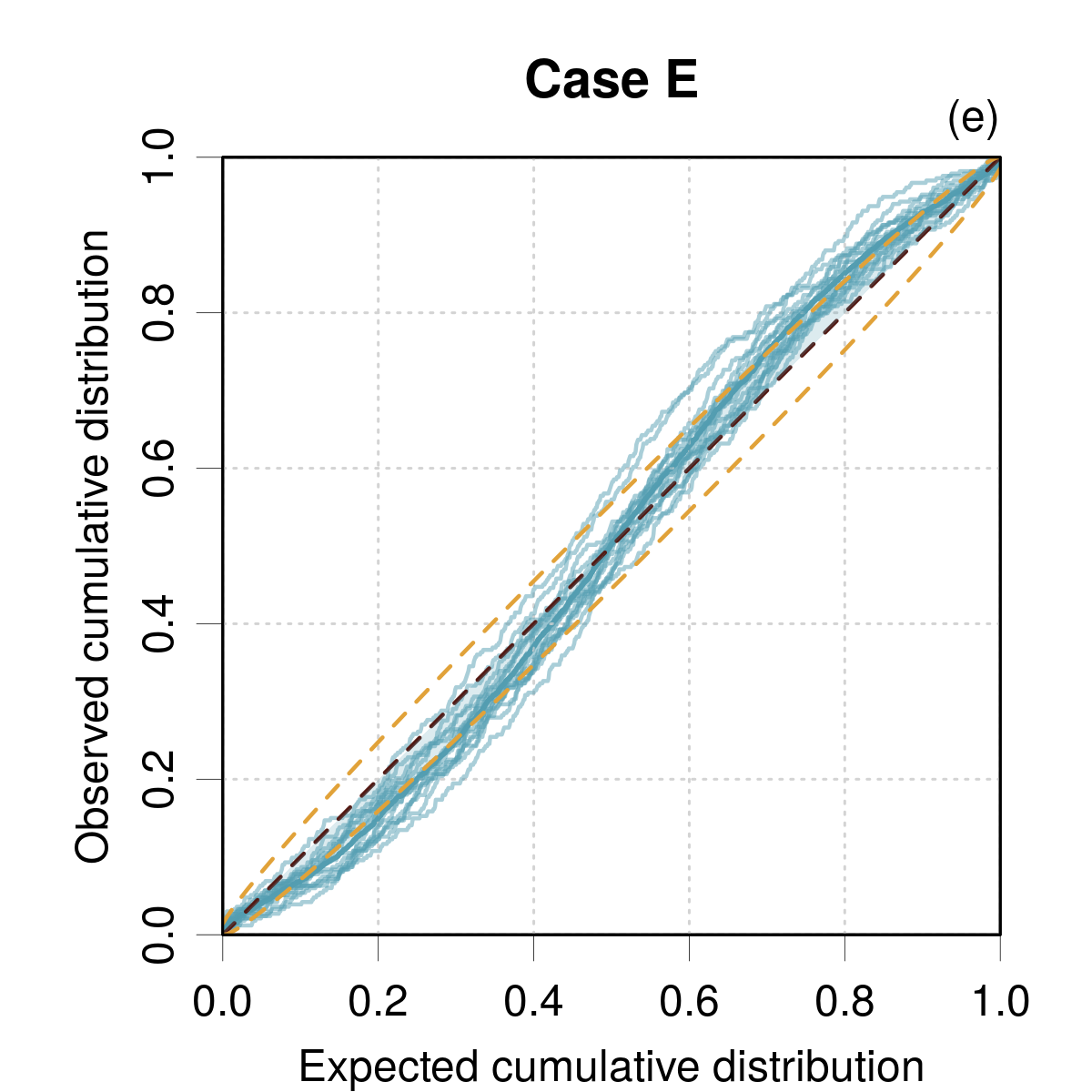} & \includegraphics[width=0.33\textwidth]{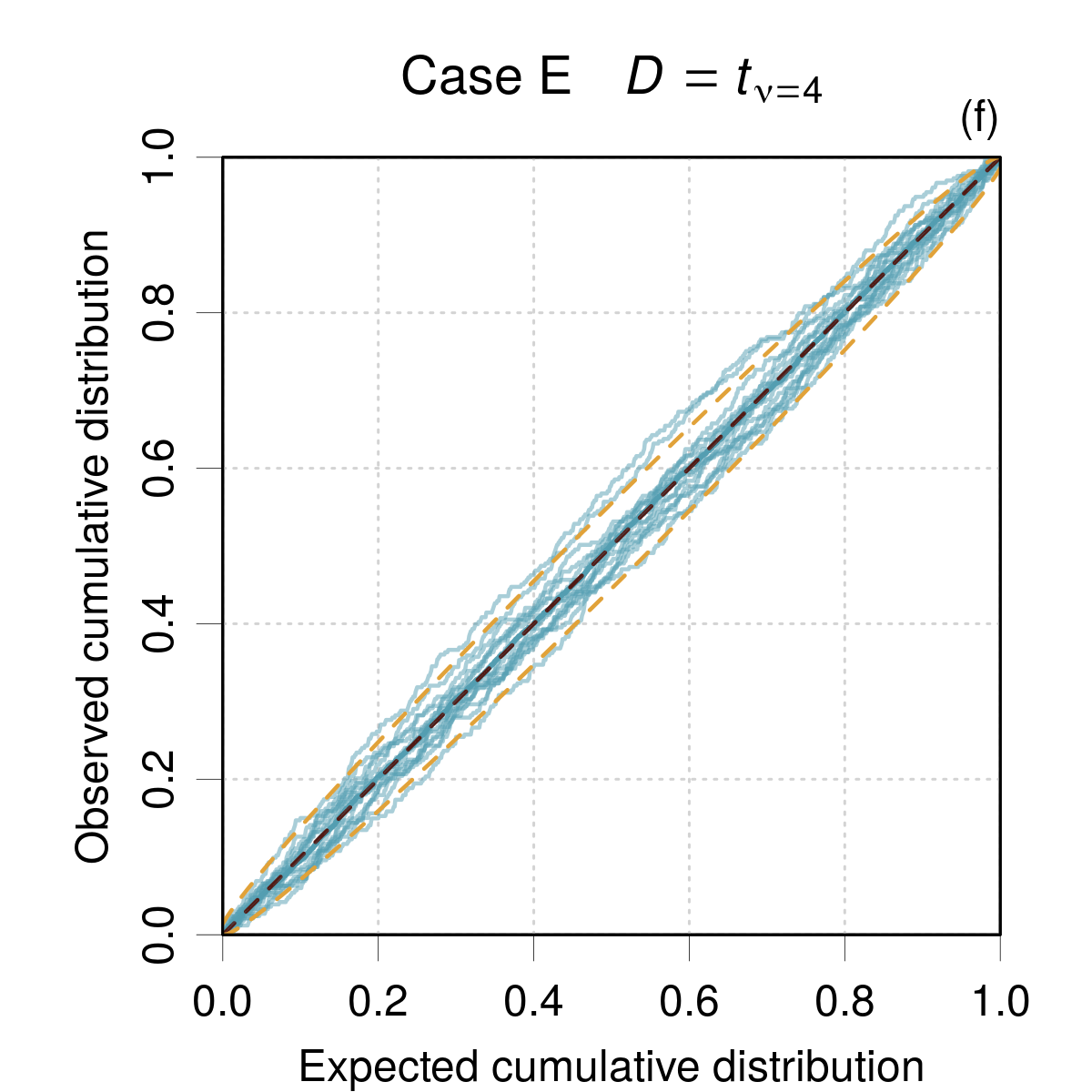}\tabularnewline
\end{tabular}
\par\end{centering}
\caption{\label{fig:Calibration-curves-cond}Conditional calibration curves
vs $u_{E}$ for cases A-E. The average calibration curve is a thick
blue line, the conditional curves are thin blue lines. The reference
line and its 95\,\% confidence interval are dashed orange lines.
The dataset has been split according to the binning of $u_{E}$ into
15 equal-counts sets.}
\end{figure*}

Examples are given in Fig.\,\ref{fig:Calibration-curves-cond} using
$u_{E}$ as the conditioning variable with 15 equal-counts bins. In
the ideal case (Case A), all the conditional curves lie within the
confidence interval estimated for a sample size corresponding to a
single bin. Case B shows that the anomaly observed on the average
calibration curve is identical at all uncertainty levels, pointing
to a consistency problem or a wrong choice of generative distribution.
Case C is interesting, as it shows that the small oscillations of
the average calibration curve around the identity line are attenuated
compared to the much larger deviations observed for conditional curves.
This points to the fact that a seemingly good calibration curve might
hide a non-consistent case. The large dispersion of conditional curves
is less likely to be due to a poor choice of generative distribution
(unless the generative distribution varies strongly with $u_{E}$)
and points to a lack of consistency. Cases D and E offer the same
diagnostic as for case B, with an ambiguity between non-consistency
and generative distribution problem. Using a Student's-$t_{\nu=4}$
generative distribution for Case E solves the problem {[}Fig.\,\ref{fig:Calibration-curves-cond}(f){]}.

\begin{figure*}[t]
\noindent \begin{centering}
\begin{tabular}{ccc}
\includegraphics[width=0.33\textwidth]{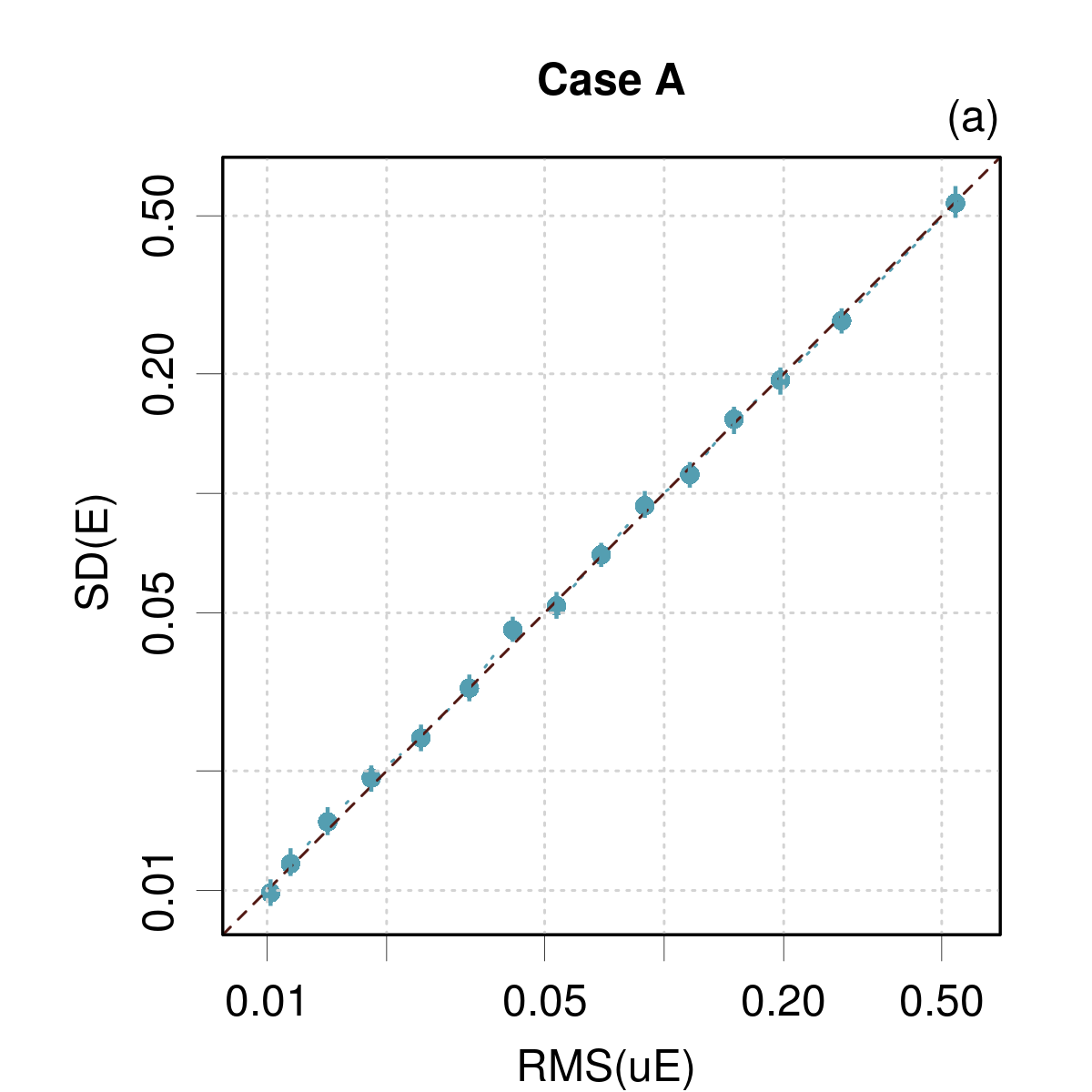} & \includegraphics[width=0.33\textwidth]{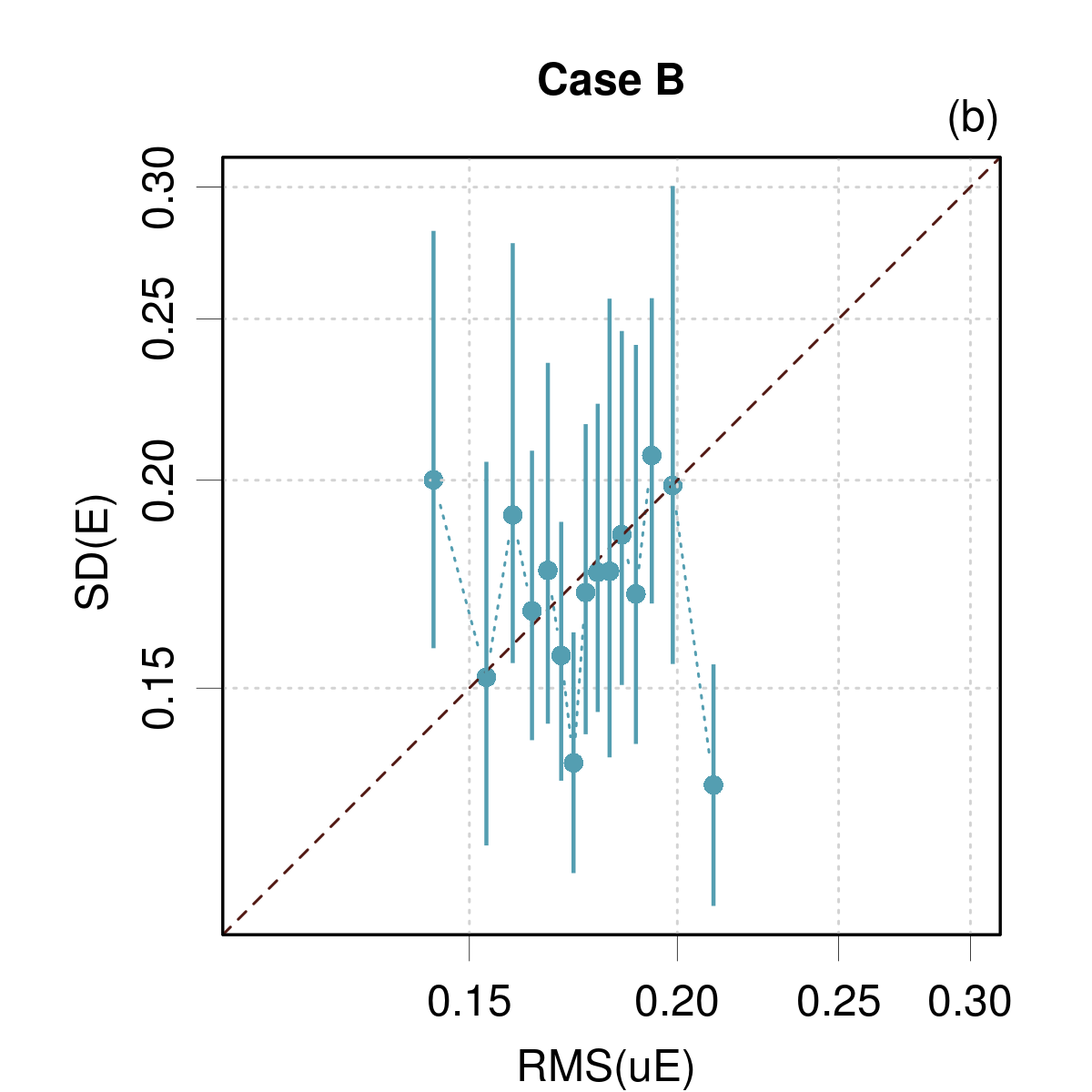} & \includegraphics[width=0.33\textwidth]{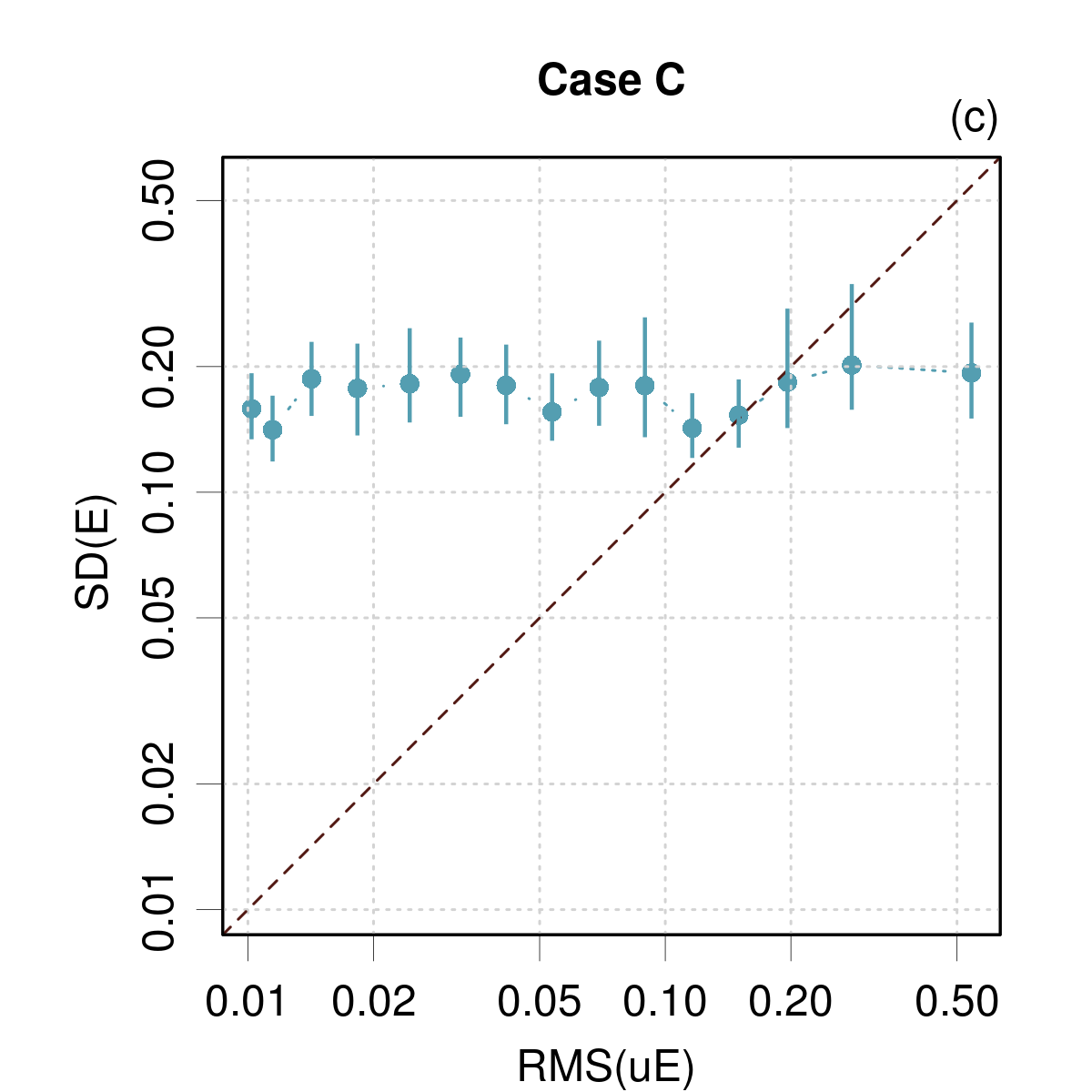}\tabularnewline
\includegraphics[width=0.33\textwidth]{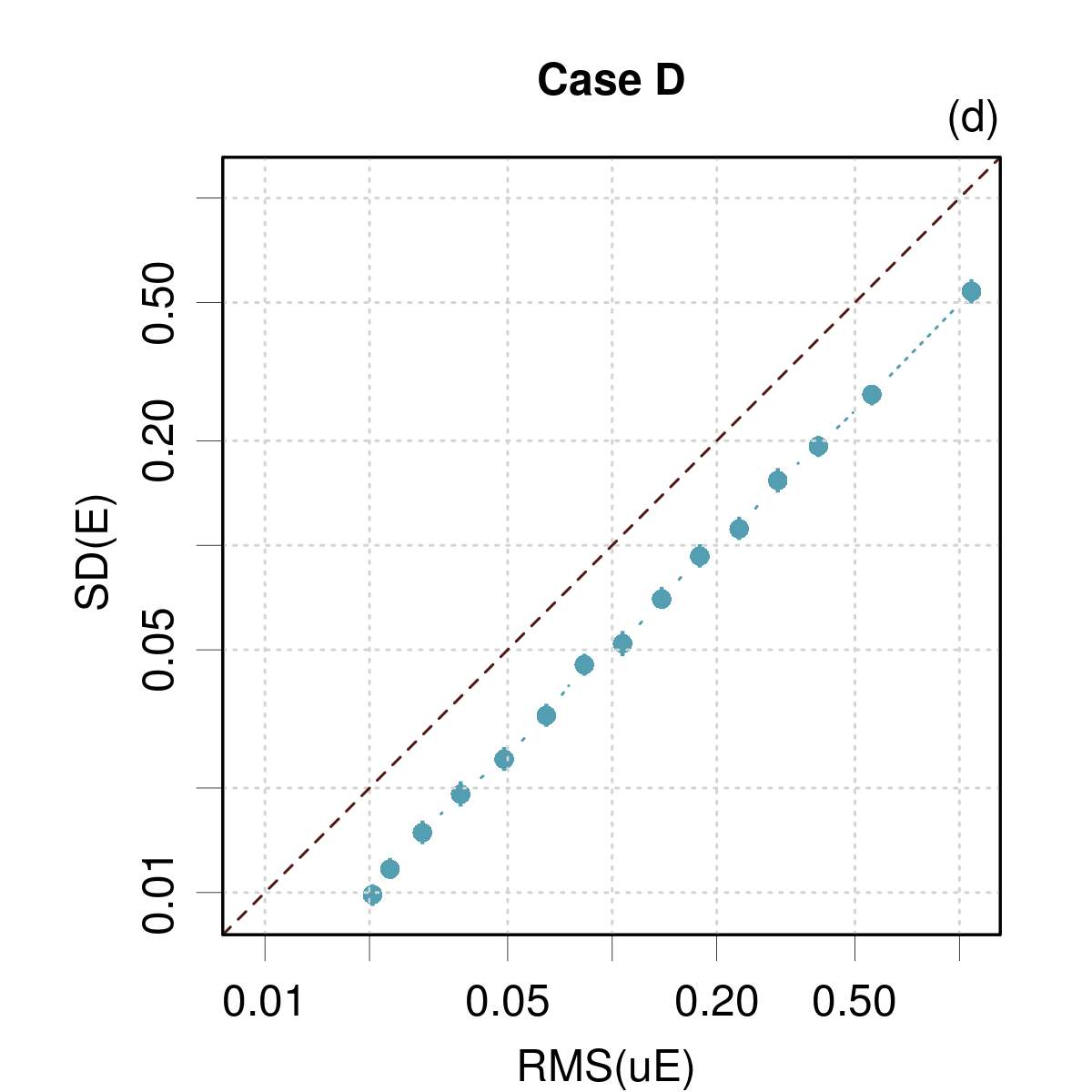} & \includegraphics[width=0.33\textwidth]{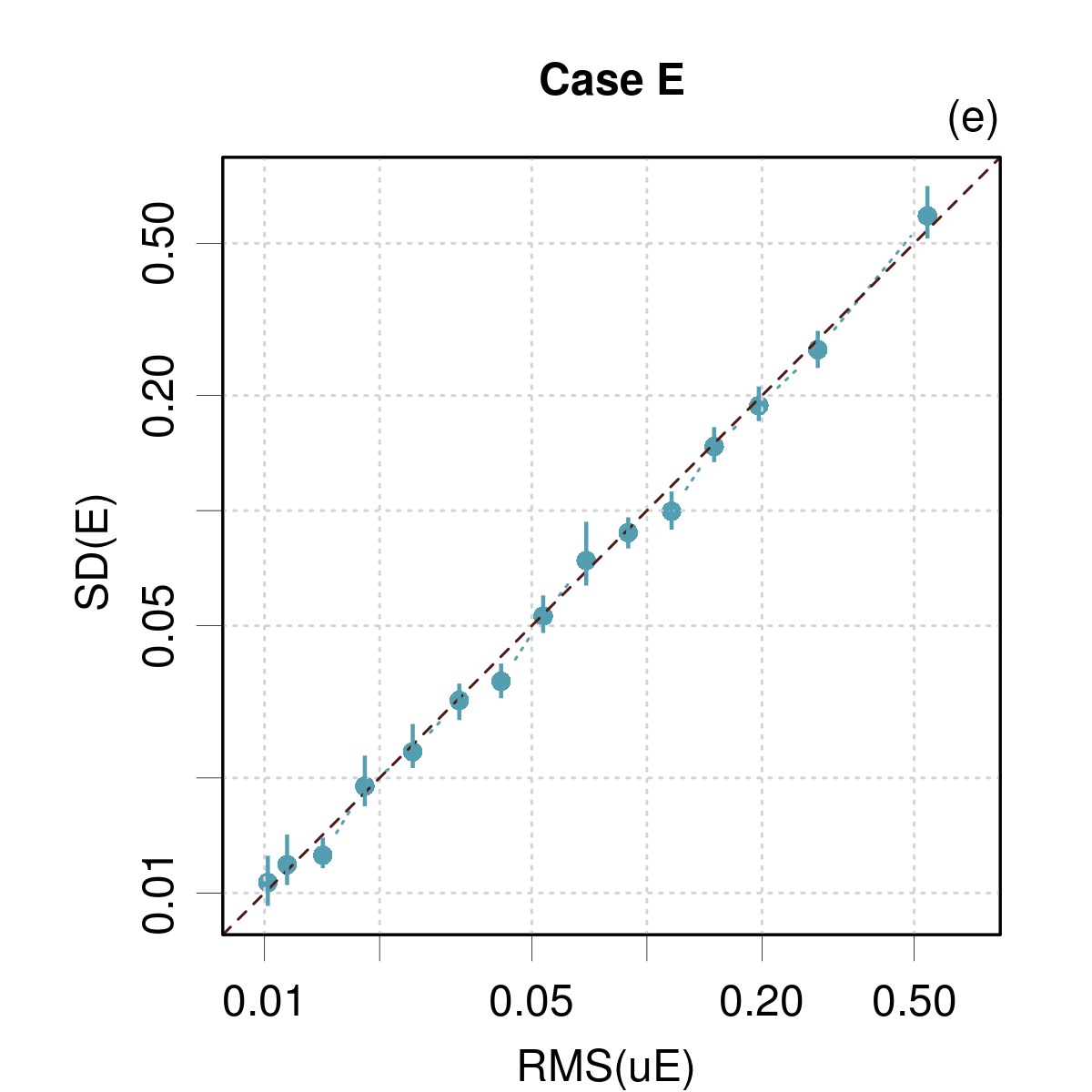} & \tabularnewline
\end{tabular}
\par\end{centering}
\caption{\label{fig:Reliability-diagrams}Reliability diagrams for cases A-E.
Not applicable to case F (homoscedastic). }
\end{figure*}

\subsubsection{Reliability diagram\label{subsec:Reliability-diagram}}

\noindent So-called reliability diagrams\citep{Guo2017,Levi2020}
test consistency through Eq.\,\ref{eq:varEvalCond} over data subsets,
where the data are ordered by increasing $u_{E}$ values and split
into bins. This method is therefore not usable for homoscedastic datasets.
Reliability diagrams can also be found in the literature as \emph{error-based
calibration plots}\citep{Scalia2020} (not to be confounded with \emph{calibration
curves}, Sect.\,\ref{subsec:Calibration-curve}), \emph{RvE} plots\citep{Palmer2022},
or \emph{RMSE vs. RMV} curves\citep{Vazquez-Salazar2022}.

For convenience, the square roots of the terms of Eq.\,\ref{eq:varEvalCond}
are used, linking the mean uncertainty in each bin (as the \emph{root
mean squared} (RMS) of $u_{E}$, also called \emph{root mean variance}
(RMV)) with the RMSE or RMSD of $E$ (both are equivalent for unbiased
errors). 

For consistent datasets, the reliability curve should lie close to
the identity line, up to statistical fluctuations due to finite bin
counts. For a conclusive analysis of deviations from the identity
line, the amplitude of these finite size effects should be estimated,
for instance by bootstrapping.\citep{Pernot2022b}

The binning strategy is important\citep{Nixon2019}: some authors
advocate for bins with identical counts,\citep{Levi2020,Scalia2020}
other for bins with identical widths\citep{Palmer2022}. Both choices
are defensible according to the distribution of uncertainties, notably
the absence or presence of heavy tails. It should also be noted that
the insensitivity of Eq.\,\ref{eq:varEVal} to the pairing between
$E$ and $u_{E}$ might still be a hindrance for large bins, but its
effect should decrease when the bins get smaller. However, small bins
are affected by large statistical fluctuations. The binning strategy
should thus be designed to offer a good compromise. An adaptive binning
scheme mixing both strategies is proposed in Appendix\,\ref{sec:PAL2022---Additional}.

For the synthetic datasets, I use a default choice of 15 equal-counts
bins, leading to about 333 points per bin (Fig.\,\ref{fig:Reliability-diagrams}).
The non-consistent cases (B, C, D) are correctly identified, although
for case B the deviation from the identity line is not noticeable
except for the two extreme bins.

\begin{figure*}[t]
\noindent \begin{centering}
\begin{tabular}{ccc}
\includegraphics[width=0.33\textwidth]{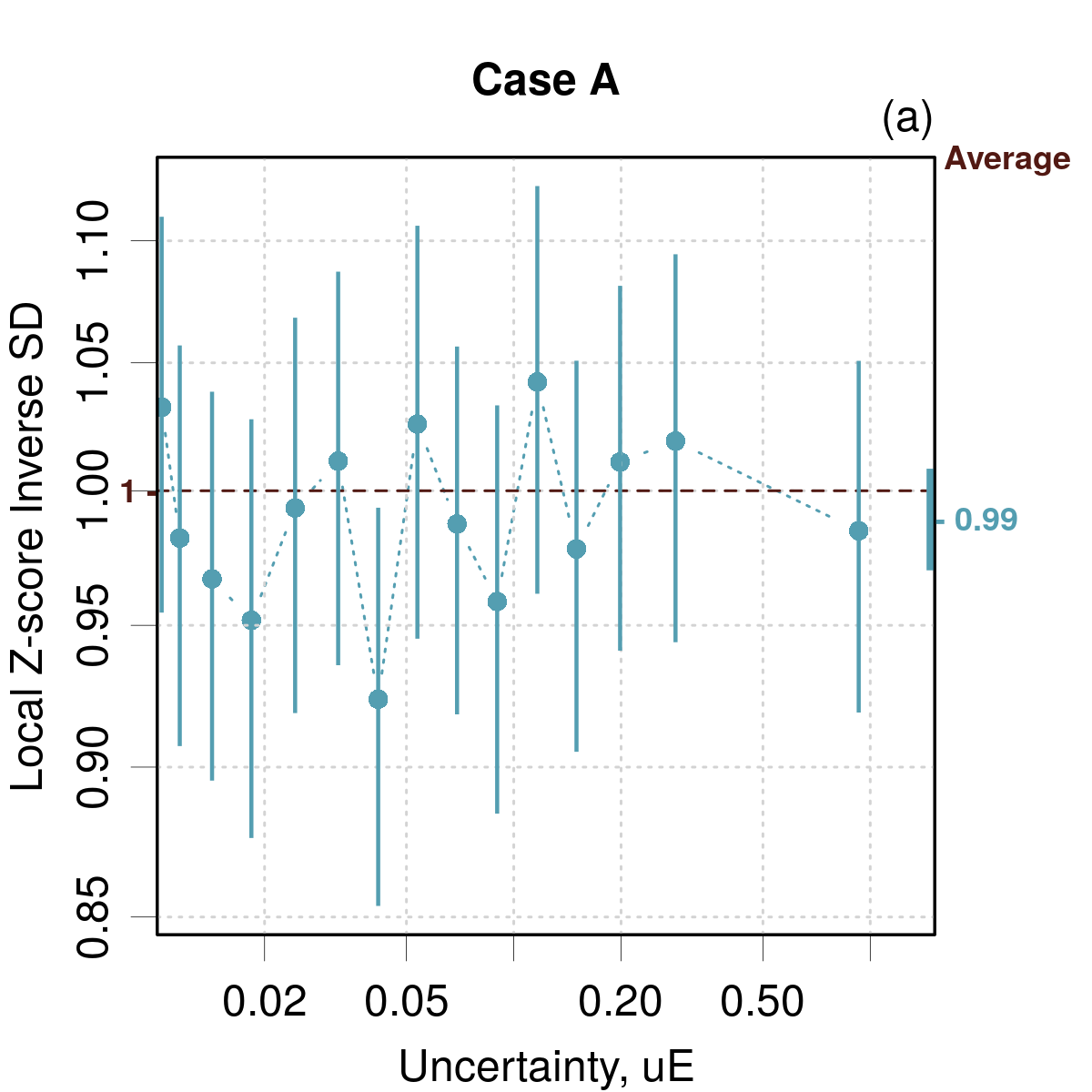} & \includegraphics[width=0.33\textwidth]{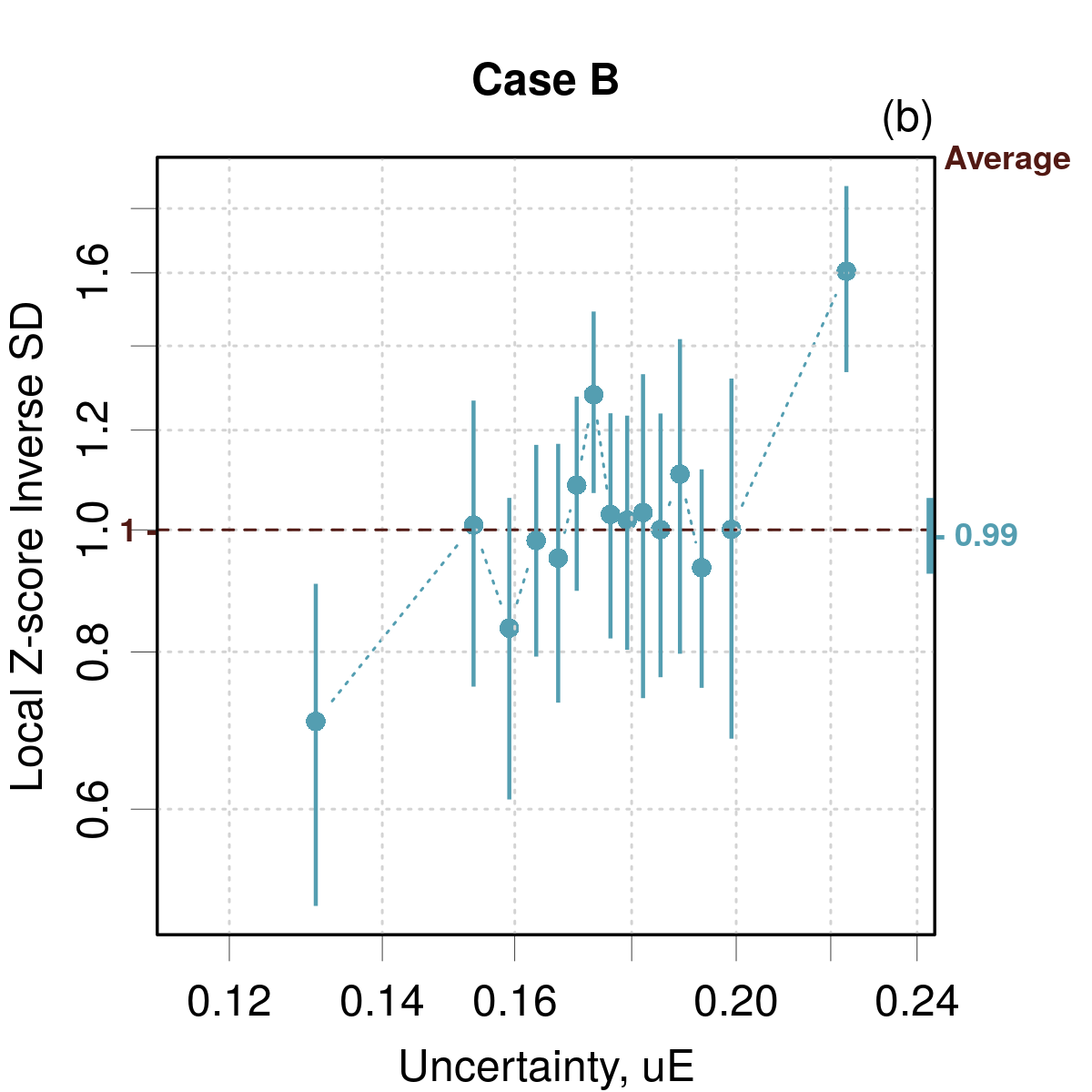} & \includegraphics[width=0.33\textwidth]{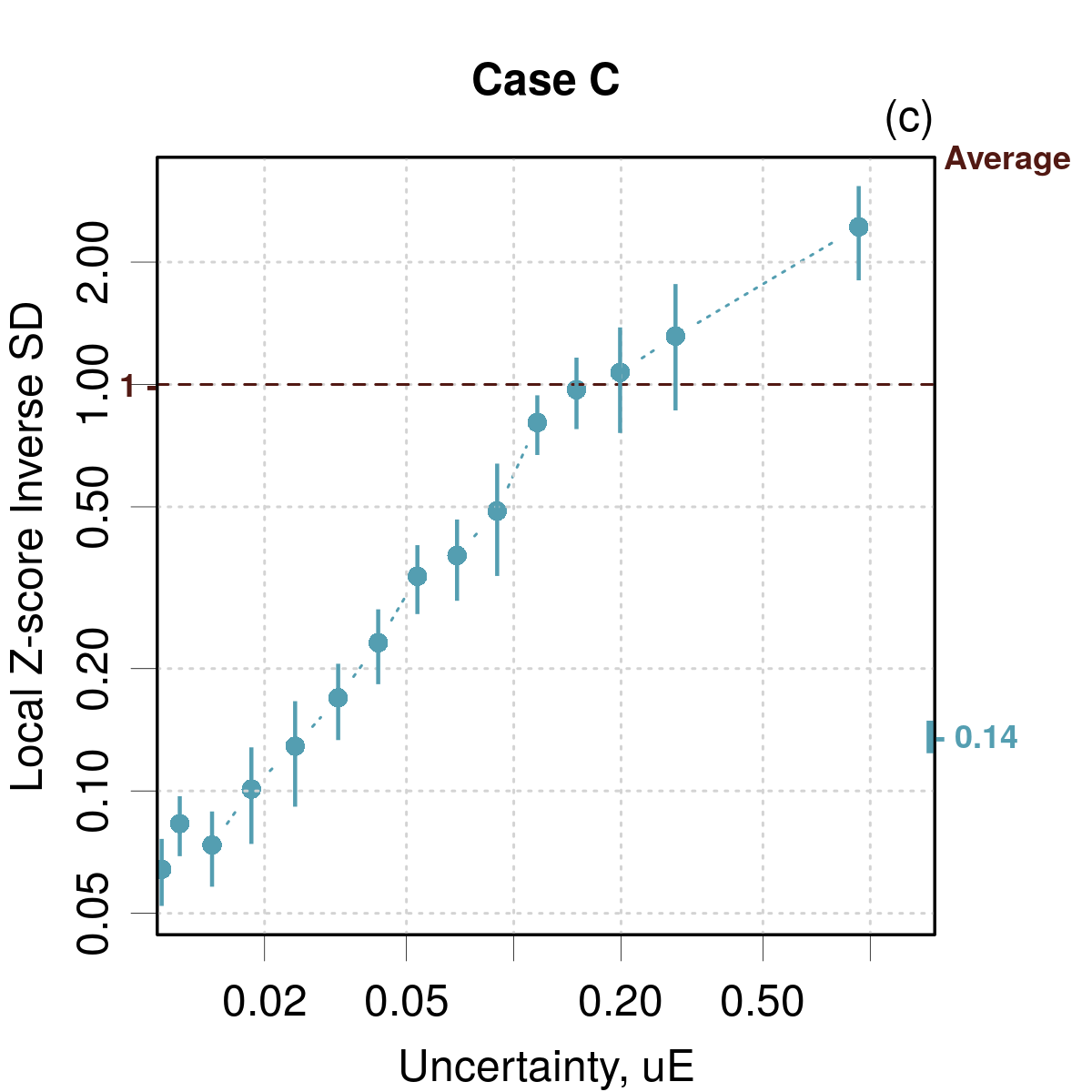}\tabularnewline
\includegraphics[width=0.33\textwidth]{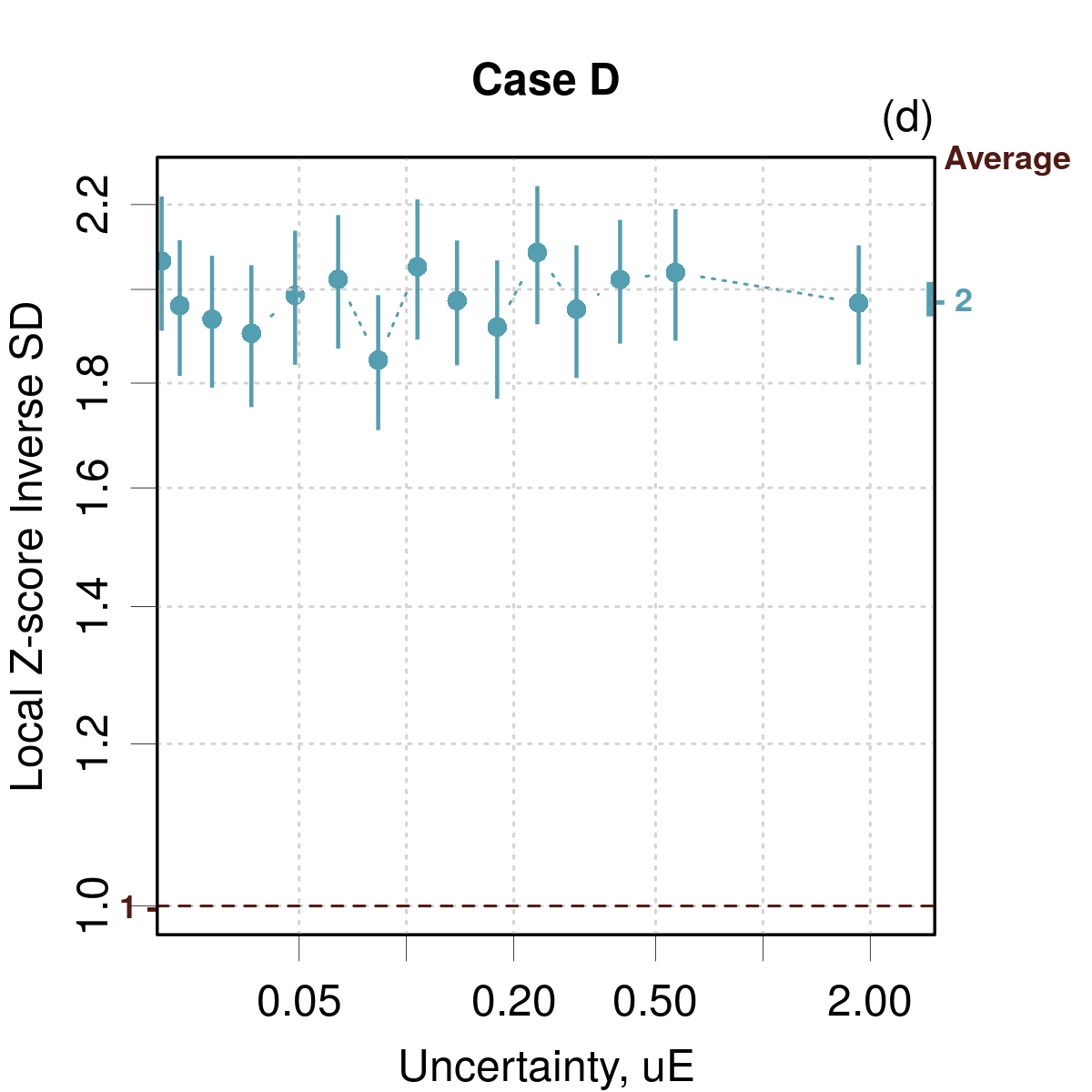} & \includegraphics[width=0.33\textwidth]{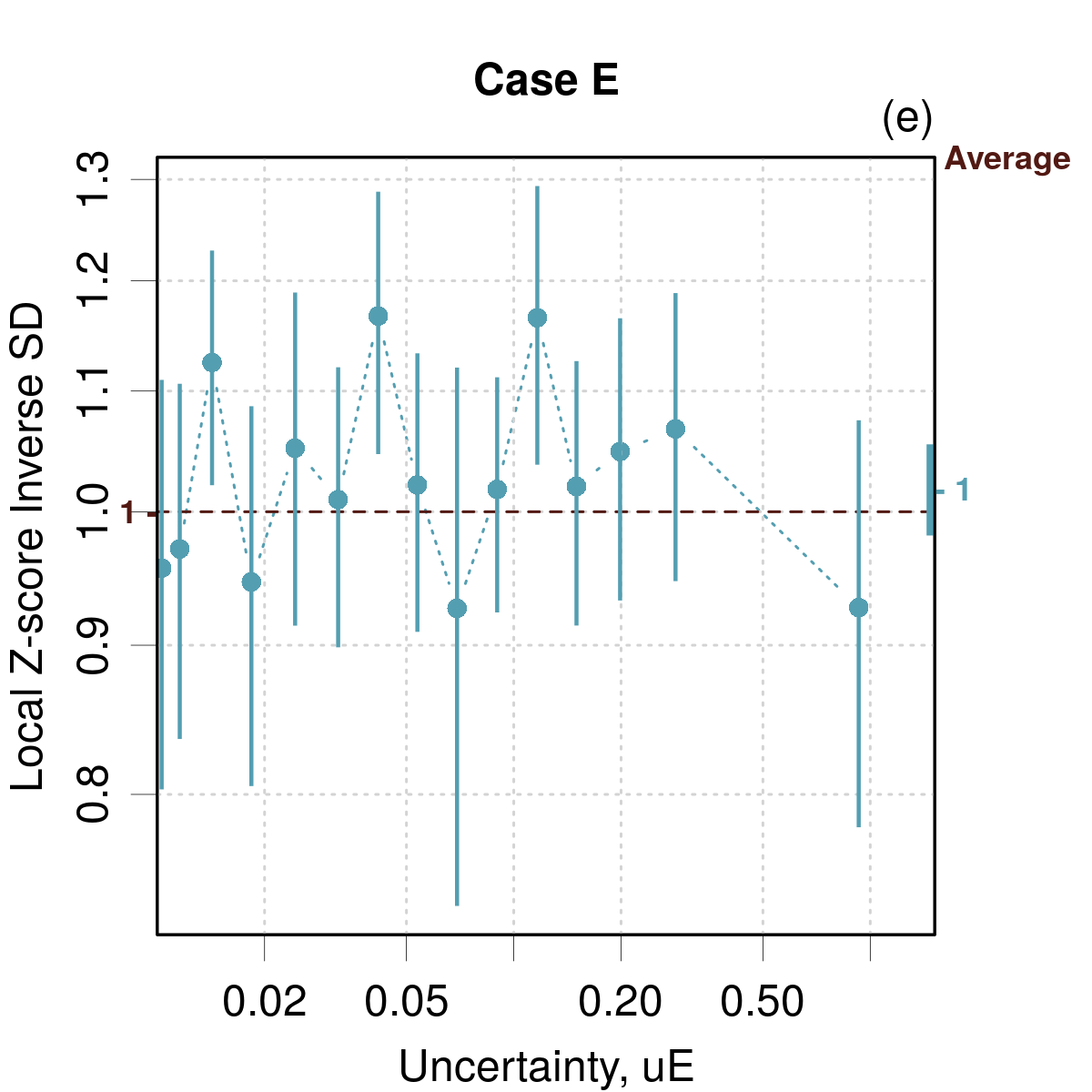} & \tabularnewline
\end{tabular}
\par\end{centering}
\caption{\label{fig:LZISD}LZISD analysis for cases A-E. Not applicable to
case F (homoscedastic). The points are reported at the center of the
bins. The error bars correspond to 95\,\% confidence intervals on
the statistic. The value and error bar reported on the right axis
corresponds to the statistic for the full dataset (Average).}
\end{figure*}

\subsubsection{Local Z-Variance analysis in uncertainty space\label{subsec:Local-Z-Variance-analysis}}

\noindent LZV analysis was introduced by Pernot\citep{Pernot2022a}
as a method to test local calibration based on Eq.\,\ref{eq:varZvalCond}.
As for reliability diagrams, it is based on a binning of the data
according to increasing uncertainties. For each bin, one estimates
$\mathrm{Var}(Z)$ and compares it to 1. Here again, error bars on
the statistic should be provided to account for finite bin counts.\citep{Pernot2022a} 

Assuming a nearly constant uncertainty value within a bin, values
of $\mathrm{Var}(Z)$ smaller/larger than 1 indicate over-/under-estimated
uncertainties (by a factor $\sqrt{\mathrm{Var}(Z)}$). This suggests
an alternative plot of the \emph{inverse of the standard deviation}
of $Z$, $\mathrm{Var}(Z)^{-1/2}$, instead of $\mathrm{Var}(Z)$,
which provides a more direct reading and quantification of the local
uncertainty deviation. This Local Z-scores Inverse Standard Deviation
(LZISD) analysis is used throughout this study.

Application of the LZISD analysis to the synthetic datasets confirms
the diagnostics of non-tightness for cases B, C and D (Fig.\,\ref{fig:LZISD}).
For Case B, the deviation from the reference is more legible than
on the corresponding reliability diagram {[}Fig.\,\ref{fig:Reliability-diagrams}(b){]},
and the largest local deviations can be directly quantified to about
30\,\% in default and 60\,\% in excess (these extreme values are
probably underestimated, as they result from a bin averaging). 

\subsubsection{Local Coverage Probability analysis in uncertainty space}

\noindent Conditional coverage with respect to an uncertainty metric
is available through the Local Coverage Probability analysis.\citep{Pernot2022a}
It can be directly built from prediction intervals if available, or
estimated from prediction uncertainties and an hypothetical generative
distribution. In both cases, the empirical coverages are drawn for
a series of data subsets based on the binning of the conditioning
variable and for the available target probabilities. The LCP analysis
of consistency is not applicable to homoscedastic datasets.

\begin{figure*}[t]
\noindent \begin{centering}
\begin{tabular}{ccc}
\includegraphics[width=0.33\textwidth]{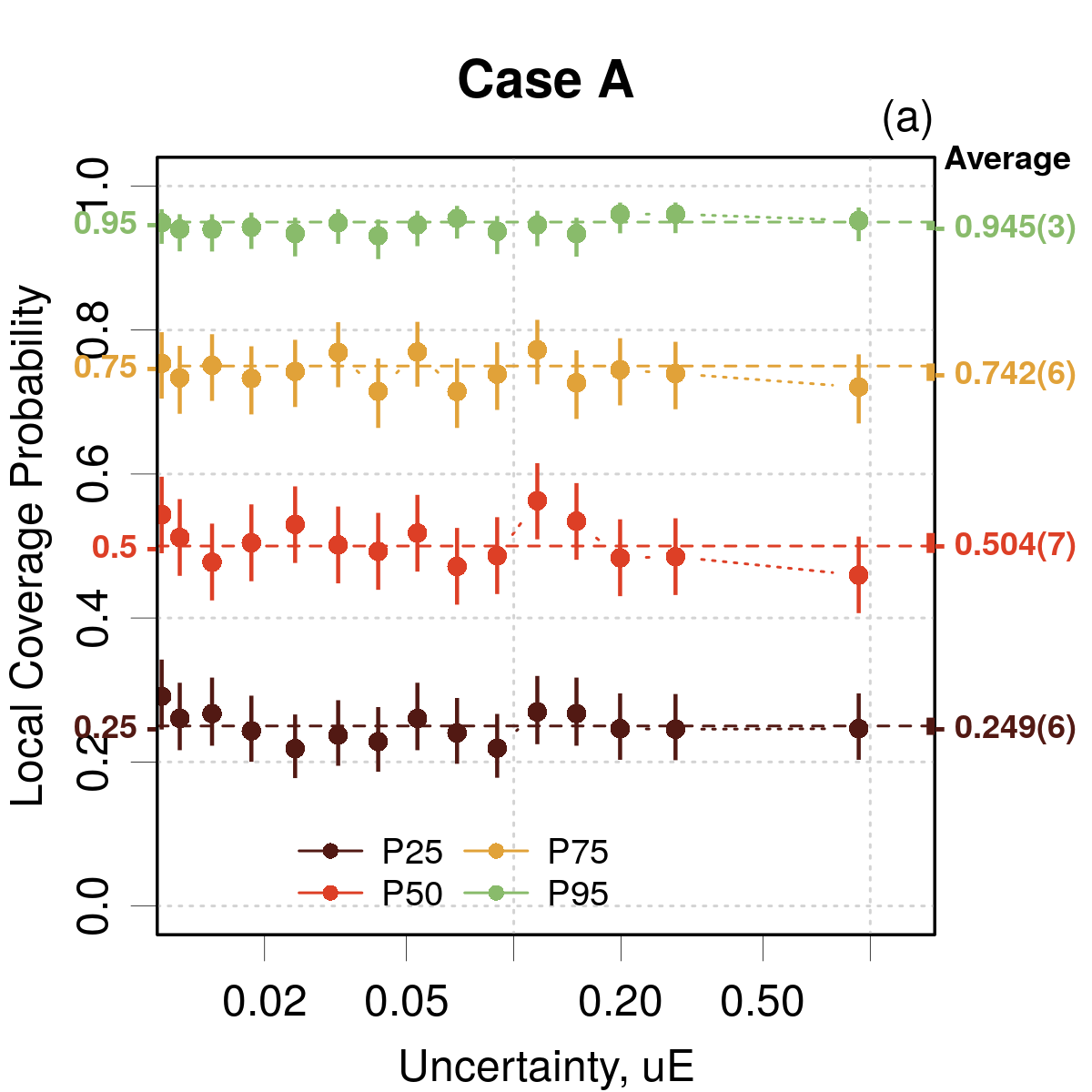} & \includegraphics[width=0.33\textwidth]{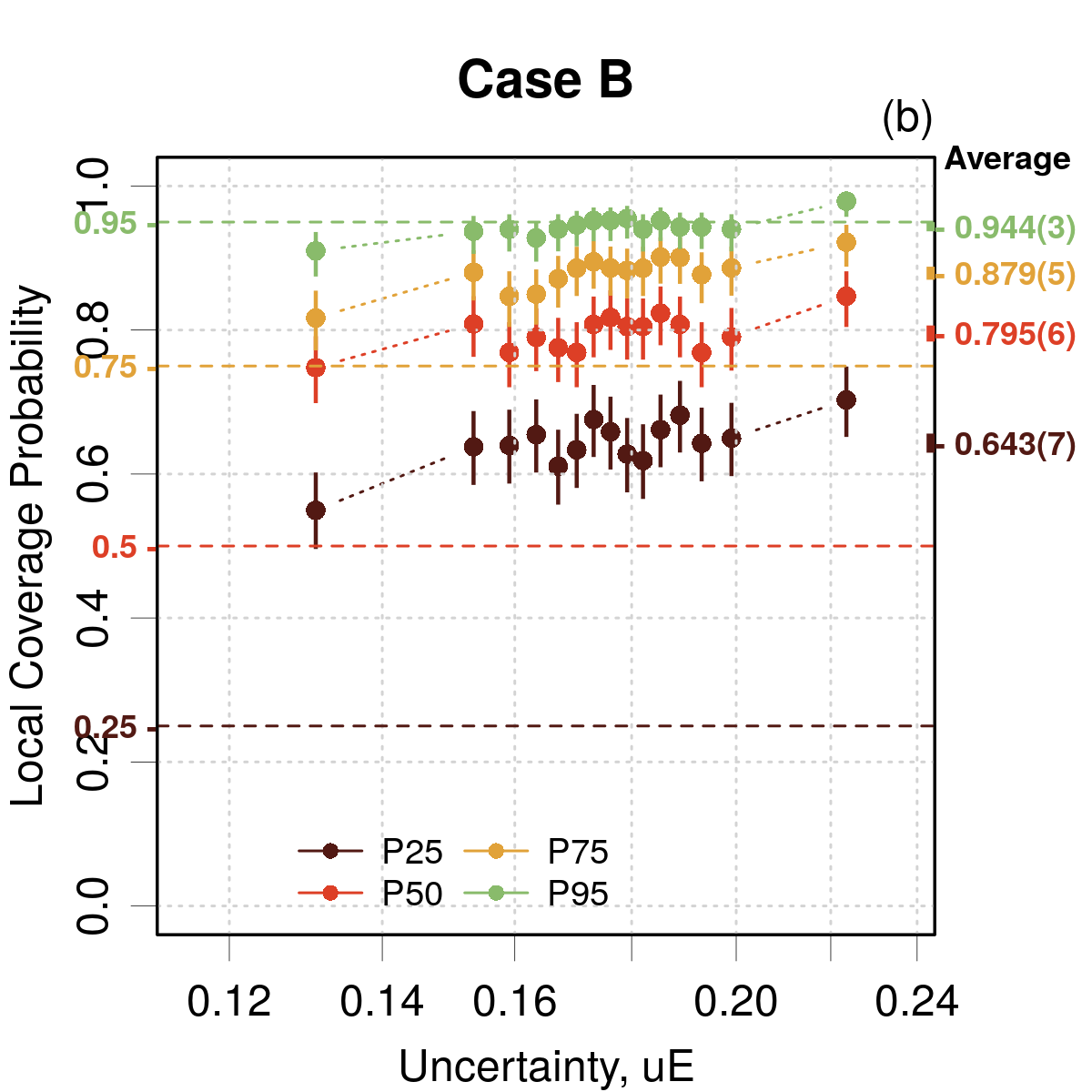} & \includegraphics[width=0.33\textwidth]{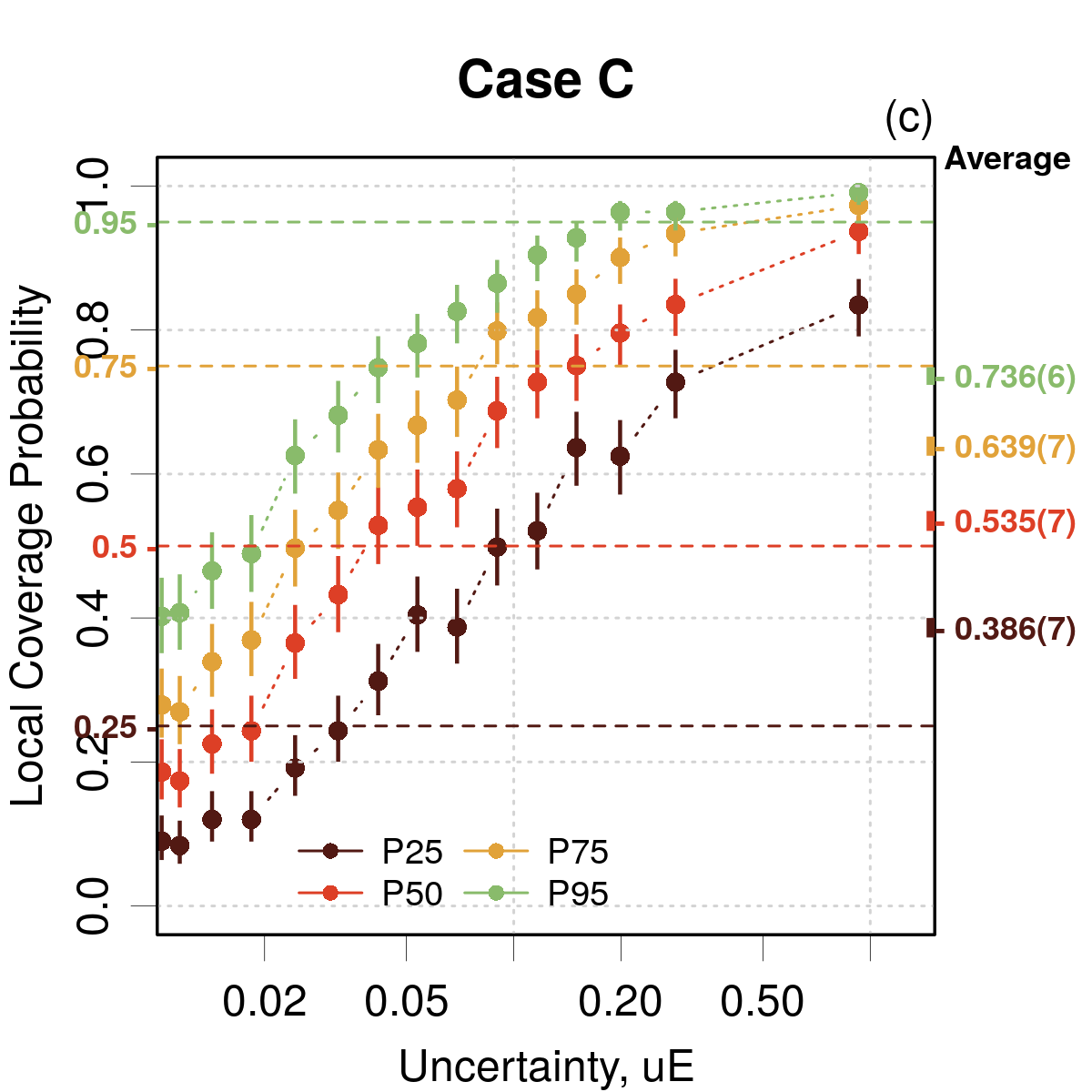}\tabularnewline
\includegraphics[width=0.33\textwidth]{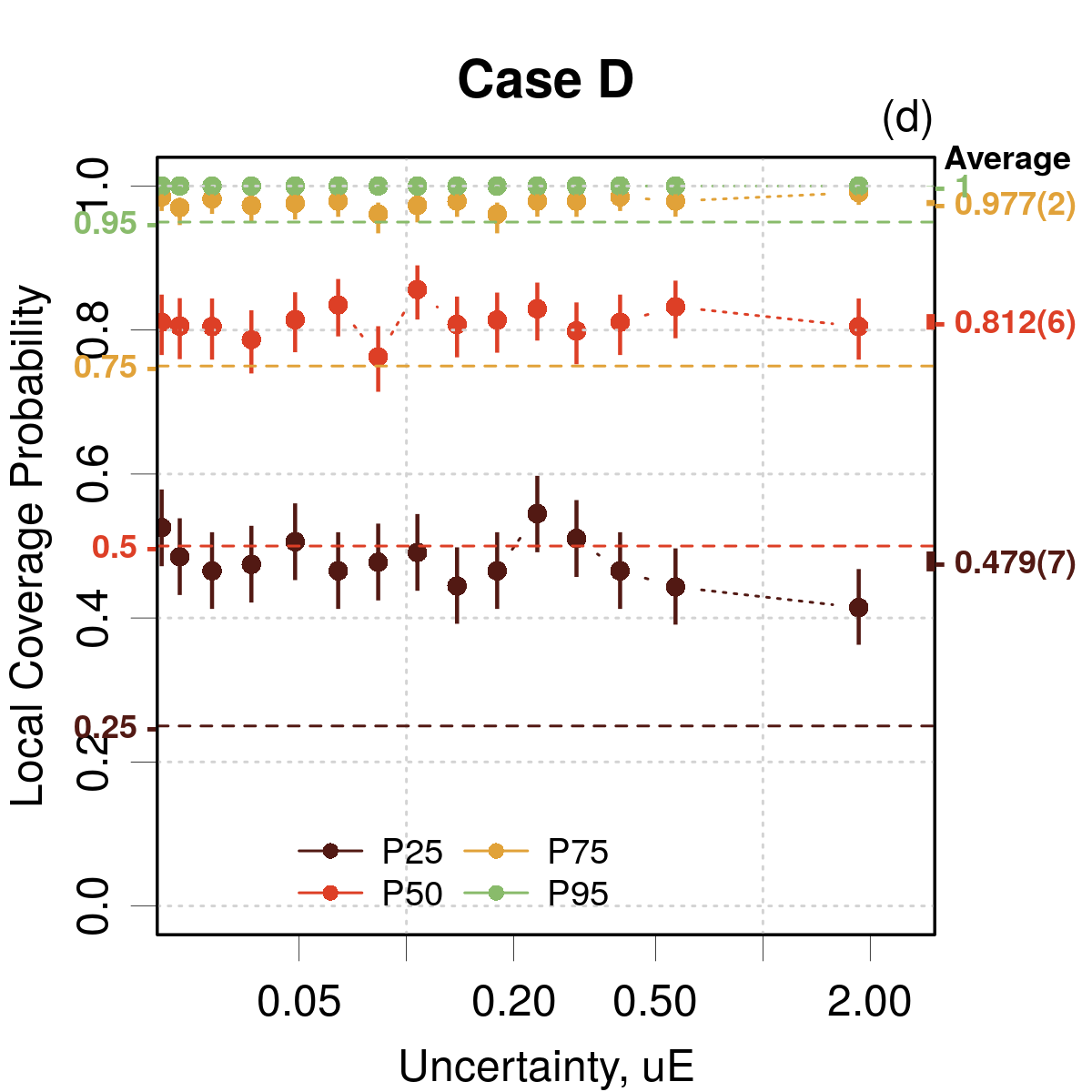} & \includegraphics[width=0.33\textwidth]{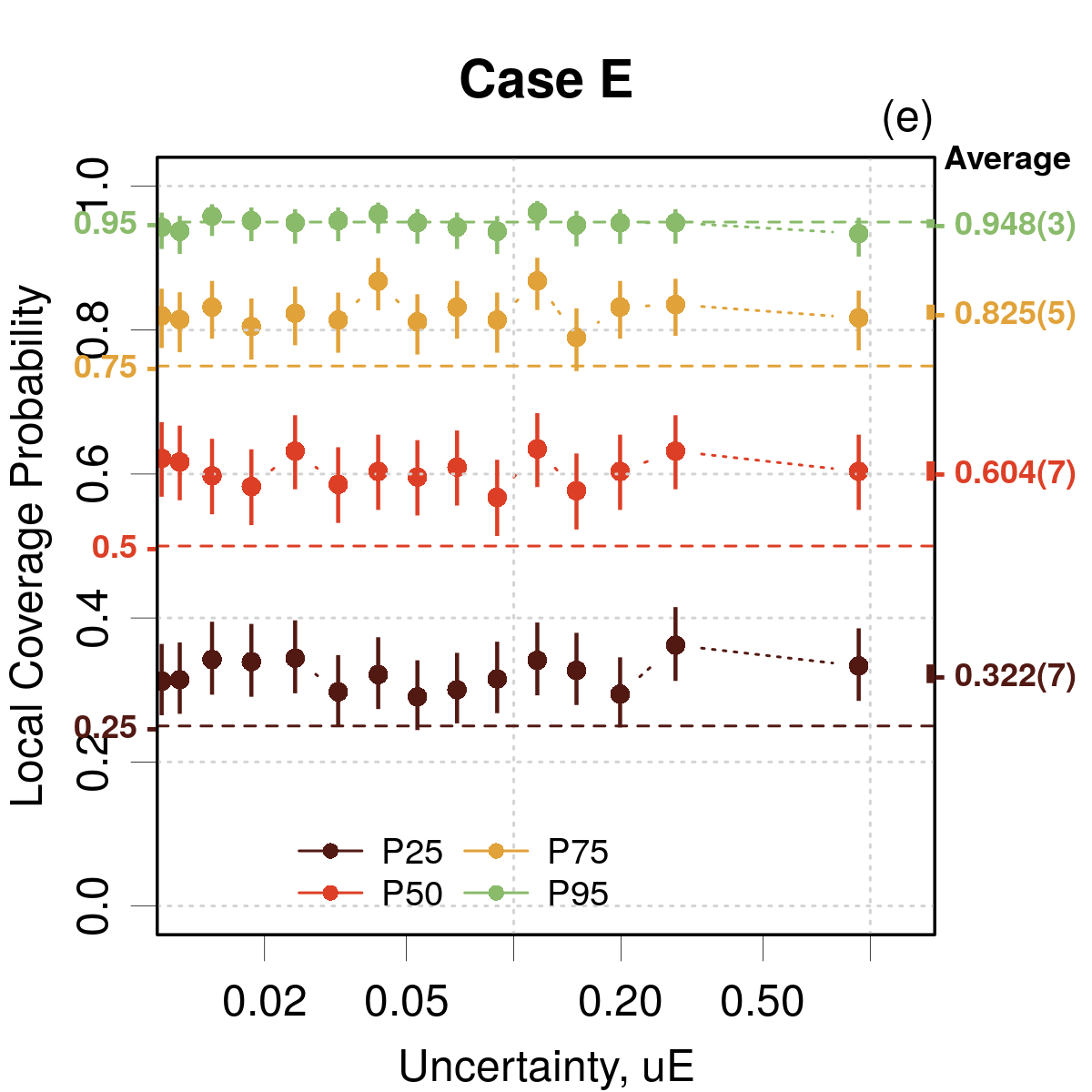} & \includegraphics[width=0.33\textwidth]{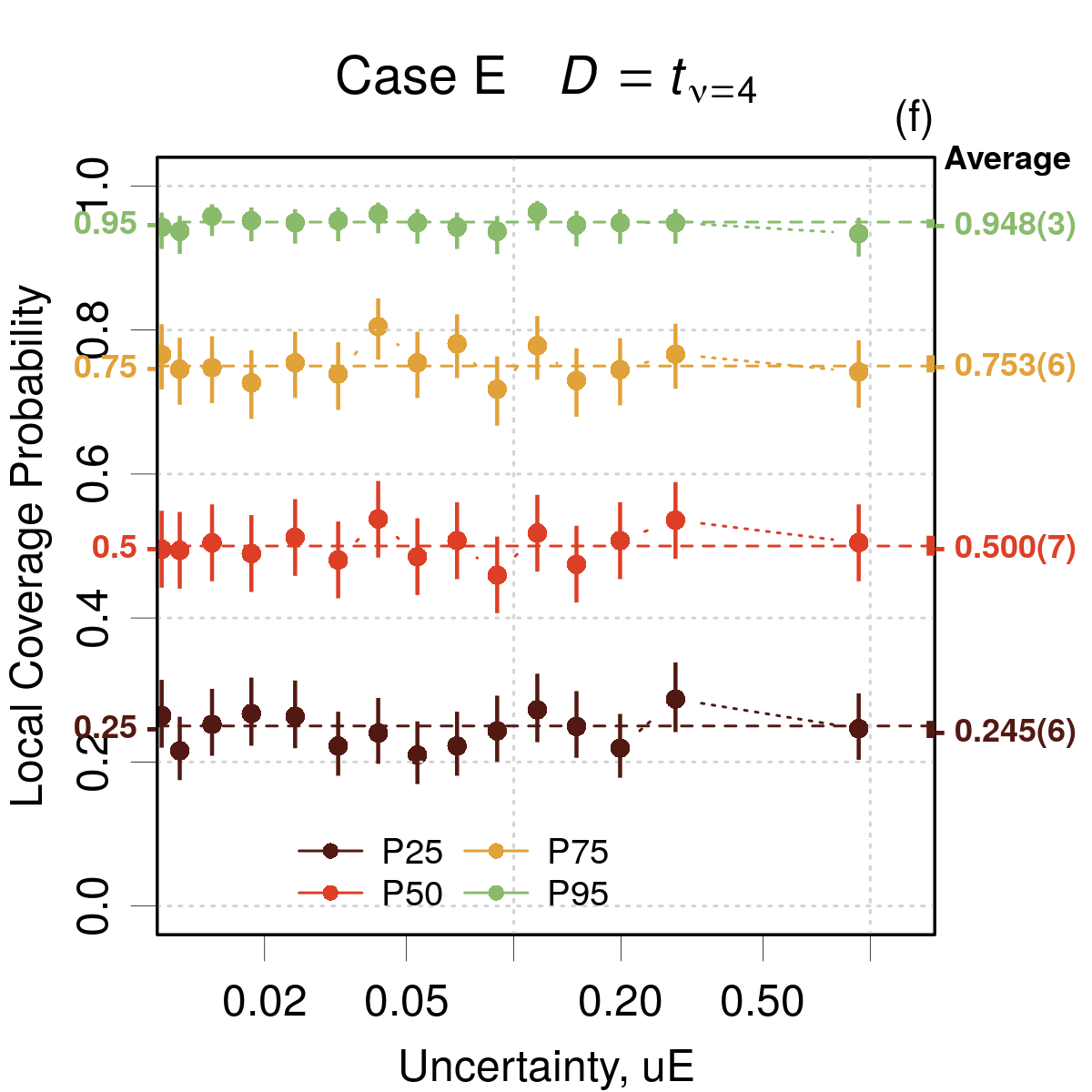}\tabularnewline
\end{tabular}
\par\end{centering}
\caption{\label{fig:LCPU}LCP analysis conditional to $u_{E}$ for cases A-E.
Not applicable to case F (homoscedastic). The points are reported
at the center of the bins. The error bars correspond to 95\,\% confidence
intervals on the statistic. The value and error bar reported on the
right axis corresponds to the statistic for the full dataset (average
coverage).}
\end{figure*}

For the synthetic datasets (Fig.\,\ref{fig:LCPU}), one makes the
hypothesis of a normal generative distribution to build prediction
intervals from the uncertainties for $\mathcal{P}=\{25,50,75,95\}$.
Case A presents perfect conditional coverages over the full uncertainty
range and probability levels. For Case B, the empirical coverages
are too large, except at the $P=95$\,\% level. The dependency along
$u_{E}$ is weak, except for the two extreme intervals, as was observed
on the LZISD analysis. A contrario, Case C presents a strong coverage
variation across $u_{E}$, with inadequate mean PICP values. For Case
D, the overestimated uncertainties produce intervals with excessive
coverage, uniformly across the uncertainty range. Finally, Case E
suffers again from the misidentification of the generative distribution,
which is solved by using the correct distribution to generate the
intervals {[}Fig.\,\ref{fig:LCPU}(f){]}. 

When a PICP value reaches 1.0, one gets no information about the mismatch
amplitude with the target probability. The Local Ranges Ratio analysis
(LRR) has been proposed by Pernot\citep{Pernot2022b} to solve this
problem. It estimates the ratio between the width of the empirical
interval and the width of the theoretical interval. This tool is not
exploited in the present study, but its availability should be kept
in mind, 

\subsubsection{Confidence curve\label{subsec:Confidence-curve}}

\noindent A confidence curve is established by estimating an error
statistic $S$ on subsets of $C$ iteratively pruned from the points
with the largest uncertainties.\citep{Scalia2020} Technically, it
is a ranking-based method, as (1) it is insensitive to the scale of
the uncertainties, and (2) the relative ranking of the errors and
uncertainties plays a determinant role. It is not applicable to homoscedastic
datasets.

If one defines $u_{k}$ as the largest uncertainty left after removing
the $k$\,\% largest uncertainties from $u_{E}$ ($k\in\{0,1,\ldots,99\}$),
a confidence statistic is defined by
\begin{equation}
c_{S}(k;E,u_{E})=S\left(E\,|\,u_{E}<u_{k}\right)\label{eq:non-normcc}
\end{equation}
where $S$ is an error statistic \textendash{} typically the Mean
Absolute Error (MAE) or Root Mean Squared Error (RMSE) \textendash{}
and $S\left(E\,|\,u_{E}<u_{k}\right)$ denotes that only those errors
$E_{i}$ paired with uncertainties $u_{E_{i}}$ smaller than $u_{k}$
are selected to compute $S$. A confidence curve is obtained by plotting
$c_{S}$ against $k$. Both normalized and non-normalized confidence
curves are used in the literature, but the RMSE-based non-normalized
version should be preferred for the validation of variance-based UQ
metrics.\citep{Pernot2022c} 

A monotonously decreasing confidence curve reveals a desirable association
between the largest errors and the largest uncertainties, an essential
feature for \emph{active learning}. It might also enable to detect
unreliable predictions. But, in order to test consistency, one needs
to compare $c_{S}$ to a reference.

\paragraph*{Reference curves.\label{subsec:Reference-curves}}

\noindent As already mentioned, only the order of $u_{E}$ values
is used to build a confidence curve, and any change of scale of $u_{E}$
leaves $c_{S}$ unchanged. Without a proper reference, $c_{S}$ cannot
inform us on calibration or consistency. 

An \emph{oracle} curve can be generated by reordering $u_{E}$ to
match the order of absolute errors. This can be expressed as
\begin{equation}
O(k;E)=c_{s}(k;E,|E|)
\end{equation}
It is evident from the above equation that the oracle is independent
of $u_{E}$ and therefore useless for calibration testing. Recast
in the probabilistic framework introduced above, the oracle would
correspond to a very implausible error distribution $D$, such that
$E_{i}=\pm u_{E_{i}}$. Although it is offered as the default reference
curve in some validation libraries, I strongly recommend against its
use for variance- or interval-based UQ metrics. 

Using Eq.\,\ref{eq:probmod}, a \emph{probabilistic reference} curve
$P$ can be generated by sampling pseudo-errors $\widetilde{E}_{i}$
for each uncertainty $u_{E_{i}}$ and calculating a confidence curve
for $\left\{ \tilde{E},u_{E}\right\} $, i.e.,
\begin{equation}
P(k;u_{E})=\left\langle c_{S}(k;\tilde{E},u_{E})\right\rangle _{\tilde{E}}
\end{equation}
where a Monte Carlo average is taken over samples of
\begin{equation}
\widetilde{E}_{i}\sim D(0,u_{E_{i}})
\end{equation}
The sampling is repeated to have converged mean and confidence band
at the 95\,\% level, typically 500 times. 

In contrast to the oracle, which depends exclusively on the errors,
the probabilistic reference depends on $u_{E}$ and a choice of distribution
$D$, but it does not depend on the actual errors $E$. Comparison
of the data confidence curve to $P$ enables to test if $E$ and $u_{E}$
are correctly linked by the probabilistic model, Eq.\,\ref{eq:probmod},
i.e. to test consistency. For RMSE-based confidence curves, $P$ does
not depend on the choice of generative distribution (only the width
of the confidence band does). This is not the case for MAE-based confidence
curves, which makes them less suitable.\citep{Pernot2022c} Given
the choice of a generative distribution $D$, interval-based metrics
can be transformed to variance-based metrics and used to build a confidence
curve and probabilistic reference. 

\begin{figure*}[t]
\noindent \begin{centering}
\begin{tabular}{ccc}
\includegraphics[width=0.33\textwidth]{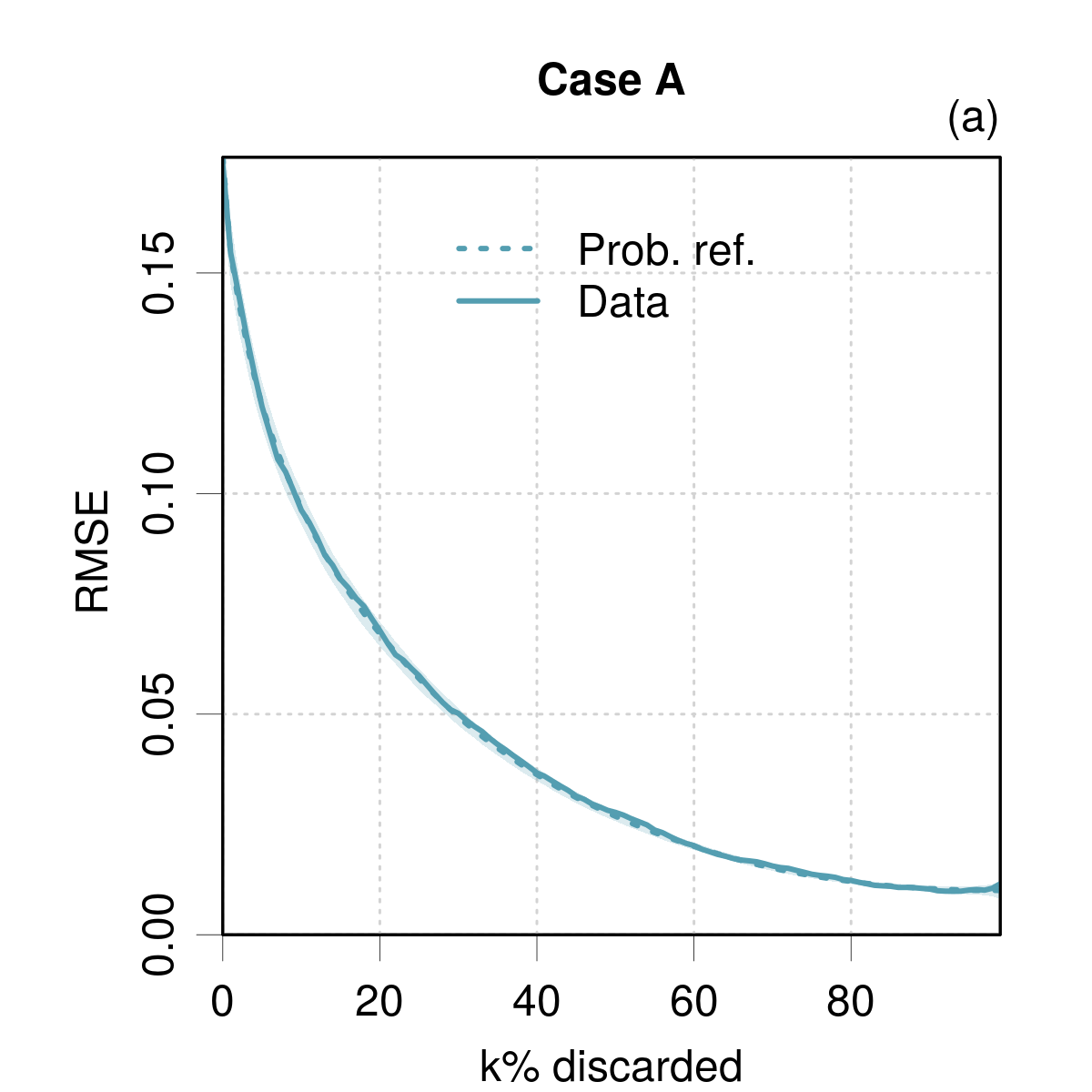} & \includegraphics[width=0.33\textwidth]{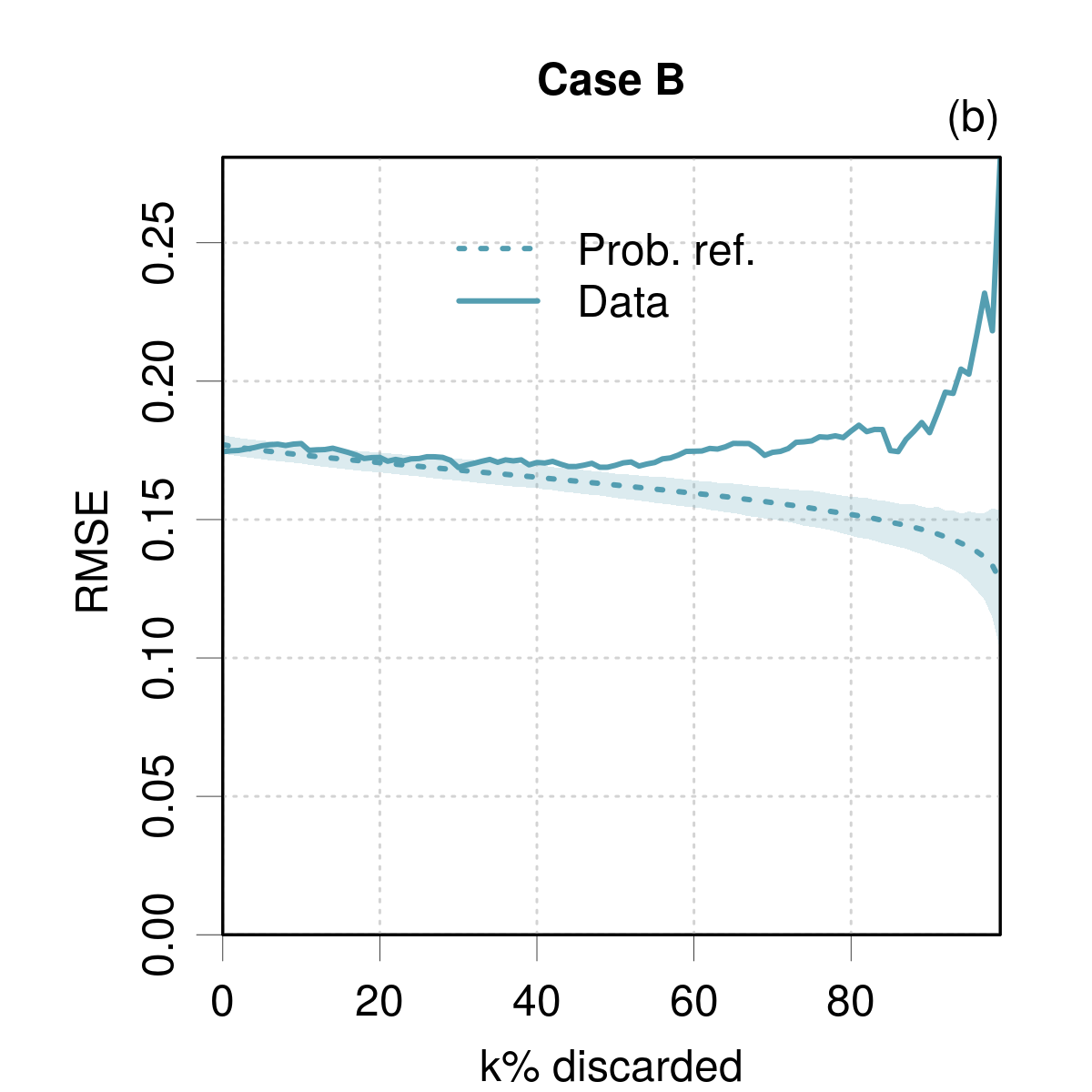} & \includegraphics[width=0.33\textwidth]{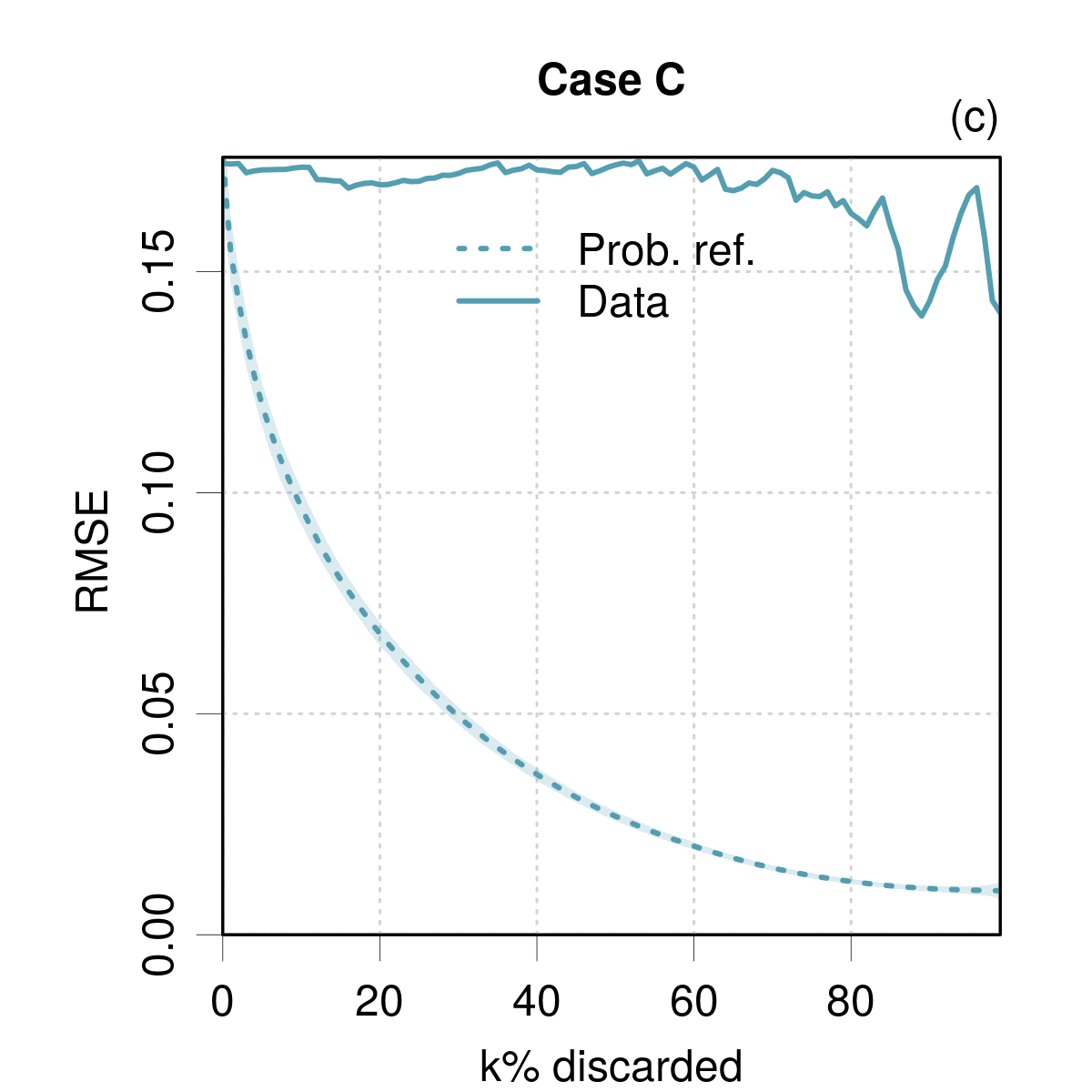}\tabularnewline
\includegraphics[width=0.33\textwidth]{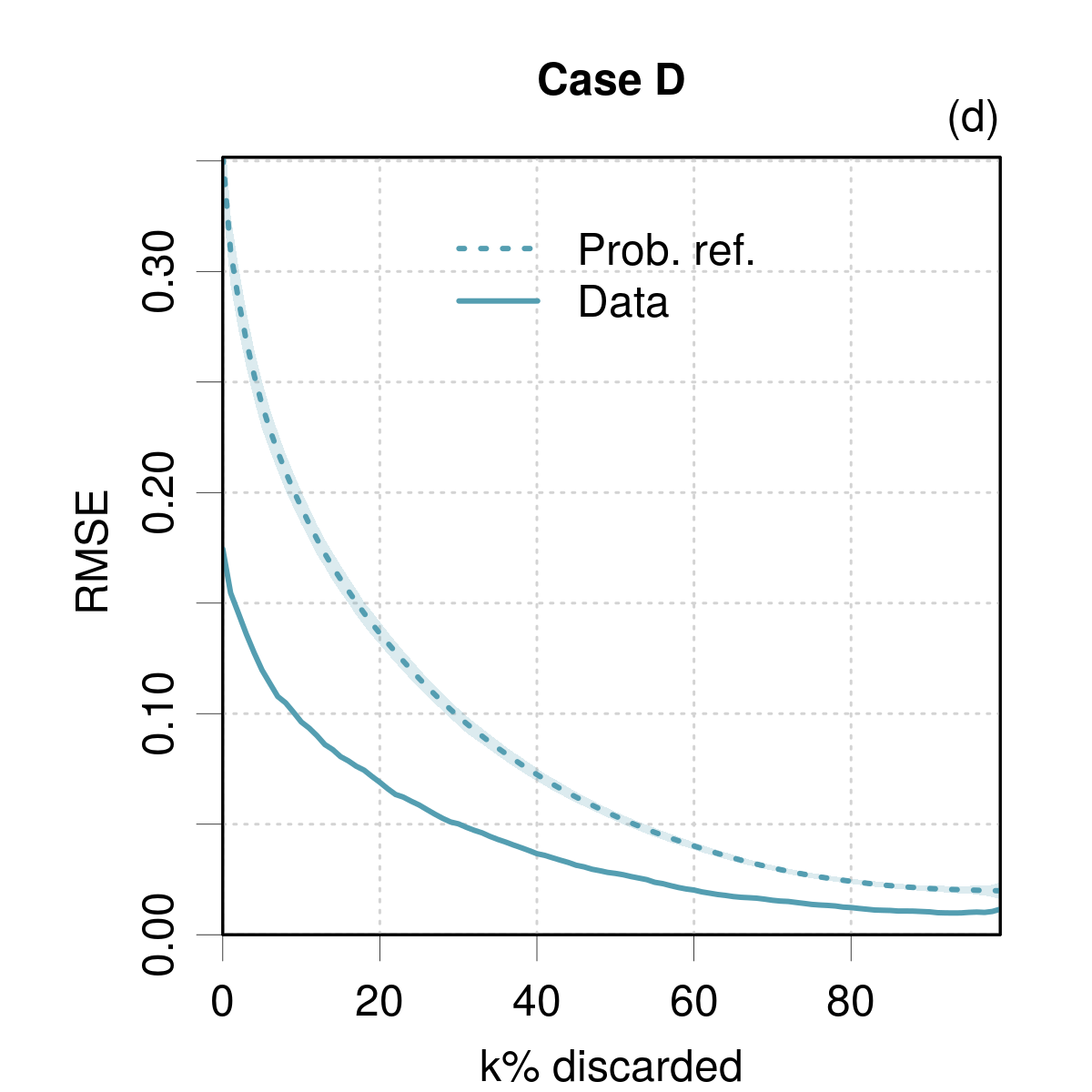} & \includegraphics[width=0.33\textwidth]{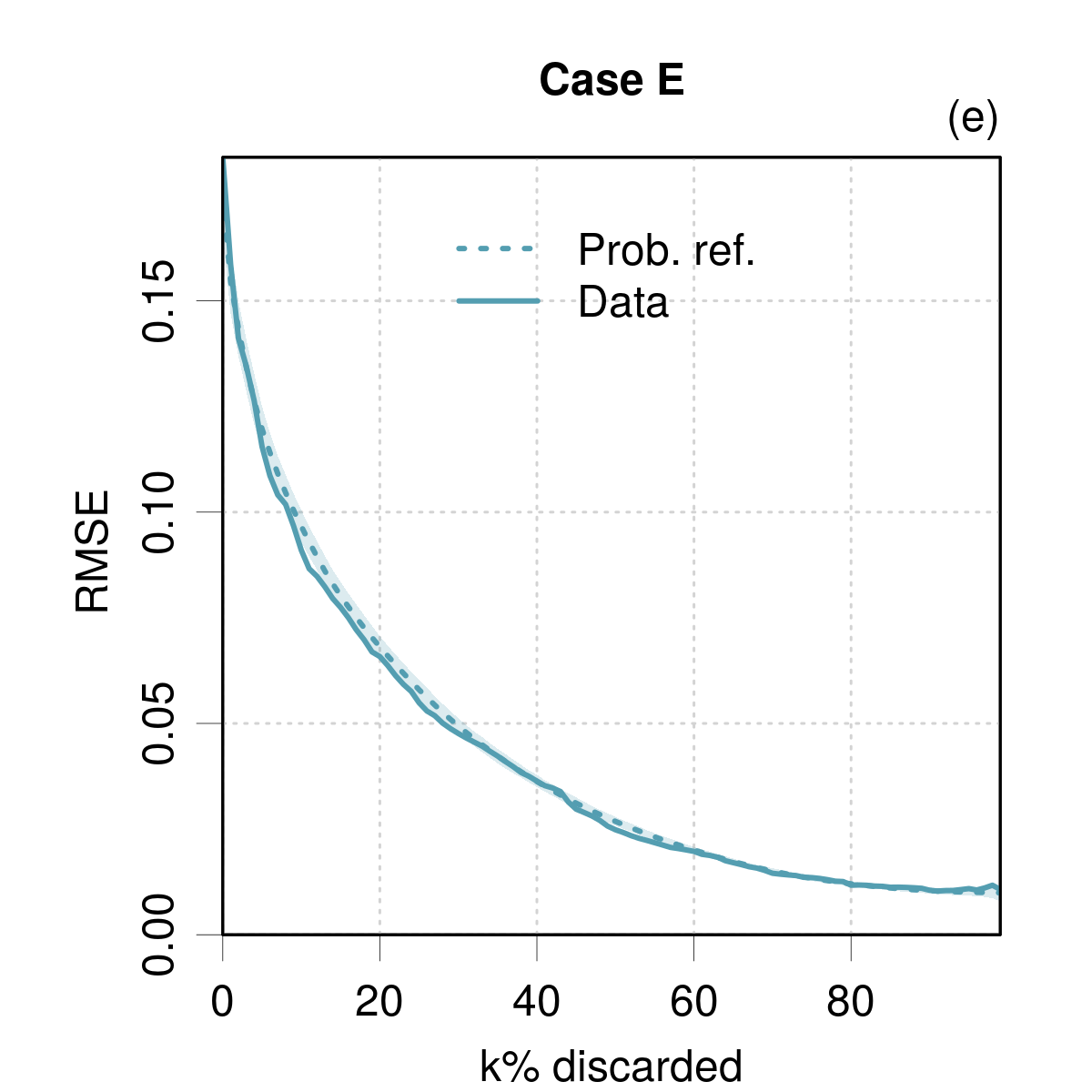} & \includegraphics[width=0.33\textwidth]{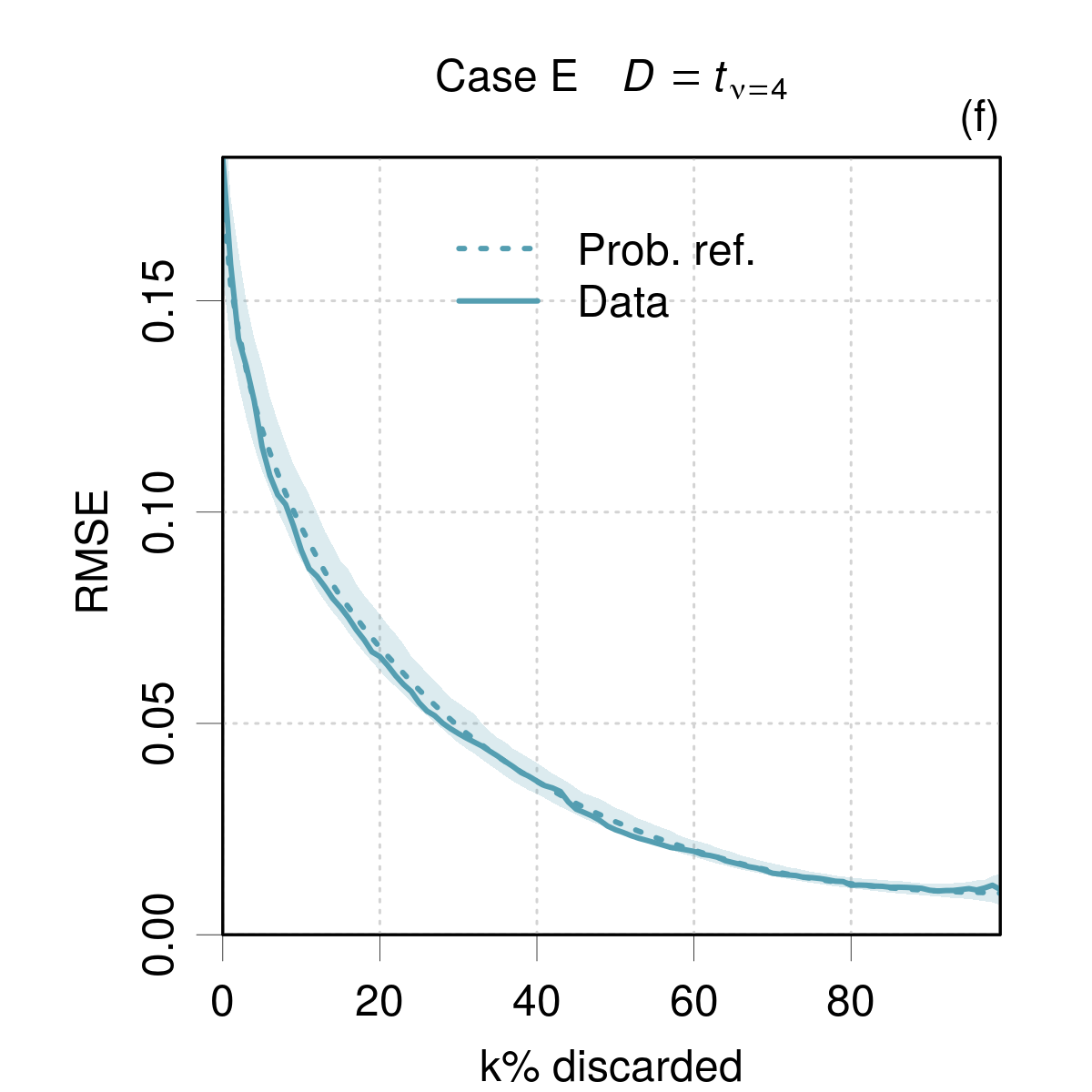}\tabularnewline
\end{tabular}
\par\end{centering}
\caption{\label{fig:Confidence-curves}Confidence curves for cases A-E. Not
applicable to case F (homoscedastic). The solid blue line is the data
confidence curve, the dashed blue line is the probabilistic reference
line and the light blue area defines its 95\,\% confidence interval.}
\end{figure*}

One can check in Fig.\,\ref{fig:Confidence-curves} that consistency
problems are well identified by the comparison of a confidence curve
to the associated probabilistic reference (Cases~B-D). When small
deviations are observed, it is however difficult to discriminate between
a genuine consistency problem and a bad choice of generative distribution
for $P$ (Case~E). The choice of the appropriate distribution for
Case~E leaves the confidence and reference curve unchanged, but widens
the confidence band of the probabilistic reference, ensuring validation
of consistency. 

\subsection{Testing adaptivity\label{subsec:Testing-adaptivity}}

\noindent Adaptivity is not commonly tested in the ML-UQ literature,
except maybe through a binary approach discerning in-distribution
and out-of-distribution predictions.\citep{Scalia2020,Hu2022}

\subsubsection{$Z$ vs Input feature plot}

\noindent The analog of the ``$E$ vs $u_{E}$'' plot for consistency
is to plot the \emph{z}-scores $Z$ as a function of a relevant input
feature $X$ to check adaptivity. The distribution of $Z$ should
be homogeneous along $X$ and symmetric around $Z=0$. Here again,
one can use guide lines $Z=\pm2$ and compare with running quantiles
for a 95\% confidence interval of $Z$ values. 

\begin{figure*}[t]
\noindent \begin{centering}
\begin{tabular}{ccc}
\includegraphics[width=0.33\textwidth]{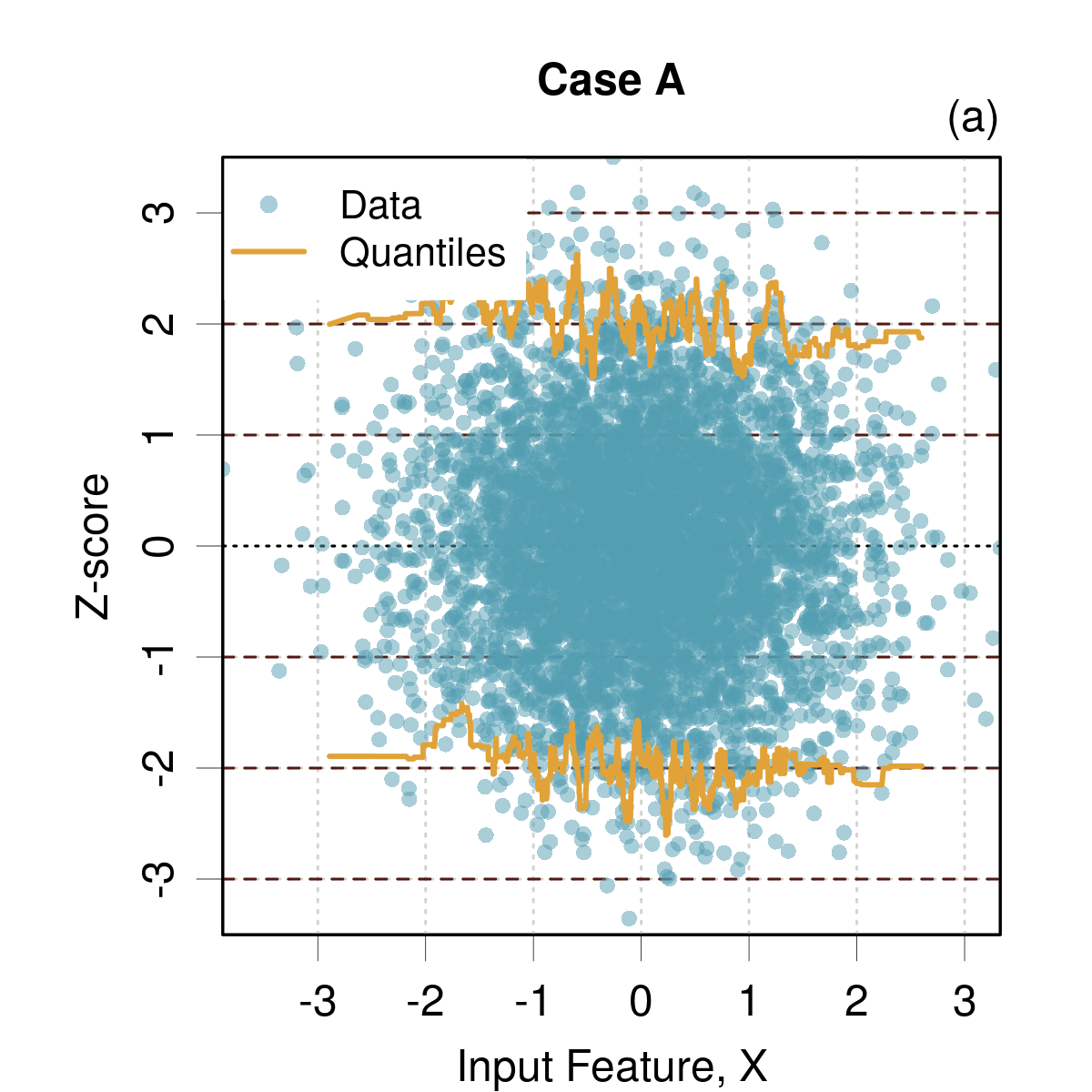} & \includegraphics[width=0.33\textwidth]{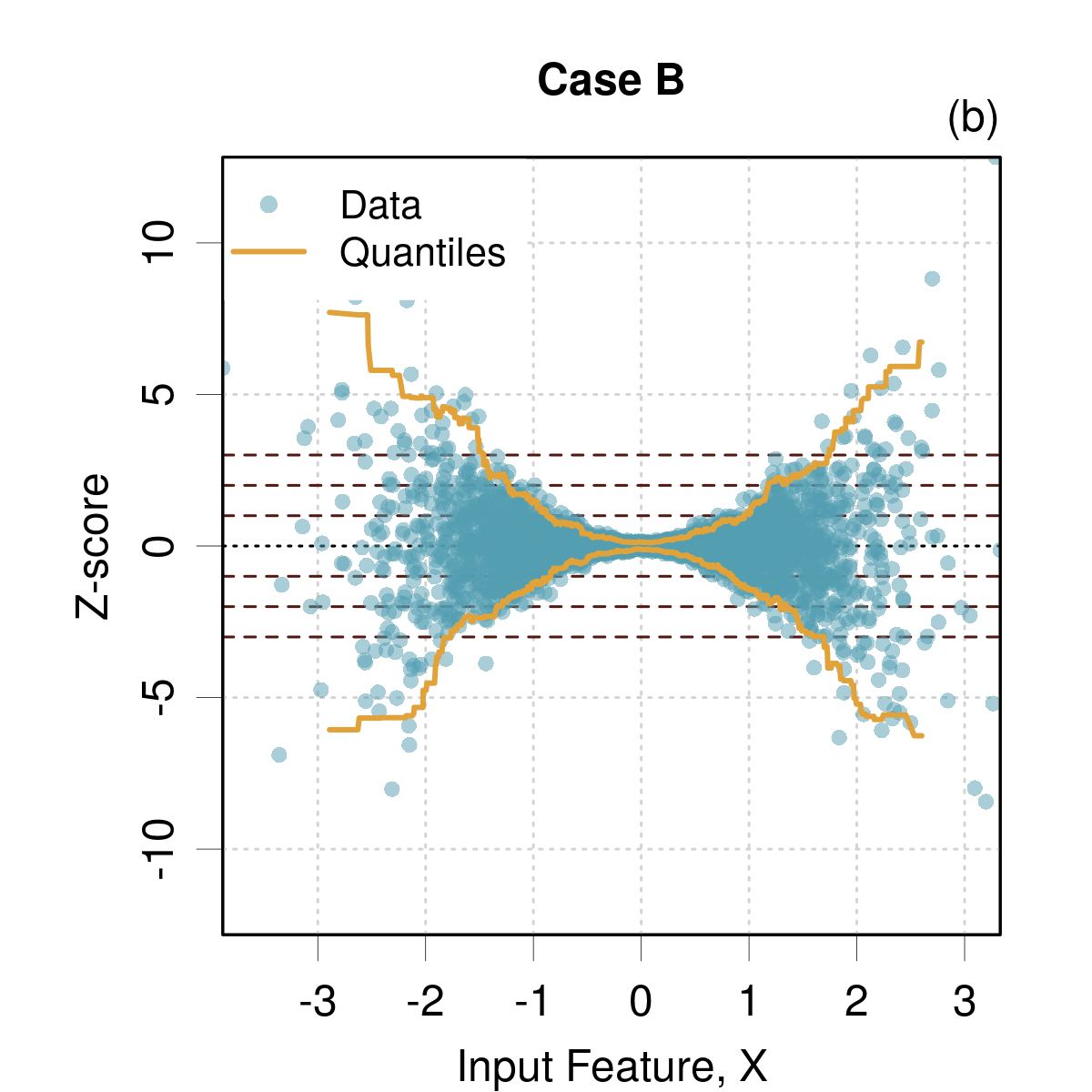} & \includegraphics[width=0.33\textwidth]{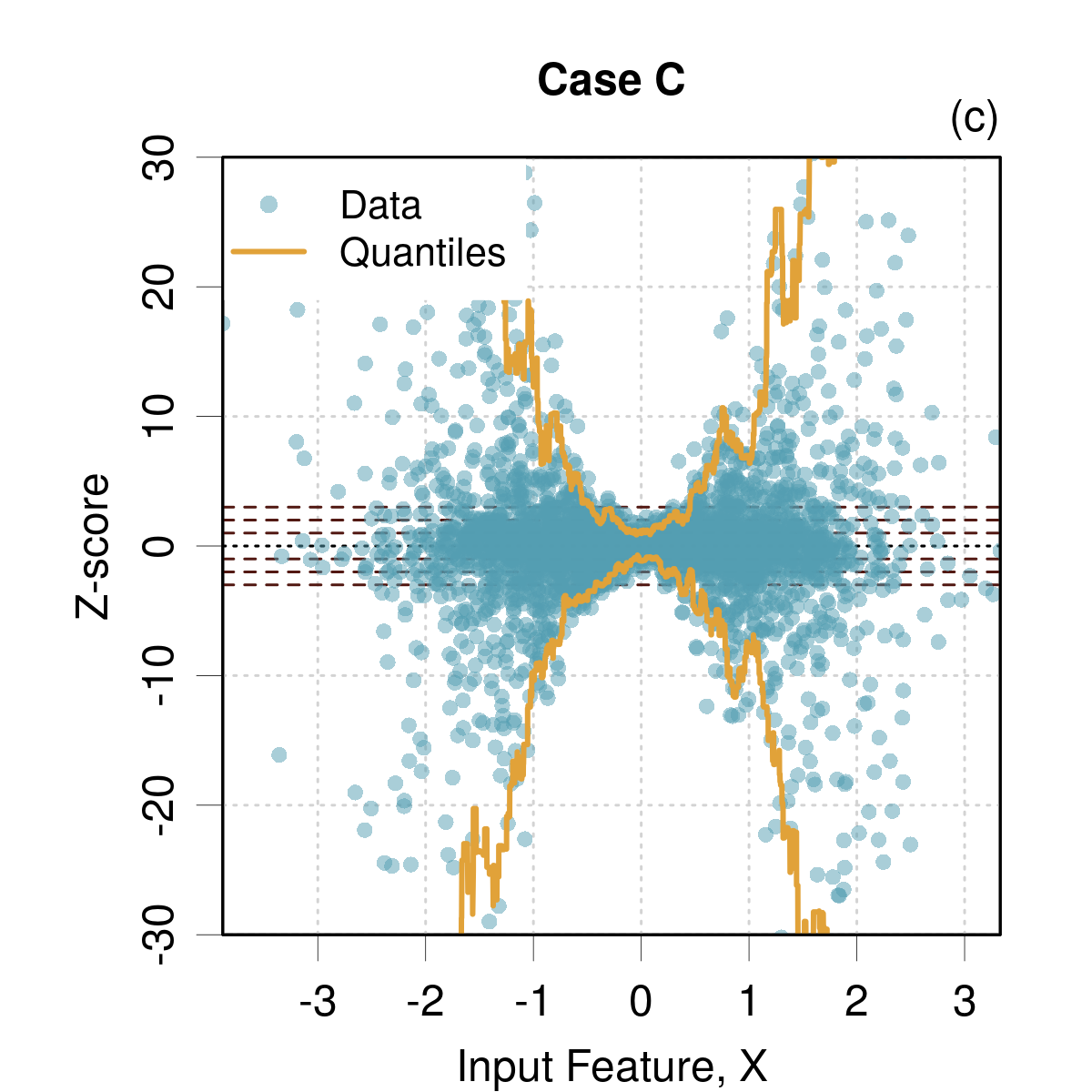}\tabularnewline
\includegraphics[width=0.33\textwidth]{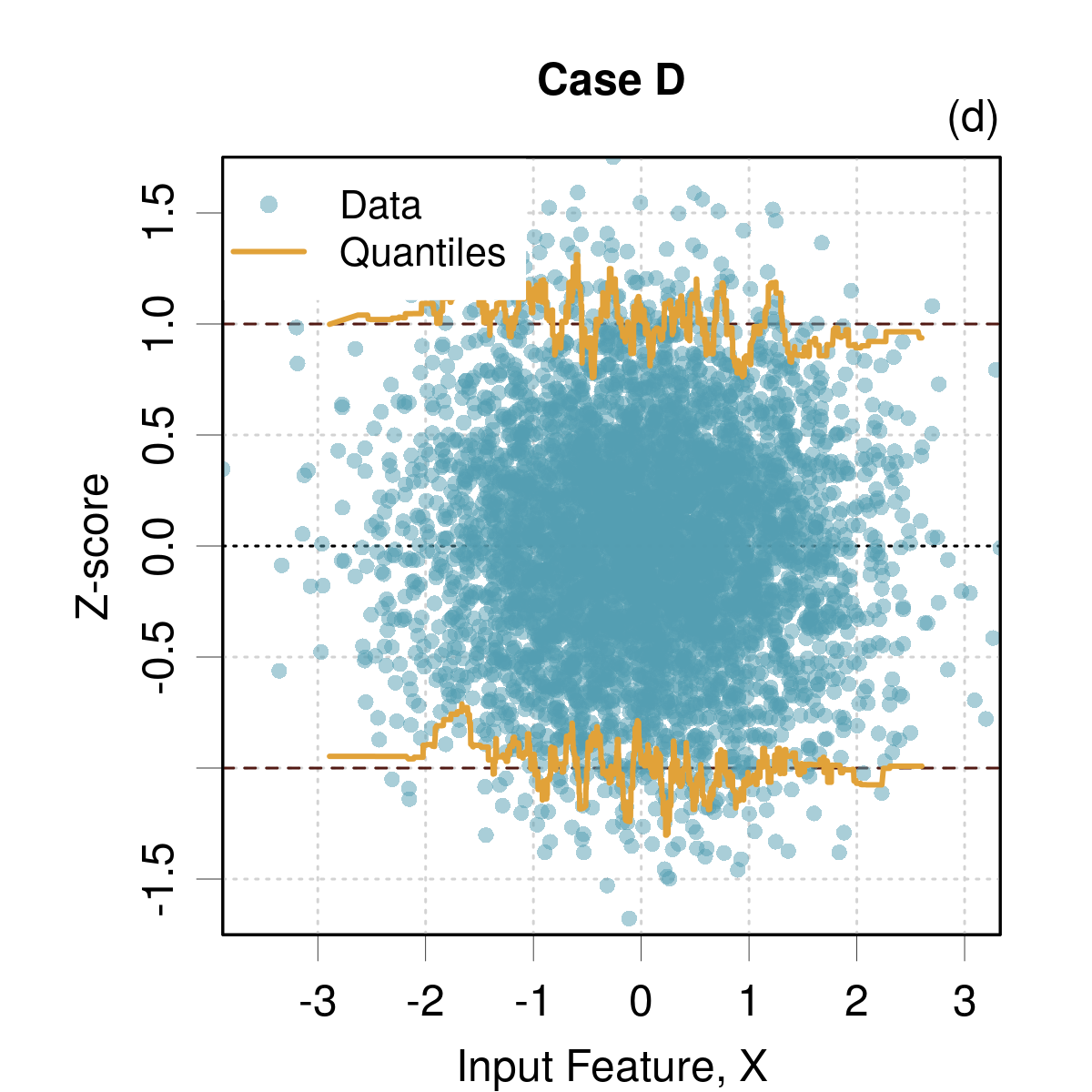} & \includegraphics[width=0.33\textwidth]{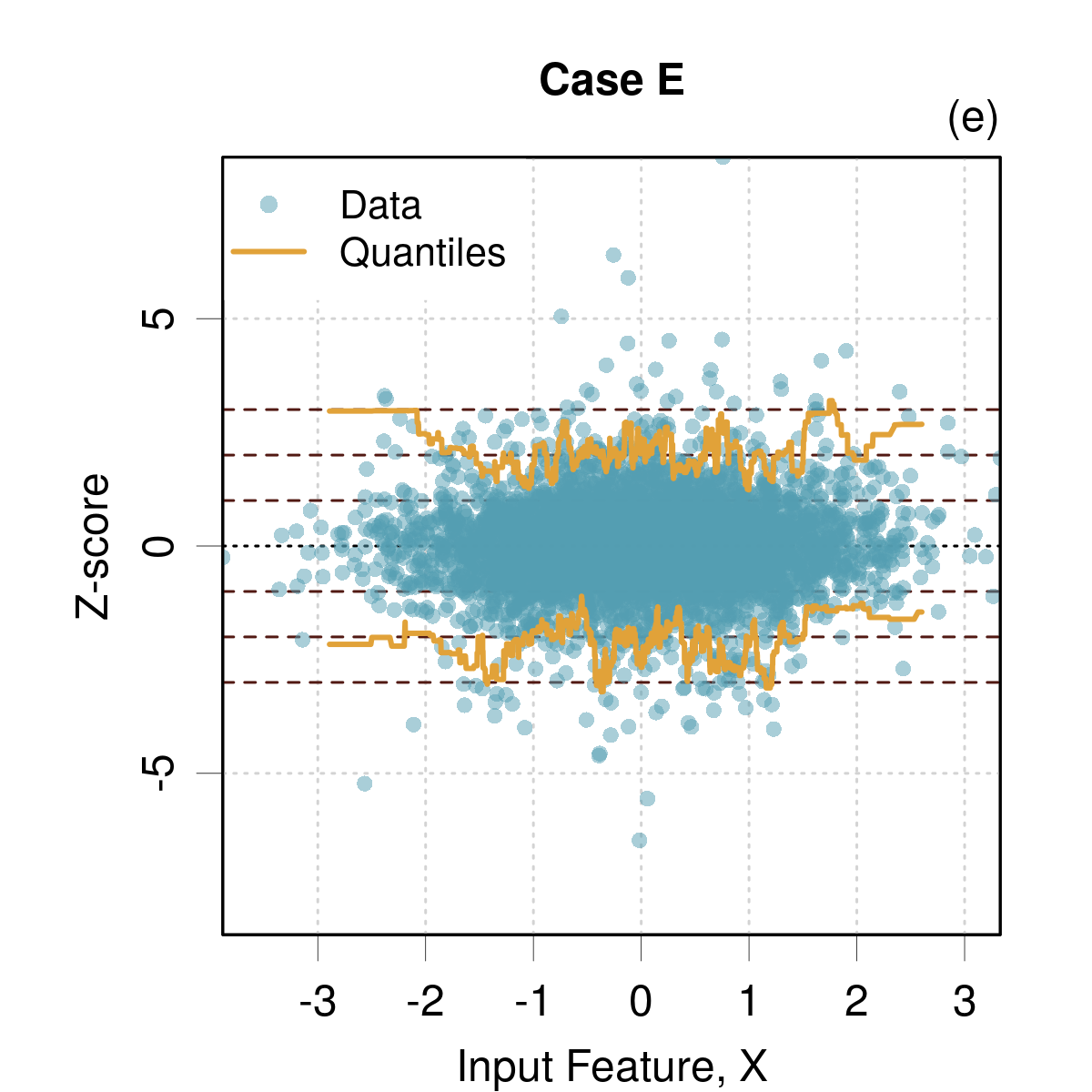} & \includegraphics[width=0.33\textwidth]{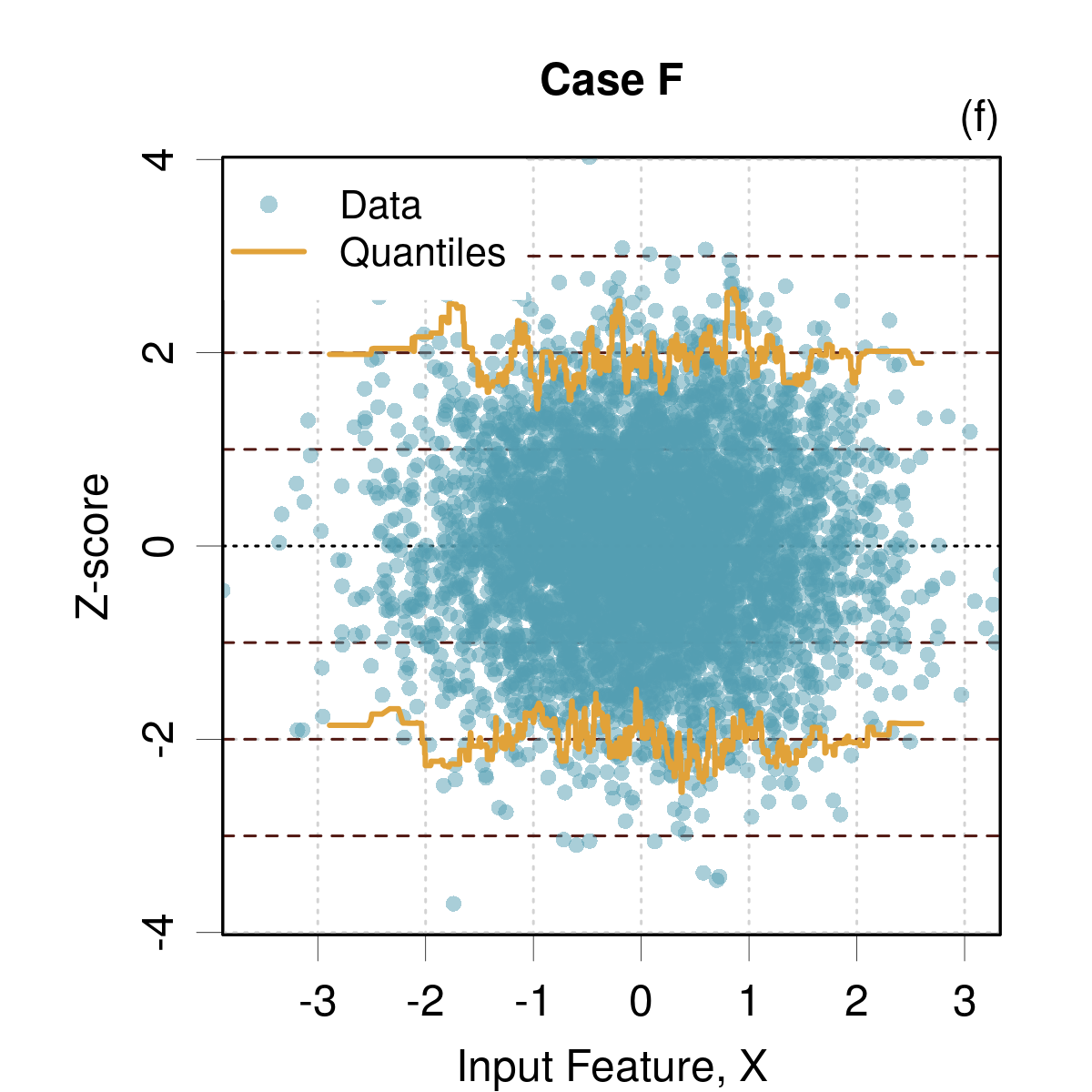}\tabularnewline
\end{tabular}
\par\end{centering}
\caption{\label{fig:validSynth-1-1}''$Z$ vs $X$'' validation plots for
cases A-F. }
\end{figure*}

Conditioning on $X$ is shown in Fig.\ref{fig:validSynth-1-1} for
the synthetic datasets. Considering the shape of the data clouds,
adaptivity can be clearly rejected in cases B and C, while the dispersion
of the \emph{z}-scores is insufficient in Case D. Validation of adaptivity
in the other cases require a more quantitative approach. 

\subsubsection{Conditional calibration curves in input feature space}

The same principle as used for consistency testing (Sect.\,\ref{subsec:Conditional-calibration-curves})
can be applied to adaptivity by grouping the data according to the
binning of a relevant input feature $X$. The method can now be applied
to homoscedastic datasets.

\begin{figure*}[t]
\noindent \begin{centering}
\begin{tabular}{ccc}
\includegraphics[width=0.33\textwidth]{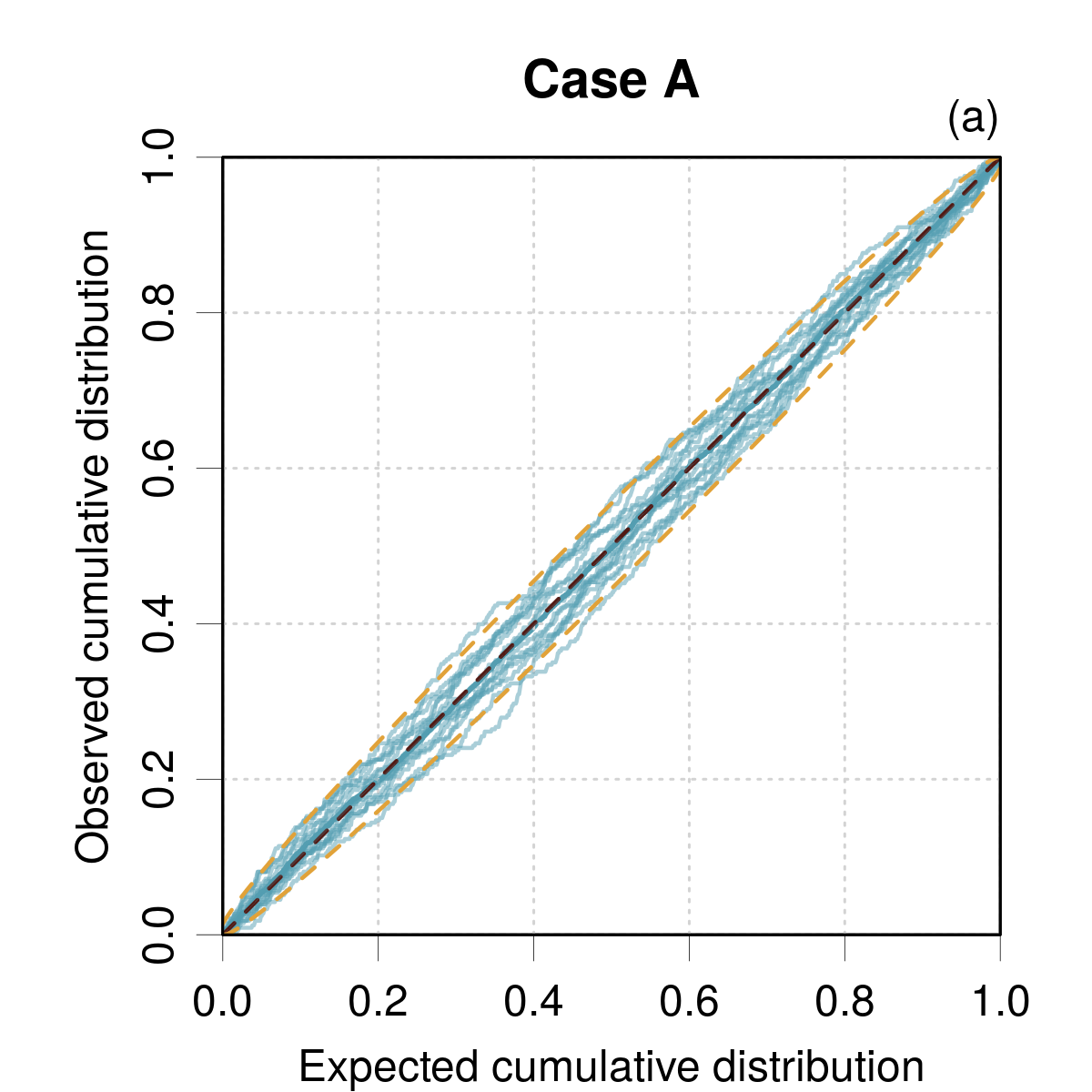} & \includegraphics[width=0.33\textwidth]{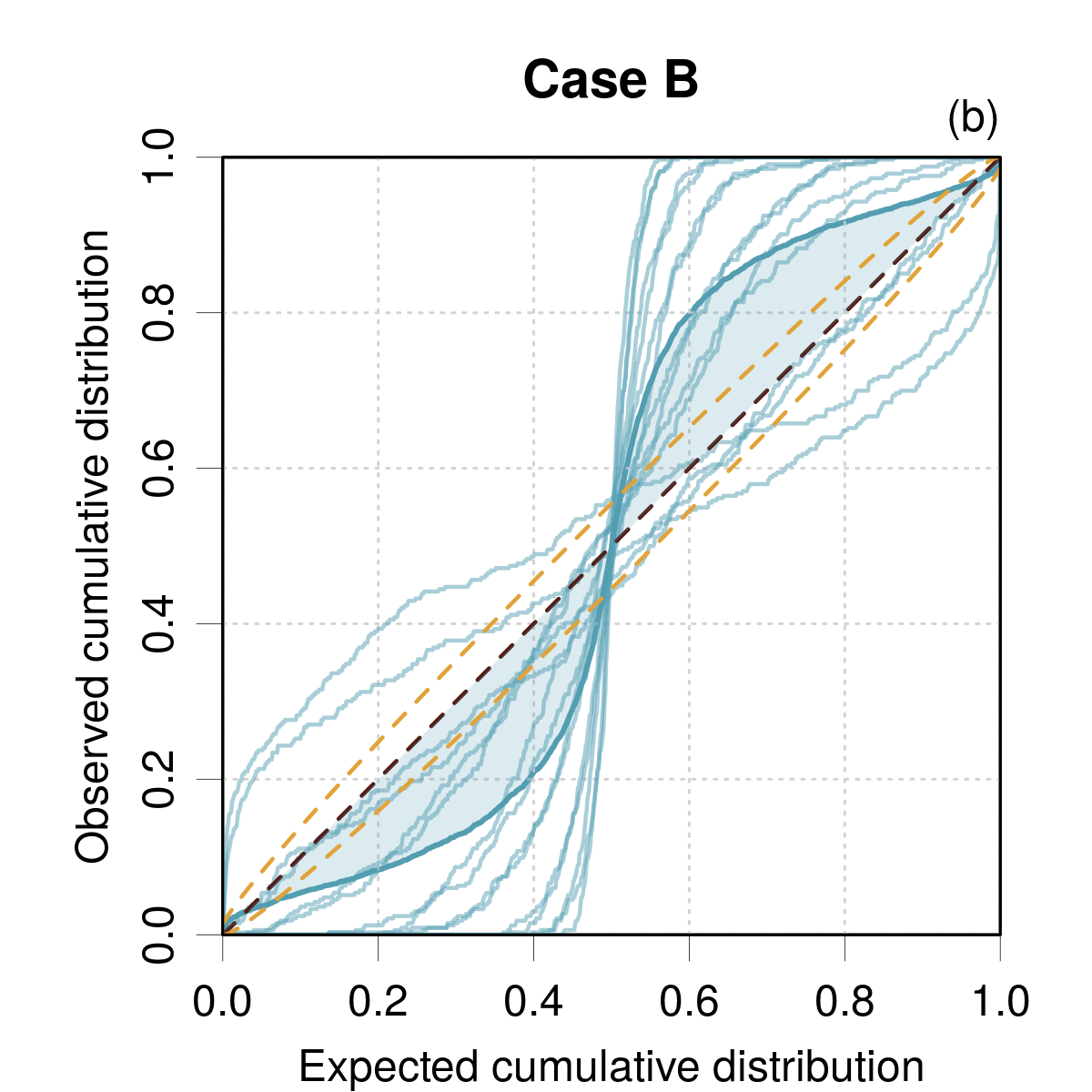} & \includegraphics[width=0.33\textwidth]{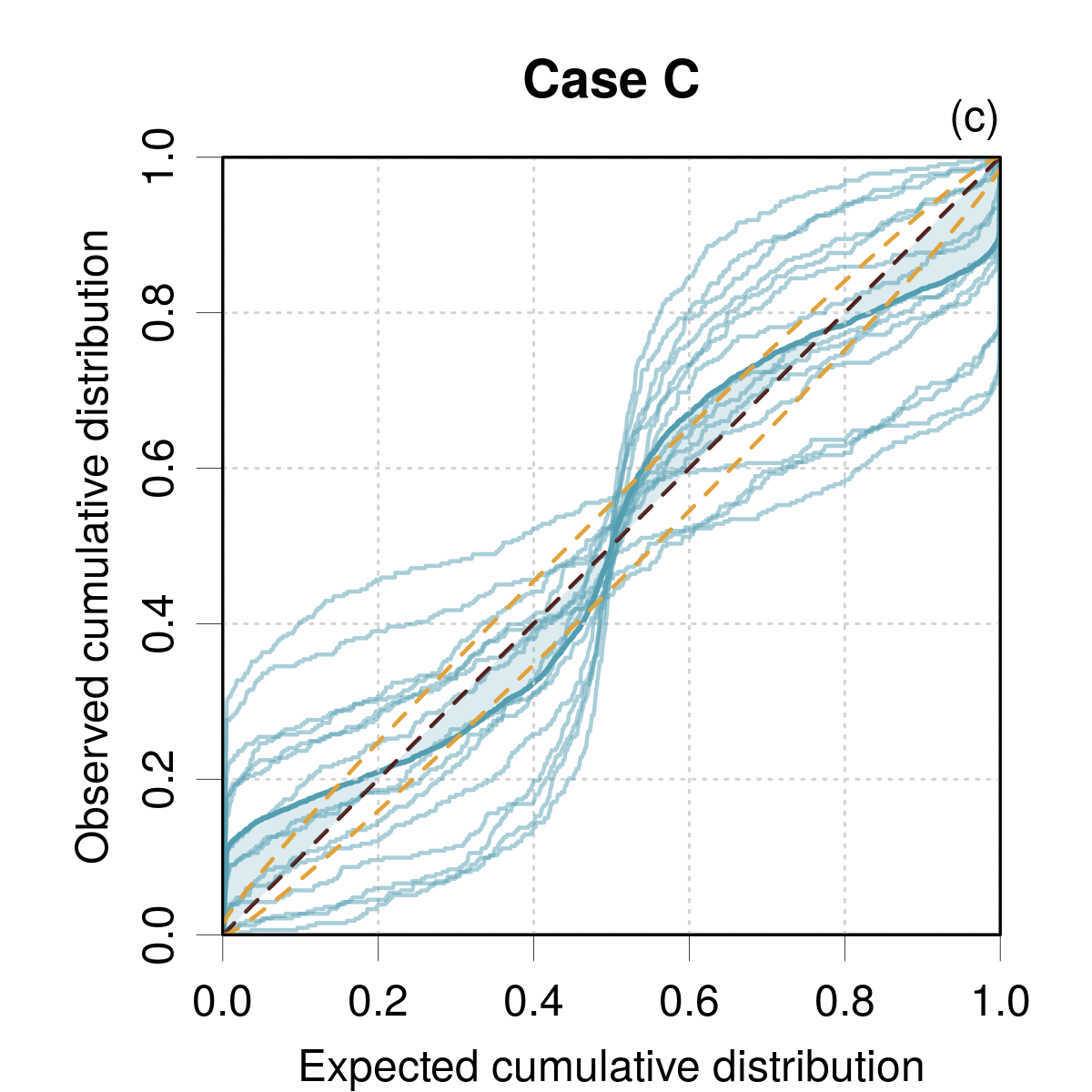}\tabularnewline
\includegraphics[width=0.33\textwidth]{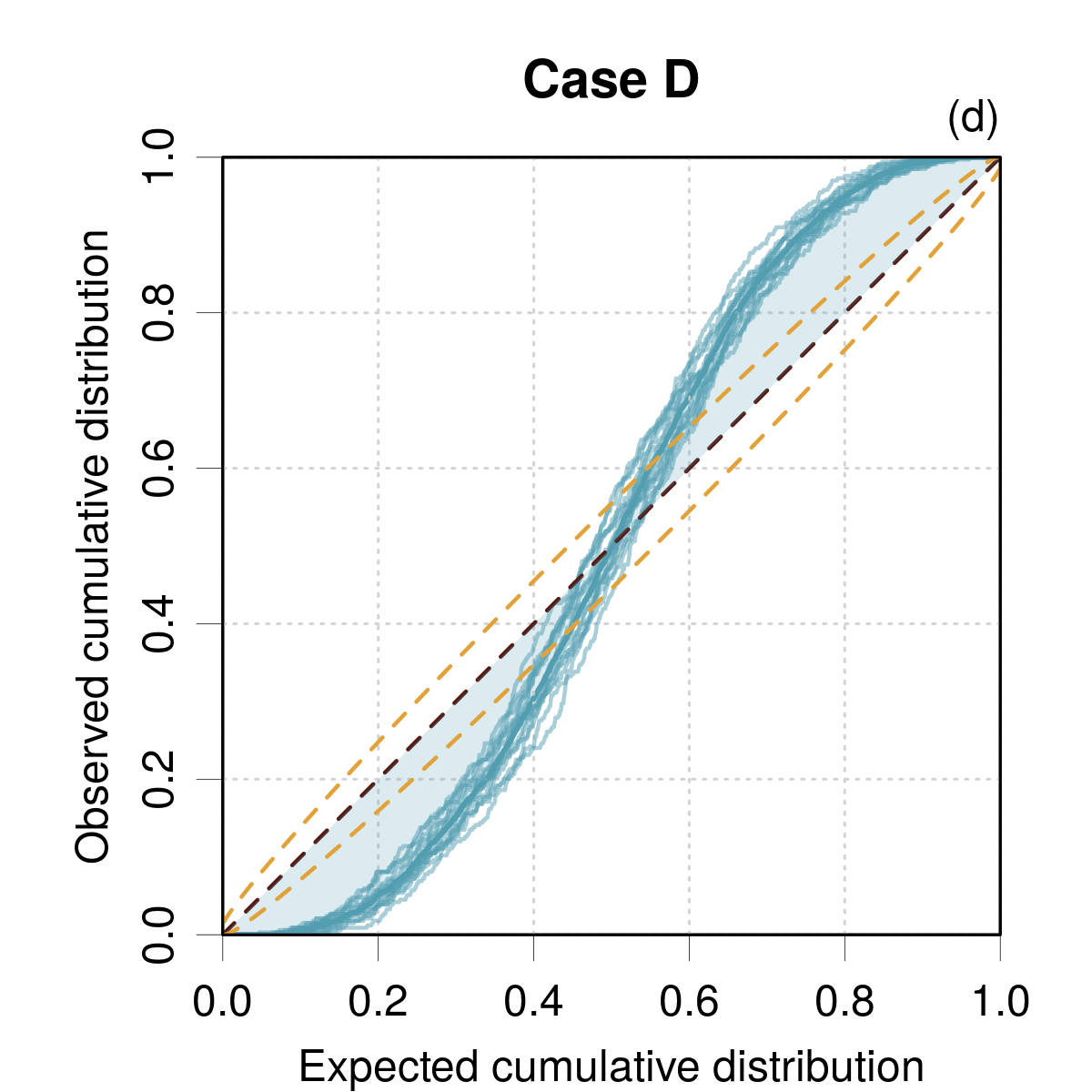} & \includegraphics[width=0.33\textwidth]{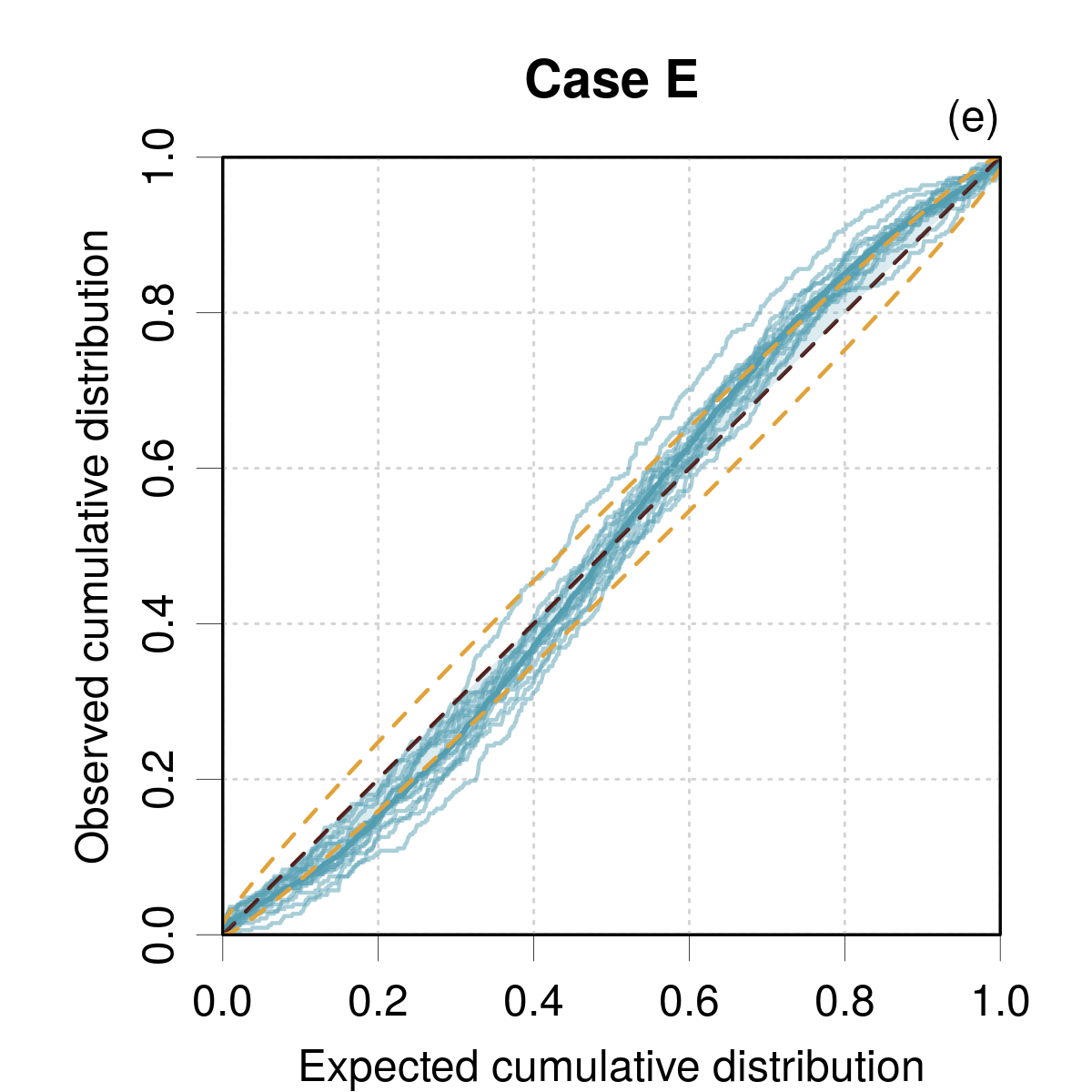} & \includegraphics[width=0.33\textwidth]{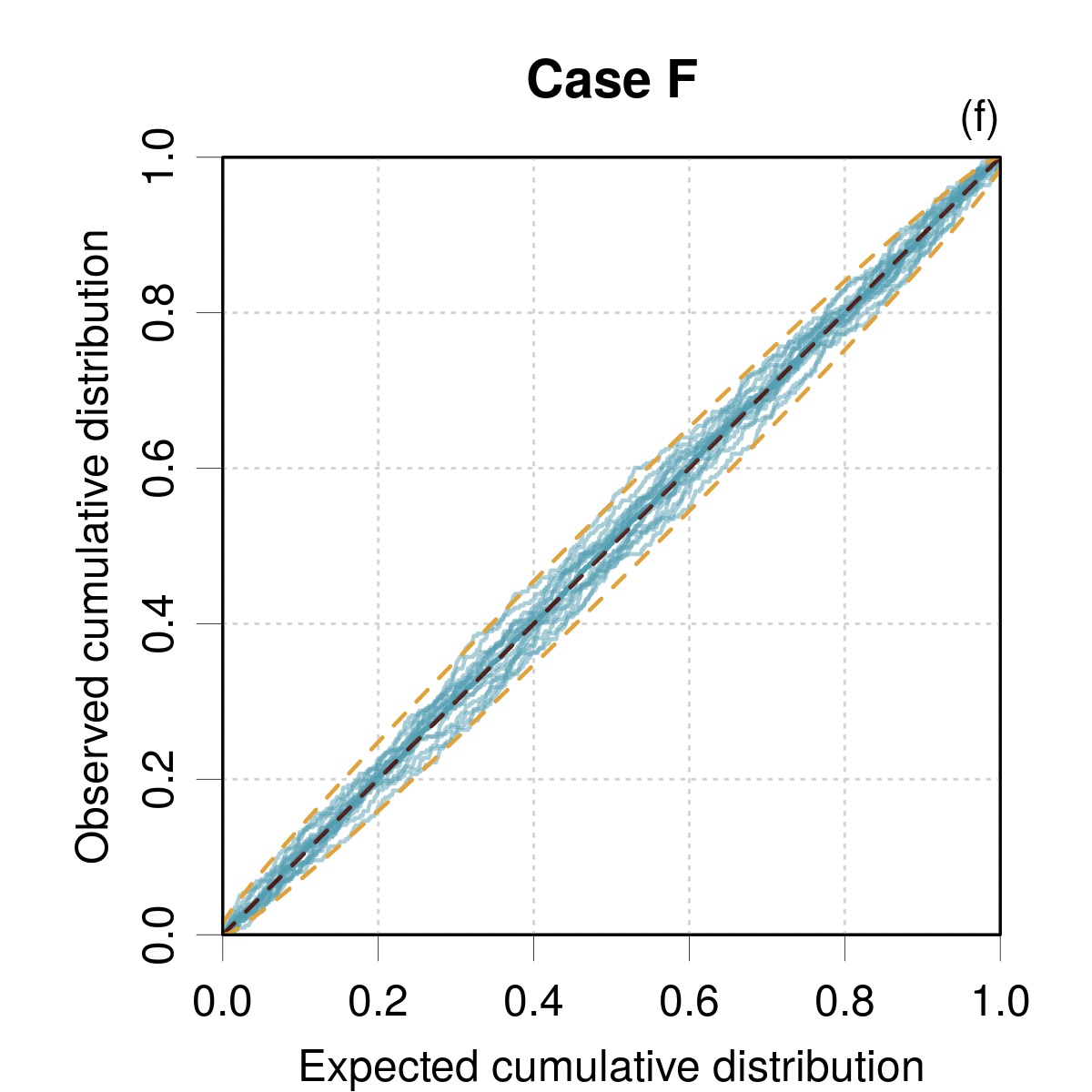}\tabularnewline
\end{tabular}
\par\end{centering}
\caption{\label{fig:Calibration-curves-1-2}Conditional calibration curves
in $X$ space for cases A-F. The average calibration curve is a thick
blue line, the conditional curves are thin blue lines. The reference
line and its 95\,\% confidence interval are dashed red lines. The
dataset has been split according to the binning of $X$ into 15 equal-counts
sets.}
\end{figure*}

Application to the synthetic datasets is presented in Fig.\,\ref{fig:Calibration-curves-1-2}.
For Cases A and F, one would conclude to a good adaptivity along $X$
. For the other cases, the deviations from the identity line are similar
to those observed when testing consistency, except for Case B, where
the conditional calibration curves are much more dispersed. 

\subsubsection{Local Z-Variance analysis in input feature space\label{subsec:Local-Z-Variance-analysis-1}}

\noindent As the LZV/LZISD analysis in $u_{E}$ space was used to
validate consistency, one can perform a LZV of LZISD analysis in $X$
space to validate adaptivity. Here again, the LZISD analysis offers
a more direct quantification of deviations. Both methods can be used
for homoscedastic uncertainties.

\begin{figure*}[t]
\noindent \begin{centering}
\begin{tabular}{ccc}
\includegraphics[width=0.33\textwidth]{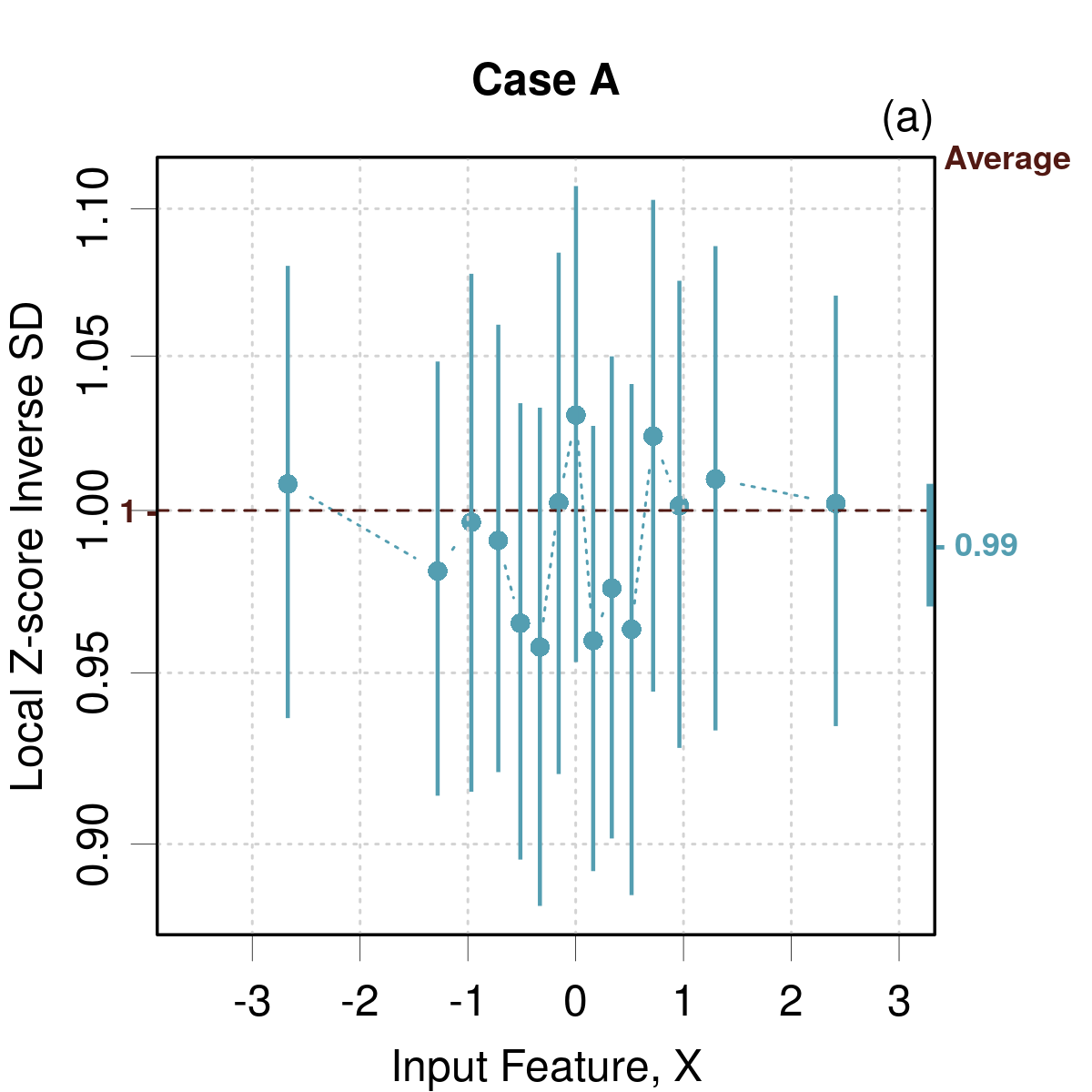} & \includegraphics[width=0.33\textwidth]{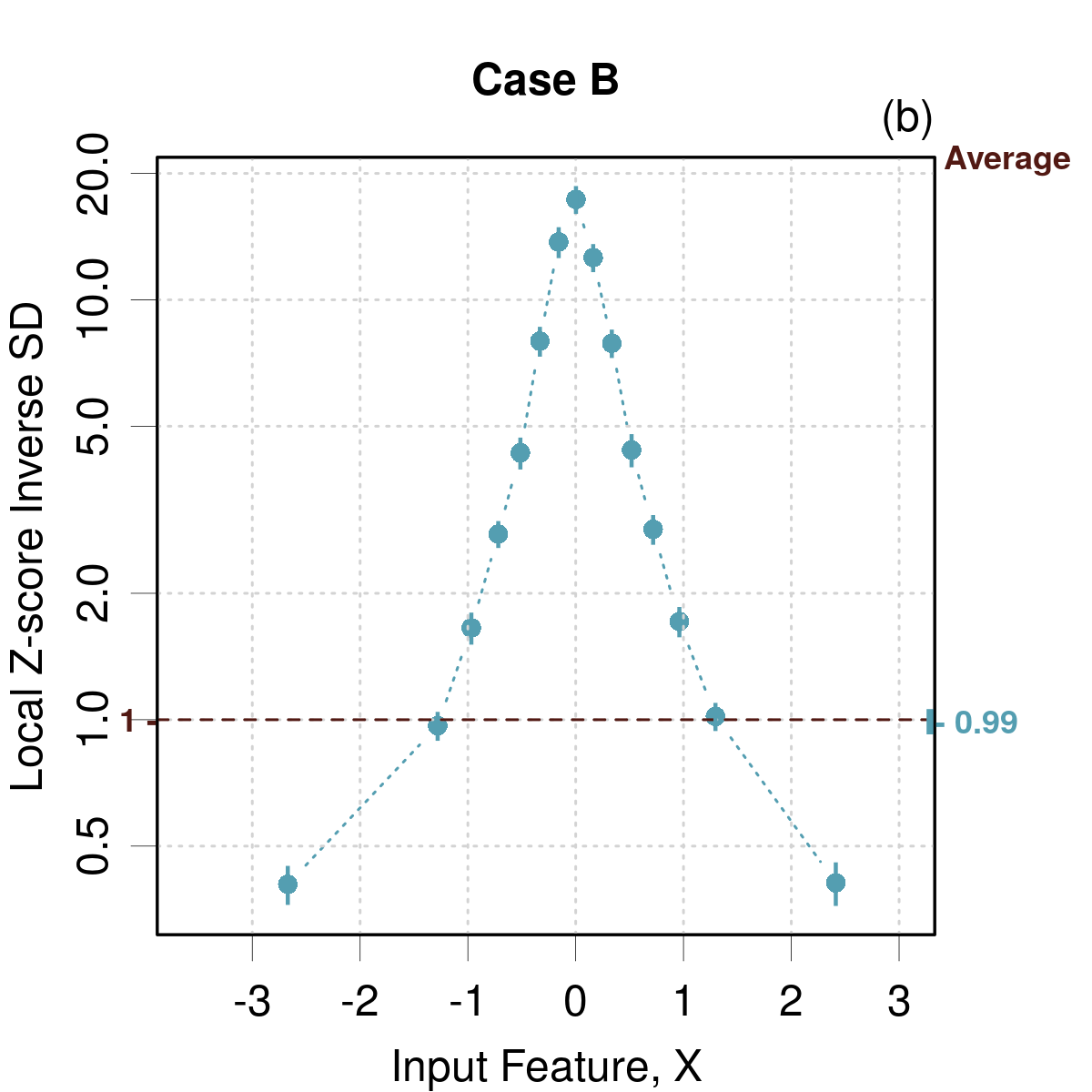} & \includegraphics[width=0.33\textwidth]{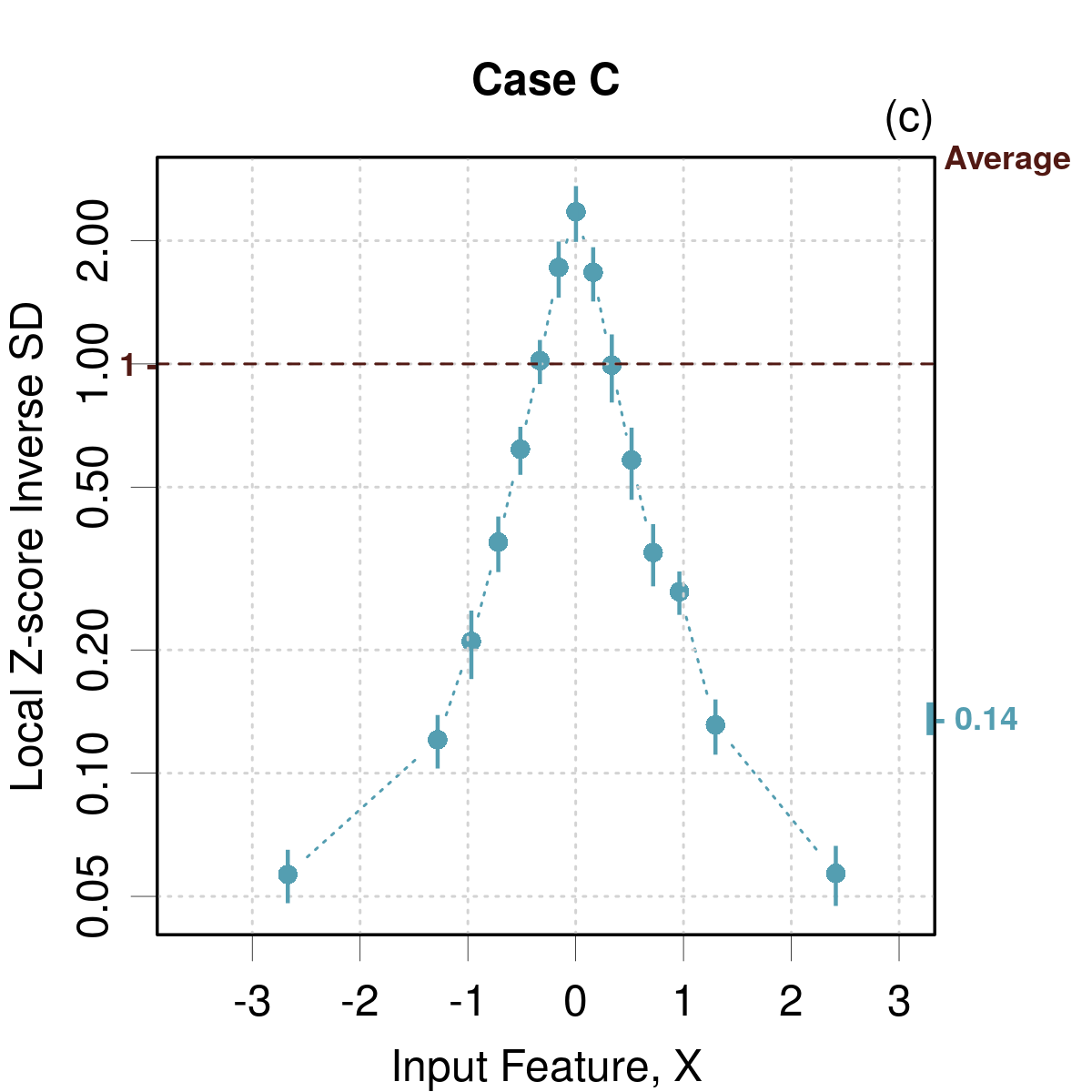}\tabularnewline
\includegraphics[width=0.33\textwidth]{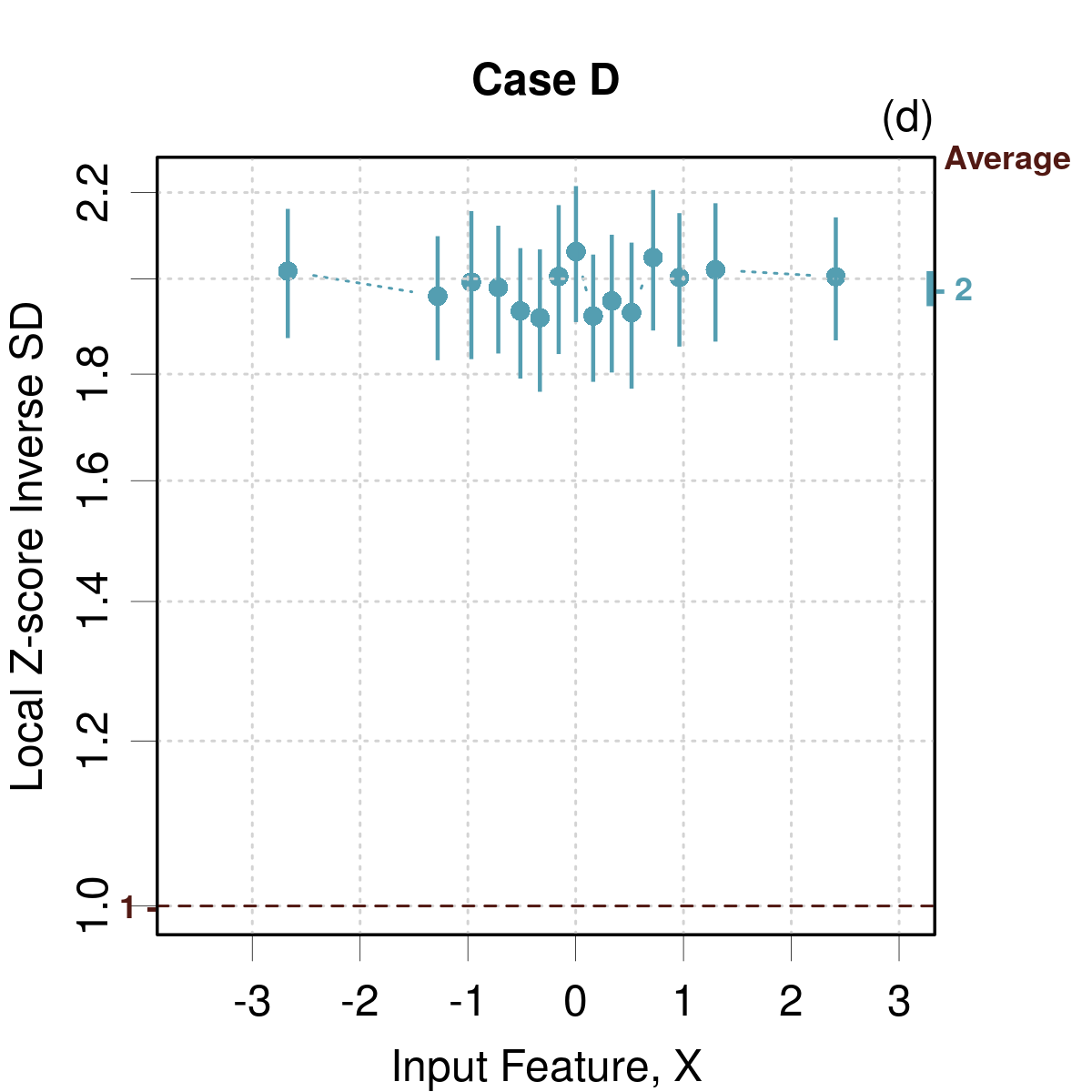} & \includegraphics[width=0.33\textwidth]{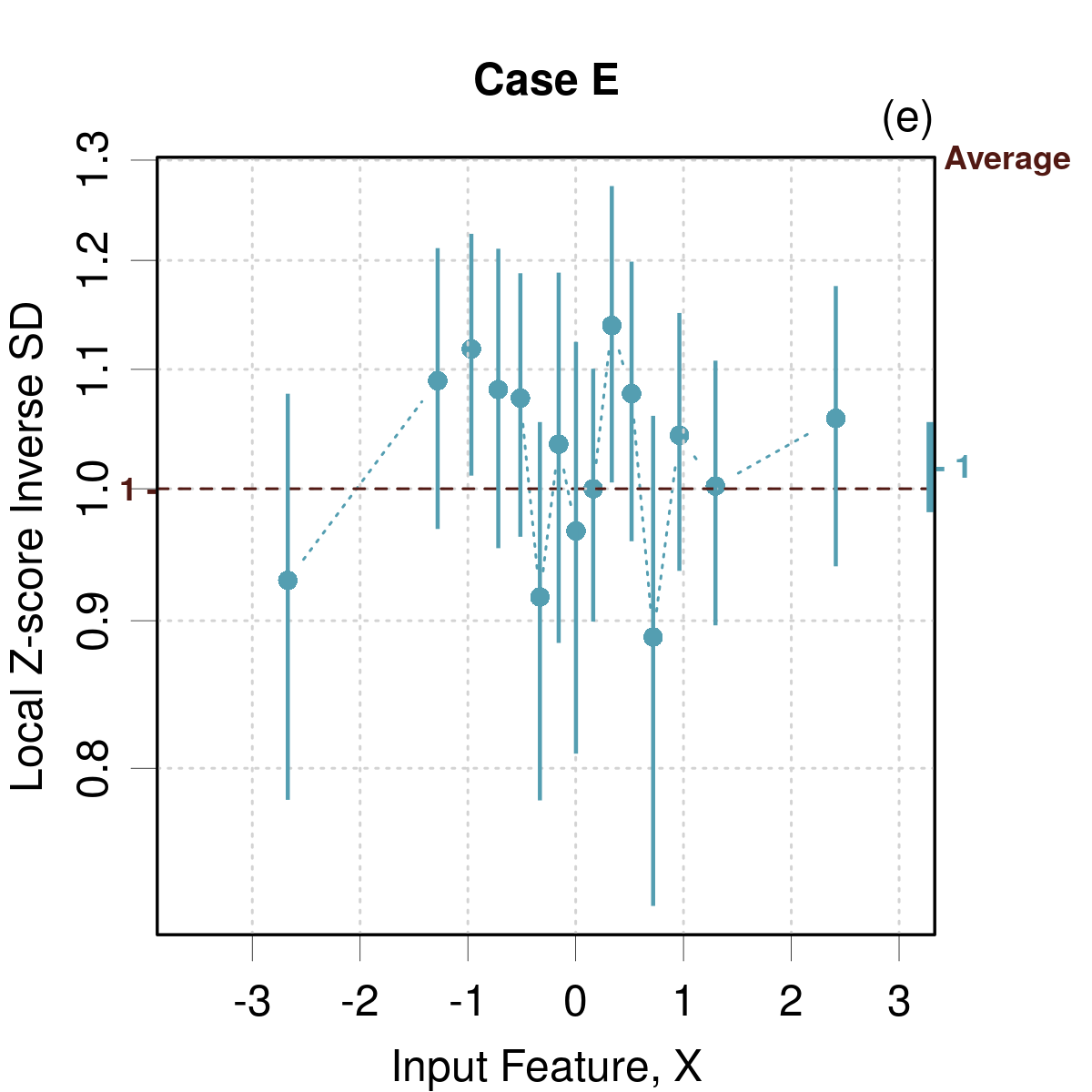} & \includegraphics[width=0.33\textwidth]{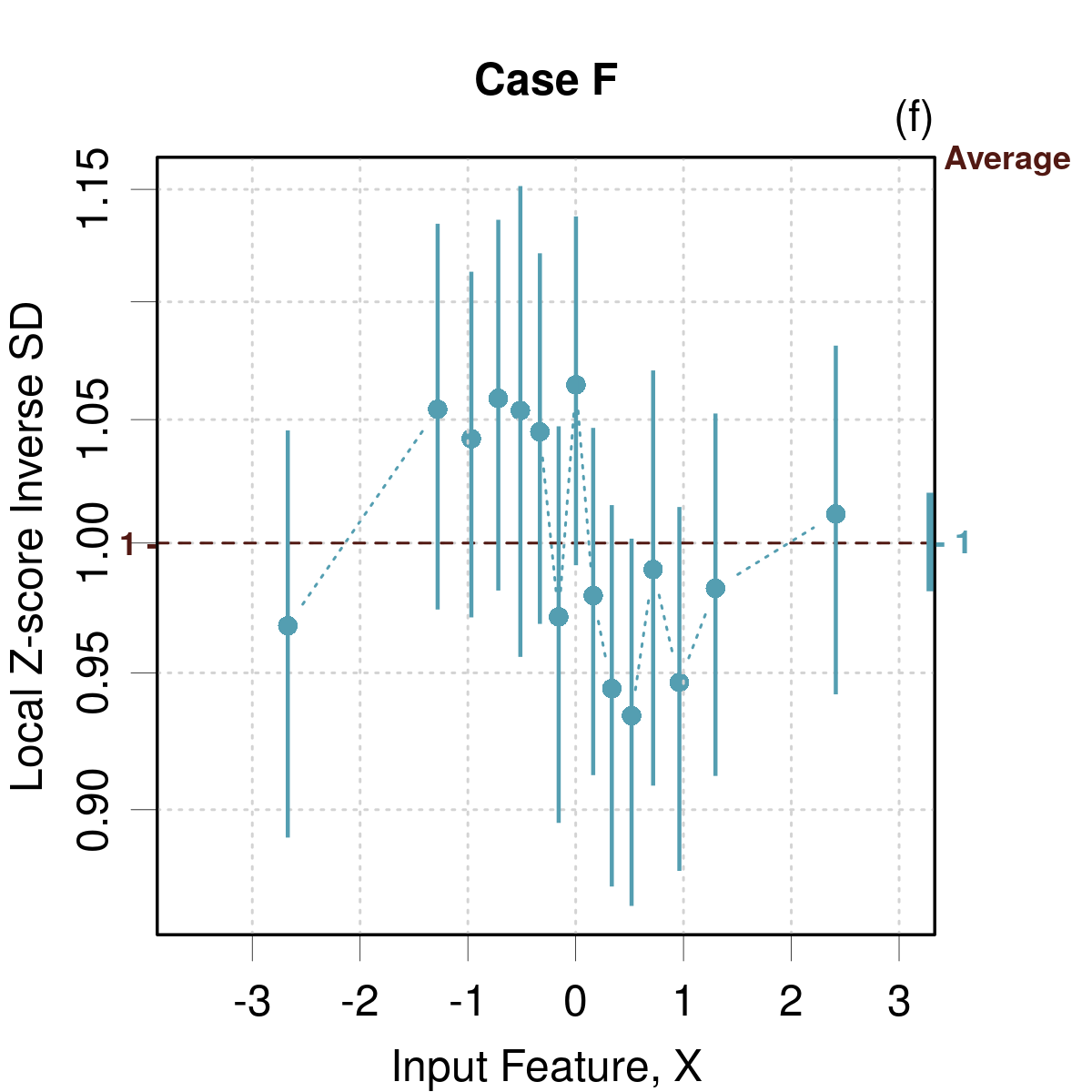}\tabularnewline
\end{tabular}
\par\end{centering}
\caption{\label{fig:LZISDV-1}LZISD analysis against $X$ for cases A-F. The
points are reported at the center of the bins. The error bars correspond
to 95\,\% confidence intervals on the statistic. The value and error
bar reported on the right axis corresponds to the statistic for the
full dataset (Average).}
\end{figure*}

The diagnostics provided by the LZISD analysis on the synthetic datasets
are non-ambiguous (Fig.\,\ref{fig:LZISDV-1}): Cases B, C and D stand
out as having strong deviations from the reference line, while Cases
A, E and F present a good adaptivity, as observed on the ``$Z$ vs
$X$'' plots. 

\subsubsection{Local Coverage Probability analysis in input feature space}

\begin{figure*}[t]
\noindent \begin{centering}
\begin{tabular}{ccc}
\includegraphics[width=0.33\textwidth]{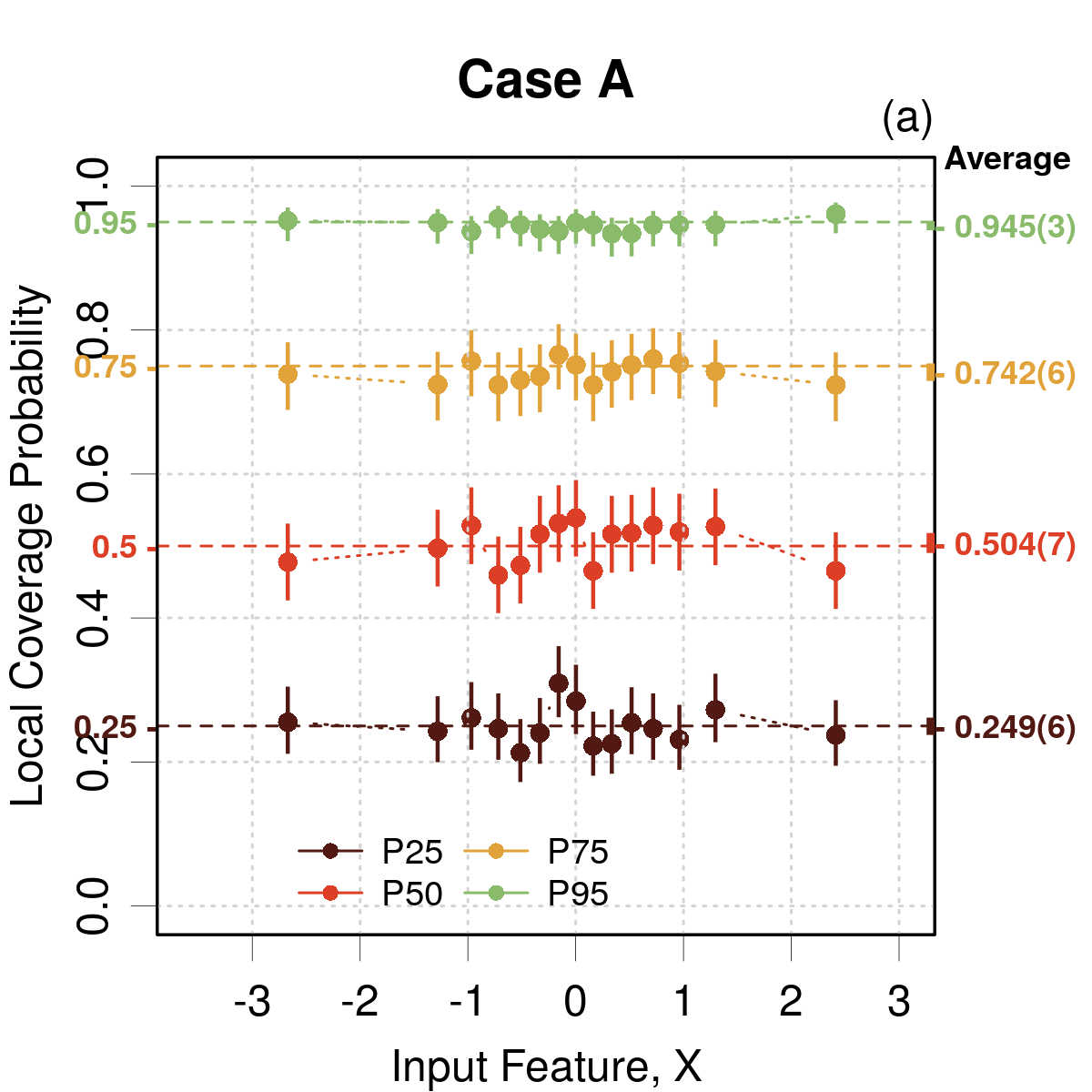} & \includegraphics[width=0.33\textwidth]{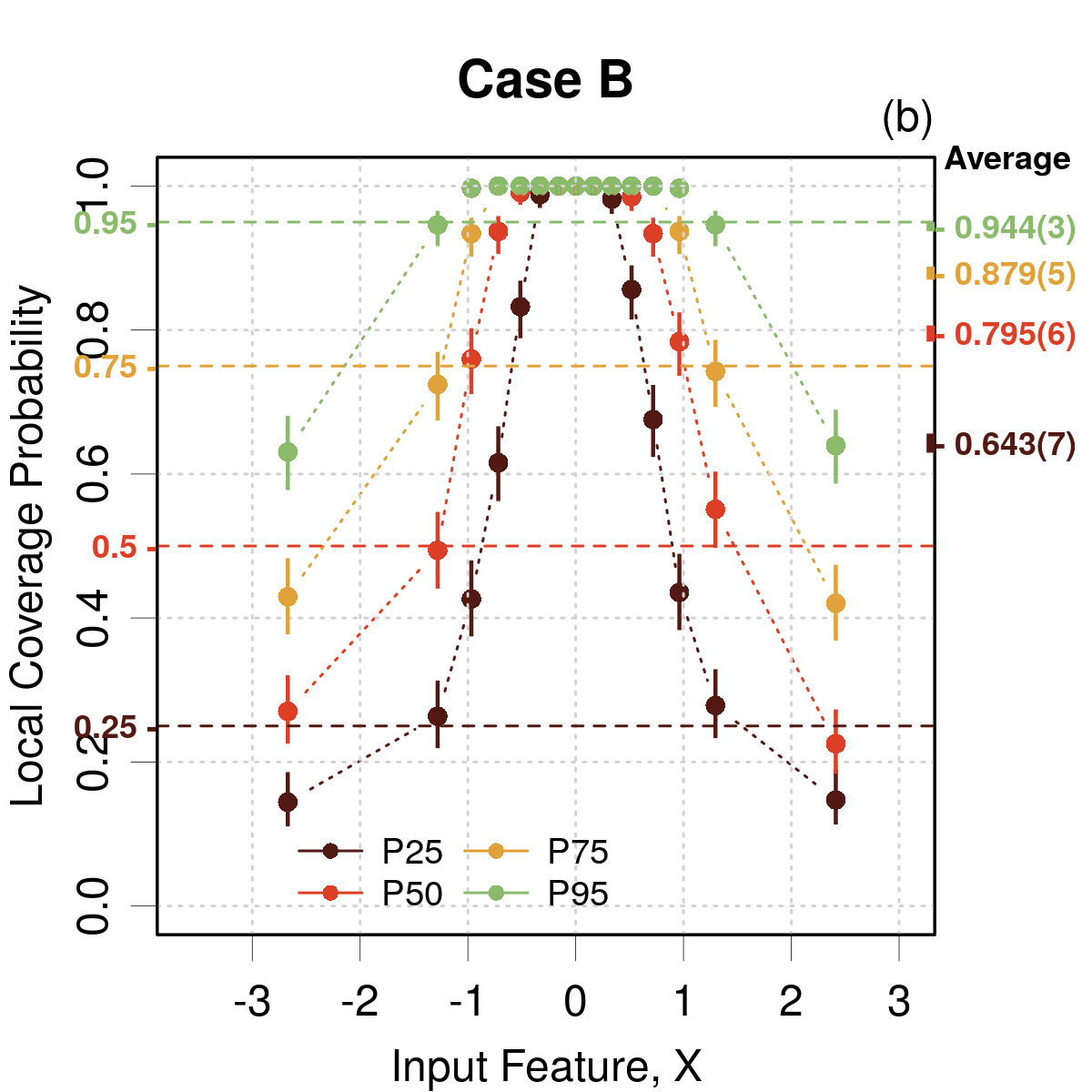} & \includegraphics[width=0.33\textwidth]{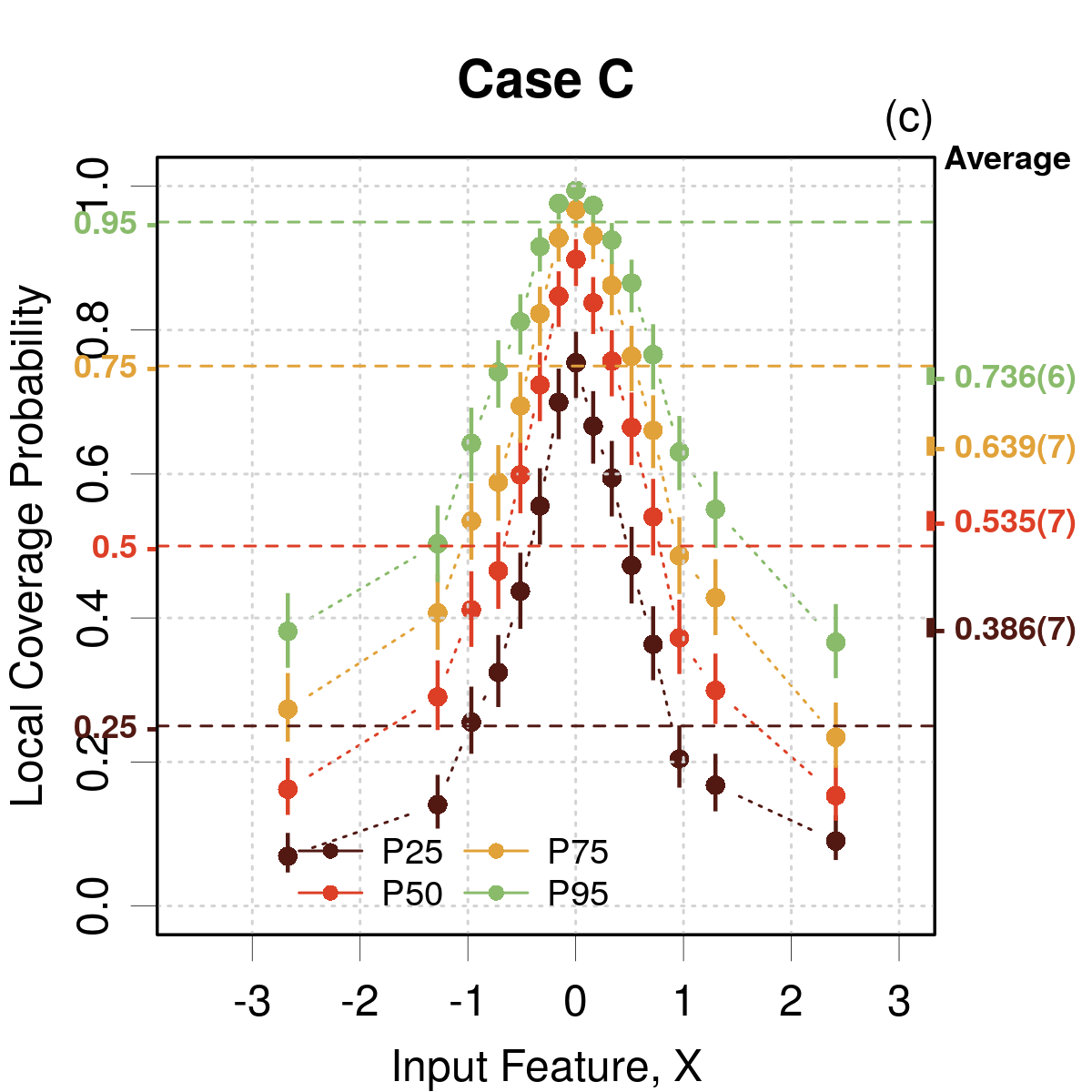}\tabularnewline
\includegraphics[width=0.33\textwidth]{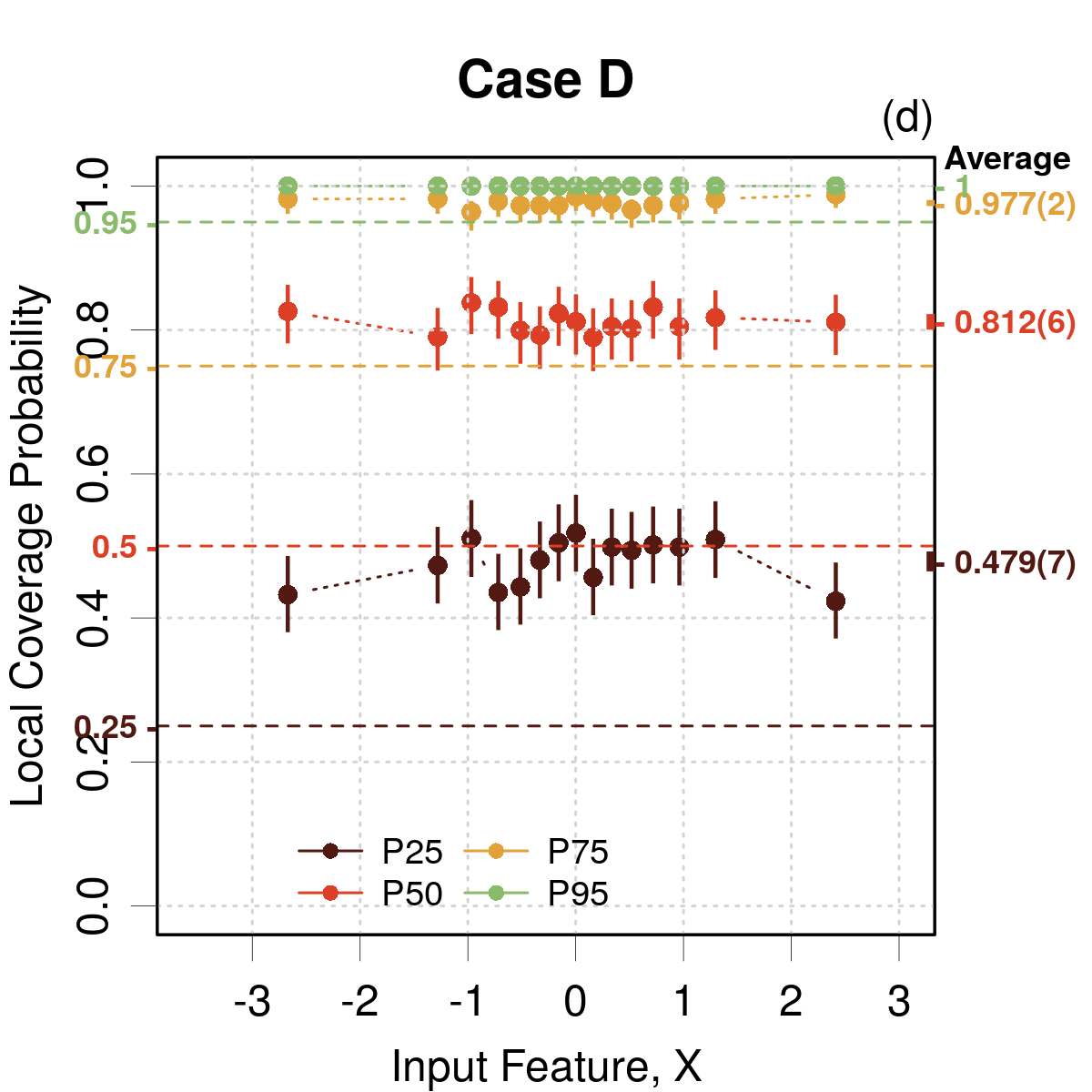} & \includegraphics[width=0.33\textwidth]{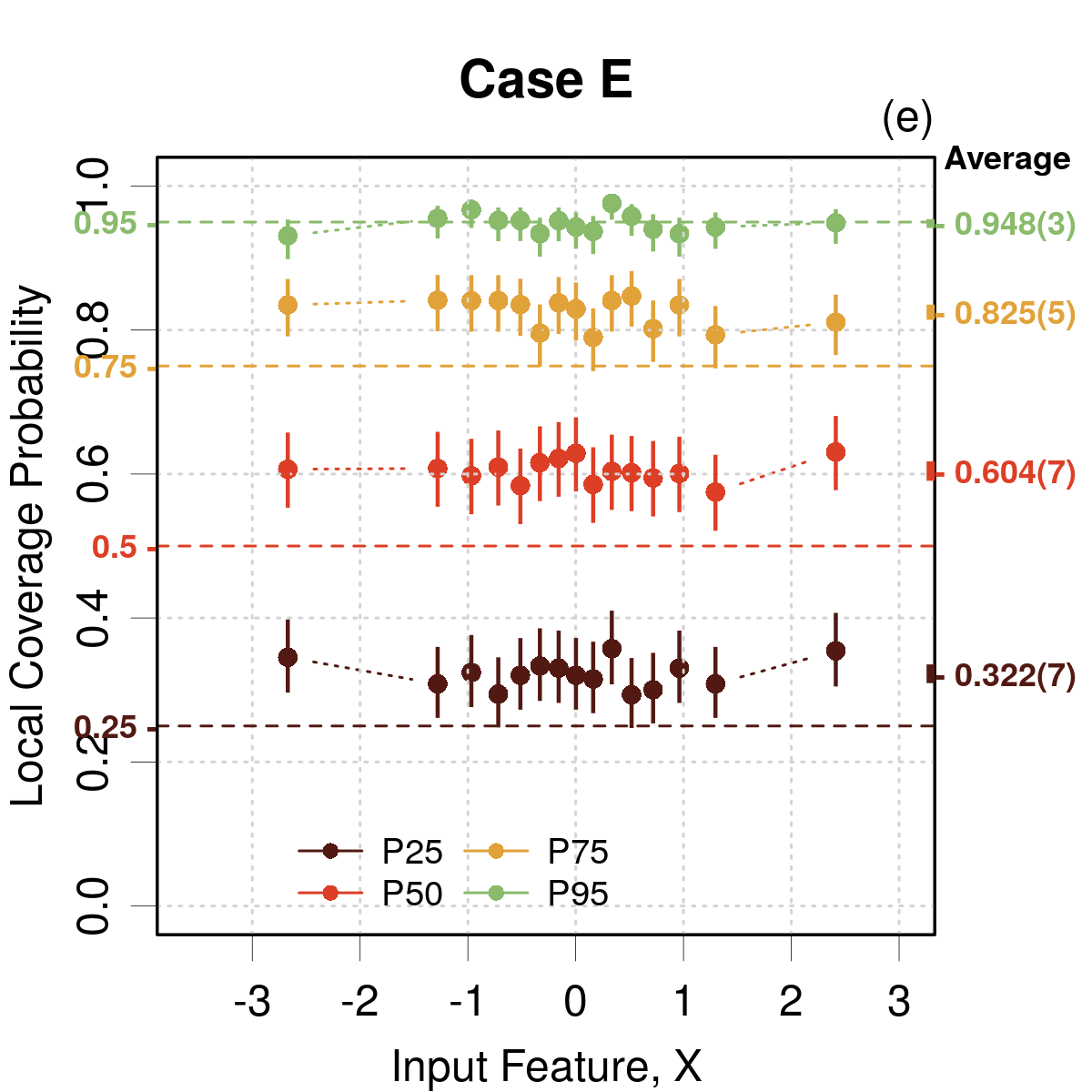} & \includegraphics[width=0.33\textwidth]{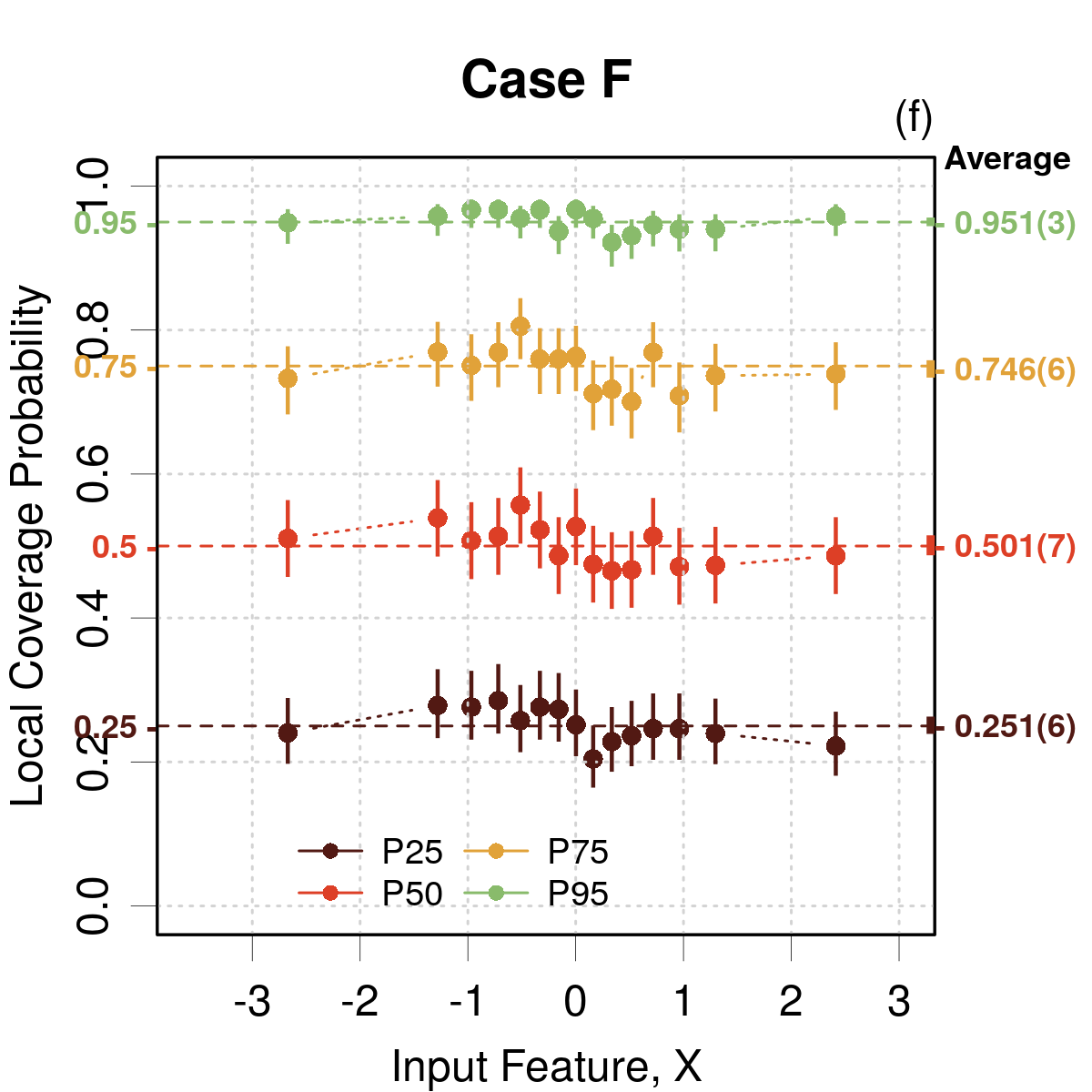}\tabularnewline
\end{tabular}
\par\end{centering}
\caption{\label{fig:LCPX}LCP analysis conditional to $X$ for cases A-F. The
points are reported at the center of the bins. The error bars correspond
to 95\,\% confidence intervals on the statistic. The value and error
bar reported on the right axis corresponds to the statistic for the
full dataset (Average).}
\end{figure*}

\noindent Using the same binning for $X$ as used in the other methods,
the LCP analysis (Fig.\,\ref{fig:LCPX}) provides results conform
with those of the conditional calibration curves (Fig.\,\ref{fig:Calibration-curves-1-2}).
Here also, the hypothesis of a normal generative distribution used
to build the probability intervals penalizes Case E, which can be
solved by an appropriate choice of distribution (not shown). 

\subsubsection{Using latent distances or $V$ as conditioning variables\label{subsec:Using-latent-distances}}

\noindent Although using one or several input features $X$ as conditioning
variable is the most direct way to test adaptivity, it might not always
be practical, for instance when input features are strings, graphs
or images. In such cases, one might use instead a latent variable
or the predicted property value $V$. 

$X$ and $V$ are interchangeable only if the latter is a monotonous
transformation of the former, but they might provide different adaptivity
diagnostics if this is not the case. Using $V$ answers to the question:
Are uncertainties reliable over the full range of predictions ? 

The adaptivity analysis of the synthetic datasets using $V$ as a
conditioning variable is presented in Appendix\,\ref{sec:Using-the-predicted}.
It shows that the non-monotonous shape of the $V=f(X)$ model might
induce artifacts in the adaptivity analysis that can be difficult
to interpret if the functional form of the model is unknown or complex.

\subsection{Validation metrics}

\noindent The present study focuses on graphical validation tools,
but validation metrics\citep{Maupin2018,Vishwakarma2021} are widely
used in the ML-UQ literature. For instance, metrics have been designed
for calibration curves\citep{Kuleshov2018,Tran2020}, reliability
diagrams\citep{Levi2020}, and confidence curves\citep{Scalia2020}.
These metrics are generally used to rank UQ methods, but they do not
provide a validation setup accounting for the statistical fluctuations
due to finite-sized datasets or bins. This is essential for small
validation datasets and is not without interest for the conditional
analysis of large ML datasets. 

Pernot\citep{Pernot2022a} introduced confidence bands for calibration
curves that enables to identify significant deviations from the identity
line, and also error bars for reliability diagrams, LZV and LCP analyses.
This provides a basis to define upper limits for calibration metrics
that can be used for validation purpose. In the present state of affairs,
calibration metrics such as the \emph{expected normalized calibration
error} (ENCE)\citep{Levi2020,Scalia2020,Vazquez-Salazar2022} or the\emph{
area under the confidence-oracle} (AUCO)\citep{Scalia2020} cannot
be used to validate consistency nor adaptivity and deserve further
consideration beyond the purpose of this article. 

\section{Applications\label{sec:Application}}

\noindent The validation methods introduced in the previous sections
are now applied to ``real life'' data coming from the materials
science and physico-chemical ML-UQ literature. The choice of datasets
enables to explore various aspects of \emph{a posteriori} validation.
It is not my intent in these reanalyses to criticize the analyzes
nor the UQ methods presented in the original studies, but simply to
show how the augmented set of validation methods proposed above might
facilitate and complete the calibration diagnostics.

\subsection{Case PAL2022\label{subsec:Case-PAL2022}}

\noindent The data have been gathered from the supplementary information
of a recent article by Palmer \emph{et al}.\citep{Palmer2022}. I
retained 8 sets of errors and uncertainties before and after calibration
by a bootstrap method, resulting from the combination of two materials
datasets (Diffusion, $M=2040$ and Perovskites, $M=3836$) and several
ML methods (RF, LR, GPR....). The datasets are tagged by the combination
of both elements: for instance, Diffusion\_RF is the dataset resulting
from the use of the Random Forest method on the Diffusion dataset.
The reader is referred to the original article for more details on
the methods and datasets. 

As only the errors $E$ and uncertainties $u_{E}$ are available,
it is not possible to test adaptivity for this dataset. My aim here
is mainly to compare the performance of reliability diagrams and LZISD
analysis against the binning strategy, and to compare their results
to confidence curves with the probabilistic reference. 

\subsubsection{Average calibration}

\begin{figure*}[t]
\noindent \begin{centering}
\includegraphics[width=0.45\textwidth]{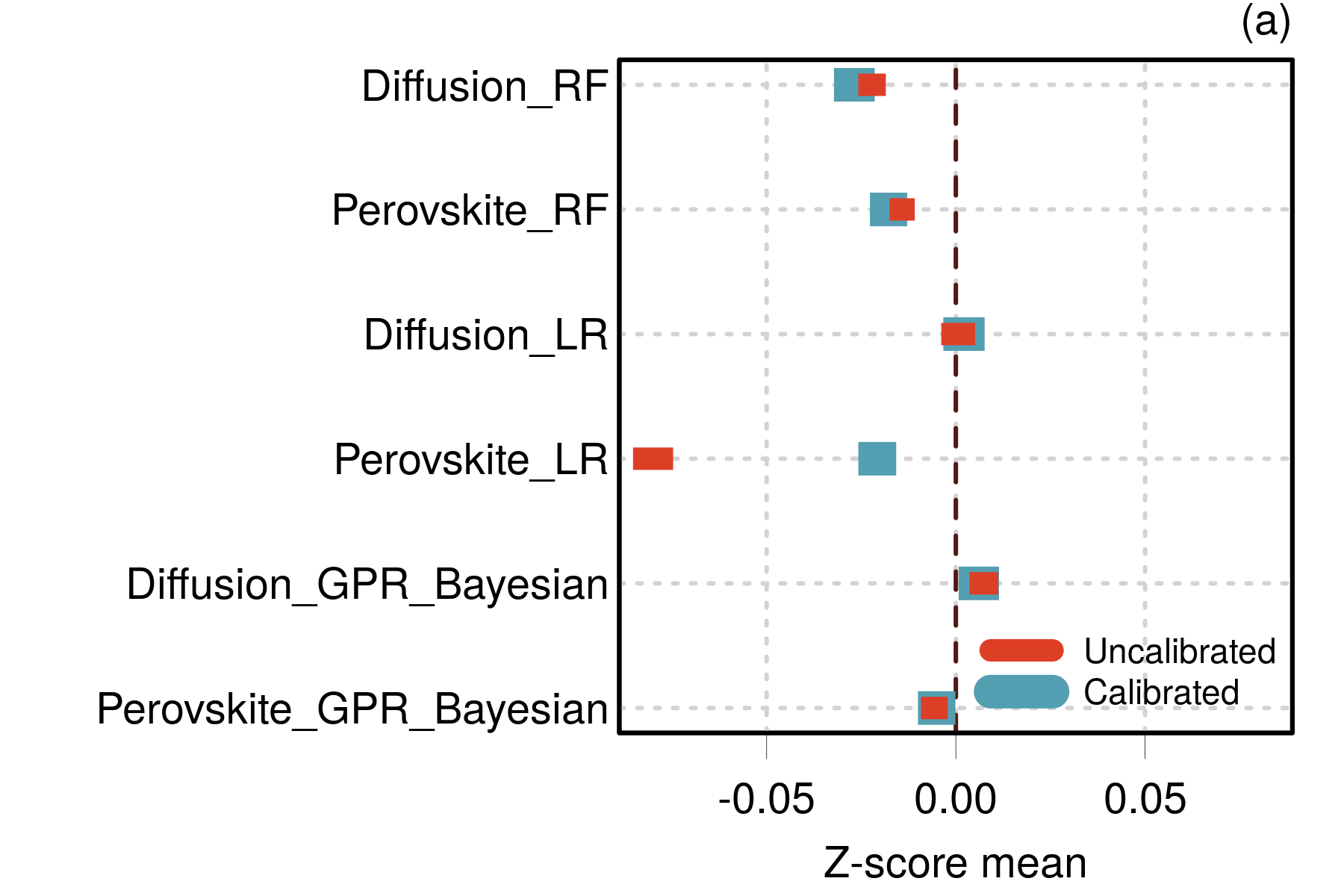}\includegraphics[width=0.45\textwidth]{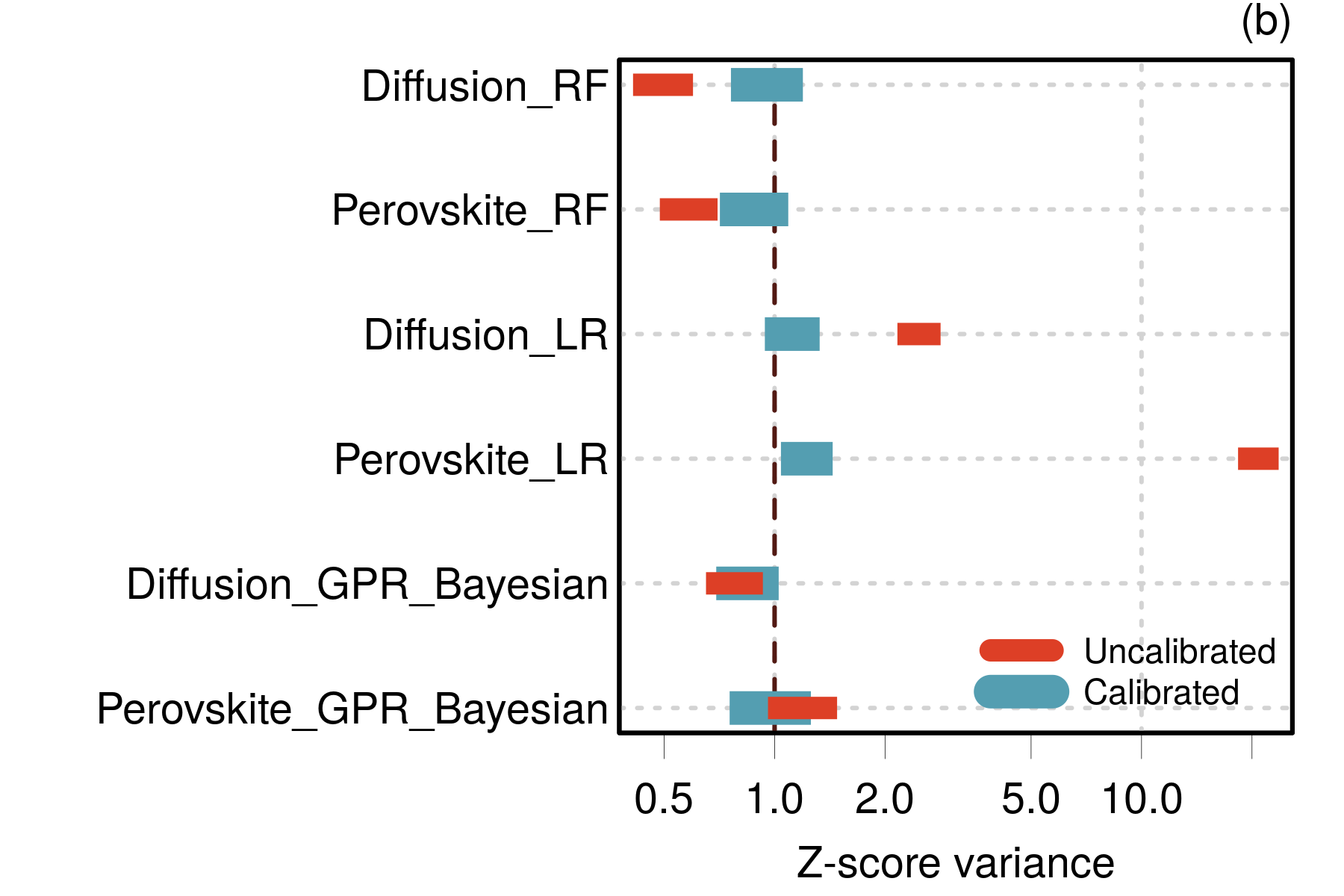}
\par\end{centering}
\caption{\label{fig:PAL2022-1}Case PAL2022: mean and variance of z-scores
for all datasets before and after calibration. The horizontal bars
represent 95\,\% confidence intervals.}
\end{figure*}

\noindent As a first step, I estimated for each dataset the mean of
\emph{z}-scores and their variance to appreciate the average impact
of the calibration method. Fig.\,\ref{fig:PAL2022-1} summarizes
the results. It can be seen that calibration leaves the mean z-score
unchanged or improves it (the calibrated value is closer to zero).
After calibration, the mean z-scores are not always null, but small
enough to consider the error sets to be unbiased. The calibration
effect is more remarkable for $\mathrm{Var}(Z)$: all the 95\,\%
confidence intervals for $\mathrm{Var}(Z)$ after calibration overlap
the target value ($1.0$), except for Perovskite\_LR with a very small
residual gap. 

\subsubsection{Consistency}

\noindent Consistency can be checked by reliability diagrams, LZISD
analysis and confidence curves with probabilistic reference. In absence
of information on the generative distribution(s), it is best to avoid
conditional calibration curves in this example. Before analyzing the
datasets, it is important to point out the impact of the binning strategy
on the conclusions drawn from reliability diagrams and LZISD analysis.\textcolor{orange}{{}
}In their study, Palmer \emph{et al.} define 15 intervals forming
a regular grid in $u_{E}$ space, with some adaptations described
in Appendix\,\ref{sec:PAL2022---Additional}. This is different from
the usual strategy consisting in designing bins with identical counts.\citep{Scalia2020,Pernot2022a}
Besides, considering the size of the datasets, 15 bins might not be
enough to reveal local consistency anomalies. The impact of the binning
strategy is explored in Appendix\,\ref{sec:PAL2022---Additional},
where an adaptive binning strategy is designed to combine both approaches.
When using the log-scale for uncertainty binning, this adaptive strategy
is efficient to reveal local consistency anomalies that do not always
appear in the original reliability diagrams.

\paragraph{Calibrated Random Forest.}

\begin{figure*}[t]
\noindent \begin{centering}
\begin{tabular}{ccc}
\includegraphics[width=0.33\textwidth]{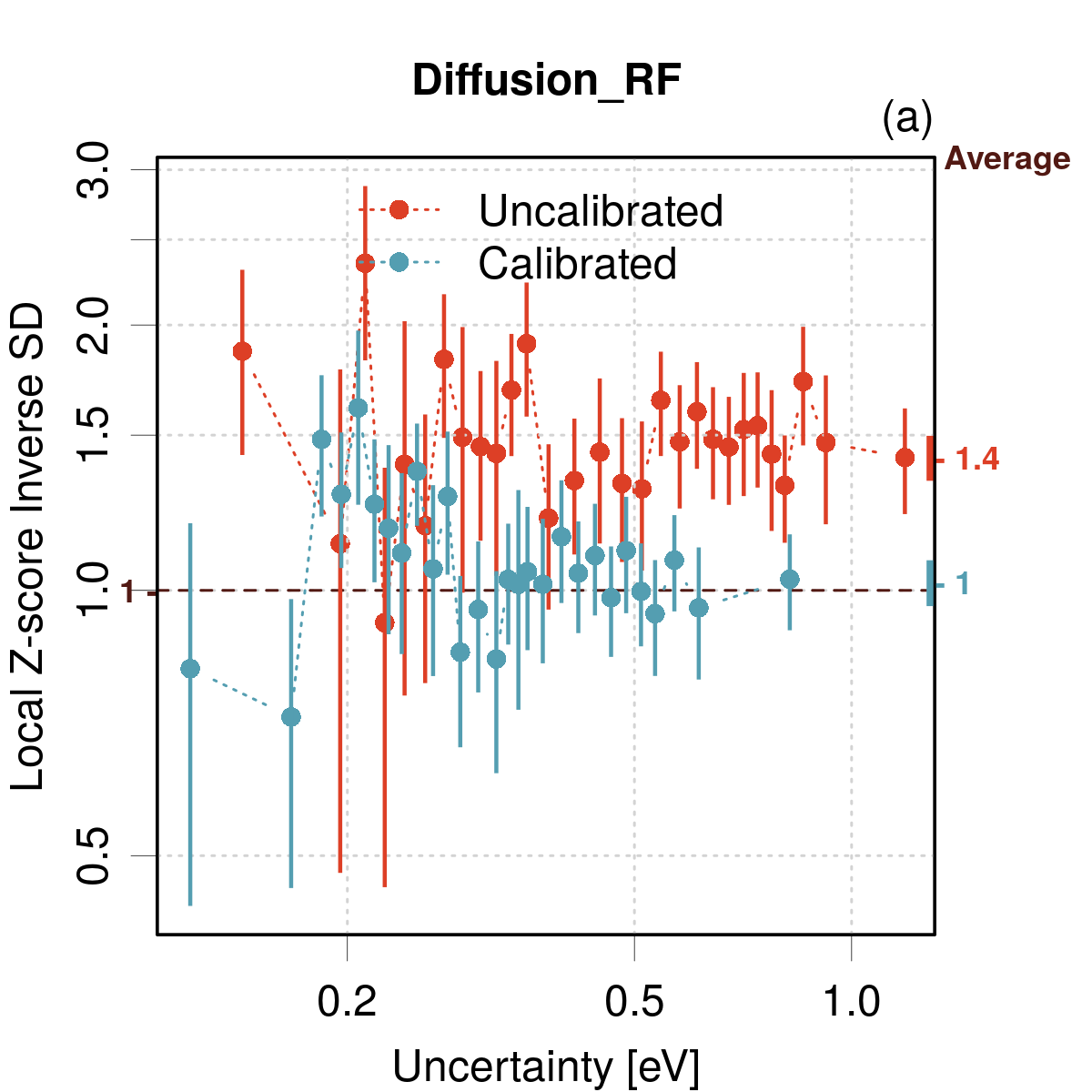} & \includegraphics[width=0.33\textwidth]{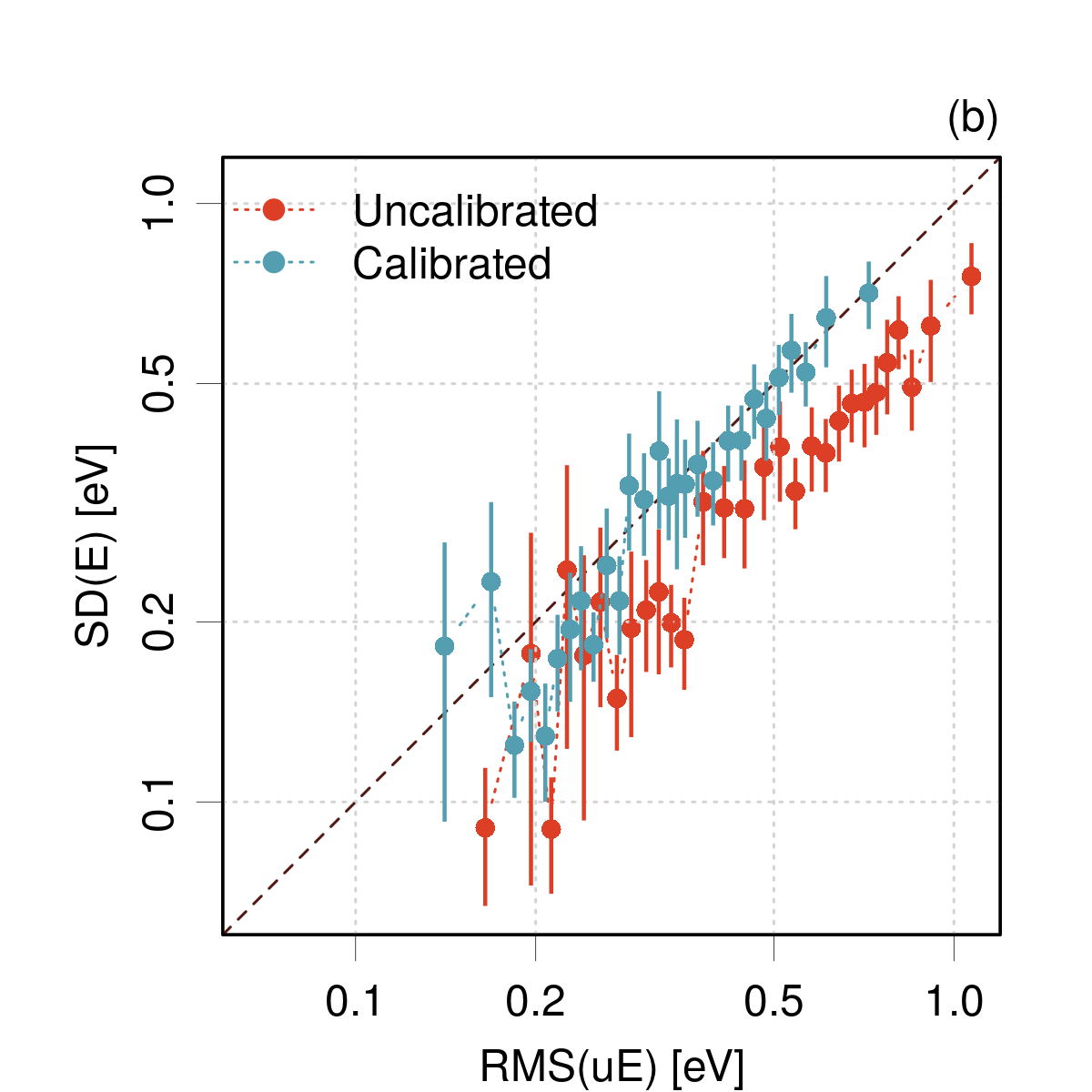} & \includegraphics[width=0.33\textwidth]{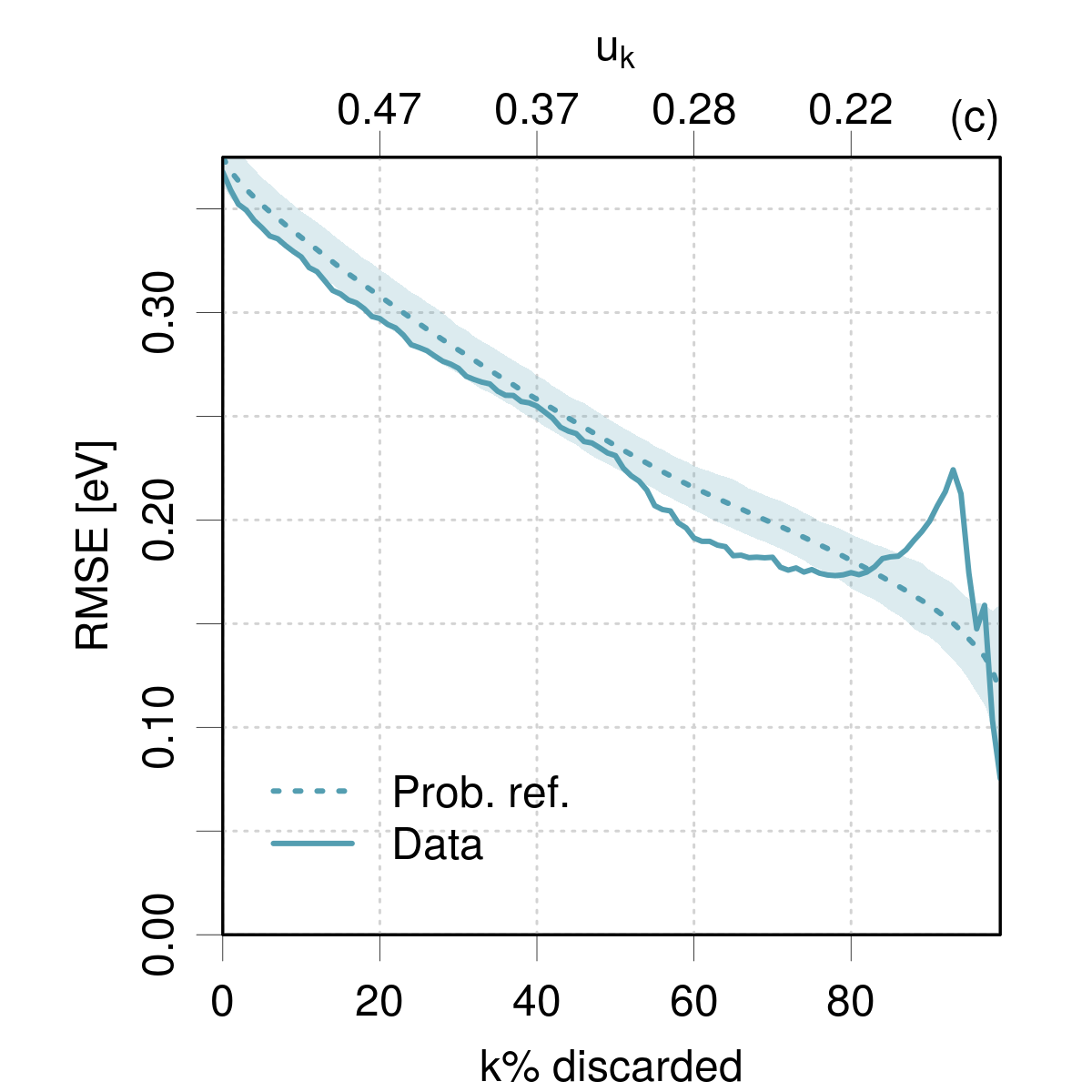}\tabularnewline
\includegraphics[width=0.33\textwidth]{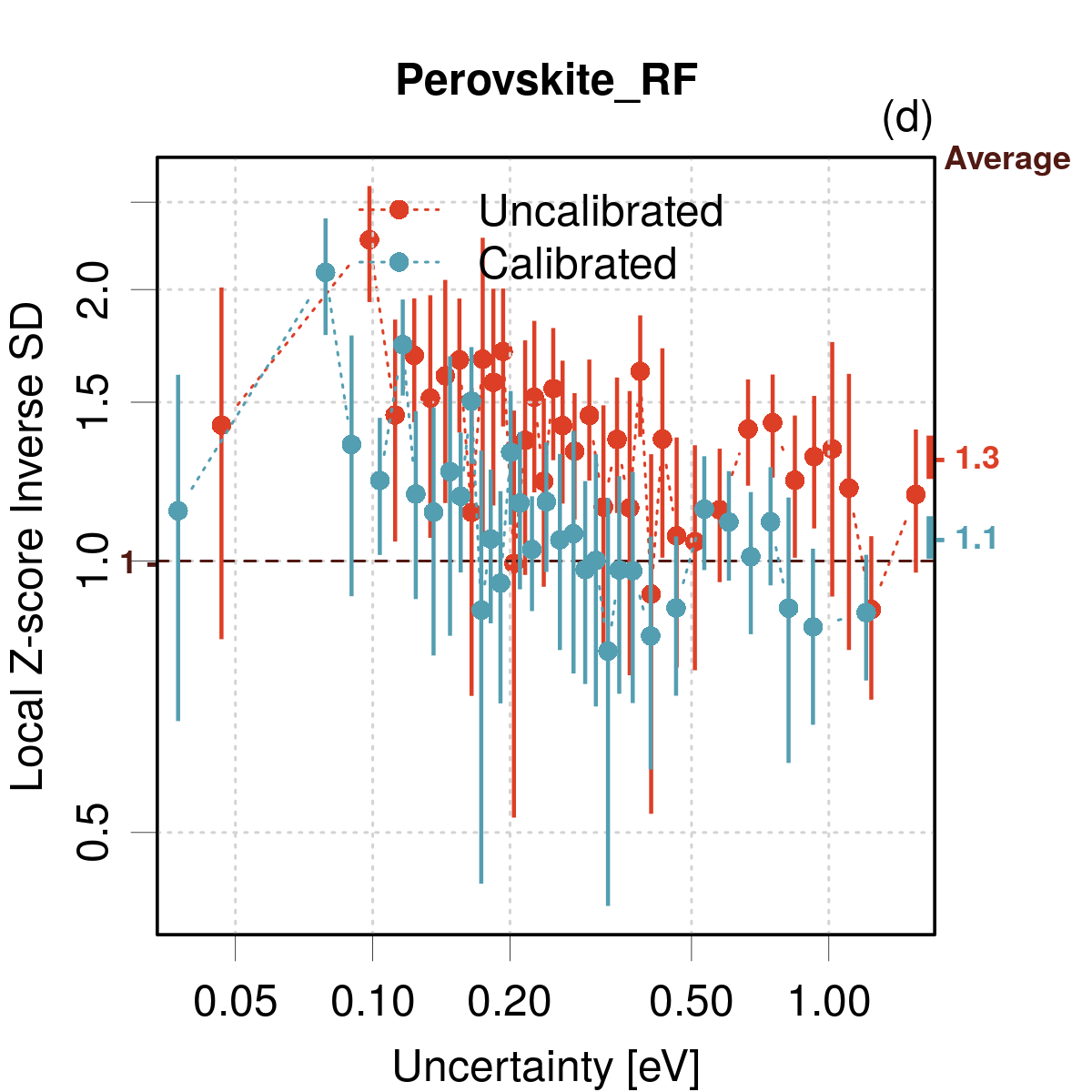} & \includegraphics[width=0.33\textwidth]{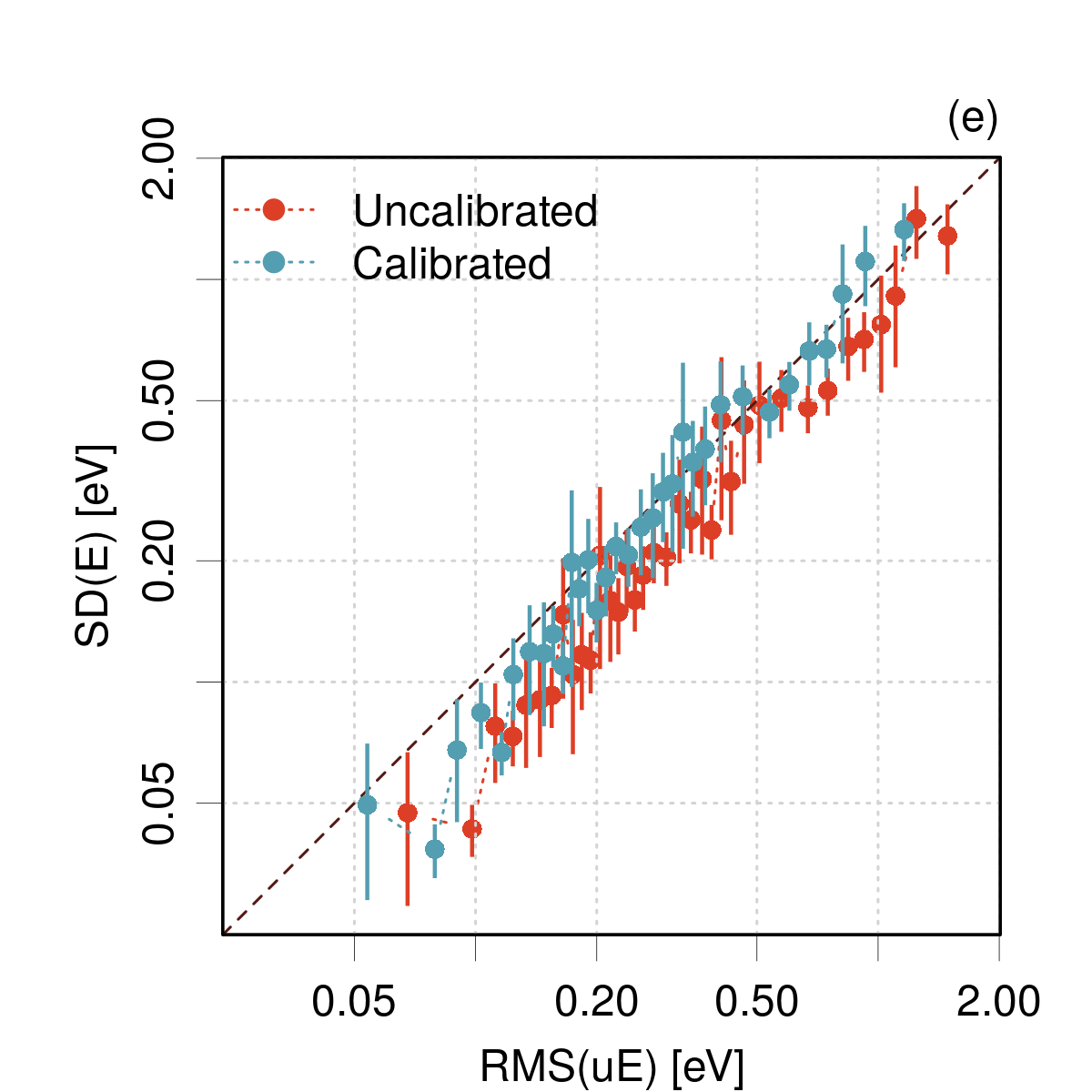} & \includegraphics[width=0.33\textwidth]{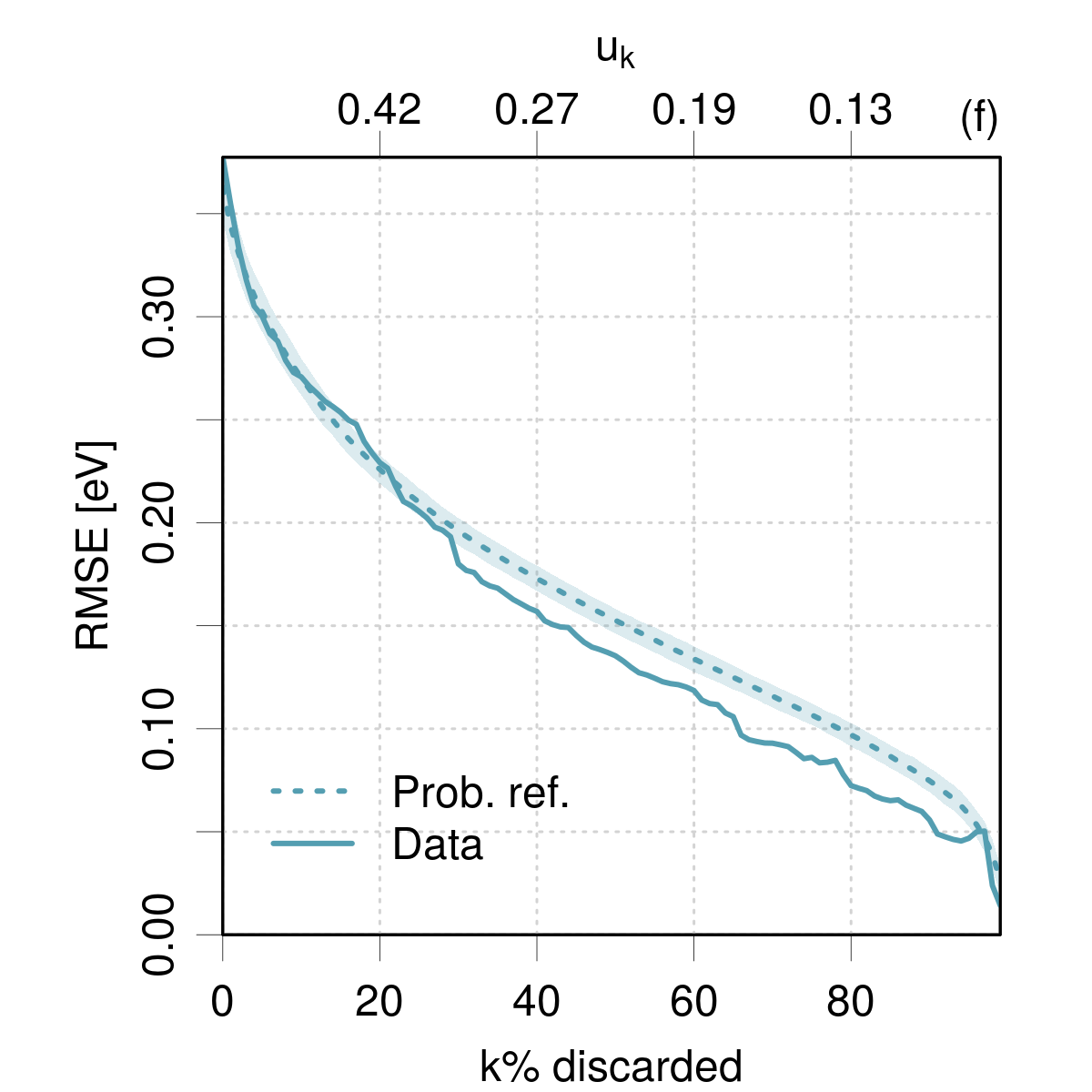}\tabularnewline
\end{tabular}
\par\end{centering}
\caption{\label{fig:PAL2022_RF}Case PAL2022: UQ validation for random forests
examples.}
\end{figure*}

The LZISD analysis and reliability diagrams are shown in Fig.\,\ref{fig:PAL2022_RF}
for both datasets. The effect of calibration is well visible on those
graphs when comparing the red dots (uncalibrated) to the blue ones
(calibrated). For both datasets, consistency after calibration is
rather good, except for the small uncertainty values (below 0.25\,eV
for Diffusion and 0.15\,eV for Perovskite). The legibility is better
on the LZISD analysis, which provides directly an overestimation value
of 50\,\% of the uncertainties around 0.2\,eV for Diffusion and
up to 100\,\% around 0.08\,eV for Perovskite. 

The confidence curves in Fig.\,\ref{fig:PAL2022_RF} confirm this
diagnostic, as they lie close to the reference for large uncertainties
and start to deviate from it at smaller uncertainties. The non-linear
scale of threshold values $u_{k}$ at the top of the plot helps to
locate the anomalies in uncertainty space. 

\paragraph{Calibrated Linear Ridge regression.}

\begin{figure*}[t]
\noindent \begin{centering}
\begin{tabular}{ccc}
\includegraphics[width=0.33\textwidth]{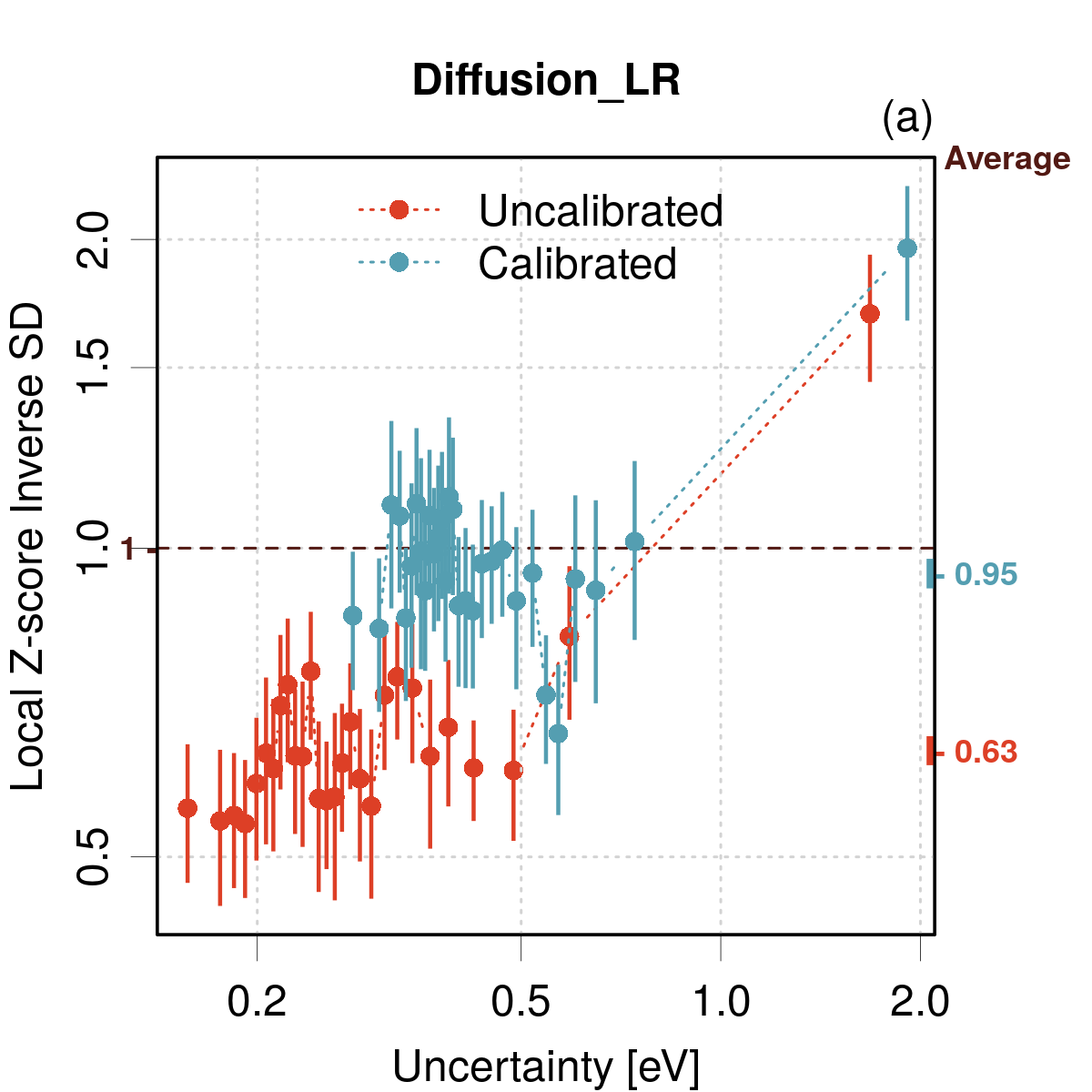} & \includegraphics[width=0.33\textwidth]{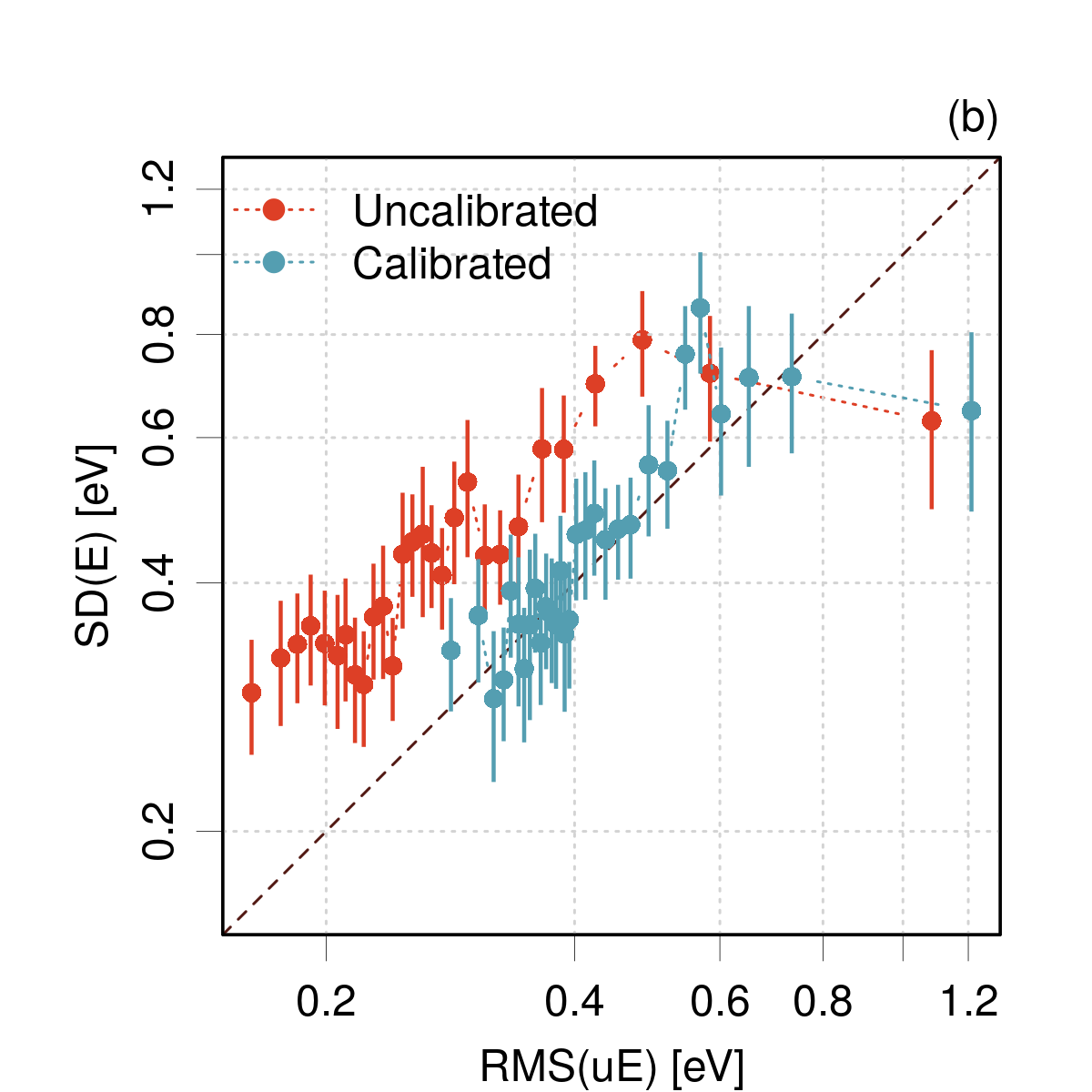} & \includegraphics[width=0.33\textwidth]{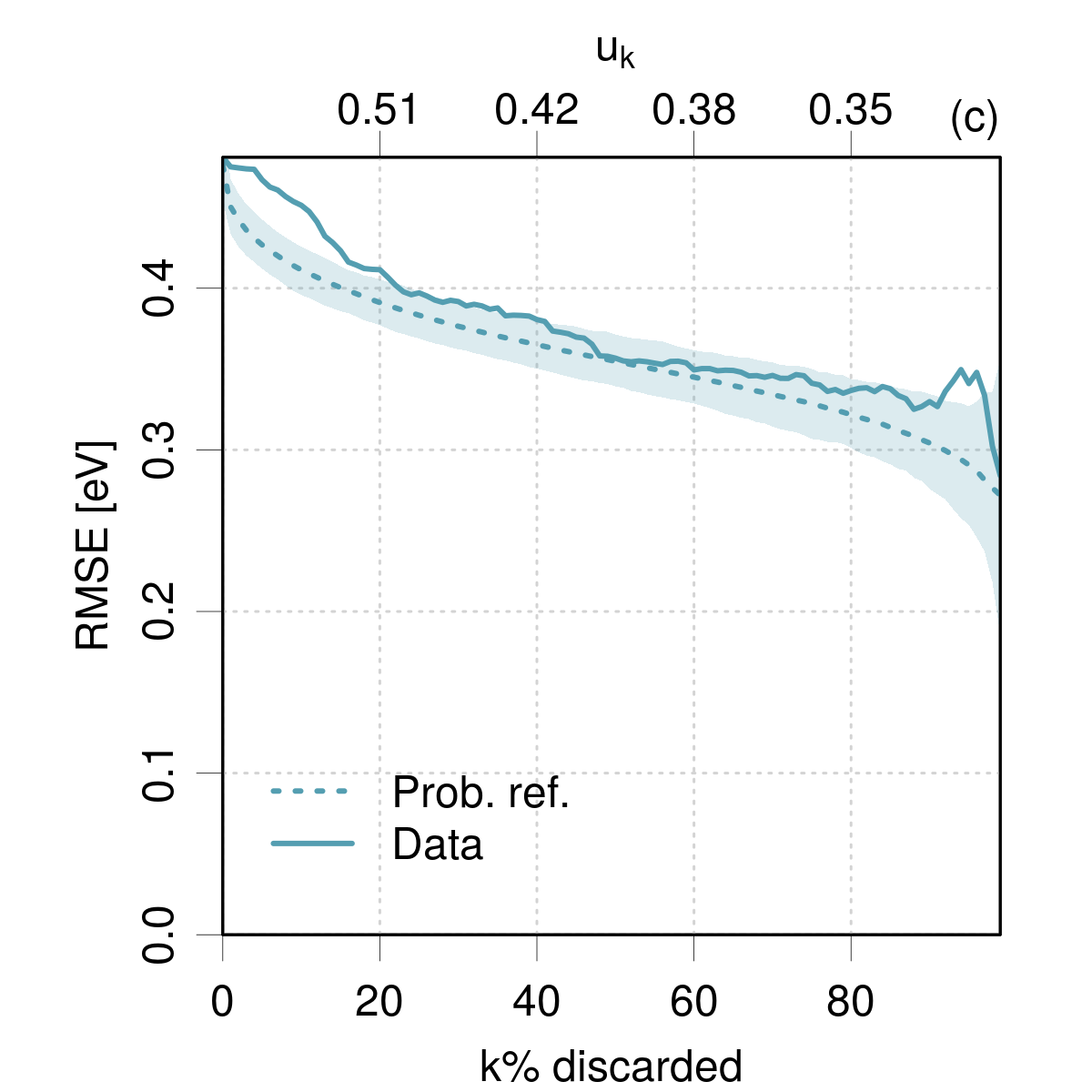}\tabularnewline
\includegraphics[width=0.33\textwidth]{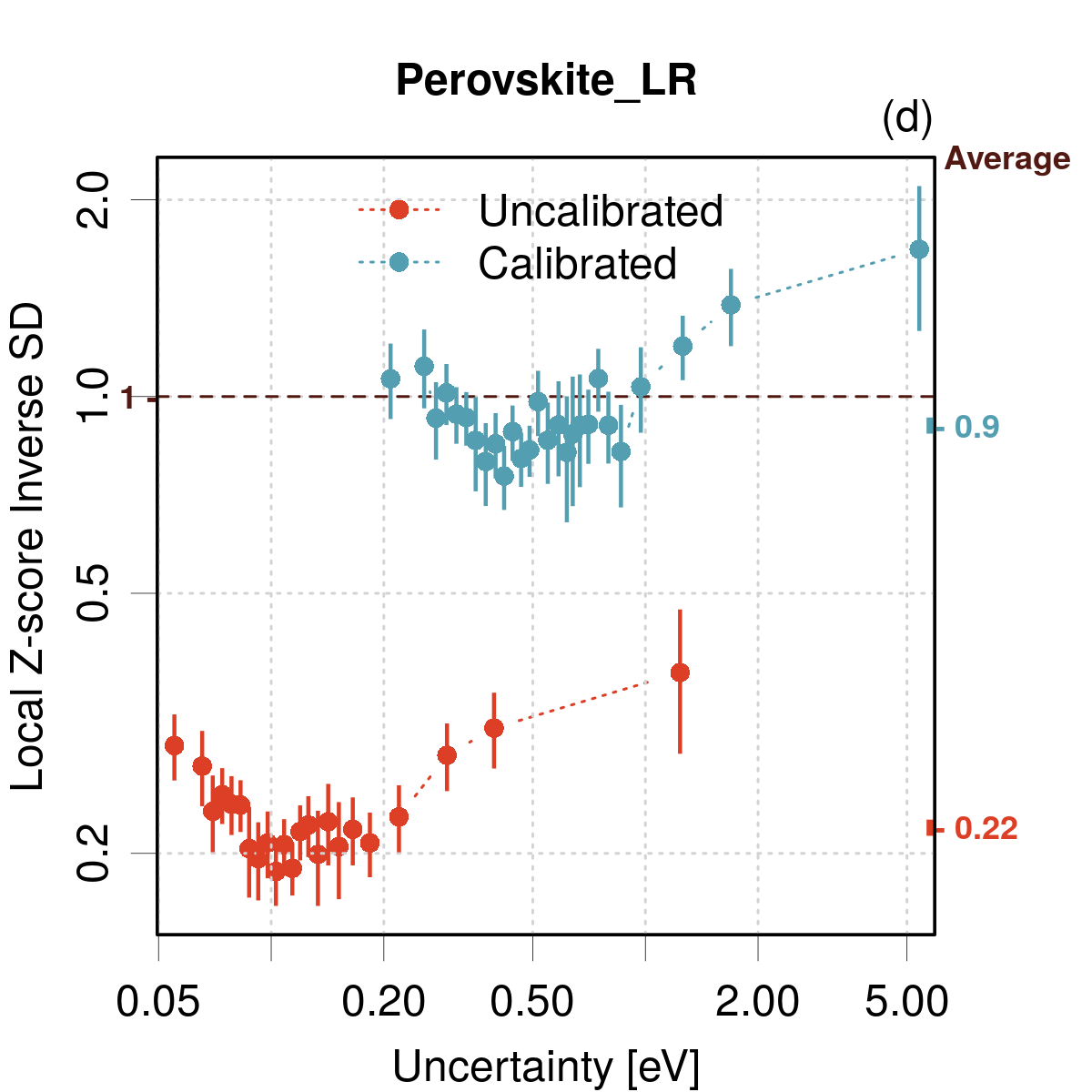} & \includegraphics[width=0.33\textwidth]{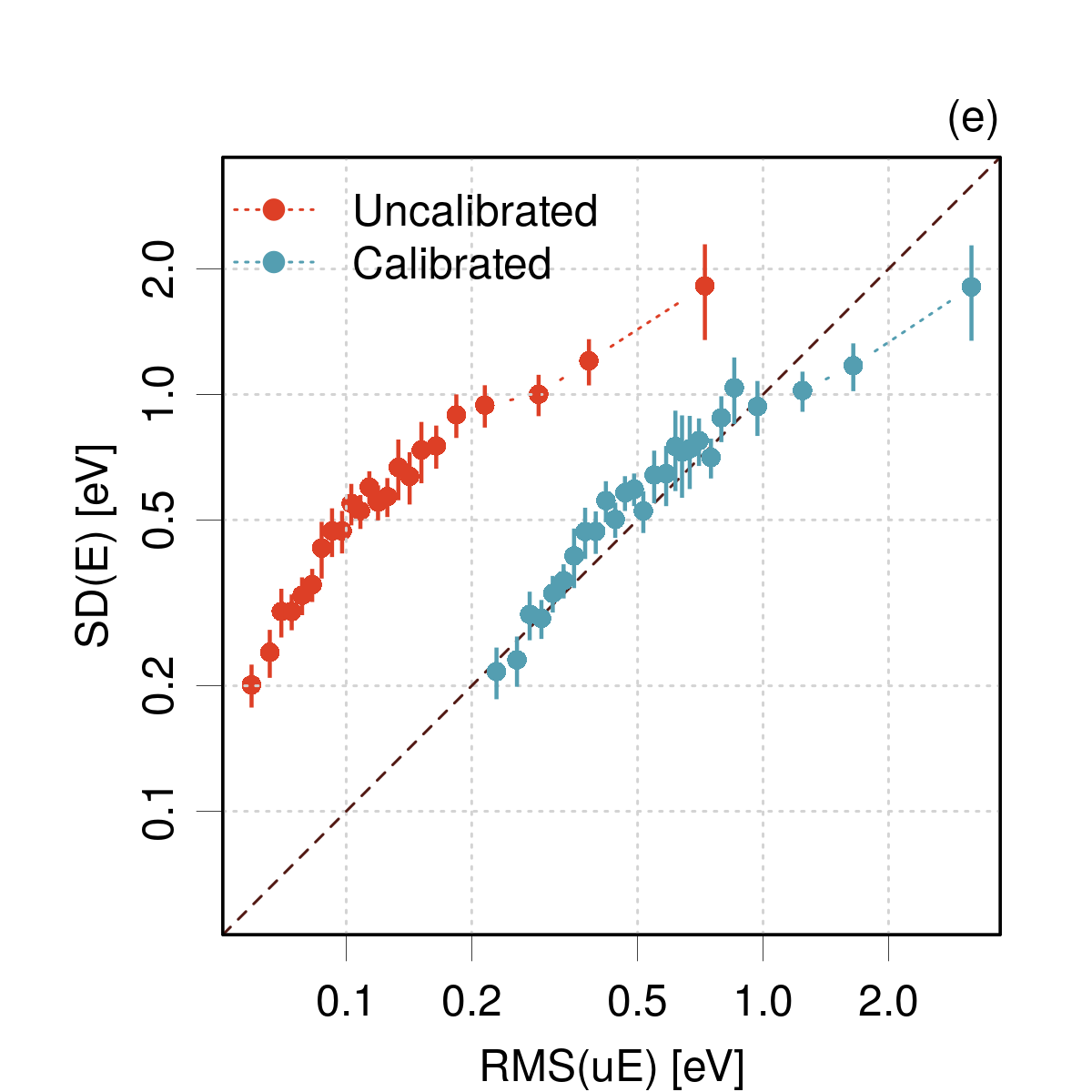} & \includegraphics[width=0.33\textwidth]{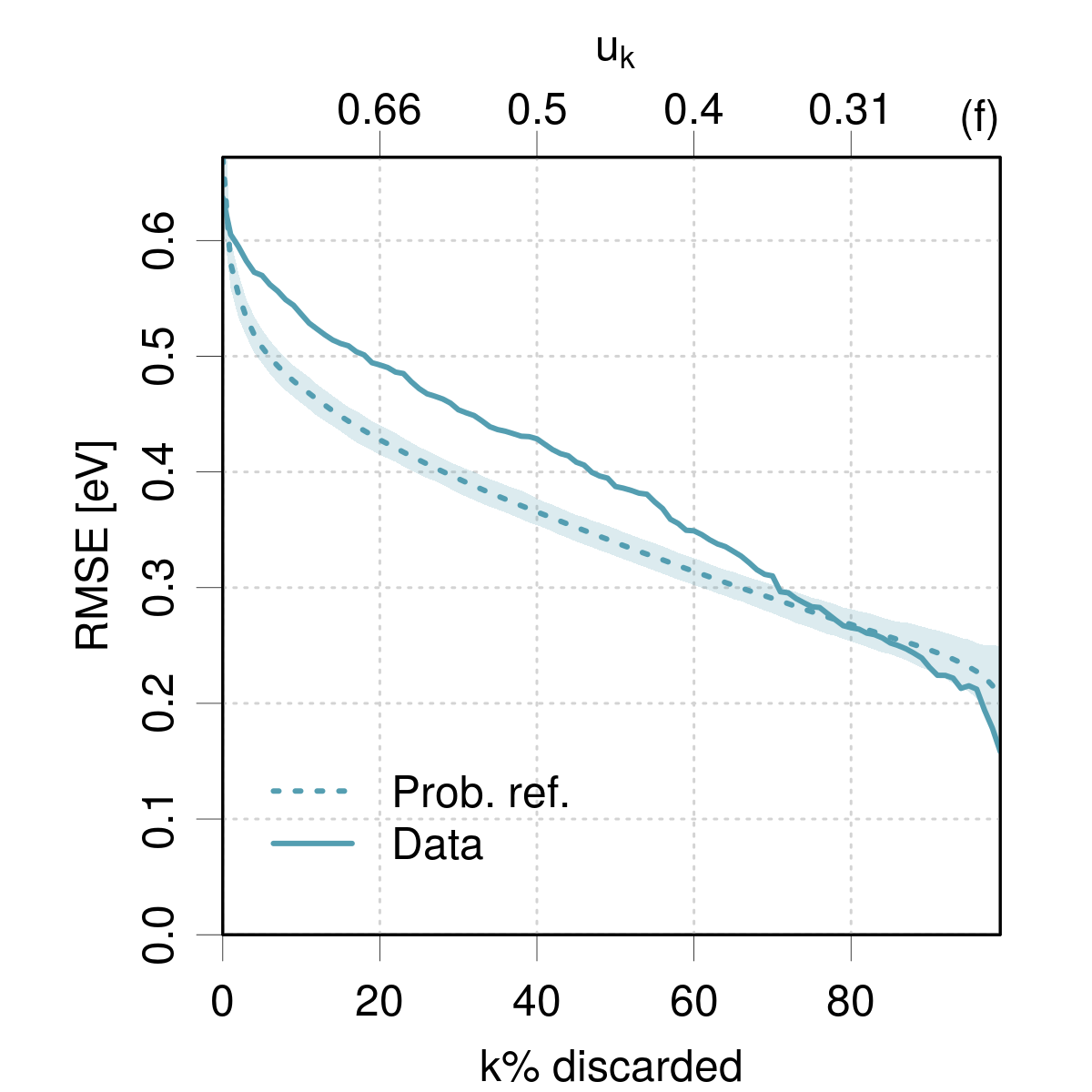}\tabularnewline
\end{tabular}
\par\end{centering}
\caption{\label{fig:PAL2022_LR}Case Pal2022: UQ validation for linear ridge
regression examples.}
\end{figure*}

Calibration and consistency of linear ridge regression results are
assessed in Fig.\,\ref{fig:PAL2022_LR}. The effect of calibration
is noticeable over the whole uncertainty range, except at large uncertainties
for the Diffusion dataset, where the situation is worsened by calibration.
In this area, the uncertainties are overestimated by a factor about
2. At smaller uncertainties, the consistency is far from perfect,
with areas of underestimated uncertainties around 0.3\,eV and 0.5\,eV,
compensating for the overestimated values at large uncertainty. The
confidence curve displays the problem at large uncertainty values,
but is less legible at smaller uncertainties. 

For the Perovskite dataset, consistency is not reached either, with
a compensation between overestimated and underestimated uncertainty
areas. The confidence curve is notably deviating from the reference
curve.

\paragraph{Calibrated Gaussian Process Regression.}

One considers here the calibration results of a Gaussian Process regression
with its Bayesian UQ estimate. It is possible to go quickly to the
conclusion that consistency is not achieved. 

\begin{figure*}[t]
\noindent \begin{centering}
\begin{tabular}{ccc}
\includegraphics[width=0.33\textwidth]{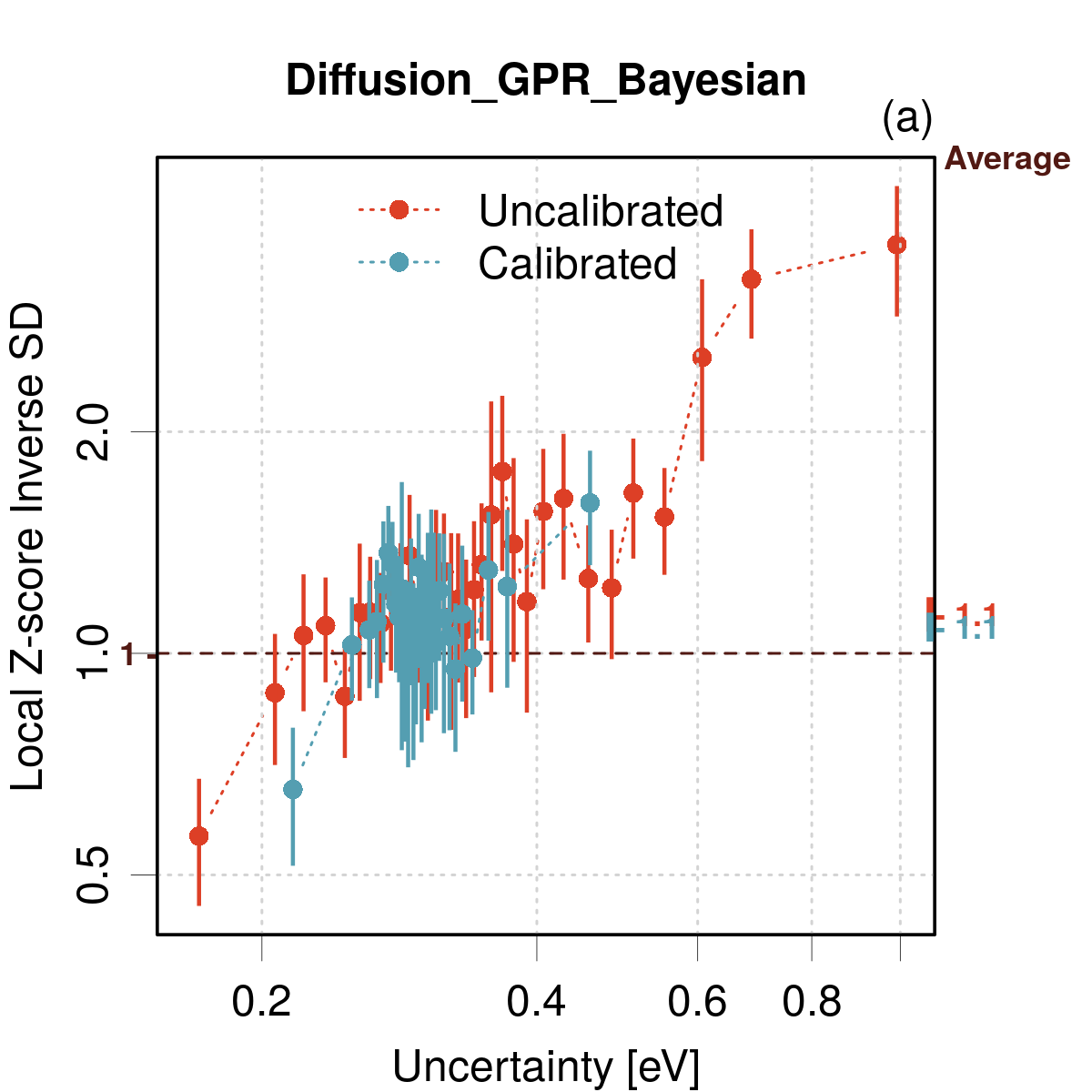} & \includegraphics[width=0.33\textwidth]{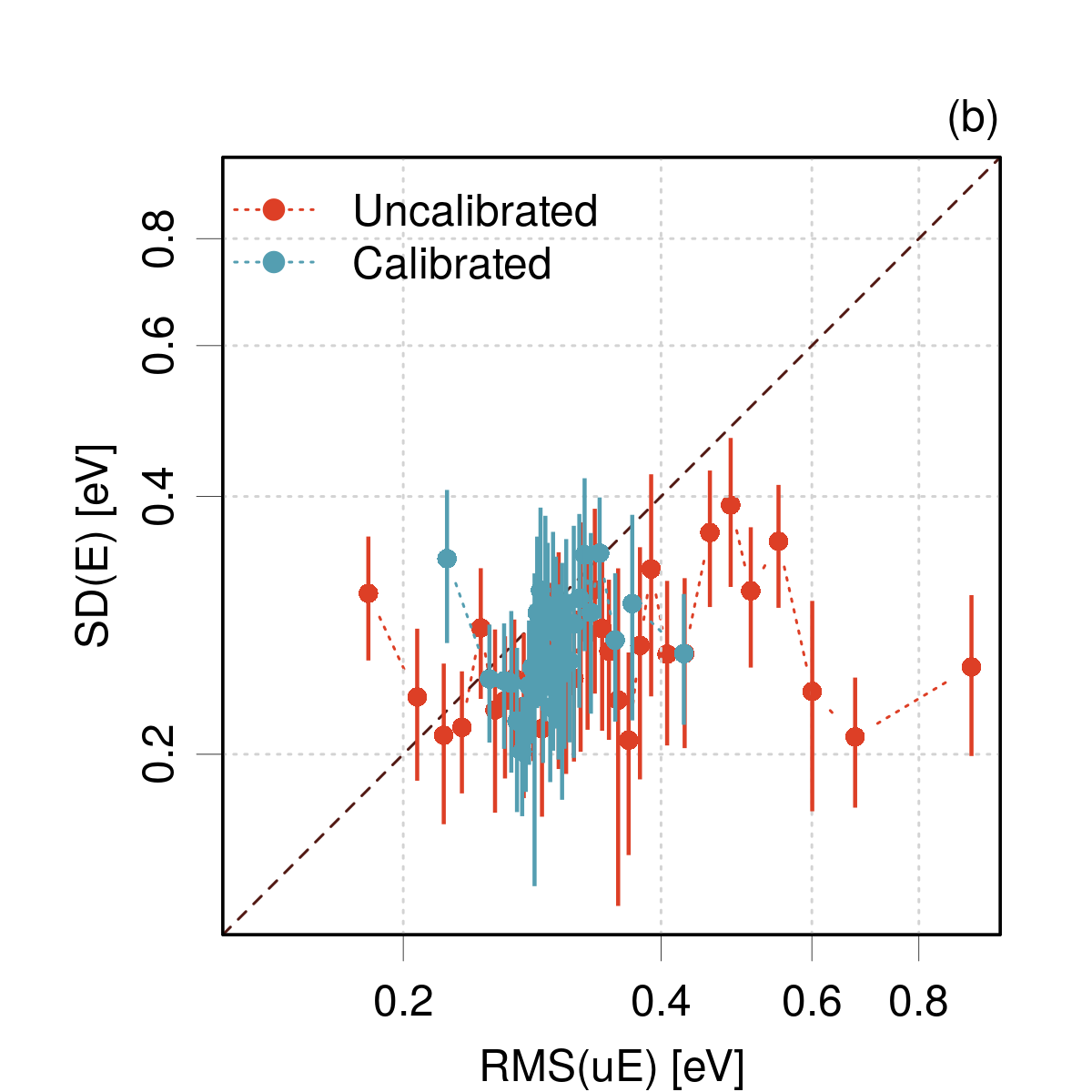} & \includegraphics[width=0.33\textwidth]{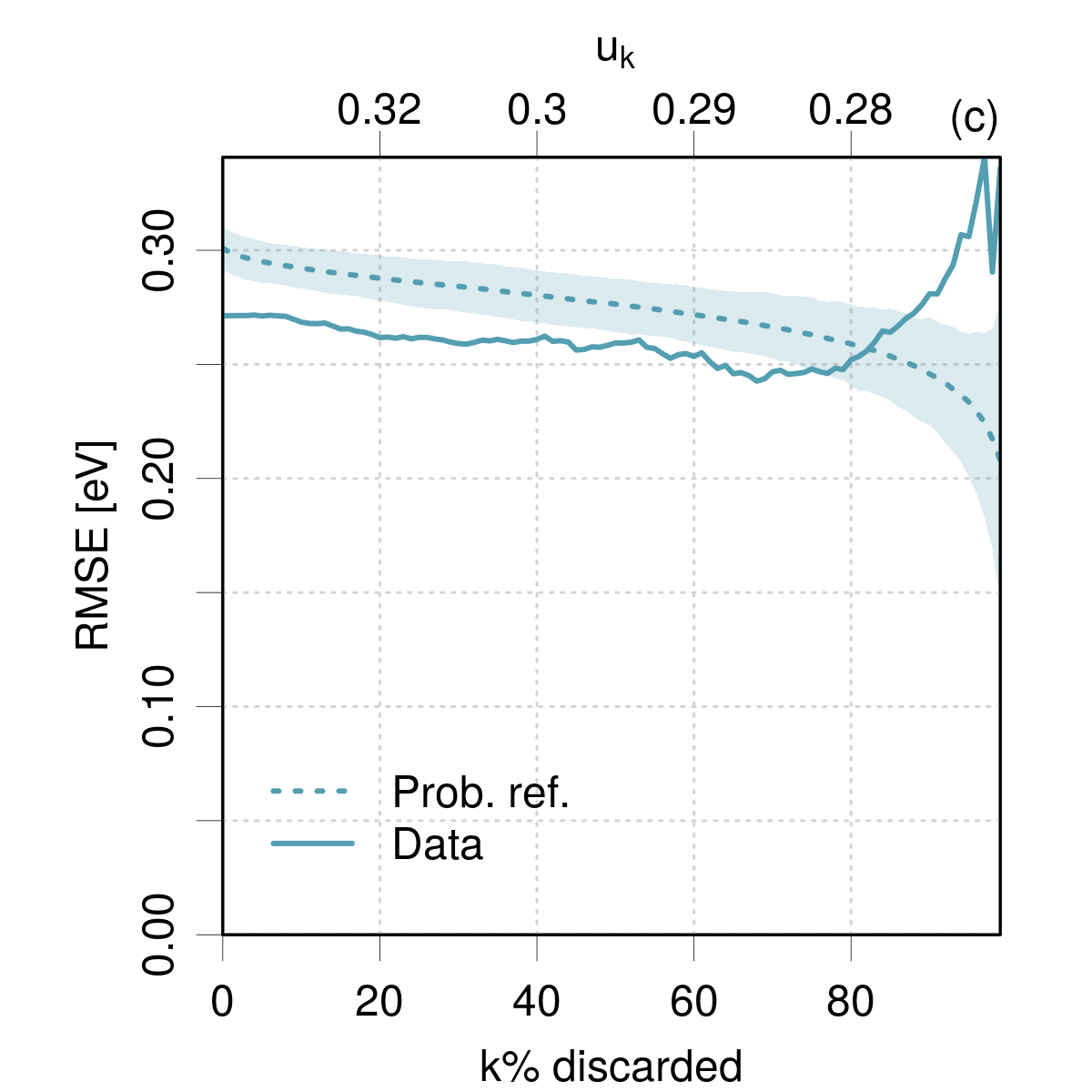}\tabularnewline
\includegraphics[width=0.33\textwidth]{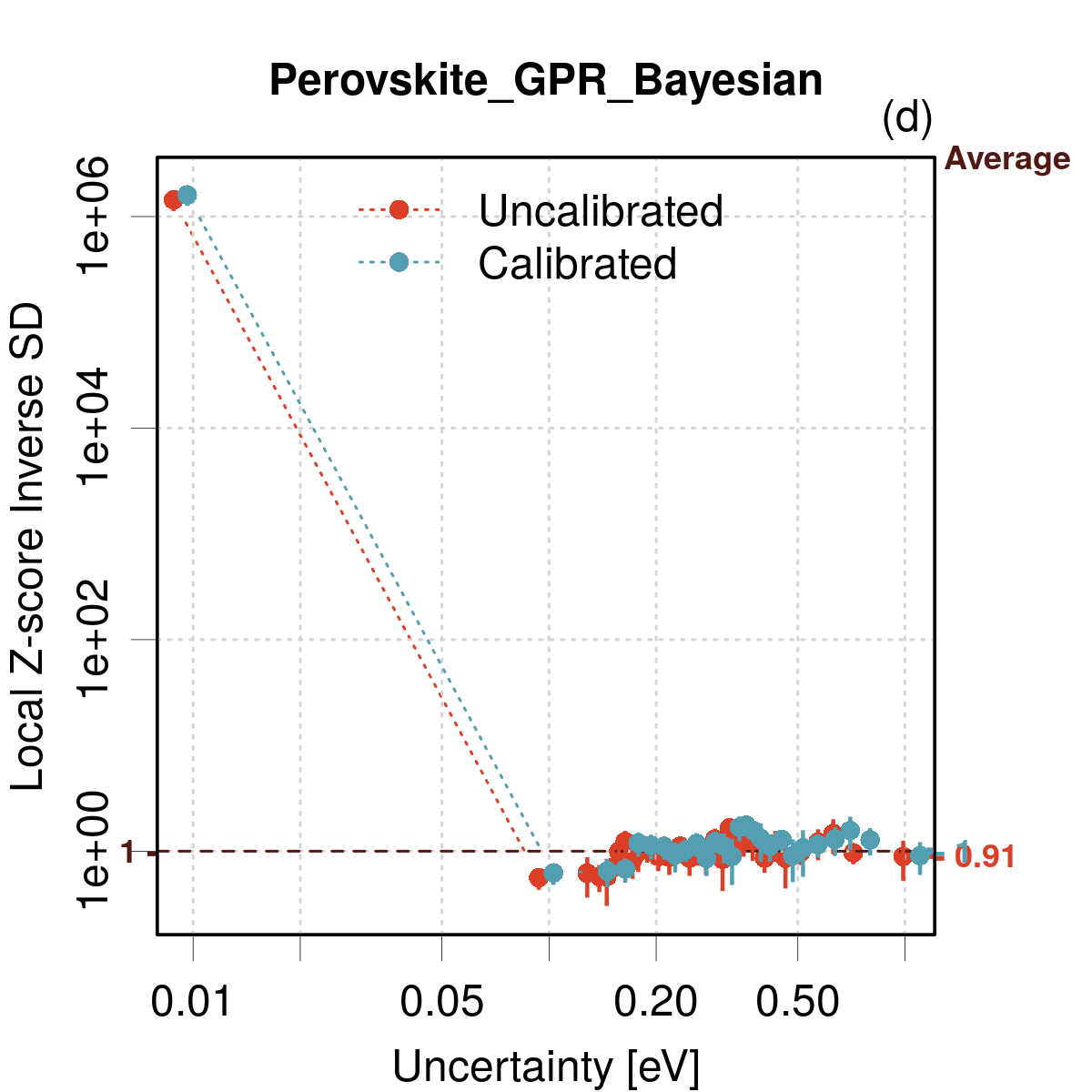} & \includegraphics[width=0.33\textwidth]{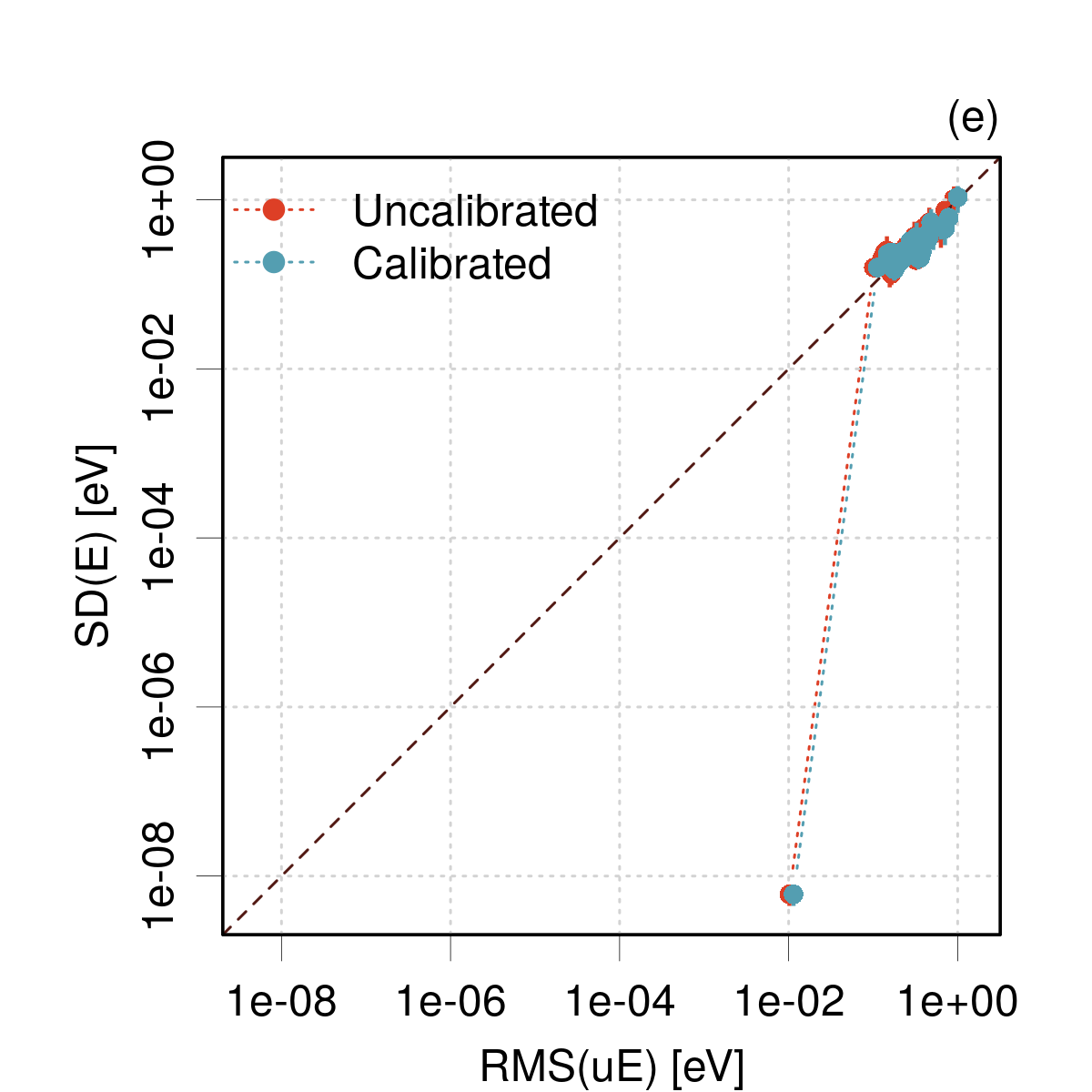} & \includegraphics[width=0.33\textwidth]{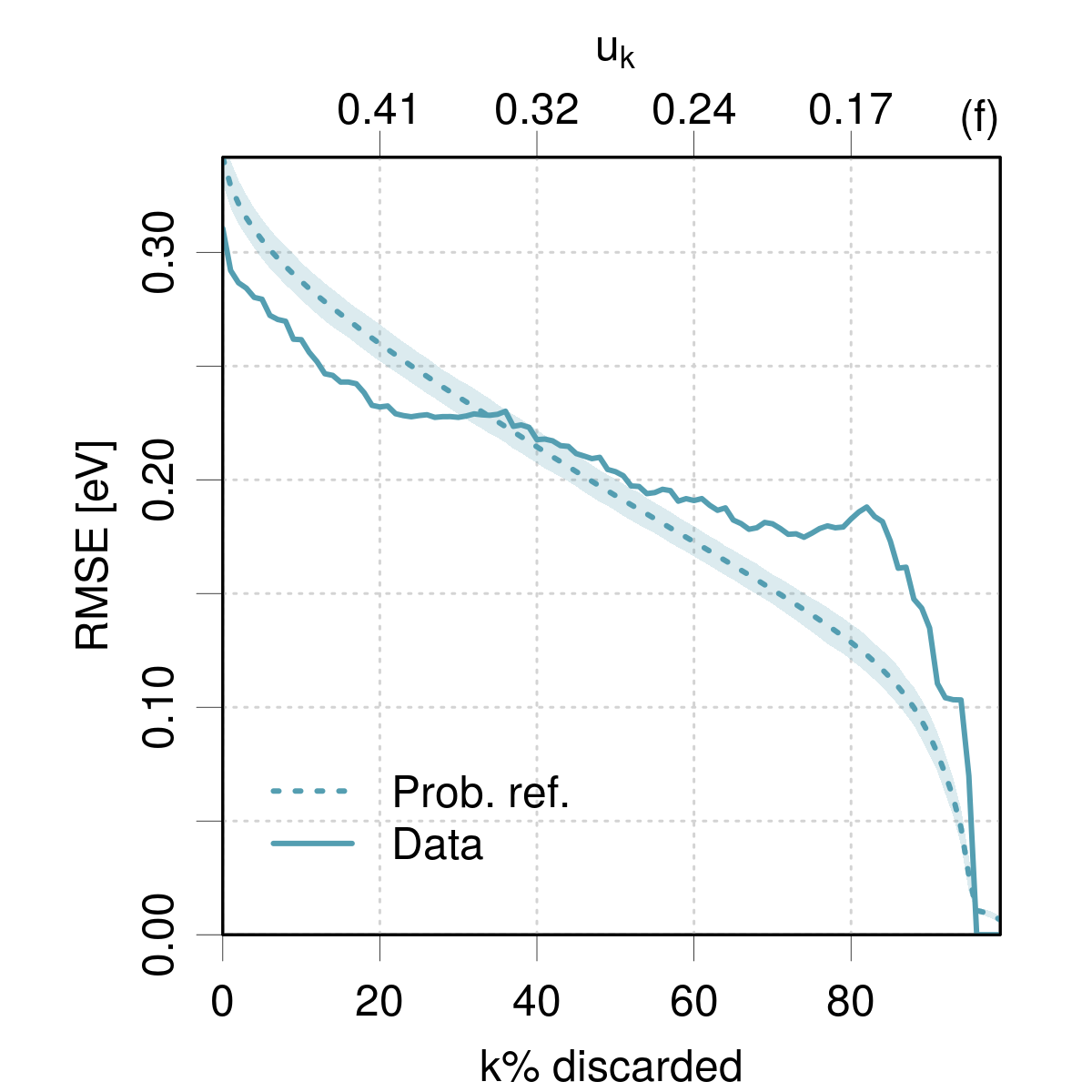}\tabularnewline
\end{tabular}
\par\end{centering}
\caption{\label{fig:PAL2022_GPRB}Case PAL2022: UQ validation for gaussian
process regression examples.}
\end{figure*}

For the Diffusion dataset, all three graphs (Fig.\,\ref{fig:PAL2022_GPRB})
concur to reject consistency. The confidence curve even suggests that
it would not be reasonable to base an active learning strategy on
these uncertainties. 

In the case of Perovskite one observes a nugget of data with tiny
errors and largely overestimated uncertainties, that was identified
by the adaptive binning strategy (Appendix\,\ref{sec:PAL2022---Additional}).
A zoom over the remaining data (not shown) reveals an heterogeneous
situations with small areas of underestimation (below 0.2\,eV) and
overestimation (around 0.3\,eV and 0.6\,eV). The confidence curve
deviates considerably from the reference and falls unexpectedly to
zero at small uncertainties, a consequence of the aforementioned data
nugget.

\paragraph{Conclusion.}

The reanalysis of these data shows that the reliability diagrams presented
in the original article with a small number of bins and no error bars
are often overoptimistic. Using an adaptive binning strategy in log-uncertainty,
conditional calibration appears heterogeneous for all datasets, meaning
that consistency is not achieved despite a rather good average calibration. 

The uncertainties are at worst within a factor two of their ideal
value, a level that has to be contrasted with their intended application.
The confidence curves show that not all ML methods provide reliable
uncertainties for active learning. 

\subsection{Case BUS2022}

\noindent The data presented in Busk \emph{et al.}\citep{Busk2022}
have been kindly collated by Jonas Busk for the present study. In
their article, Busk \emph{et al.} extend a message passing neural
network in order to predict properties of molecules and materials
with a calibrated probabilistic predictive distribution. An a posteriori
isotonic regression on data unseen during the training is used to
ensure calibration. For UQ validation, these authors used the more
or less standard trio of reliability diagrams, calibration curves
(named \emph{quantile-calibration} plot) and MAE-based confidence
curves with the oracle reference. Note that they duly express a reserve
about using the oracle: ``\emph{However, we do not expect a perfect
ranking...}''\citep{Busk2022}. 

Considering the validation results published for the QM9 dataset in
Fig.\,2 of the original article, a few questions arise: (1) what
is causing the imperfect calibration curve ?; (2) if this is a wrong
distribution hypothesis, how does it affect the MAE-based confidence
curve ?; (3) how is the confidence curve analysis modified if one
uses the RMSE statistic and the probabilistic reference ?; and, (4)
what is the diagnostic for adaptivity? The present reanalysis aims
to answers these questions. 

\subsubsection{Reanalysis of the QM9 dataset}

\noindent The QM9 dataset consists of 13\,885 predicted ($V,u_{V}$),
reference ($R$) atomization energies, and molecular formulas. These
data are transformed to\textcolor{violet}{{} }$C=\left\{ X,E,u_{E}\right\} $
according to Sect.\,\ref{subsec:Validation-datasets}, where $X$
is the molecular mass generated from the formulas. One should thus
be able to test calibration, consistency and adaptivity. Average calibration
is readily assessed ($Var(Z)=0.96(2)$).

\begin{figure*}[t]
\noindent \begin{centering}
\includegraphics[viewport=0bp 0bp 1200bp 1100bp,clip,width=0.33\textwidth]{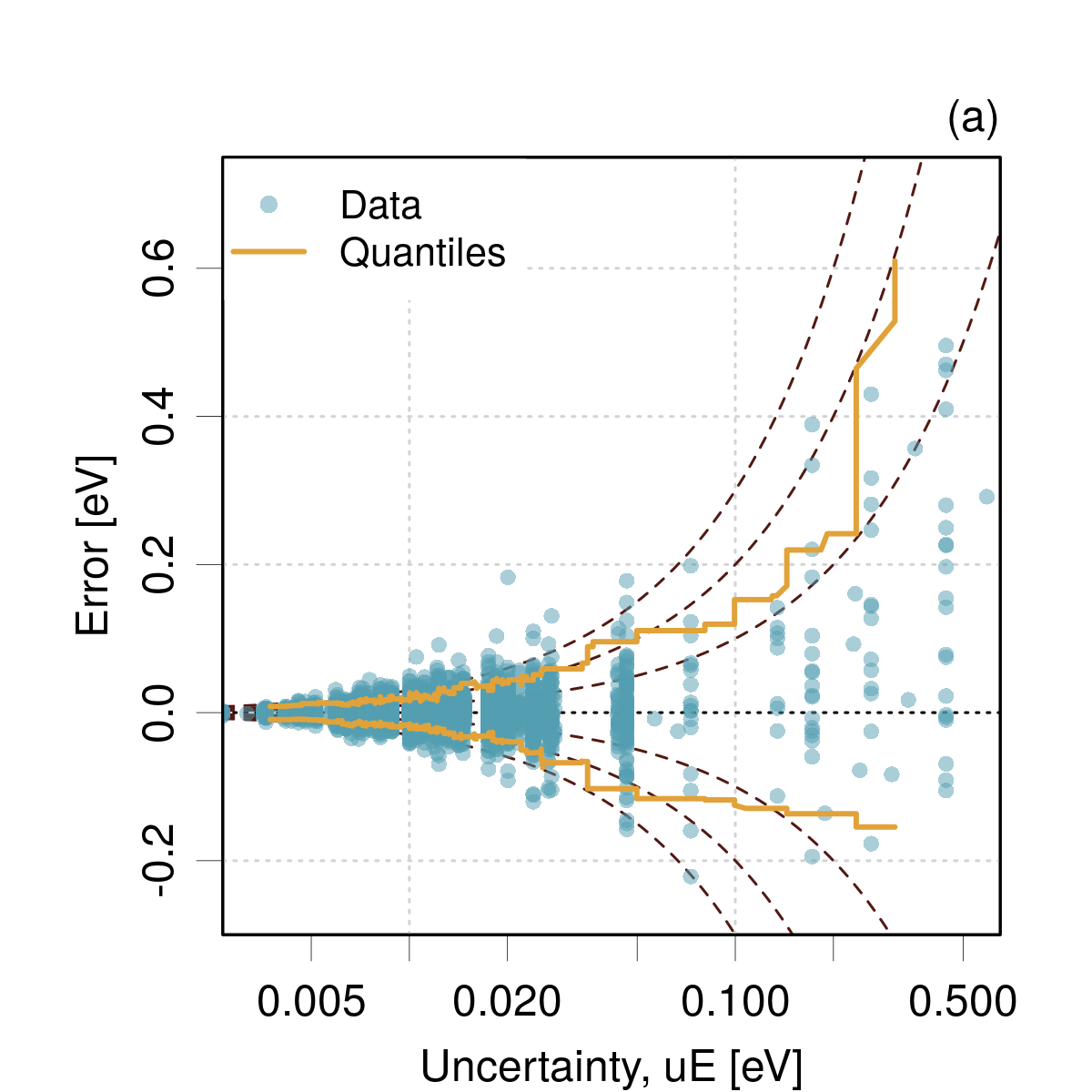}\includegraphics[viewport=0bp 0bp 1200bp 1100bp,clip,width=0.33\textwidth]{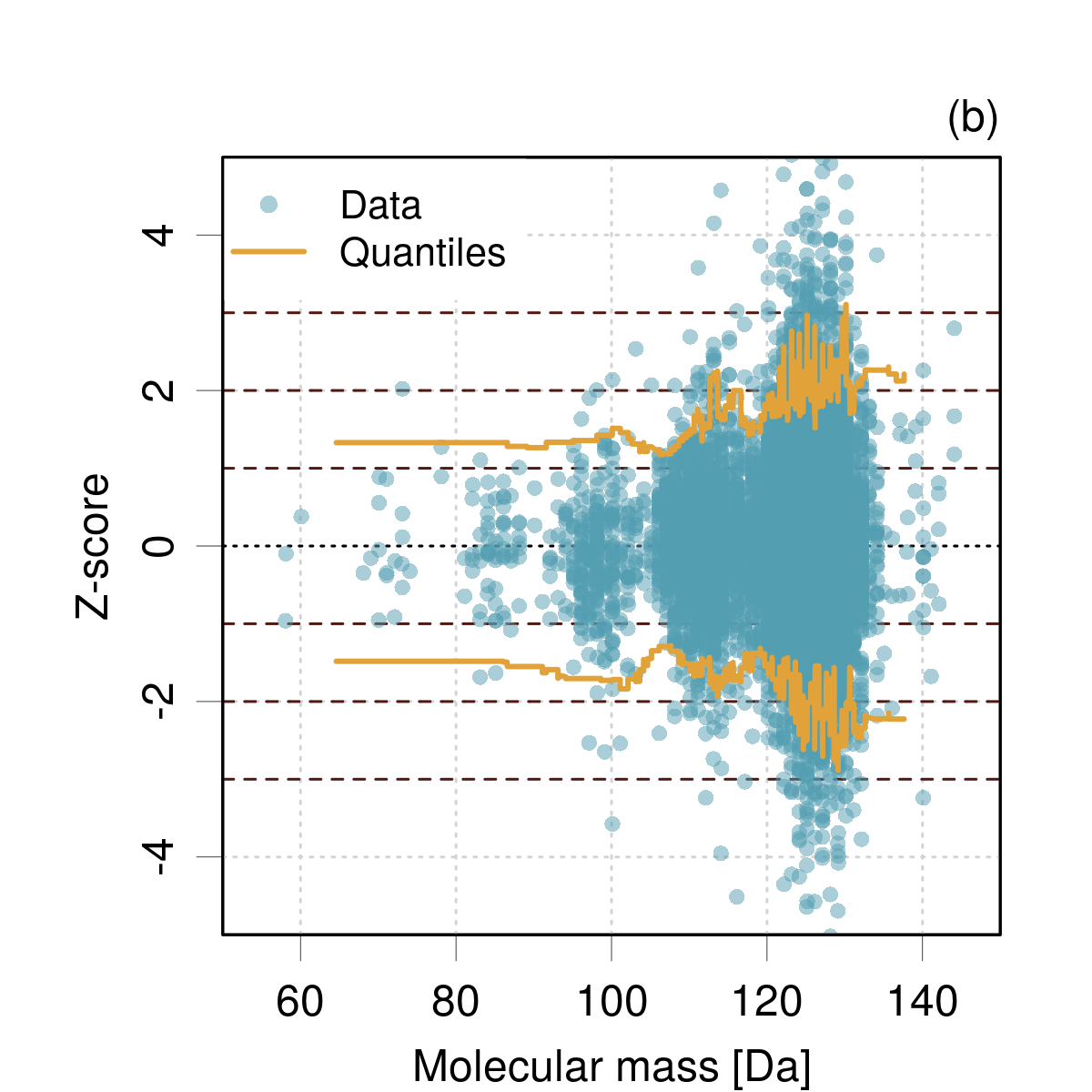}
\par\end{centering}
\caption{\label{fig:Plot-of-errors}Case BUS2022: (a) errors vs. uncertainty;
(b) \emph{z}-scores vs. prediction.}
\end{figure*}

It is always instructive to inspect the raw data through ``$E$ vs
$u_{E}$'' and ``$Z$ vs $X$'' plots to get a global appreciation
of the link between these quantities (Fig.\,\ref{fig:Plot-of-errors}).
One sees in Fig.\,\ref{fig:Plot-of-errors}(a) that the data points
are neatly distributed between the $E=\pm3u_{E}$ guiding lines, and
the 0.025 and 0.975 running quantile lines seem to follow closely
the $E=\pm2u_{E}$ lines up to $uE\simeq0.05$. Above this value,
the uncertainties seem overestimated and the errors are somehow biased
toward the positive values. However, this concerns a very small population
(about 95 points) and filtering them out does not affect significantly
the calibration statistics presented below. The problem might be simply
due to the sparsity of the data in this uncertainty range. 

The ``$Z$ vs $X$'' plot in Fig.\,\ref{fig:Plot-of-errors}(b)
using molecular masses enables to check if calibration is homogeneous
in mass space. The shape of the running quantile lines indicate that
uncertainties are probably overestimated for masses smaller than the
main cluster around 125-130\,Da, evolving to a slight underestimation
above this peak. Depending on its amplitude, this systematic effect
might be problematic, and it hints at a lack of adaptivity. A quantitative
analysis of this feature is presented below.

\begin{figure*}[t]
\noindent \begin{centering}
\begin{tabular}{ccc}
\includegraphics[width=0.33\textwidth]{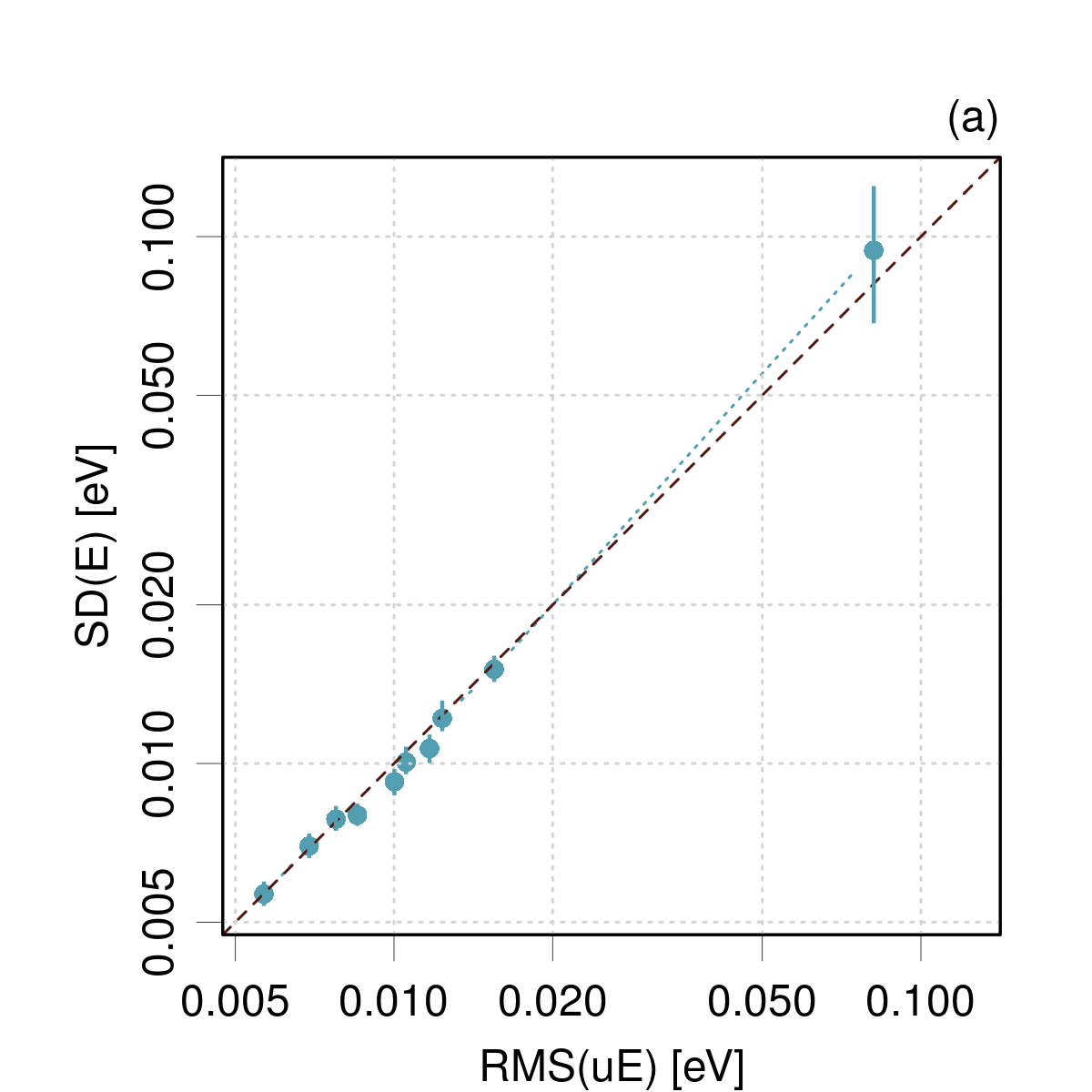} & \includegraphics[width=0.33\textwidth]{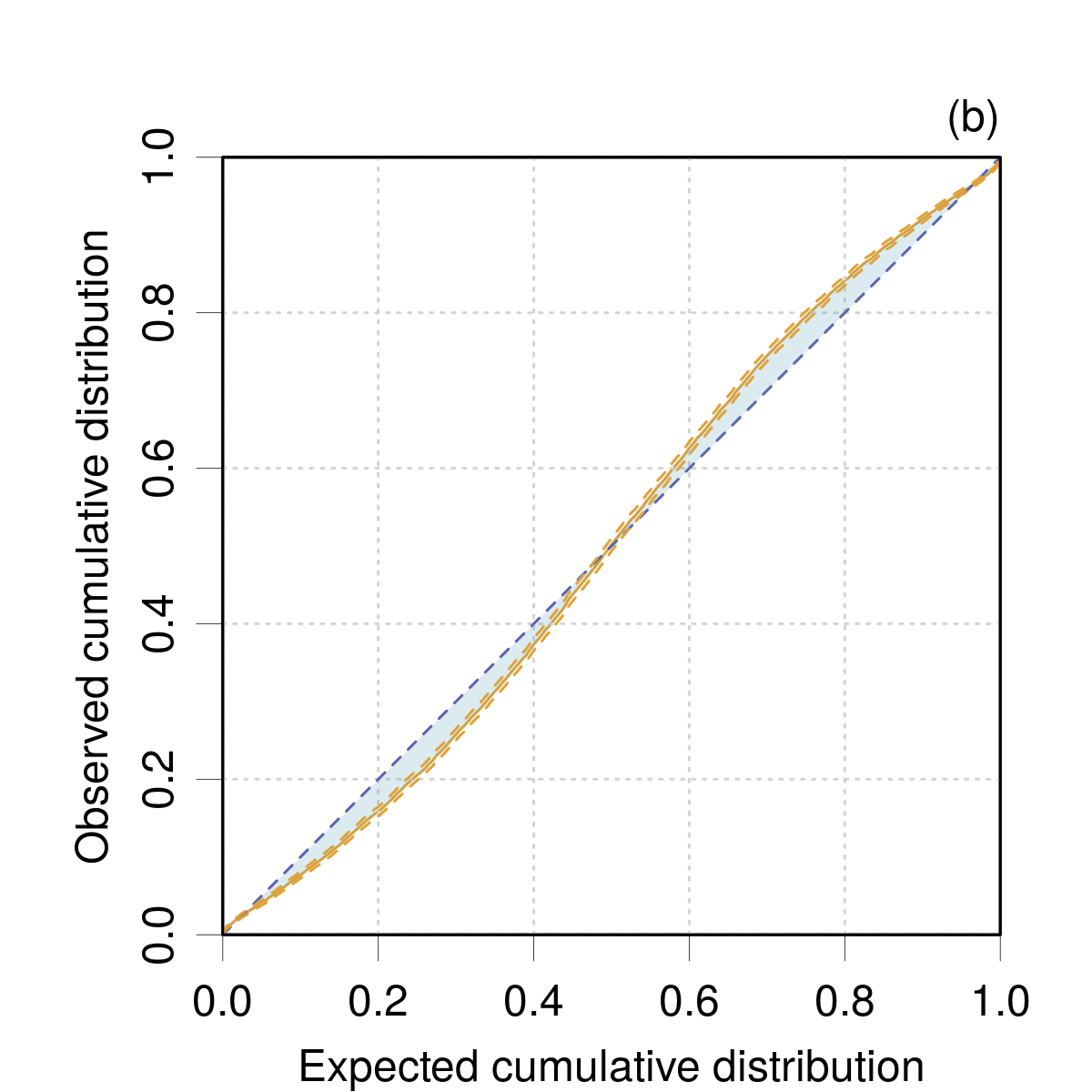} & \includegraphics[width=0.33\textwidth]{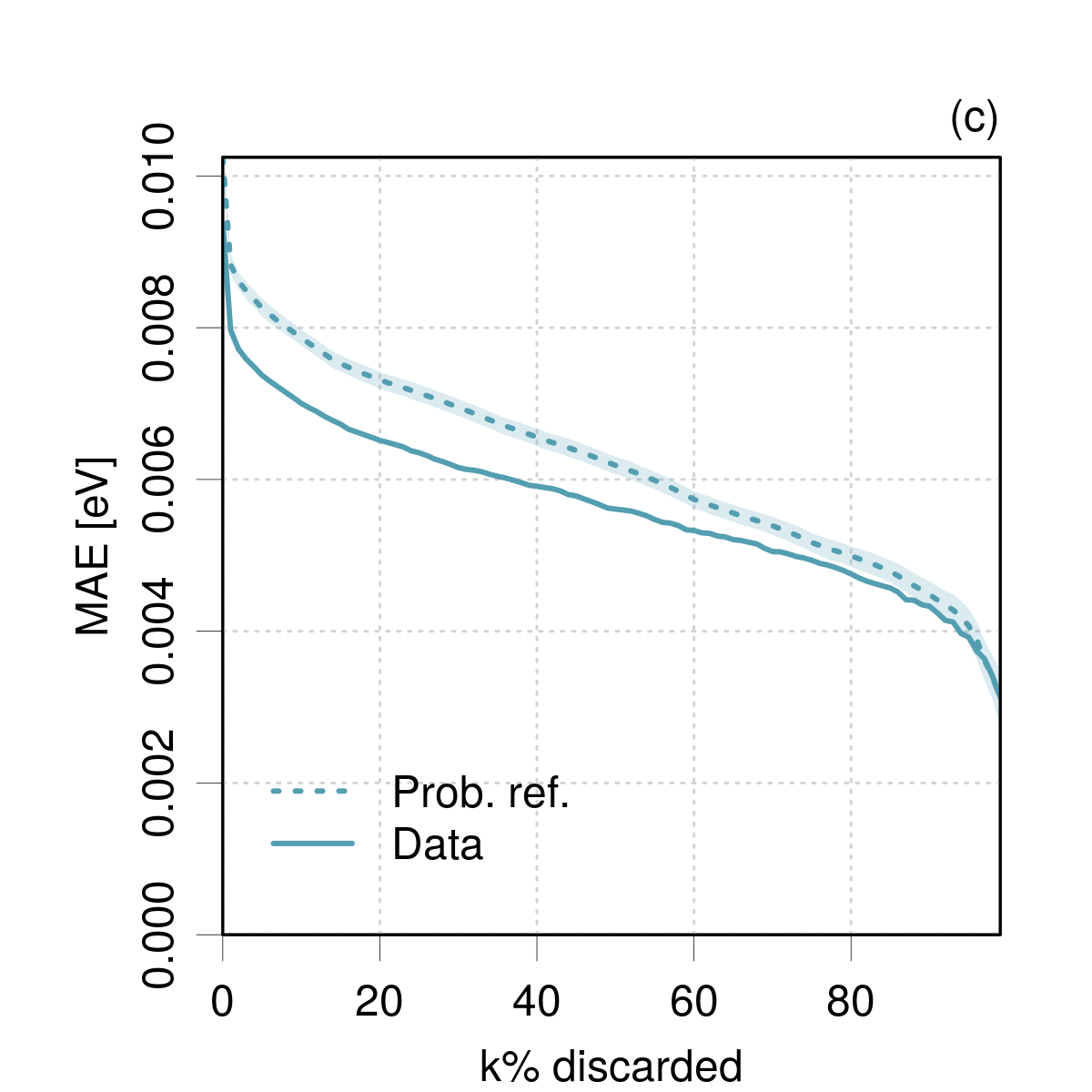}\tabularnewline
\includegraphics[width=0.33\textwidth]{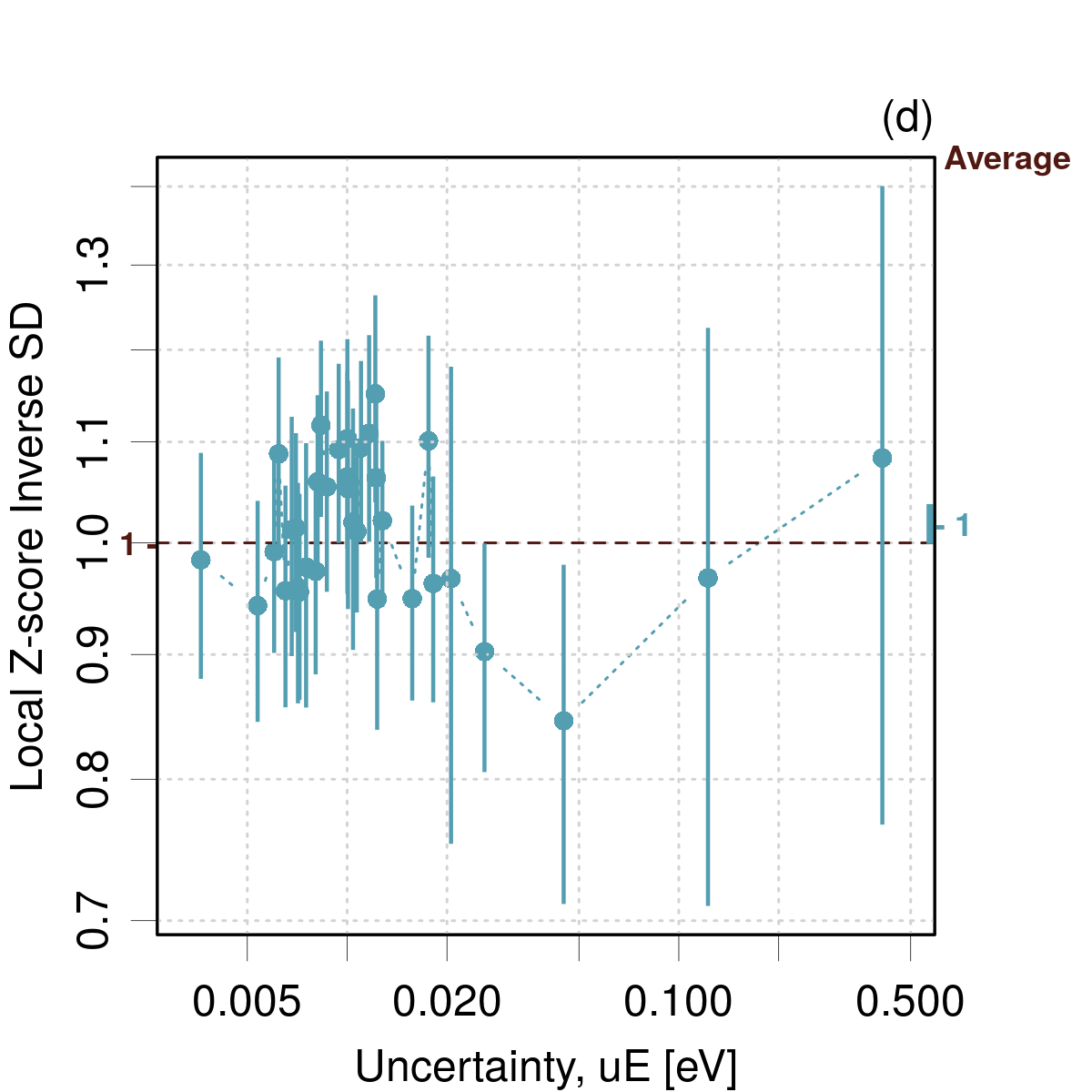} & \includegraphics[width=0.33\textwidth]{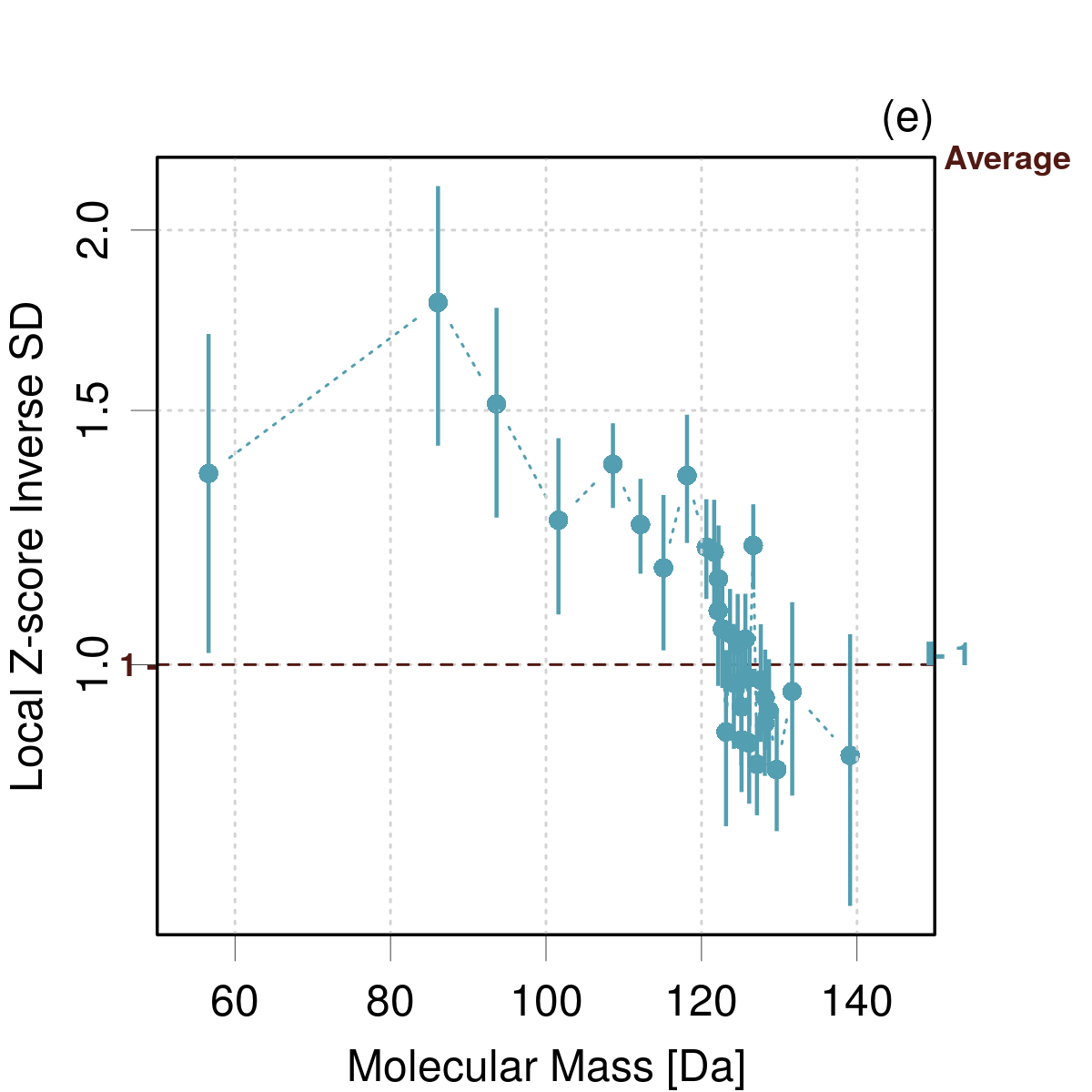} & \includegraphics[width=0.33\textwidth]{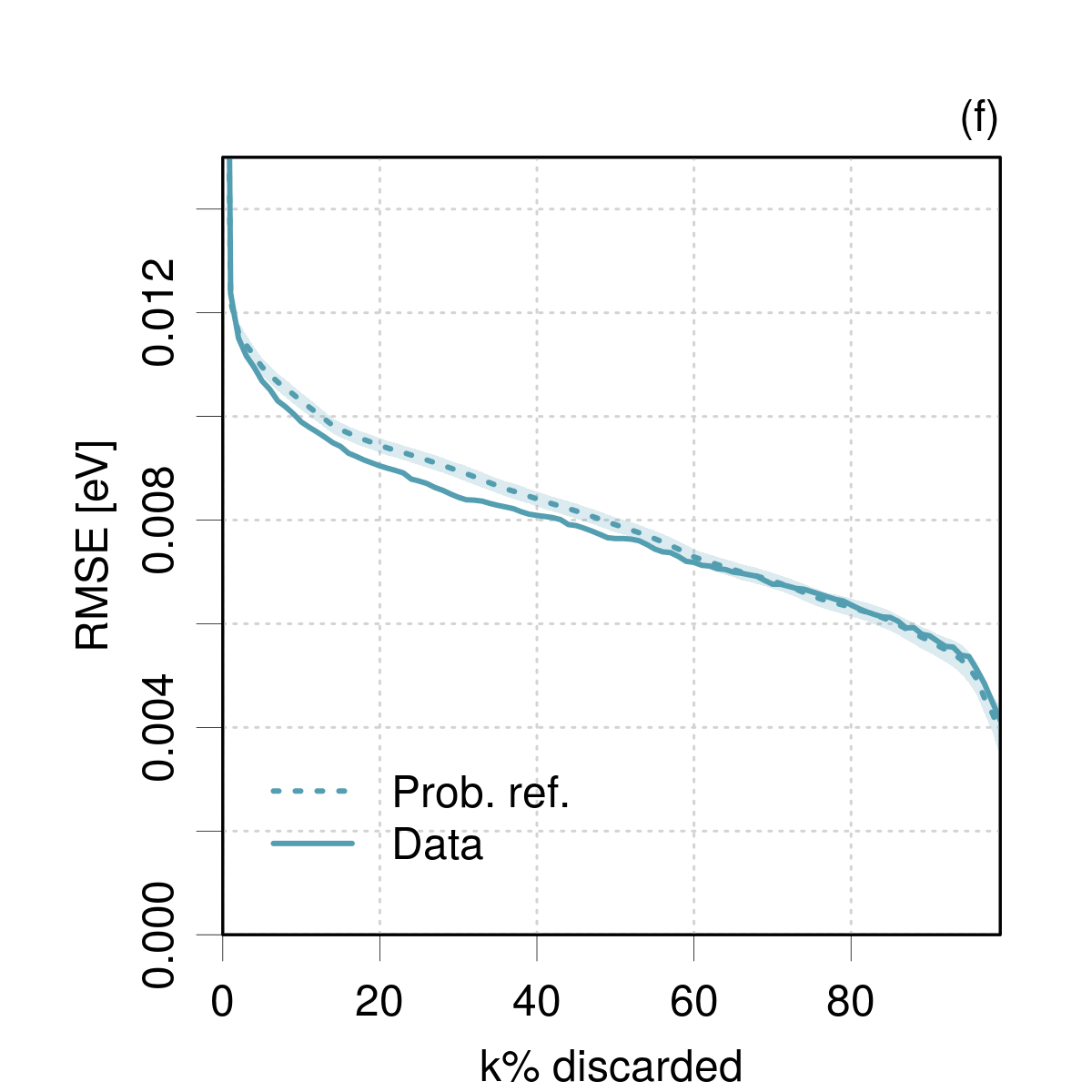}\tabularnewline
\end{tabular}
\par\end{centering}
\caption{\label{fig:LZISD-1}Case BUS2022: validation plots. (a) reliability
diagram with 10 equal-counts bins; (b) calibration curve; (c) MAE-based
confidence curve; (d) LZISD analysis vs. $u_{E}$, with adaptive binning
strategy; (e) LZISD analysis vs. $V$, with adaptive binning strategy;
(f) RMSE-based confidence curve (the data curve has been truncated
at 0.015\,eV, but it decreases sharply from 0.031\,eV). }
\end{figure*}

In a first step, calibration analyses are done to reproduce the Fig.\,2
of the original article: reliability diagram with 10 equal-counts
bins {[}Fig.\,\ref{fig:LZISD-1}(a){]}, calibration curve with the
normal hypothesis {[}Fig.\,\ref{fig:LZISD-1}(b){]} and MAE-based
confidence curve (replacing the unsuitable oracle by the probabilistic
reference) {[}Fig.\,\ref{fig:LZISD-1}(c){]}. Then, further analyses
are performed to complement the information provided by the first
set: LZISD analysis in uncertainty and $V$ space {[}Fig.\,\ref{fig:LZISD-1}(d,e){]},
and RMSE-based confidence curve {[}Fig.\,\ref{fig:LZISD-1}(f){]}.

The reliability diagram {[}Fig.\,\ref{fig:LZISD-1}(a){]} does not
enable to see any major problem and would lead us to conclude that
the data are consistent. However, the calibration curve {[}Fig.\,\ref{fig:LZISD-1}(b){]}
is not perfect. Considering the good consistency provided by the reliability
diagram, one should conclude that there is probably a distribution
problem, i.e., the generative distribution of errors that is used
to build the probabilistic reference curve should not be normal. Moreover,
the confidence curve {[}Fig.\,\ref{fig:LZISD-1}(c){]} is not very
close to the probabilistic reference, which, assuming a good consistency,
points also to a problem of distribution, as the MAE-based probabilistic
reference is sensitive to the choice of distribution. This distribution
problem is analyzed more specifically below (Sect.\,\ref{subsec:Non-normality-of-the}).

A LZISD analysis with an adaptive binning strategy is performed to
compare with the original reliability diagram {[}Fig.\,\ref{fig:LZISD-1}(d){]}.
Indeed, one observes some discrepancies: overestimation and underestimation
of $u_{E}$ go locally up to 15\,\%. There is notably an overestimation
trend around 0.01\,eV. This feature is also present in the reliability
diagram, but it is more difficult to visualize small deviations on
a parity plot. 

The LZISD analysis against the molecular mass {[}Fig.\,\ref{fig:LZISD-1}(e){]}
confirms and quantifies what has been observed in {[}Fig.\,\ref{fig:Plot-of-errors}(b){]}:
an overestimation between 40 and 80\,\% at masses below 120\,Da,
and a slight underestimation by about 20\,\% above 130\,Da (although
there are too few data in this range to conclude decisively). Uncertainties
are therefore mostly reliable for the main mass peak between 120 and
130\,Da, but not outside of this range. This heterogeneity of calibration
and lack of adaptivity is not observable in the other validation plots.

Finally, a RMSE-based confidence curve is reported in Fig.\,\ref{fig:LZISD-1}(f).
It has been truncated at 0.015\,eV, but drops sharply from 0.031\,eV
because of the small set of large positive errors corresponding to
the largest uncertainties {[}Fig.\,\ref{fig:Plot-of-errors}(a){]}.
This confidence curve shows a much better agreement with the probabilistic
reference than the MAE-based one, albeit there is still a mismatch
due to the consistency defects observed in the LZISD analysis and/or
to the distribution problem mentioned above.

\subsubsection{Non-normality of the generative probabilistic model\label{subsec:Non-normality-of-the}}

\begin{figure*}[t]
\noindent \begin{centering}
\begin{tabular}{ccc}
\includegraphics[width=0.33\textwidth]{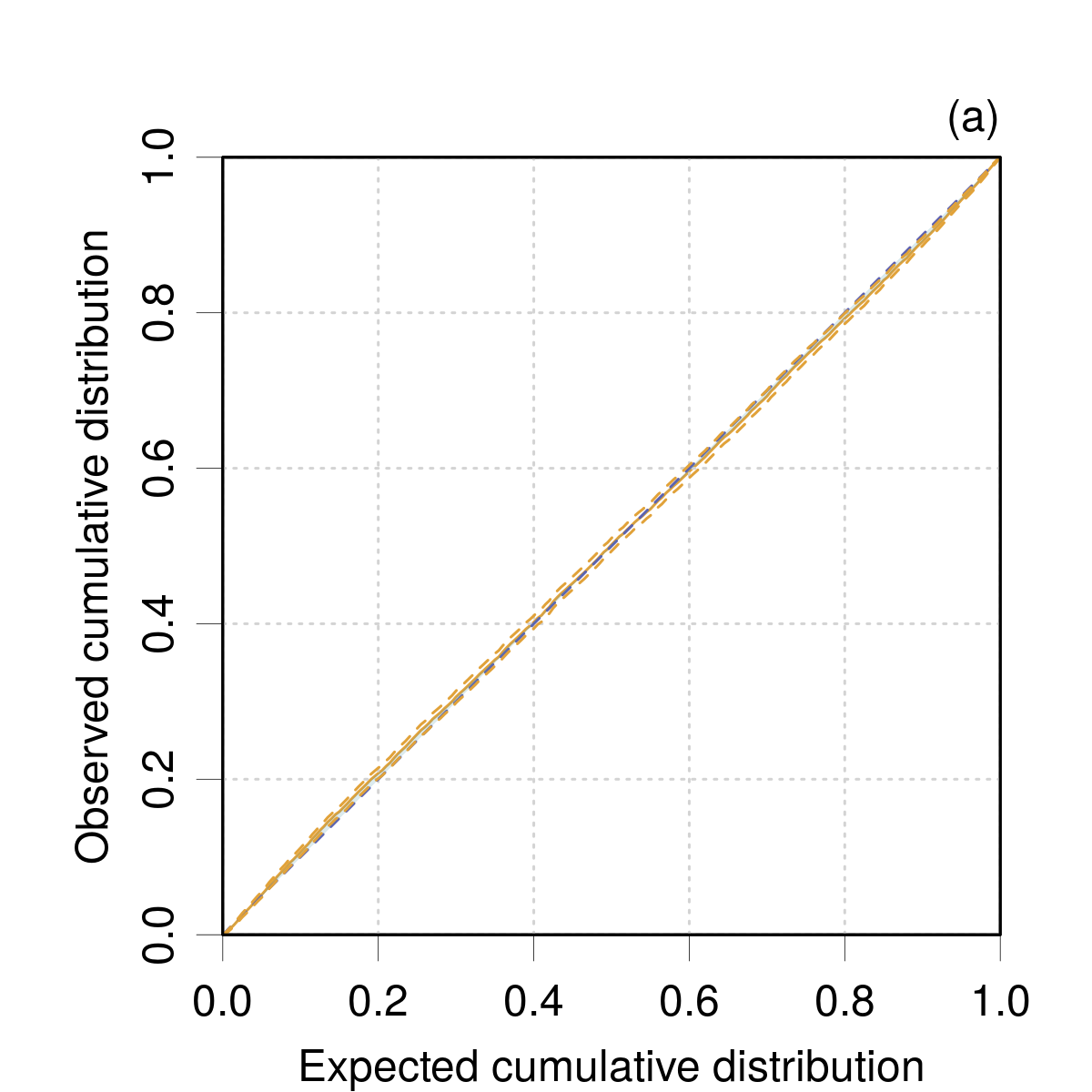} & \includegraphics[width=0.33\textwidth]{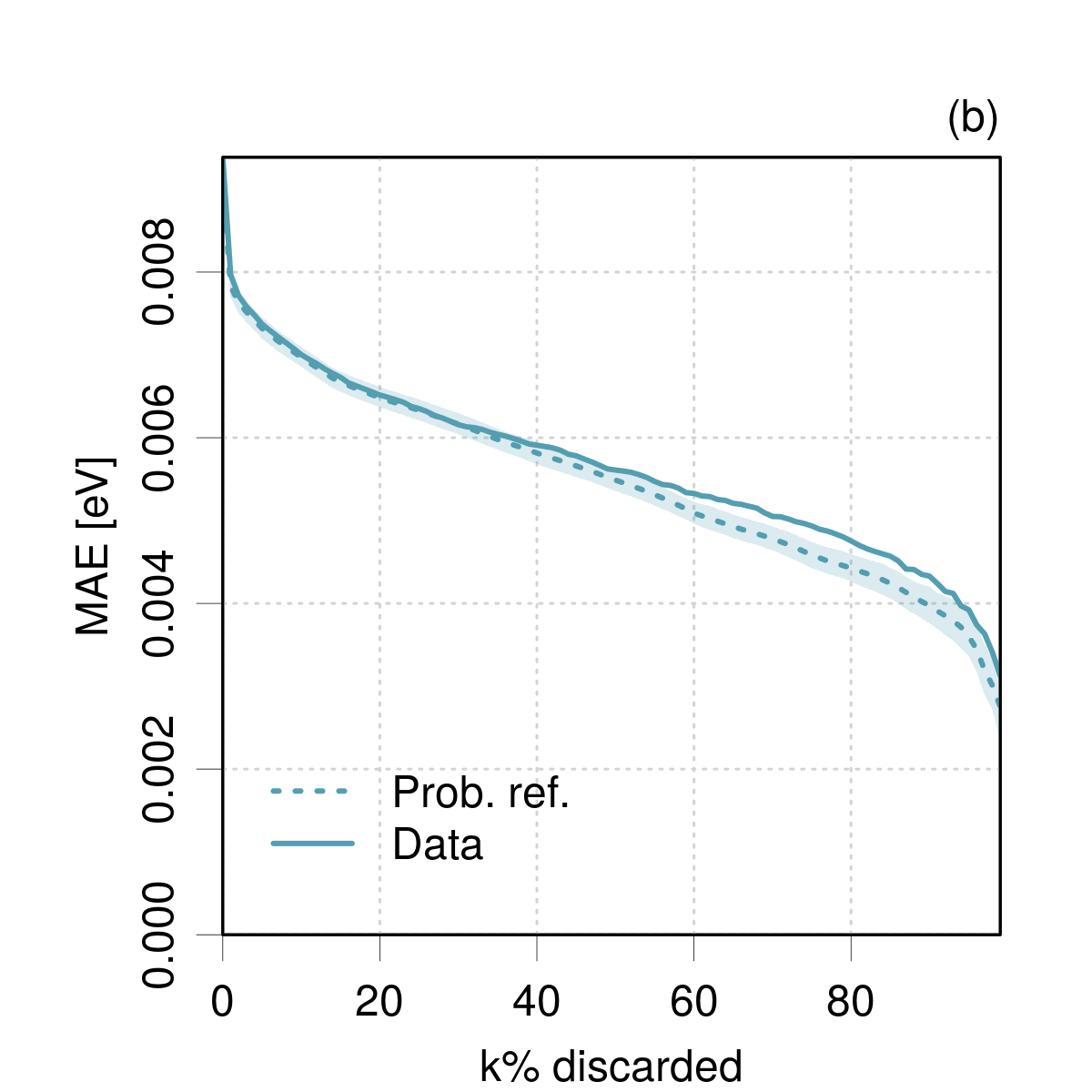} & \includegraphics[width=0.33\textwidth]{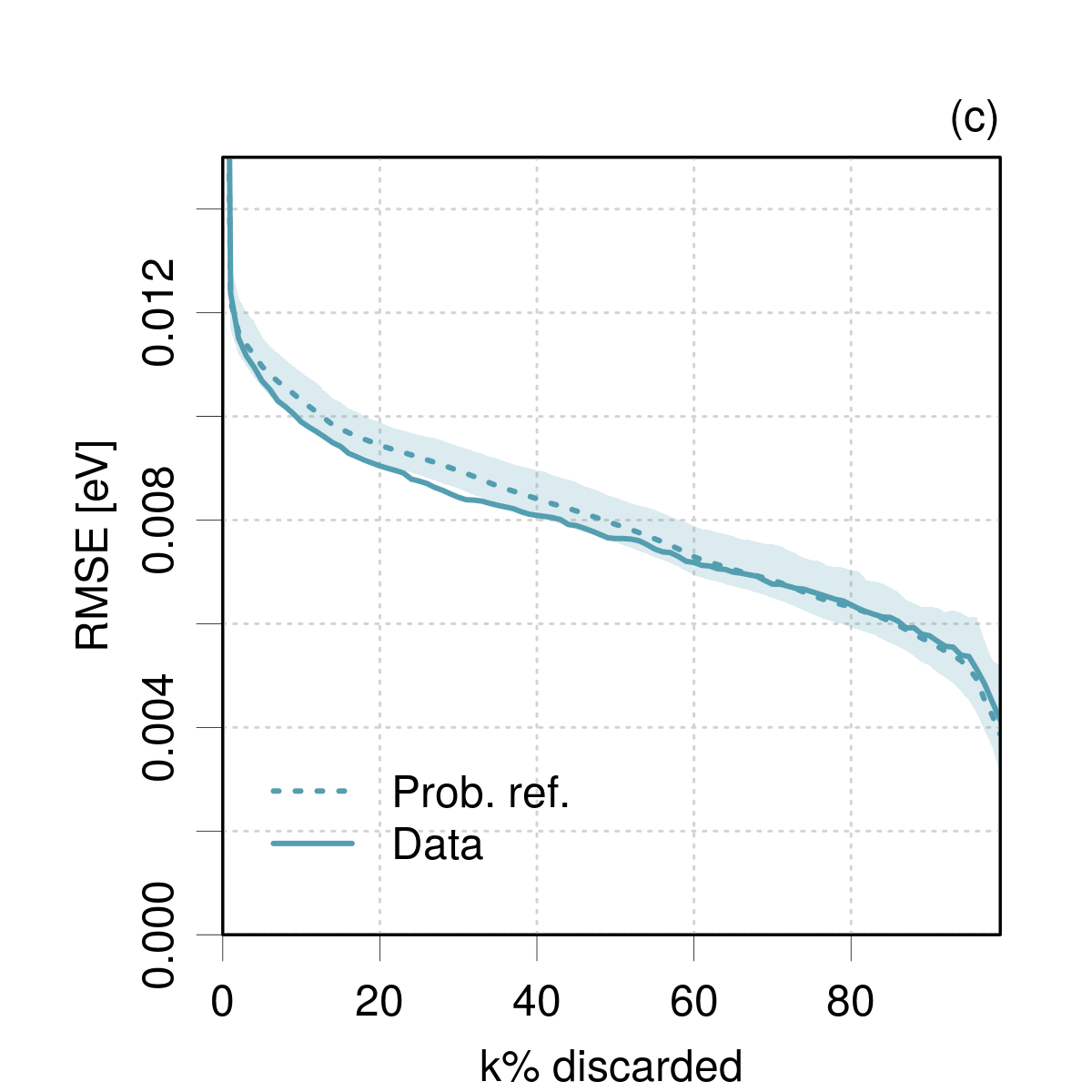}\tabularnewline
\end{tabular}
\par\end{centering}
\caption{\label{fig:LZISD-2}Case BUS2022: validation plots with a Student's-\emph{$t_{\nu=4}$}
generative distribution. (a) calibration curve; (b) MAE-based confidence
curve; (c) RMSE-based confidence curve (the data curve has been truncated
at 0.015\,eV, but it decreases sharply from 0.031\,eV). }
\end{figure*}

\noindent The deviations of the calibration curve observed in Fig.\,\ref{fig:LZISD-1}(b)
correspond neatly to observations made for one of the synthetic datasets
(Case~E) when drawing error from a Student's-\emph{t} distribution
with four degrees of freedom and comparing the percentiles with a
normal reference {[}Fig.\,\ref{fig:Calibration-curves}(e){]}. Indeed,
if one substitutes the normal reference in the calibration plot by
a $t_{\nu=4}$ reference, one obtains a perfect calibration curve
{[}Fig.\,\ref{fig:LZISD-2}(a){]}. 

One can consider two major sources of non-normality in this example:
the Bayesian model using an inverse-gamma prior on the uncertainty
variable\citep{Soleimany2021,Busk2022}, and/or the small size of
the ensemble of neural networks ($n=5$).\citep{Busk2022} The latter
would explain perfectly the observed four degrees of freedom (see
Sect.\,\ref{par:Average-calibration.}). Note that averaging over
small ensembles is expected to affect also $\mathrm{Var}(Z)$, but,
in the present case, this is corrected by the \emph{a posteriori}
calibration stage, leaving only a distribution effect.

Assuming a $t_{\nu=4}$ generative distribution for the probabilistic
reference of confidence curves improves also considerably the agreement
with the data curve {[}Fig.\,\ref{fig:LZISD-2}(b,c){]}, both for
the MAE- and RMSE-based approaches. Note the widening of the confidence
area of the probabilistic reference in both cases. However, there
remains small deviations indicating that consistency is not perfect.

\subsubsection{Conclusions}

\noindent All diagnostics based on $E$ and $u_{E}$ conclude to a
good calibration and consistency, notably if one accounts for a non-normal
generative model ($t_{\nu=4}$). The main feature revealed by this
reanalysis is the lack of adaptivity seen in the LZISD analysis in
molecular mass space. It leads to a significant underestimation of
the quality of predictions for the smaller molecules in the QM9 dataset.
By contrast, the confidence curves indicate that the uncertainties
can be used for active learning without a second thought. 

\subsection{Case HU2022}

\noindent These data were kindly collated and provided by Y. Hu to
facilitate this reanalysis of the results presented in Figs. 3e and
3f of the paper by Hu \emph{et al.}\citep{Hu2022}.\textcolor{orange}{{}
}These figures present plots of the errors on interatomic potentials
for the QM9 dataset vs two distances: a distance in feature space,
$X_{F}$ and a distance in latent space\citep{Janet2019}, $X_{L}$.
Both distances have been calibrated to interval-based UQ metrics:
the half-ranges of prediction intervals, $U_{F,P}$ and $U_{L,P}$
have been obtained for $P\in\{68,95\}$, by multiplication of the
corresponding distances by probability-specific scaling factors optimized
by conformal calibration to ensure average coverage.\textcolor{orange}{{}
}The percentage of points within the resulting intervals are reported
in the original figures and show a successful average calibration
in both cases. However, the distribution of points in these figures
hints at different conditional calibration statistics for both metrics,
which I propose to quantify in this reanalysis. The reader is referred
to the original article for details on the dataset and ML-UQ methods.
\begin{figure*}[t]
\noindent \begin{centering}
\begin{tabular}{ccc}
\includegraphics[width=0.33\textwidth]{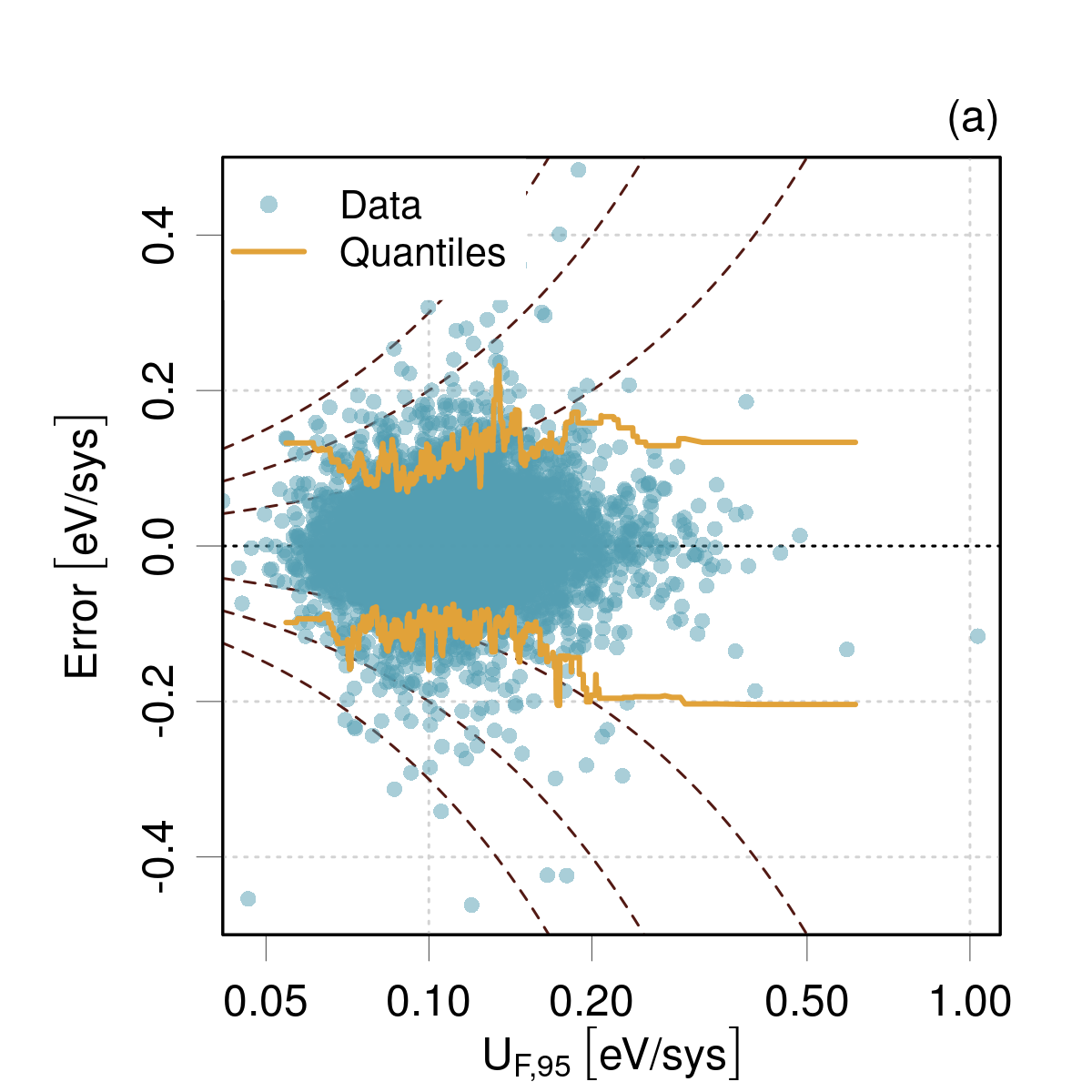} & \includegraphics[width=0.33\textwidth]{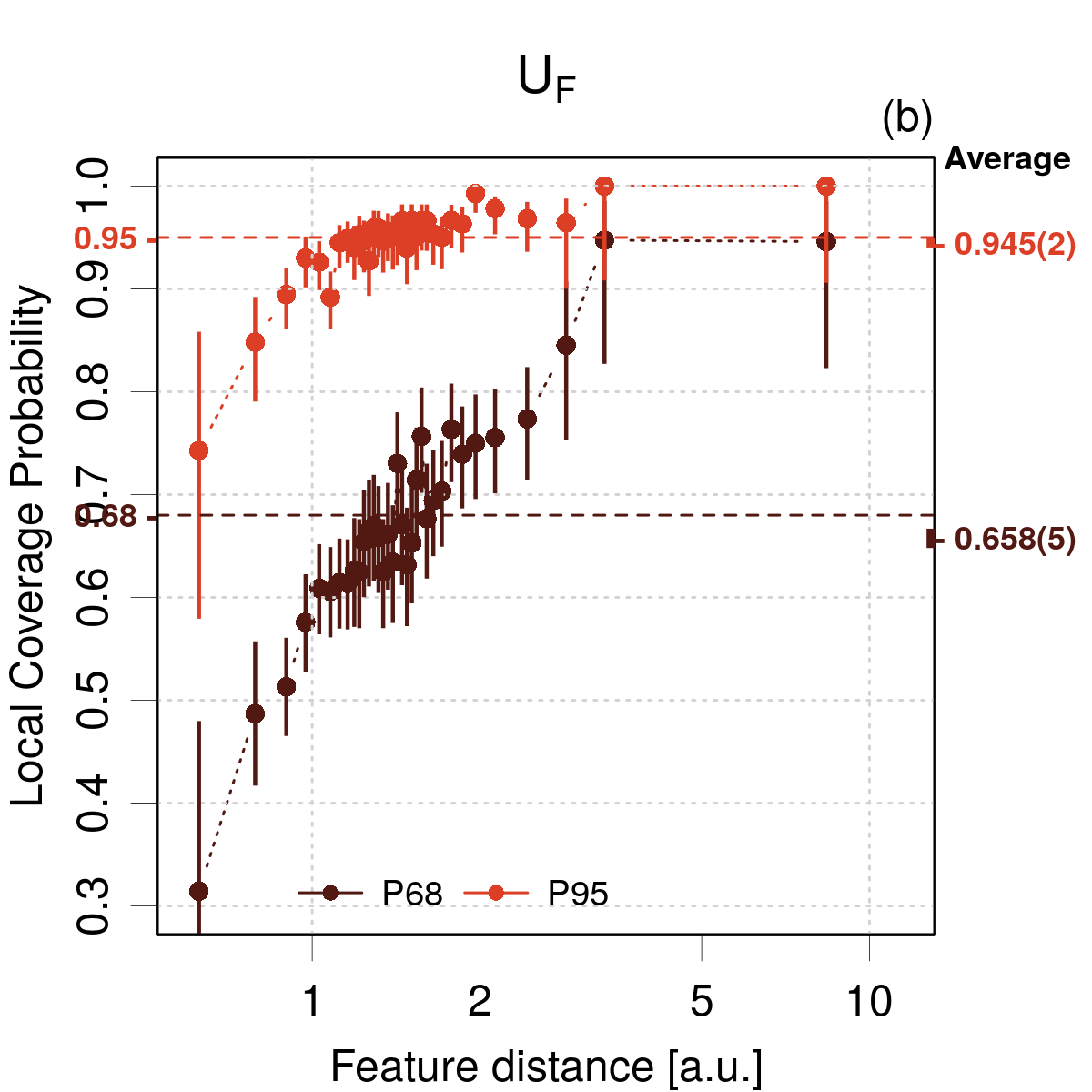} & \includegraphics[width=0.33\textwidth]{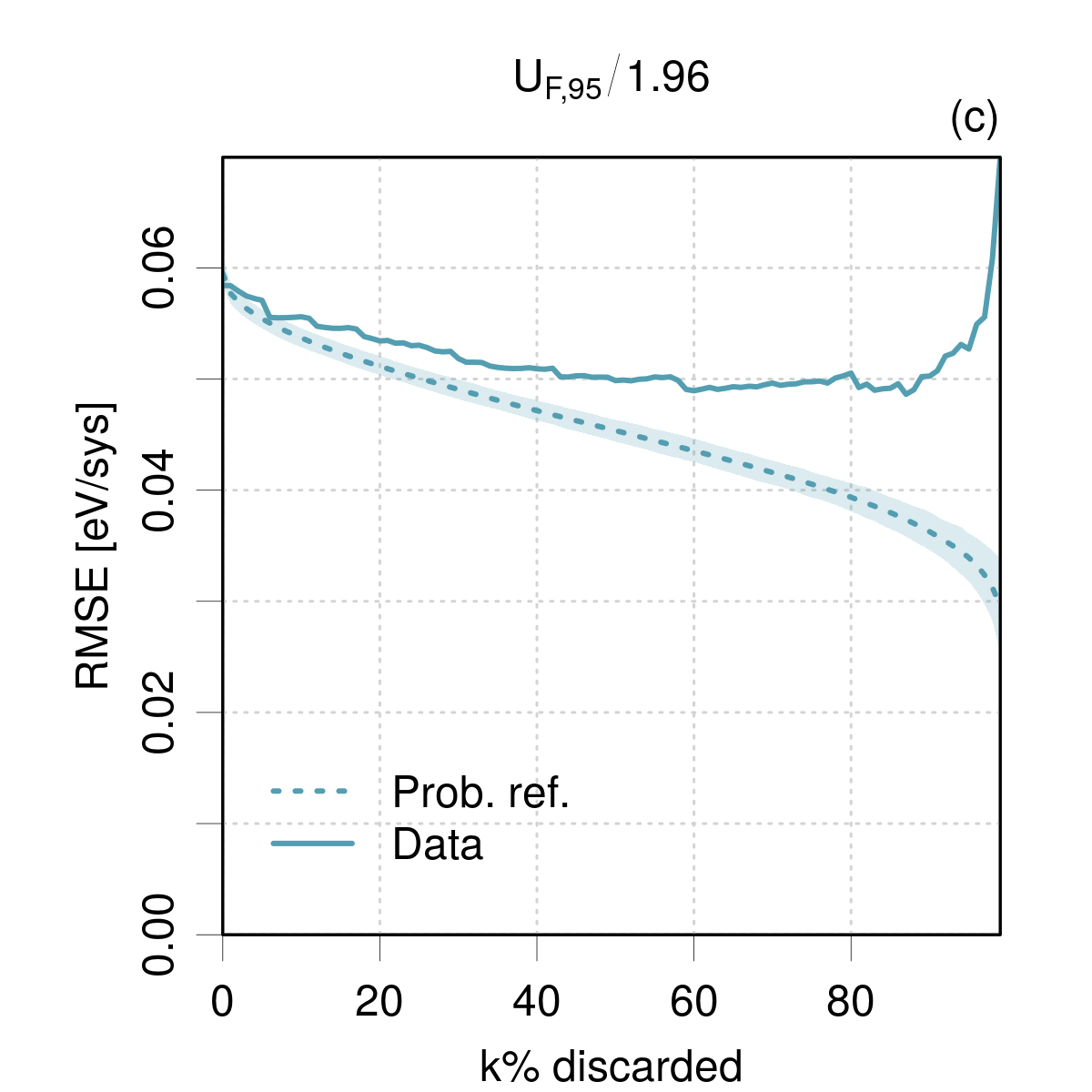}\tabularnewline
\includegraphics[width=0.33\textwidth]{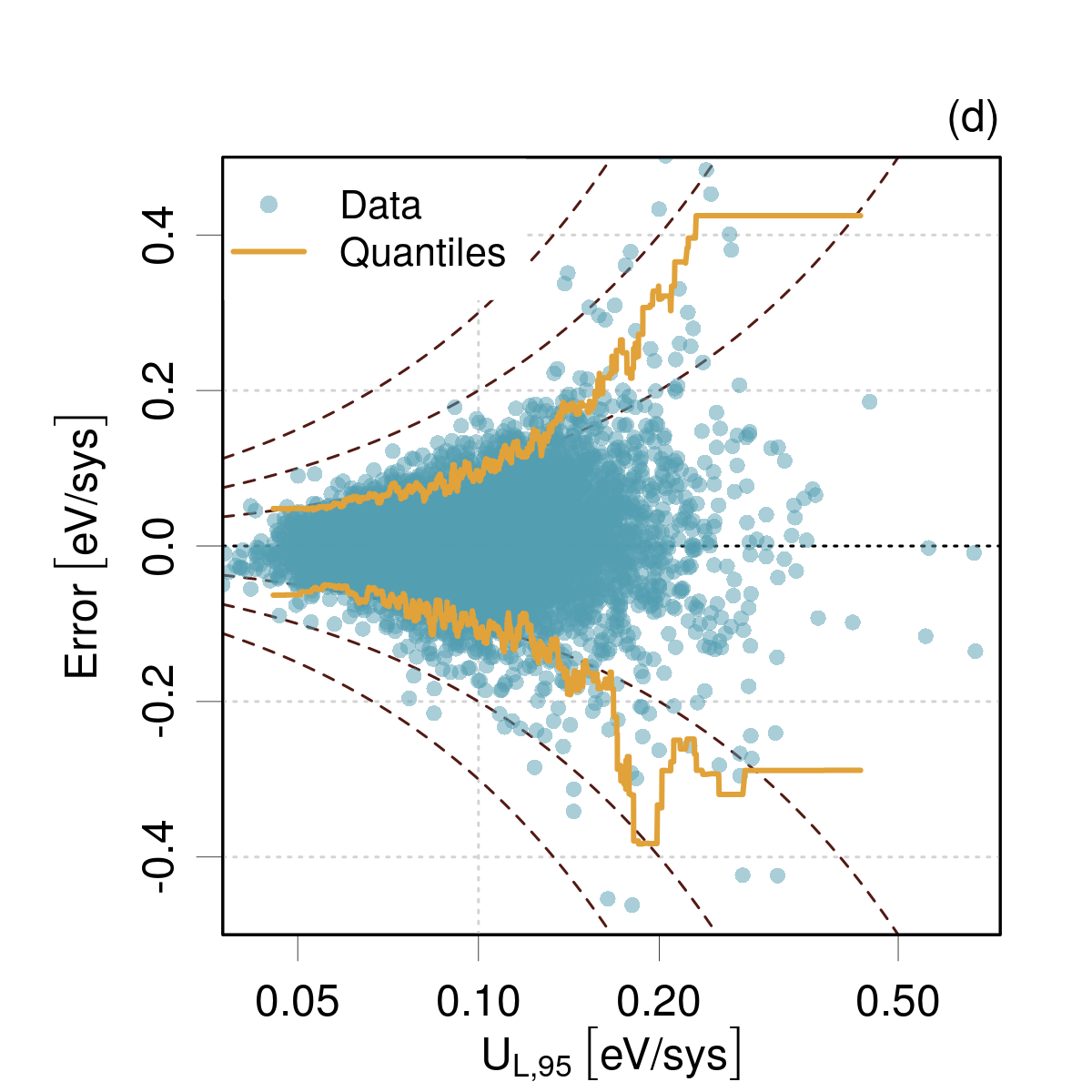} & \includegraphics[width=0.33\textwidth]{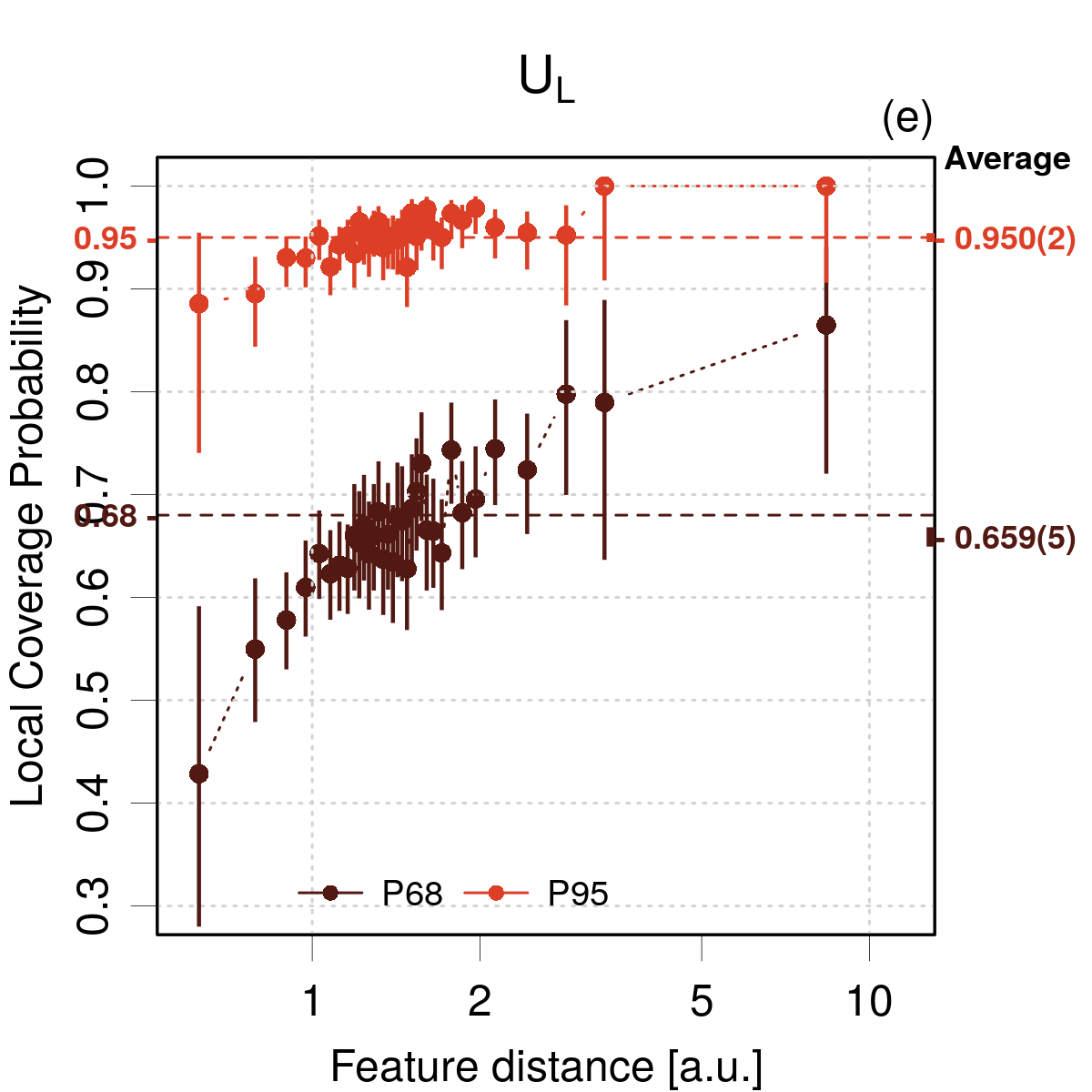} & \includegraphics[width=0.33\textwidth]{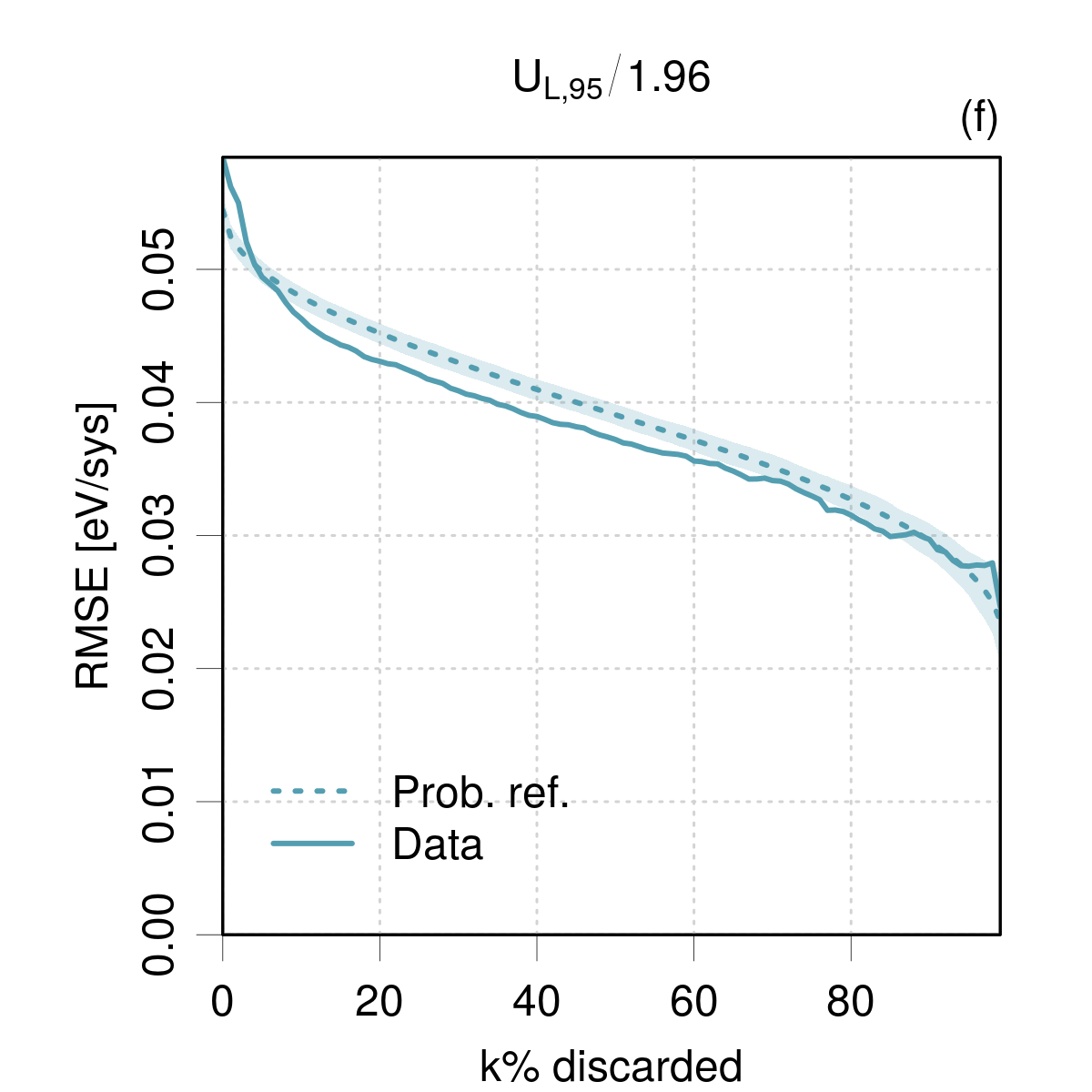}\tabularnewline
\end{tabular}
\par\end{centering}
\caption{\textcolor{orange}{\label{fig:LCP-analysis-of}}Case HU2022: (a) Error
vs $U_{F,95}$; (b) LCP analysis for $U_{F,68}$ and $U_{F,95}$;
(c) confidence curve for $U_{F,95}$; (d-f) idem with $U_{L,P}$.
The LCP analysis uses an adaptive binning with a stating point of
20 bins).}
\end{figure*}

A plot of the errors vs the expanded uncertainties at the 95\,\%
level {[}Fig.\,\ref{fig:LCP-analysis-of}(a,d){]} confirms the features
observed in the original figures: in this plot, the running quantiles
should follow the $y=\pm x$ guide lines, which is not the case for
$U_{F,95}$, while this seems to be much better for $U_{L,95}$, at
least below 0.2\,eV/sys.

To quantify this difference, the data are then analyzed with the LCP
analysis in feature distance space {[}Fig.\,\ref{fig:LCP-analysis-of}(b,e){]}.
As the data are conditioned over the feature distance, one is testing
the adaptivity of the conformal calibration for individual predictions.
Note that for $U_{F,95}$ the distances and uncertainties are proportional,
and the adaptivity diagnostic is also a consistency diagnostic! A
satisfying average coverage is reached in all cases, in agreement
with the conclusions of the original article. However, the local coverage
of $U_{L,P}$ is more homogeneous than for $U_{F,P}$ at both probability
levels. The latter presents strong under- and over-coverage areas.
The adaptivity of $U_{L,P}$ is not far from perfect at the 95\,\%
level, except for the extreme bins, but is not good at the 68\,\%
level. 

It is also worthwhile to look at confidence curves{[}Fig.\,\ref{fig:LCP-analysis-of}(c,f){]}.
The original article presents calibration curves which are very close
to the identity line for both metrics. One can thus confidently transform
the expanded uncertainties at the 95\,\% level to standard uncertainties
using a normal distribution hypothesis. The resulting confidence curves
display a much better behavior for $U_{L}$ than for $U_{F}$. For
active learning, uncertainty built on latent distance looks therefore
much more reliable than when built on feature distance. However, consistency
is not perfect for $U_{L}$. 

A simple scaling of the distance metrics by conformal inference results
in a good average calibration, but it appears unable to ensure consistency/adaptivity.
A better calibration would require to use \emph{conditional conformal
calibration}, which is reputed for its difficulty,\citep{Vovk2012,Barber2019,Angelopoulos2021}
and, as far as I know, has not yet been implemented in the ML-UQ context
considered here. 

\section{Conclusions\label{sec:Conclusion}}

\noindent This article introduces a principled framework for the validation
of UQ metrics. The studied examples target mostly machine learning
in a materials science and chemistry context, but the applicability
is more general. The concept of conditional calibration enables to
define two aspects of local calibration: consistency, which assesses
the reliability of UQ metrics across the range of uncertainty values,
and adaptivity, which assesses the reliability of UQ metrics across
the range of relevant input features. 

The more or less standard UQ validation methods (calibration curves,
reliability diagrams and confidence curves) were recast in the proposed
framework, showing that they are not designed to test adaptivity.
In consequence, adaptivity is presently a blind spot in the ML-UQ
validation practice. It was shown on a few examples that consistency
and adaptivity are scarcely reached by modern ML-UQ methods, and that
a positive consistency diagnostic does not augur of a positive adaptivity
diagnostic. Both aspects of local calibration should be tested. 

\medskip{}

\noindent Let us summarize the main points arising from this study.

\paragraph{Consistency testing.}

\noindent Beside the exploratory ``$E$ vs $u_{E}$'' plots, three
validation methods have some pertinence to test consistency of variance-based
UQ metrics:
\begin{itemize}
\item Error vs Uncertainty plots offer a very informative preliminary analysis,
devoid of any artifacts due to a binning strategy or the choice of
a generative distribution. However, they do not enable to conclude
positively on consistency, which requires more quantitative diagnostics.
\item Reliability diagrams and LZV/LZISD analysis are direct implementation
of the conditional calibration tests. It was shown that the LZV/LZISD
analysis offers a visually more discriminant approach, along with
a direct quantification of possible uncertainty misestimations. Both
methods are sensitive to the choice of a binning scheme, and an adaptive
binning scheme is proposed to compensate for the drawbacks of equal-counts
or equal-width binning strategies.
\item The RMSE-based confidence curve is not designed to check consistency,
but, augmented with a probabilistic reference, it offers an interesting
approach, combining two diagnostics: the eligibility of the UQ method
for a reliable active learning and the consistency of the dataset.
Moreover, it is less dependent on a binning strategy than the reliability
diagrams or LZV/LZISD analysis. An inconvenience is the dependence
of the statistical validation reference curve (more precisely its
confidence band) on the choice of a generative distribution.
\end{itemize}
\noindent I have also shown that conditional calibration curves could
be considered to assess consistency. As for confidence curves, their
disadvantage for variance-based UQ metrics is the difficulty to discern
between a lack of consistency or a bad choice of generative distribution
hypothesis. For the latter point, care has to be taken in particular
for UQ metrics based on small ensembles. 

I would then recommend the Error vs Uncertainty plot, the LZISD analysis
and the RMSE-based confidence curve with its probabilistic reference
as complementary diagnostics for the consistency of variance-based
UQ metrics.

For interval-based UQ metrics the LCP analysis seems to be the best
choice, as it implements directly the conditional calibration test.
Calibration curves have been shown to have possibly ambiguous diagnostics,
which can be improved by using conditional calibration curves. In
order to avoid undue hypotheses about an error generative distribution,
interval-based validation methods should be reserved to interval-based
UQ metrics.

\paragraph{Adaptivity testing.}

Adaptation of the standard methods to test conditional calibration
over some input feature is not ideal but feasible, and one could envision
conditional reliability diagrams, conditional calibration curves,
and even conditional confidence curves. This approach requires large
validation datasets and leads to complex graphs that are not easily
legible. In addition to the Z vs Input feature plots, I would then
recommend the LZISD analysis and the LCP analysis as general tools
to test adaptivity of variance- and interval-based UQ metrics, respectively. 

It was also noted that using the predicted values $V$ as a conditioning
variable to test adaptivity is not without risks, as it might produce
spurious features that complicate the diagnostic. This should probably
be generalized to any property that is susceptible to be strongly
correlated with the errors. Further studies are required to clear
this up.

\bigskip{}

\noindent I hope that the proposed framework, articulated around conditional
calibration, will help ML-UQ researchers to approach UQ validation
with a renewed interest. Up to now, ML-UQ validation studies were
mainly focused on conditional calibration with respect to uncertainty,
i.e. consistency. There is a strong need that adaptivity be also considered,
notably for final users who expect UQ metrics to be reliable throughout
input features space. 

\section*{Acknowledgments}

\noindent I warmly thank J. Busk for providing me the data for the
BUS2022 case, and Y. Hu and A.\,J. Medford for the data in the HU2022
case.

\section*{Author Declarations}

\subsection*{Conflict of Interest}

The authors have no conflicts to disclose.

\section*{Code and data availability\label{sec:Code-and-data}}

\noindent The code and data to reproduce the results of this article
are available at \url{https://github.com/ppernot/2023_Primer/releases/tag/v1.0}
and at Zenodo (\url{https://doi.org/10.5281/zenodo.7731669}). The
\texttt{R},\citep{RTeam2019} \href{https://github.com/ppernot/ErrViewLib}{ErrViewLib}
package implements the validation functions used in the present study,
under version \texttt{ErrViewLib-v1.7.0} (\url{https://github.com/ppernot/ErrViewLib/releases/tag/v1.7.0}),
also available at Zenodo (\url{https://doi.org/10.5281/zenodo.7729100}).\textcolor{orange}{{}
}The \texttt{UncVal} graphical interface to explore the main UQ validation
methods provided by \texttt{ErrViewLib} is also freely available (\url{https://github.com/ppernot/UncVal}).

\bibliographystyle{unsrturlPP}
\bibliography{NN}

\begin{thebibliography}{10}

\bibitem{Vishwakarma2021}
G.~Vishwakarma, A.~Sonpal, and J.~Hachmann.
\newblock \href{http://dx.doi.org/10.1016/j.trechm.2020.12.004}{Metrics for
  {Benchmarking} and {Uncertainty} {Quantification}: {Quality},
  {Applicability}, and {Best} {Practices} for {Machine} {Learning} in
  {Chemistry}}.
\newblock {\em Trends in Chemistry}, 3:146--156, 2021.

\bibitem{Gruich2023}
C.~Gruich, V.~Madhavan, Y.~Wang, and B.~Goldsmith.
\newblock \href{http://dx.doi.org/10.48550/arXiv.2302.02595}{{Clarifying Trust
  of Materials Property Predictions using Neural Networks with
  Distribution-Specific Uncertainty Quantification}}.
\newblock {\em arXiv:2302.02595}, February 2023.

\bibitem{Pearce2018}
T.~Pearce, A.~Brintrup, M.~Zaki, and A.~Neely.
\newblock \href{https://proceedings.mlr.press/v80/pearce18a.html}{{High-Quality
  Prediction Intervals for Deep Learning: A Distribution-Free, Ensembled
  Approach}}.
\newblock In {\em {International Conference on Machine Learning}}, pages
  4075--4084. PMLR, 2018.
\newblock URL: \url{https://proceedings.mlr.press/v80/pearce18a.html}.

\bibitem{Musil2019}
F.~Musil, M.~J. Willatt, M.~A. Langovoy, and M.~Ceriotti.
\newblock \href{http://dx.doi.org/10.1021/acs.jctc.8b00959}{Fast and accurate
  uncertainty estimation in chemical machine learning}.
\newblock {\em J. Chem. Theory Comput.}, 15:906--915, 2019.

\bibitem{Hirschfeld2020}
L.~Hirschfeld, K.~Swanson, K.~Yang, R.~Barzilay, and C.~W. Coley.
\newblock \href{http://dx.doi.org/10.1021/acs.jcim.0c00502}{{Uncertainty
  Quantification Using Neural Networks for Molecular Property Prediction}}.
\newblock {\em J. Chem. Inf. Model.}, 60:3770--3780, 2020.

\bibitem{Tran2020}
K.~Tran, W.~Neiswanger, J.~Yoon, Q.~Zhang, E.~Xing, and Z.~W. Ulissi.
\newblock \href{http://dx.doi.org/10.1088/2632-2153/ab7e1a}{Methods for
  comparing uncertainty quantifications for material property predictions}.
\newblock {\em Mach. Learn.: Sci. Technol.}, 1:025006, 2020.

\bibitem{Abdar2021}
M.~Abdar, F.~Pourpanah, S.~Hussain, D.~Rezazadegan, L.~Liu, M.~Ghavamzadeh,
  P.~Fieguth, X.~Cao, A.~Khosravi, U.~R. Acharya, V.~Makarenkov, and
  S.~Nahavandi.
\newblock \href{http://dx.doi.org/10.1016/j.inffus.2021.05.008}{{A review of
  uncertainty quantification in deep learning: Techniques, applications and
  challenges}}.
\newblock {\em Information Fusion}, 76:243--297, 2021.

\bibitem{Gawlikowski2021}
J.~Gawlikowski, C.~R.~N. Tassi, M.~Ali, J.~Lee, M.~Humt, J.~Feng, A.~Kruspe,
  R.~Triebel, P.~Jung, R.~Roscher, M.~Shahzad, W.~Yang, R.~Bamler, and X.~X.
  Zhu.
\newblock \href{http://arxiv.org/abs/2107.03342}{A {Survey} of {Uncertainty} in
  {Deep} {Neural} {Networks}}.
\newblock {\em arXiv:2107.03342}, 2021.
\newblock arXiv: 2107.03342.
\newblock URL: \url{http://arxiv.org/abs/2107.03342}.

\bibitem{Tynes2021}
M.~Tynes, W.~Gao, D.~J. Burrill, E.~R. Batista, D.~Perez, P.~Yang, and
  N.~Lubbers.
\newblock \href{http://dx.doi.org/10.1021/acs.jcim.1c00670}{Pairwise difference
  regression: A machine learning meta-algorithm for improved prediction and
  uncertainty quantification in chemical search}.
\newblock {\em J. Chem. Inf. Model.}, 61:3846--3857, 2021.
\newblock PMID: 34347460.

\bibitem{Zelikman2021}
E.~Zelikman, C.~Healy, S.~Zhou, and A.~Avati.
\newblock \href{http://arxiv.org/abs/2005.12496}{{CRUDE}: {Calibrating}
  {Regression} {Uncertainty} {Distributions} {Empirically}}.
\newblock {\em arXiv:2005.12496 [cs, stat]}, March 2021.
\newblock arXiv: 2005.12496.
\newblock URL: \url{http://arxiv.org/abs/2005.12496}.

\bibitem{Hu2022}
Y.~Hu, J.~Musielewicz, Z.~W. Ulissi, and A.~J. Medford.
\newblock \href{http://dx.doi.org/10.1088/2632-2153/aca7b1}{{Robust and
  scalable uncertainty estimation with conformal prediction for machine-learned
  interatomic potentials}}.
\newblock {\em Mach. Learn.: Sci. Technol.}, November 2022.

\bibitem{Varivoda2022}
D.~Varivoda, R.~Dong, S.~S. Omee, and J.~Hu.
\newblock \href{http://dx.doi.org/10.48550/arXiv.2211.02235}{{Materials
  Property Prediction with Uncertainty Quantification: A Benchmark Study}}.
\newblock {\em arXiv:2211.02235}, November 2022.

\bibitem{He2023}
W.~He and Z.~Jiang.
\newblock \href{http://dx.doi.org/10.48550/arXiv.2302.13425}{{A Survey on
  Uncertainty Quantification Methods for Deep Neural Networks: An Uncertainty
  Source Perspective}}.
\newblock {\em arXiv:2302.13425}, February 2023.

\bibitem{Liu2021}
Y.~Liu, M.~Pagliardini, T.~Chavdarova, and S.~U. Stich.
\newblock \href{http://dx.doi.org/10.48550/arXiv.2112.05000}{{The Peril of
  Popular Deep Learning Uncertainty Estimation Methods}}.
\newblock {\em arXiv:2112.05000}, 2021.

\bibitem{Pernot2022b}
P.~Pernot.
\newblock \href{http://dx.doi.org/10.1063/5.0109572}{Prediction uncertainty
  validation for computational chemists}.
\newblock {\em J. Chem. Phys.}, 157:144103, 2022.

\bibitem{GUM}
{BIPM}, {IEC}, {IFCC}, {ILAC}, {ISO}, {IUPAC}, {IUPAP}, and {OIML}.
\newblock
  \href{http://www.bipm.org/utils/common/documents/jcgm/JCGM_100_2008_F.pdf}{Evaluation
  of measurement data - {G}uide to the expression of uncertainty in measurement
  ({GUM})}.
\newblock Technical Report 100:2008, Joint Committee for Guides in Metrology,
  JCGM, 2008.
\newblock URL:
  \url{http://www.bipm.org/utils/common/documents/jcgm/JCGM_100_2008_F.pdf}.

\bibitem{Irikura2004}
K.~K. Irikura, R.~D. Johnson, and R.~N. Kacker.
\newblock \href{http://dx.doi.org/10.1088/0026-1394/41/6/003}{{U}ncertainty
  associated with virtual measurements from computational quantum chemistry
  models}.
\newblock {\em Metrologia}, 41:369--375, 2004.

\bibitem{Ruscic2004}
B.~Ruscic, R.~E. Pinzon, M.~L. Morton, G.~von Laszevski, S.~J. Bittner, S.~G.
  Nijsure, K.~A. Amin, M.~Minkoff, and A.~F. Wagner.
\newblock \href{http://dx.doi.org/10.1021/jp047912y}{Introduction to {Active
  Thermochemical Tables}: Several "key" enthalpies of formation revisited}.
\newblock {\em J. Phys. Chem. A}, 108:9979--9997, 2004.

\bibitem{Janet2019}
J.~P. Janet, C.~Duan, T.~Yang, A.~Nandy, and H.~J. Kulik.
\newblock \href{http://dx.doi.org/10.1039/C9SC02298H}{A quantitative
  uncertainty metric controls error in neural network-driven chemical
  discovery}.
\newblock {\em Chem. Sci.}, 10:7913--7922, 2019.

\bibitem{Hullermeier2021}
E.~H\"ullermeier and W.~Waegeman.
\newblock \href{http://dx.doi.org/10.1007/s10994-021-05946-3}{{Aleatoric and
  epistemic uncertainty in machine learning: an introduction to concepts and
  methods}}.
\newblock {\em Mach. Learn.}, 110:457--506, 2021.

\bibitem{Korolev2022}
V.~Korolev, I.~Nevolin, and P.~Protsenko.
\newblock \href{http://dx.doi.org/10.1038/s41598-022-19205-5}{{A universal
  similarity based approach for predictive uncertainty quantification in
  materials science}}.
\newblock {\em Sci. Rep.}, 12:1--10, 2022.

\bibitem{Angelopoulos2021}
A.~N. Angelopoulos and S.~Bates.
\newblock \href{http://dx.doi.org/10.48550/arXiv.2107.07511}{{A Gentle
  Introduction to Conformal Prediction and Distribution-Free Uncertainty
  Quantification}}.
\newblock {\em arXiv:2107.07511}, July 2021.

\bibitem{Vovk2012}
V.~Vovk.
\newblock Conditional validity of inductive conformal predictors.
\newblock In S.~C.~H. Hoi and W.~Buntine, editors, {\em Proceedings of the
  Asian Conference on Machine Learning}, volume~25 of {\em Proceedings of
  Machine Learning Research}, pages 475--490, Singapore Management University,
  Singapore, 04--06 Nov 2012. PMLR.

\bibitem{Reiher2022}
M.~Reiher.
\newblock
  \href{http://dx.doi.org/https://doi.org/10.1002/ijch.202100101}{Molecule-specific
  uncertainty quantification in quantum chemical studies}.
\newblock {\em Isr. J. Chem.}, 62(1-2):e202100101, 2022.

\bibitem{Pernot2022c}
P.~Pernot.
\newblock \href{http://dx.doi.org/10.48550/arXiv.2206.15272}{{Confidence curves
  for UQ validation: probabilistic reference vs. oracle}}.
\newblock {\em arXiv:2206.15272}, June 2022.

\bibitem{Gneiting2007a}
T.~Gneiting, F.~Balabdaoui, and A.~E. Raftery.
\newblock
  \href{http://dx.doi.org/https://doi.org/10.1111/j.1467-9868.2007.00587.x}{Probabilistic
  forecasts, calibration and sharpness}.
\newblock {\em J. R. Statist. Soc. B}, 69:243--268, 2007.

\bibitem{Gneiting2014}
T.~Gneiting and M.~Katzfuss.
\newblock
  \href{http://dx.doi.org/10.1146/annurev-statistics-062713-085831}{Probabilistic
  forecasting}.
\newblock {\em Annu. Rev. Stat. Appl.}, 1:125--151, 2014.

\bibitem{Pernot2022a}
P.~Pernot.
\newblock \href{http://dx.doi.org/10.1063/5.0084302}{The long road to
  calibrated prediction uncertainty in computational chemistry}.
\newblock {\em J. Chem. Phys.}, 156:114109, 2022.

\bibitem{Kuleshov2018}
V.~Kuleshov, N.~Fenner, and S.~Ermon.
\newblock \href{https://proceedings.mlr.press/v80/kuleshov18a.html}{Accurate
  uncertainties for deep learning using calibrated regression}.
\newblock In J.~Dy and A.~Krause, editors, {\em Proceedings of the 35th
  International Conference on Machine Learning}, volume~80 of {\em Proceedings
  of Machine Learning Research}, pages 2796--2804. PMLR, 10--15 Jul 2018.
\newblock URL: \url{https://proceedings.mlr.press/v80/kuleshov18a.html}.

\bibitem{Levi2020}
D.~Levi, L.~Gispan, N.~Giladi, and E.~Fetaya.
\newblock \href{http://arxiv.org/abs/1905.11659}{Evaluating and {Calibrating}
  {Uncertainty} {Prediction} in {Regression} {Tasks}}.
\newblock {\em arXiv:1905.11659}, 2020.
\newblock URL: \url{http://arxiv.org/abs/1905.11659}.

\bibitem{Chung2020}
Y.~Chung, W.~Neiswanger, I.~Char, and J.~Schneider.
\newblock \href{http://dx.doi.org/10.48550/ARXIV.2011.09588}{Beyond pinball
  loss: Quantile methods for calibrated uncertainty quantification}.
\newblock {\em arXiv:2011.09588}, 2020.

\bibitem{Laves2020}
M.-H. Laves, S.~Ihler, J.~F. Fast, L.~A. Kahrs, and T.~Ortmaier.
\newblock
  \href{https://proceedings.mlr.press/v121/laves20a.html}{Well-calibrated
  regression uncertainty in medical imaging with deep learning}.
\newblock In T.~Arbel, I.~Ben~Ayed, M.~de~Bruijne, M.~Descoteaux, H.~Lombaert,
  and C.~Pal, editors, {\em Proceedings of the Third Conference on Medical
  Imaging with Deep Learning}, volume 121 of {\em Proceedings of Machine
  Learning Research}, pages 393--412. PMLR, 06--08 Jul 2020.
\newblock URL: \url{https://proceedings.mlr.press/v121/laves20a.html}.

\bibitem{Luo2021}
R.~Luo, A.~Bhatnagar, Y.~Bai, S.~Zhao, H.~Wang, C.~Xiong, S.~Savarese,
  S.~Ermon, E.~Schmerling, and M.~Pavone.
\newblock \href{http://dx.doi.org/10.48550/arXiv.2102.10809}{{Local
  Calibration: Metrics and Recalibration}}.
\newblock {\em arXiv:2102.10809}, February 2021.

\bibitem{Ilg2018}
E.~Ilg, O.~\c{C}i\c{c}ek, S.~Galesso, A.~Klein, O.~Makansi, F.~Hutter, and
  T.~Brox.
\newblock \href{http://dx.doi.org/10.48550/ARXIV.1802.07095}{Uncertainty
  estimates and multi-hypotheses networks for optical flow}.
\newblock {\em arXiv:1802.07095}, 2018.
\newblock URL: \url{https://arxiv.org/abs/1802.07095}.

\bibitem{Scalia2020}
G.~Scalia, C.~A. Grambow, B.~Pernici, Y.-P. Li, and W.~H. Green.
\newblock \href{http://dx.doi.org/10.1021/acs.jcim.9b00975}{Evaluating scalable
  uncertainty estimation methods for deep learning-based molecular property
  prediction}.
\newblock {\em J. Chem. Inf. Model.}, 60:2697--2717, 2020.

\bibitem{Kacker2010}
R.~N. Kacker, R.~Kessel, and K.-D. Sommer.
\newblock \href{http://dx.doi.org/10.6028/jres.115.031}{Assessing differences
  between results determined according to the guide to the expression of
  uncertainty in measurement}.
\newblock {\em J. Res. Nat. Inst. Stand. Technol.}, 115(6):453, 2010.

\bibitem{Cauchois2021}
M.~Cauchois, S.~Gupta, and J.~C. Duchi.
\newblock \href{http://dx.doi.org/10.5555/3546258.3546339}{{Knowing what you
  know: valid and validated confidence sets in multiclass and multilabel
  prediction}}.
\newblock {\em J. Mach. Learn. Res.}, 22:3681--3722, 2021.

\bibitem{Feldman2021}
S.~Feldman, S.~Bates, and Y.~Romano.
\newblock \href{http://dx.doi.org/10.48550/arXiv.2106.00394}{{Improving
  Conditional Coverage via Orthogonal Quantile Regression}}.
\newblock {\em arXiv:2106.00394}, 2021.

\bibitem{Note1}
Note that this definition differs from the one in my previous studies, which
  were based on the existence of a predicted value and an expanded uncertainty,
  leading to zero-centered intervals that should contain the error. In the
  present case the intervals are error-centered, but should contain 0.

\bibitem{Guo2017}
C.~Guo, G.~Pleiss, Y.~Sun, and K.~Q. Weinberger.
\newblock \href{https://proceedings.mlr.press/v70/guo17a.html}{{On Calibration
  of Modern Neural Networks}}.
\newblock In {\em {International Conference on Machine Learning}}, pages
  1321--1330. 2017.
\newblock URL: \url{https://proceedings.mlr.press/v70/guo17a.html}.

\bibitem{Palmer2022}
G.~Palmer, S.~Du, A.~Politowicz, J.~P. Emory, X.~Yang, A.~Gautam, G.~Gupta,
  Z.~Li, R.~Jacobs, and D.~Morgan.
\newblock \href{http://dx.doi.org/10.1038/s41524-022-00794-8}{{Calibration
  after bootstrap for accurate uncertainty quantification in regression
  models}}.
\newblock {\em npj Comput. Mater.}, 8:1--9, 2022.

\bibitem{Vazquez-Salazar2022}
L.~I. Vazquez-Salazar, E.~D. Boittier, and M.~Meuwly.
\newblock \href{http://dx.doi.org/10.48550/arXiv.2207.06916}{{Uncertainty
  quantification for predictions of atomistic neural networks}}.
\newblock {\em arXiv:2207.06916}, July 2022.

\bibitem{Nixon2019}
J.~Nixon, M.~Dusenberry, G.~Jerfel, T.~Nguyen, J.~Liu, L.~Zhang, and D.~Tran.
\newblock \href{http://dx.doi.org/10.48550/arXiv.1904.01685}{{Measuring
  Calibration in Deep Learning}}.
\newblock {\em arXiv:1904.01685}, April 2019.

\bibitem{Maupin2018}
K.~A. Maupin, L.~P. Swiler, and N.~W. Porter.
\newblock \href{http://dx.doi.org/10.1115/1.4042443}{{Validation Metrics for
  Deterministic and Probabilistic Data}}.
\newblock {\em J. Verif. Validation Uncertainty Quantif.}, 3:031002, 2018.

\bibitem{Busk2022}
J.~Busk, P.~B. J{\o}rgensen, A.~Bhowmik, M.~N. Schmidt, O.~Winther, and
  T.~Vegge.
\newblock \href{http://dx.doi.org/10.1088/2632-2153/ac3eb3}{Calibrated
  uncertainty for molecular property prediction using ensembles of message
  passing neural networks}.
\newblock {\em Mach. Learn.: Sci. Technol.}, 3:015012, 2022.

\bibitem{Soleimany2021}
A.~P. Soleimany, A.~Amini, S.~Goldman, D.~Rus, S.~N. Bhatia, and C.~W. Coley.
\newblock \href{http://dx.doi.org/10.1021/acscentsci.1c00546}{{Evidential Deep
  Learning for Guided Molecular Property Prediction and Discovery}}.
\newblock {\em ACS Cent. Sci.}, 7:1356--1367, 2021.

\bibitem{Barber2019}
R.~F. Barber, E.~J. Cand\`{e}s, A.~Ramdas, and R.~J. Tibshirani.
\newblock \href{http://dx.doi.org/10.48550/arXiv.1903.04684}{{The limits of
  distribution-free conditional predictive inference}}.
\newblock {\em arXiv:1903.04684}, 2019.

\bibitem{RTeam2019}
{R Core Team}.
\newblock \href{http://www.R-project.org/}{{\em {R}: {A} {L}anguage and
  {E}nvironment for {S}tatistical {C}omputing}}.
\newblock R Foundation for Statistical Computing, Vienna, Austria, 2019.
\newblock URL: \url{http://www.R-project.org/}.

\end{thebibliography}

\appendix

\section*{Appendices}

\beginappendix

\section{Using the predicted value to test adaptivity\label{sec:Using-the-predicted}}

A posteriori validation of adaptivity on datasets that do not contain
input features $X$ might still be attempted by using the predicted
values $V$ as proxy. Conditional calibration on $V$ is not necessarily
identical to conditional calibration on $X$, but it might still provide
useful diagnostics.

\begin{figure*}[p]
\noindent \begin{centering}
\begin{tabular}{ccc}
\includegraphics[width=0.33\textwidth]{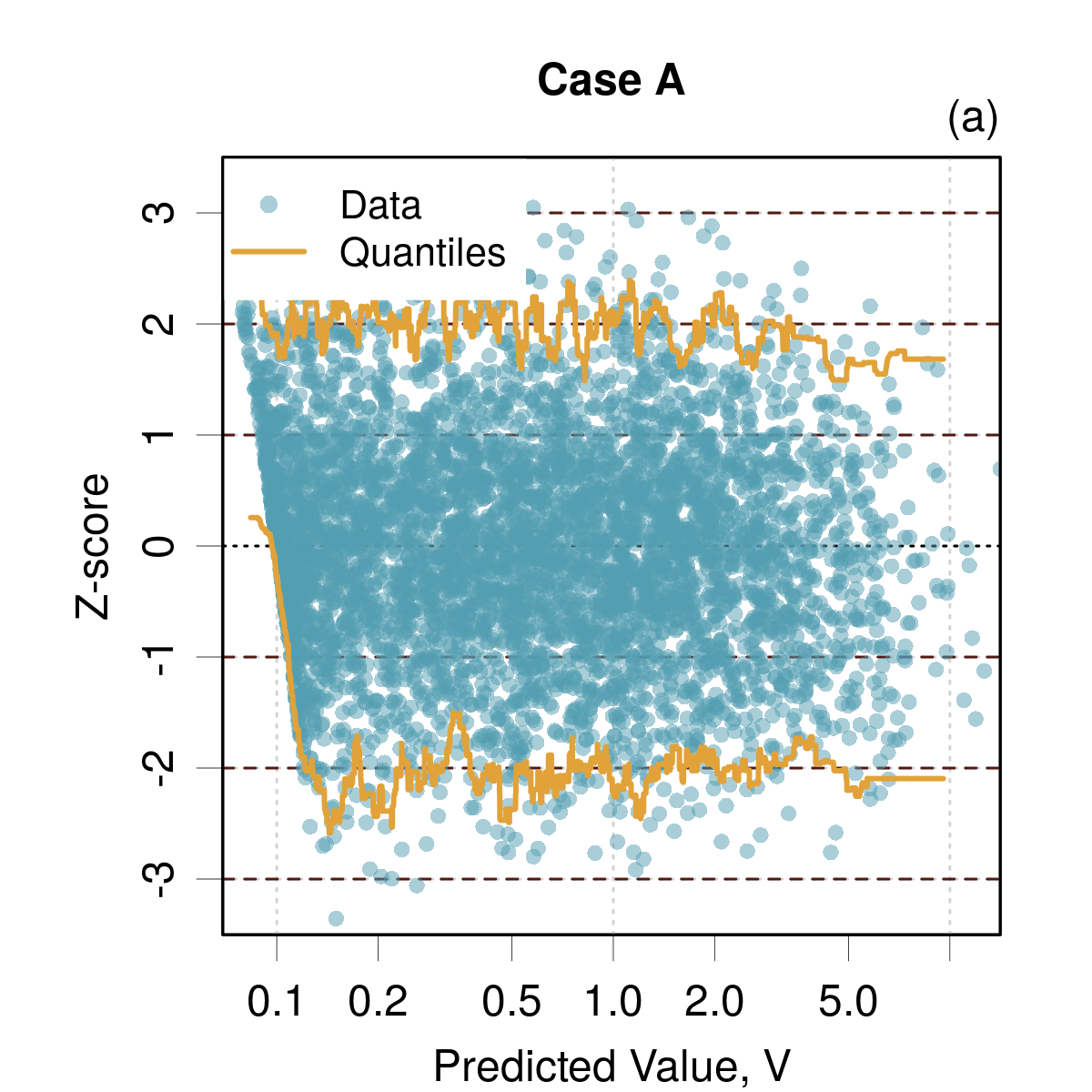} & \includegraphics[width=0.33\textwidth]{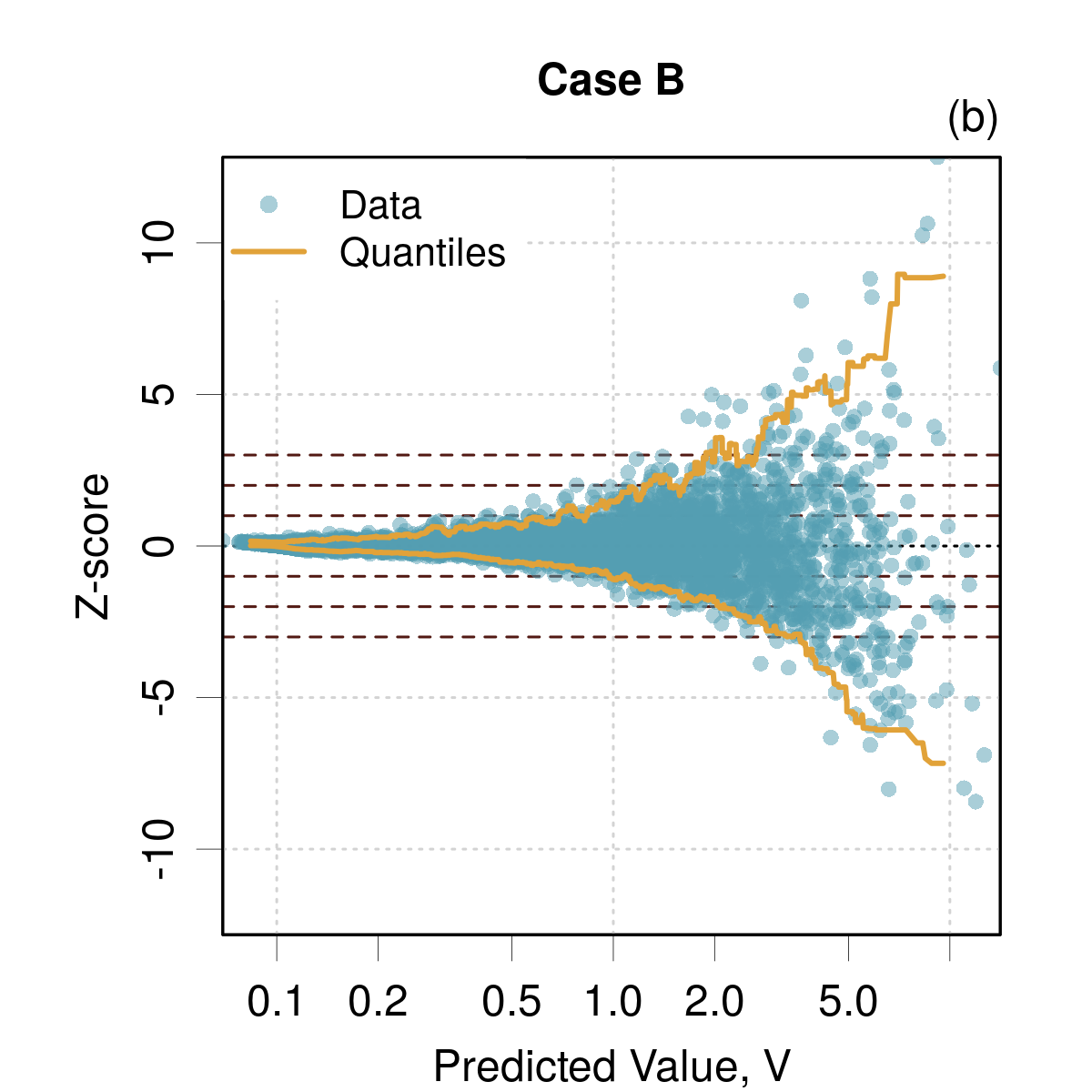} & \includegraphics[width=0.33\textwidth]{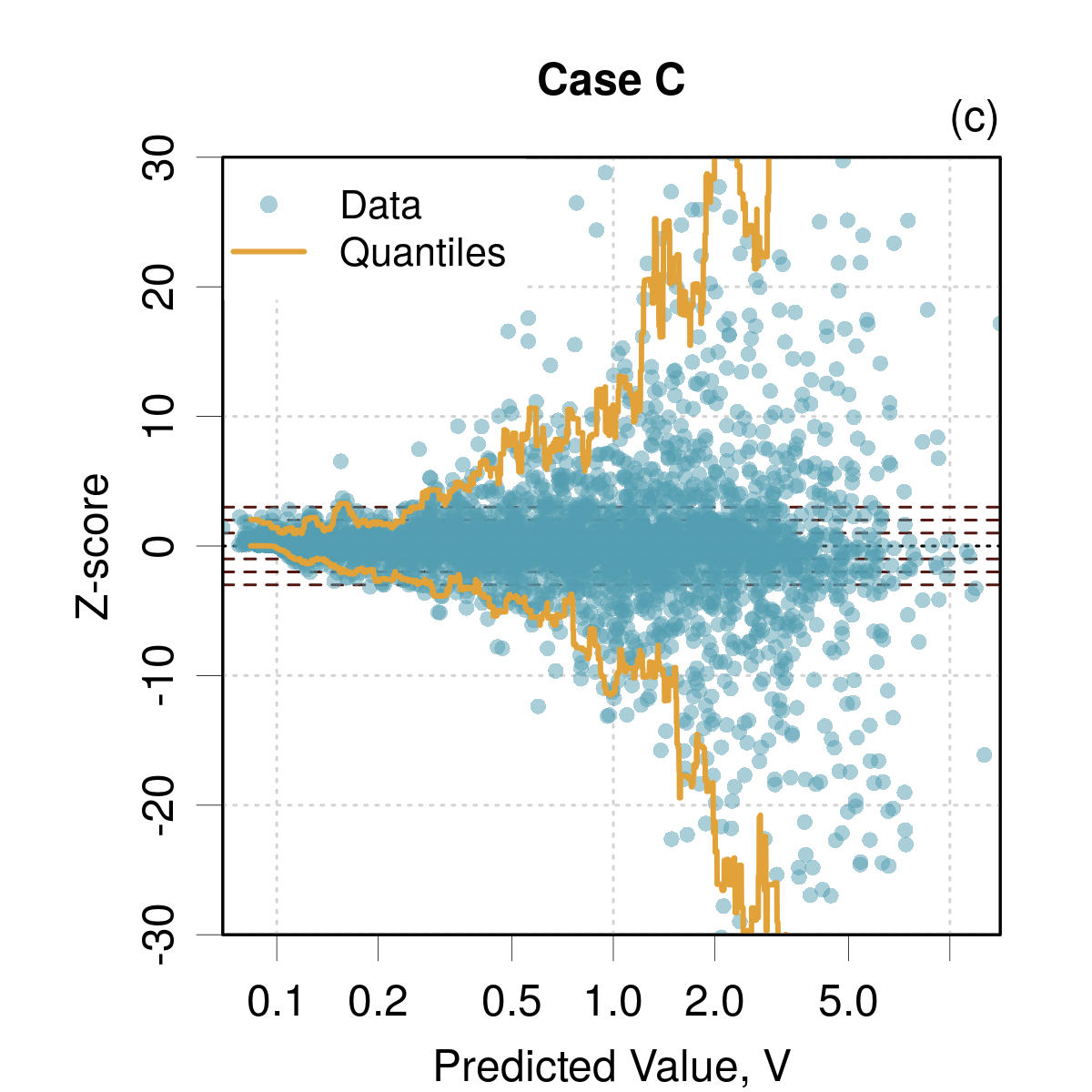}\tabularnewline
\includegraphics[width=0.33\textwidth]{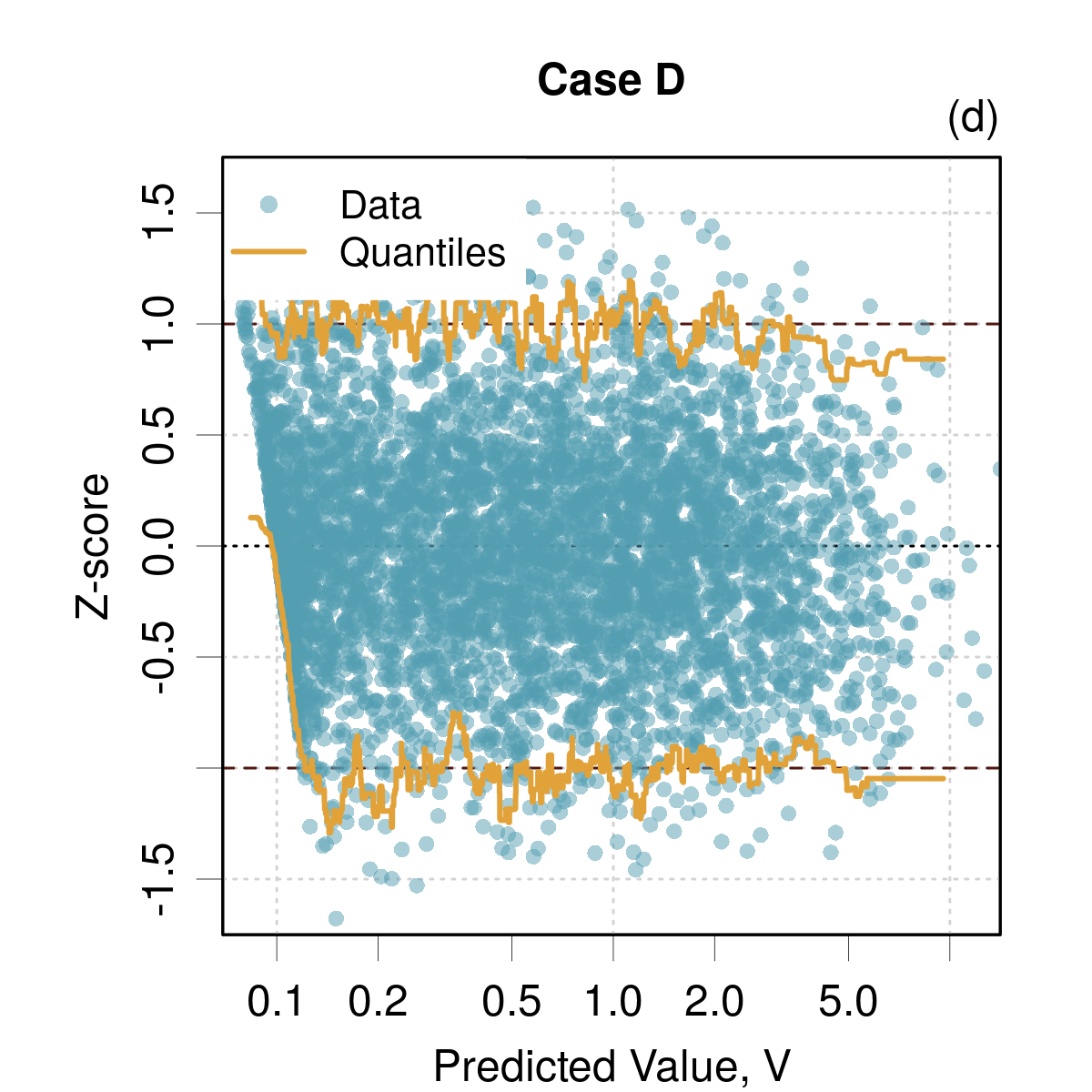} & \includegraphics[width=0.33\textwidth]{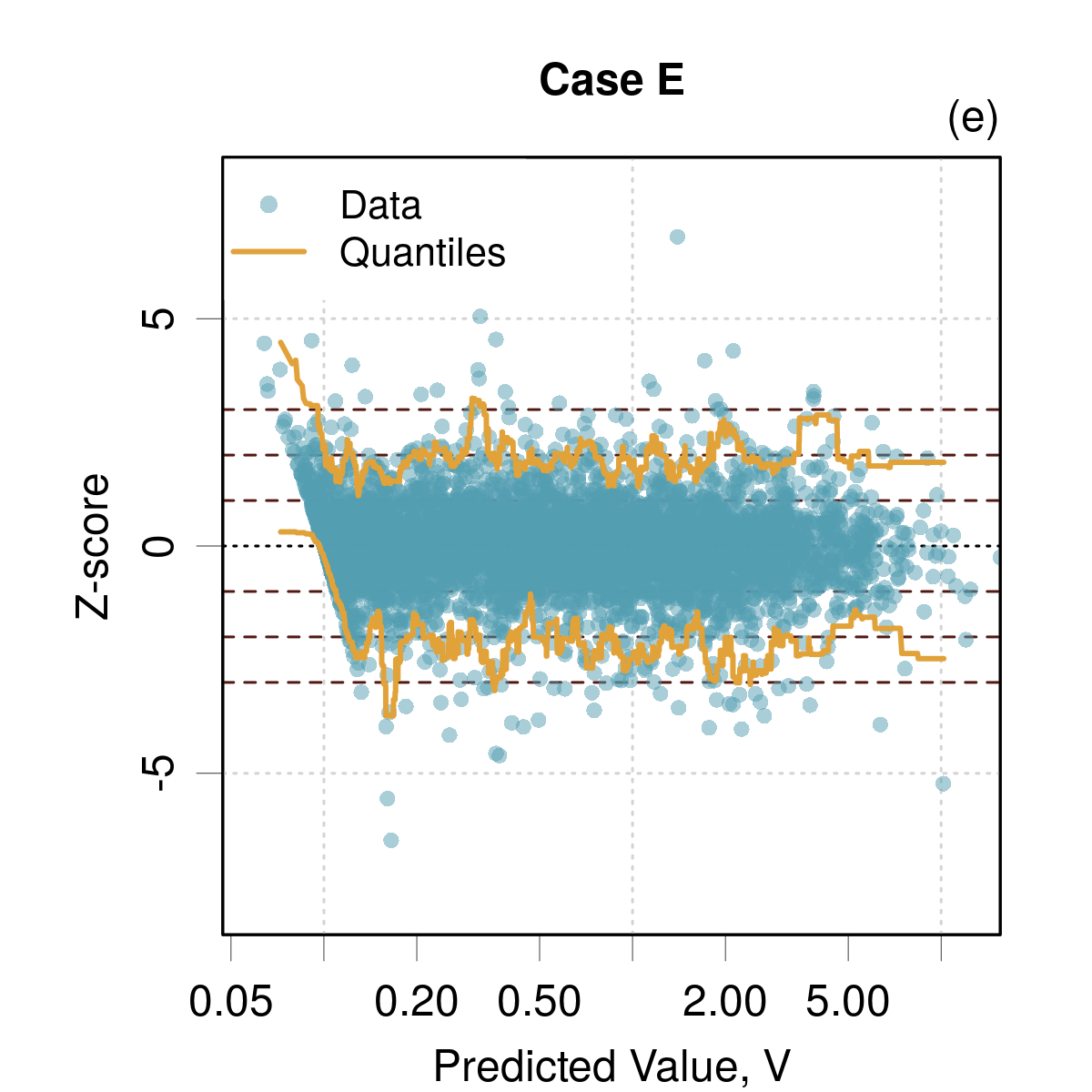} & \includegraphics[width=0.33\textwidth]{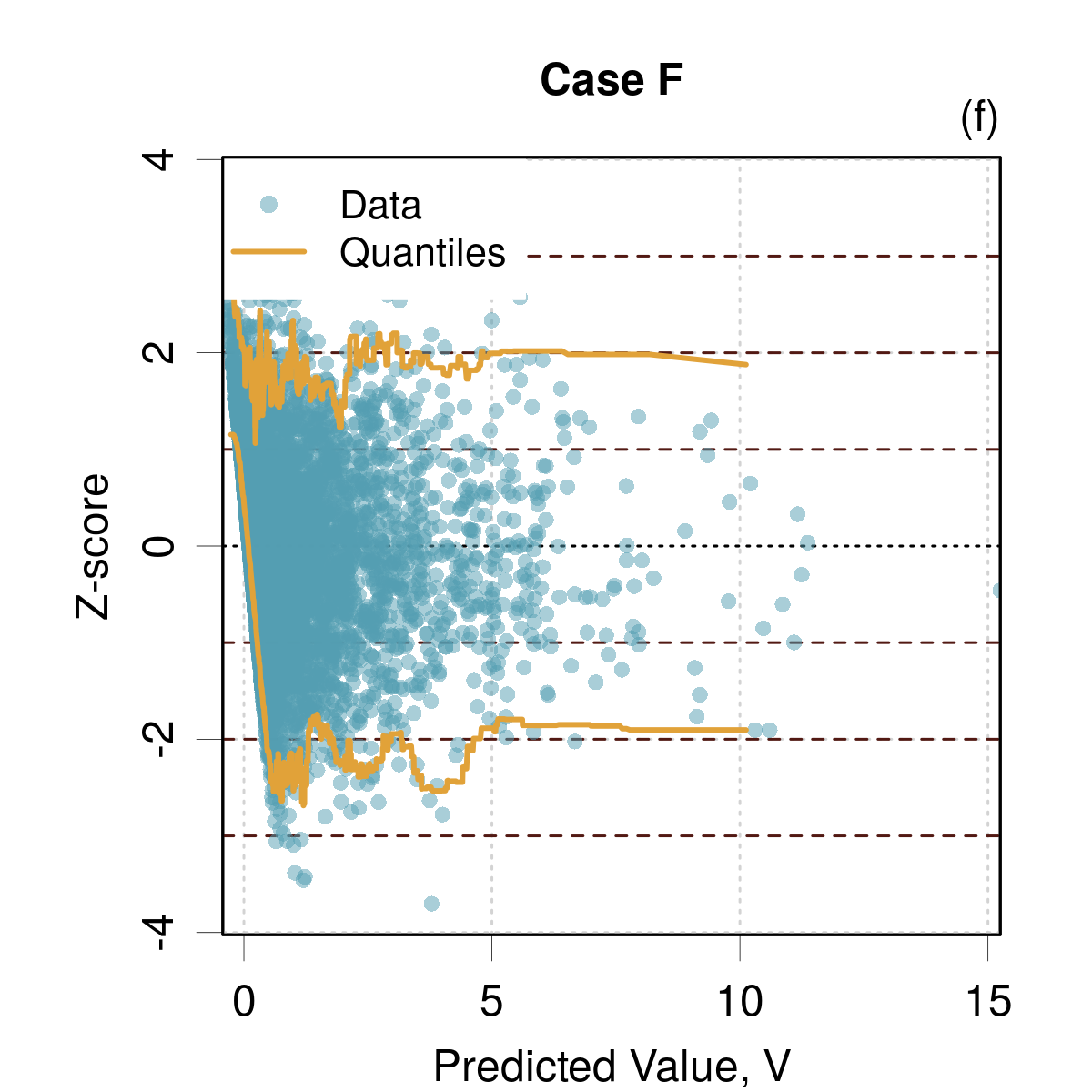}\tabularnewline
\end{tabular}
\par\end{centering}
\noindent \begin{centering}
\begin{tabular}{ccc}
\includegraphics[width=0.33\textwidth]{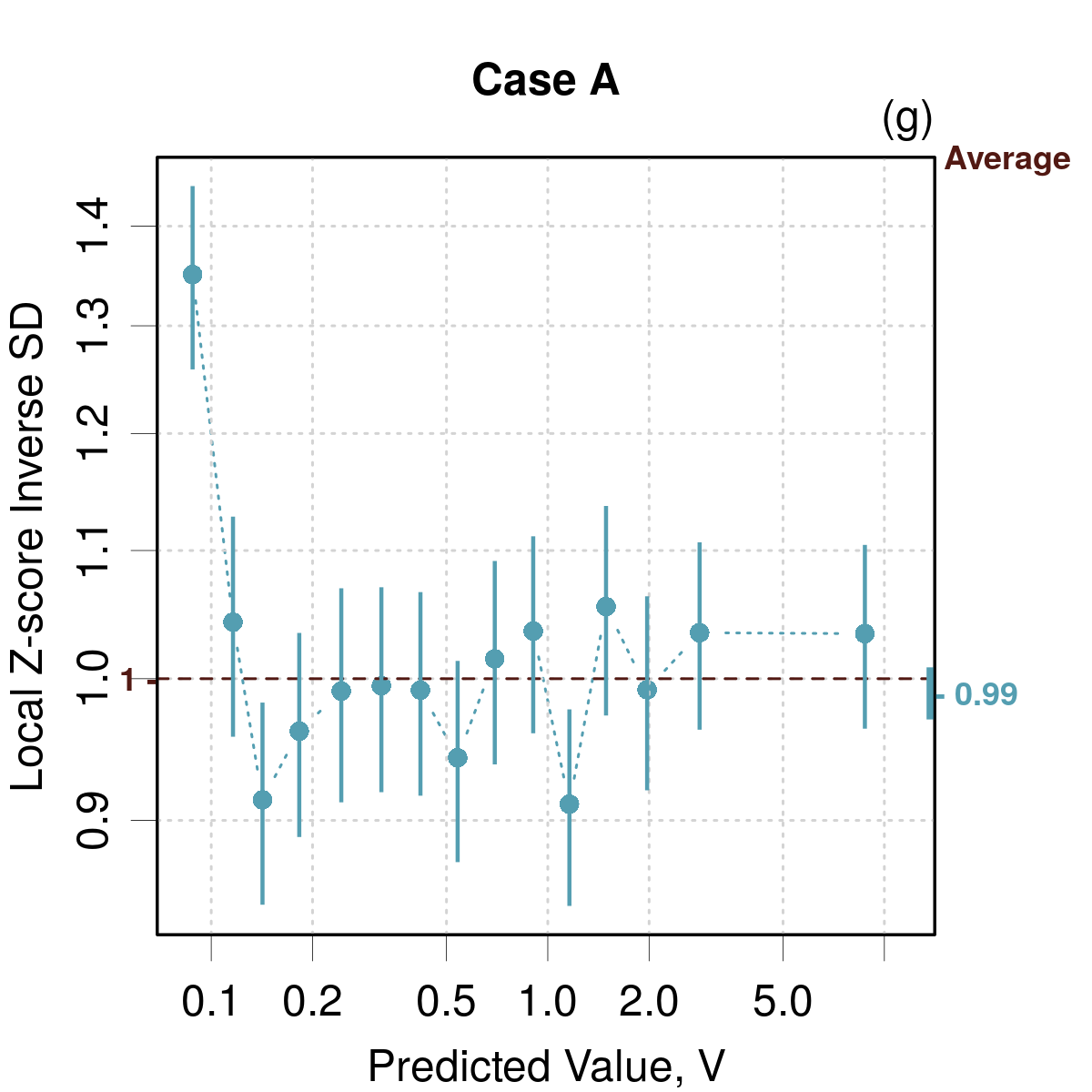} & \includegraphics[width=0.33\textwidth]{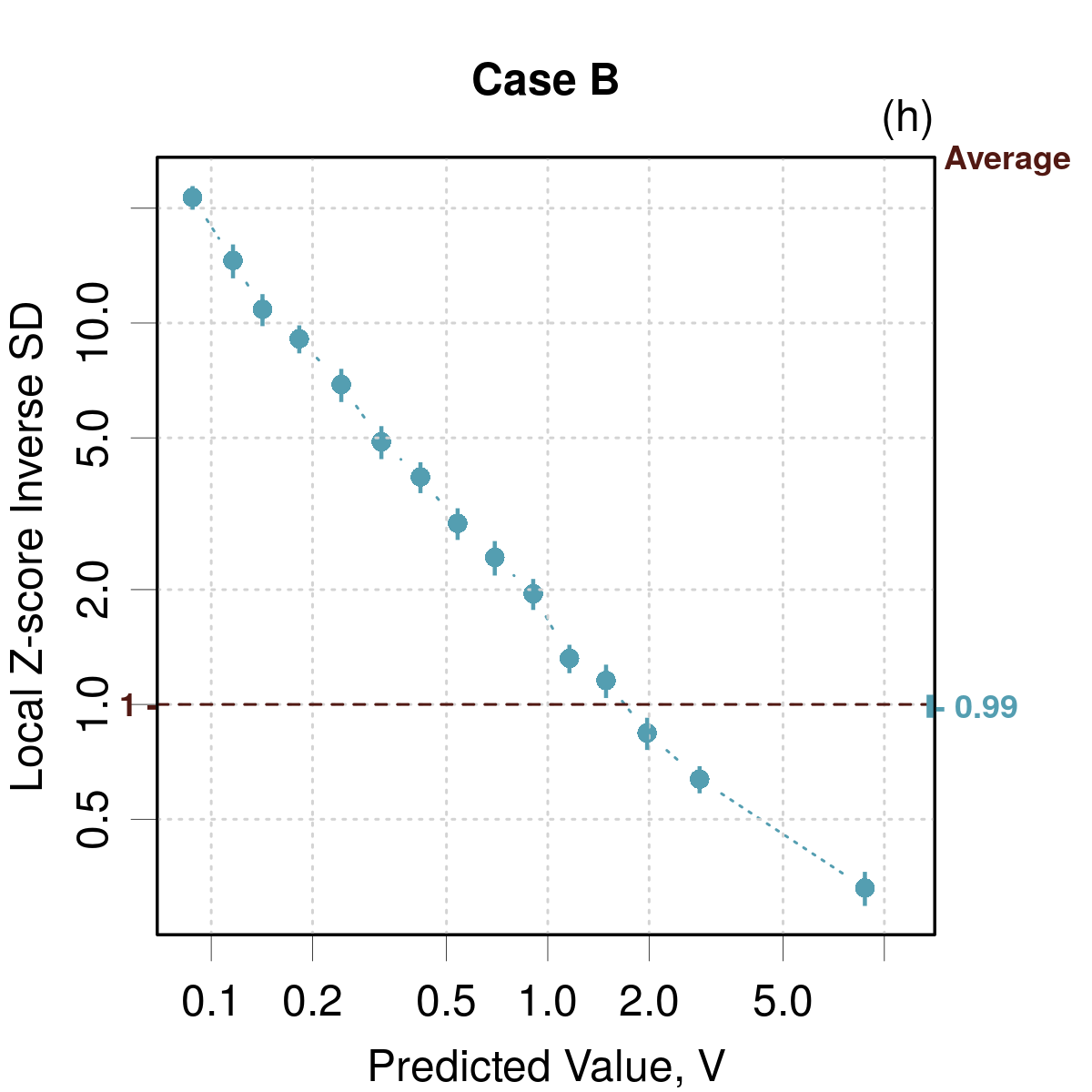} & \includegraphics[width=0.33\textwidth]{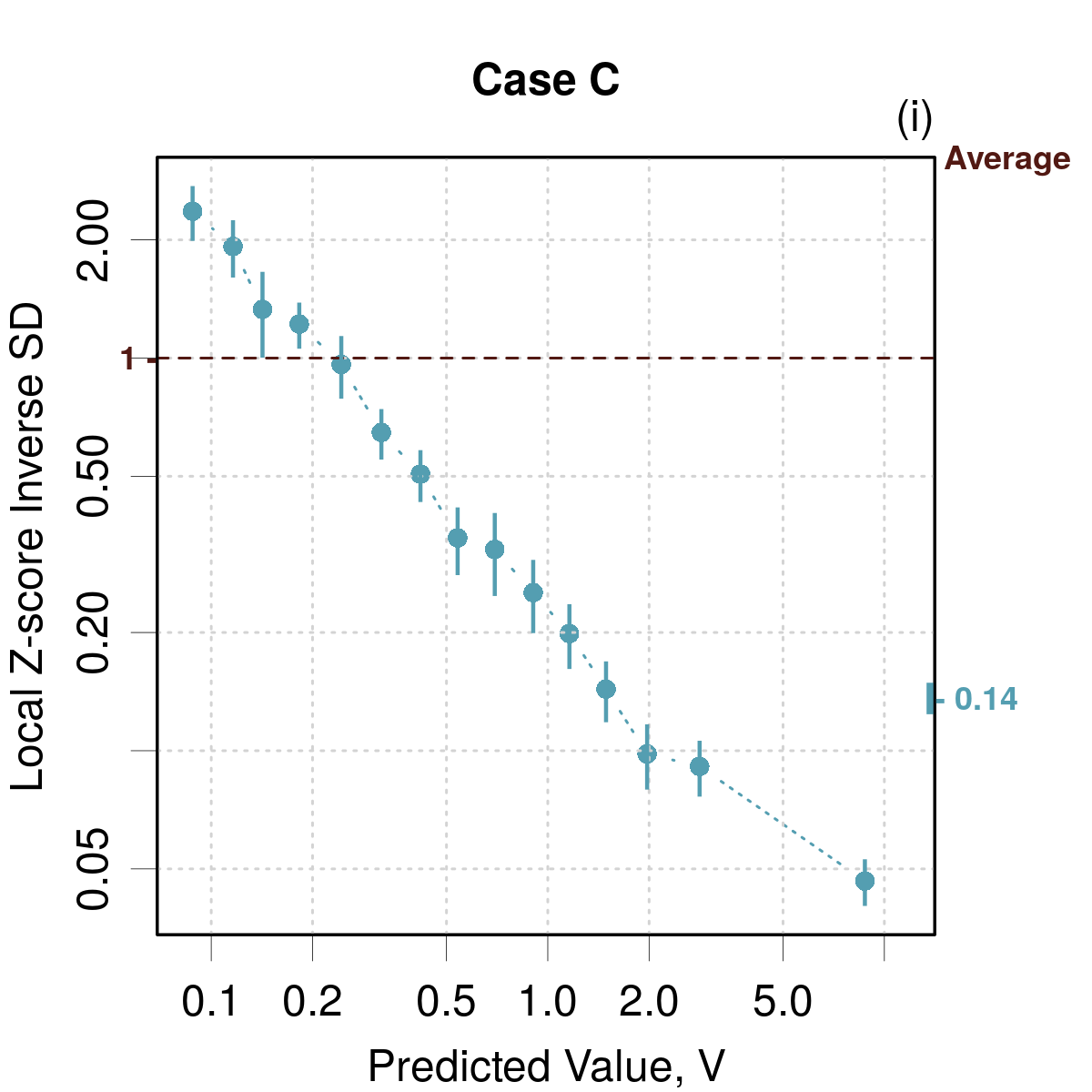}\tabularnewline
\includegraphics[width=0.33\textwidth]{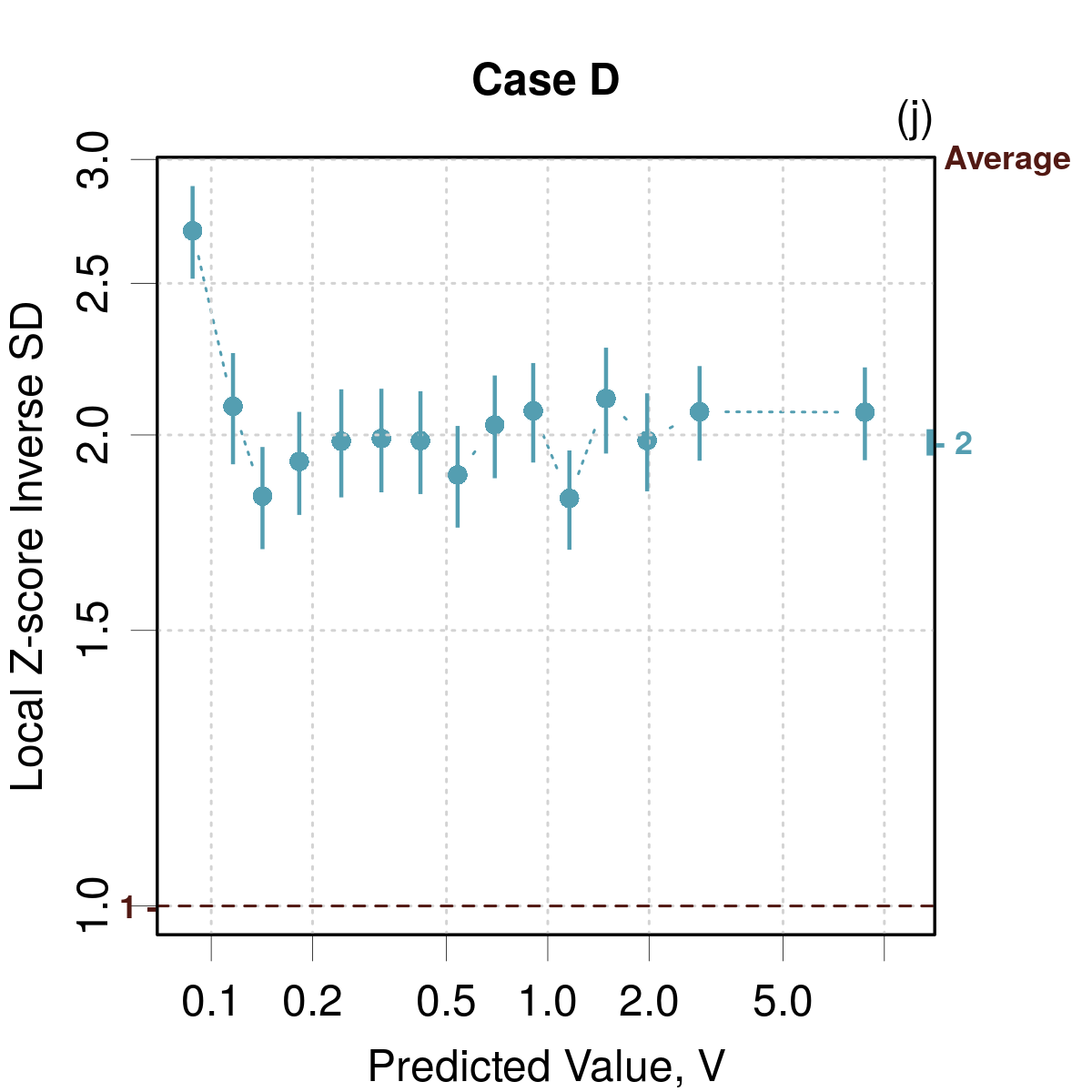} & \includegraphics[width=0.33\textwidth]{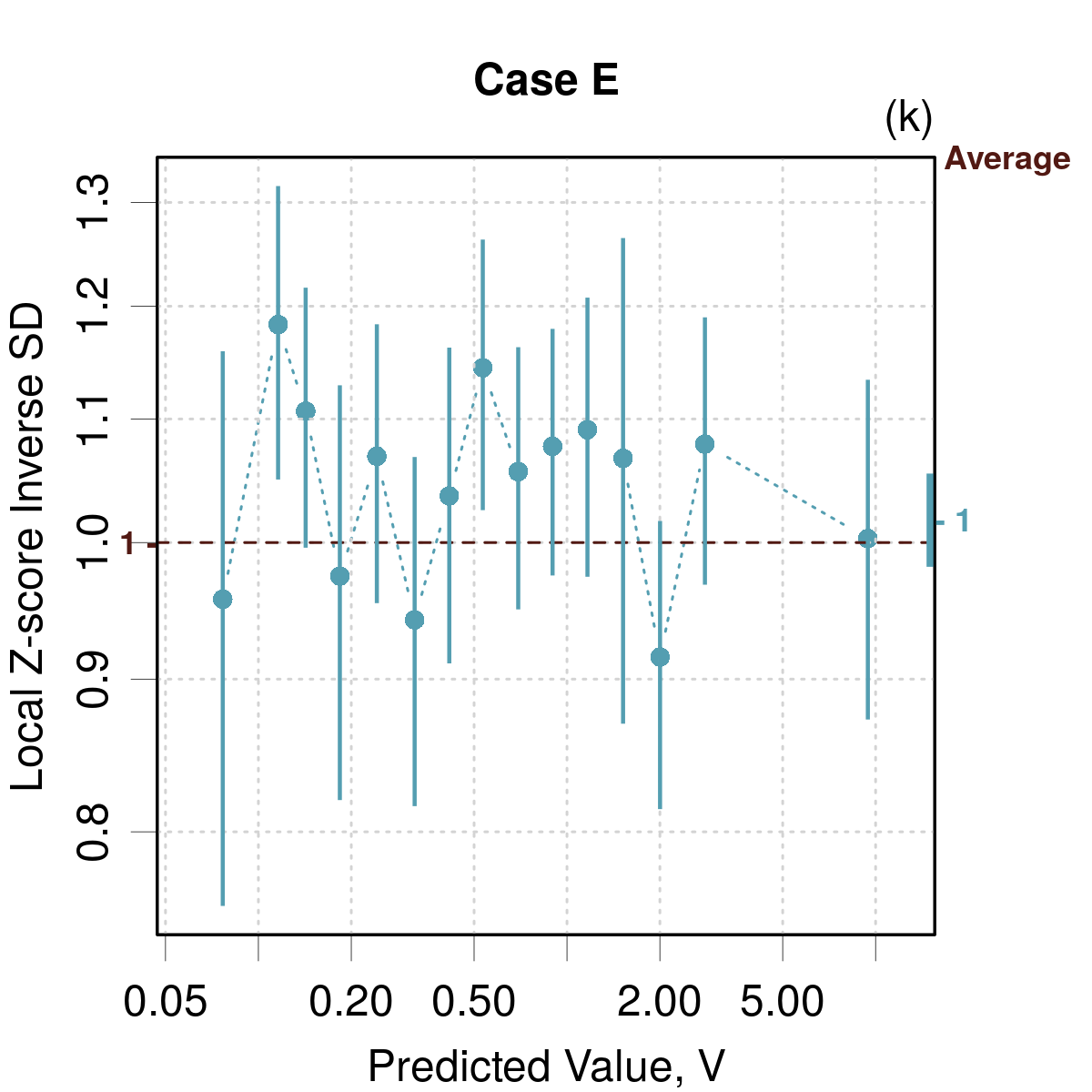} & \includegraphics[width=0.33\textwidth]{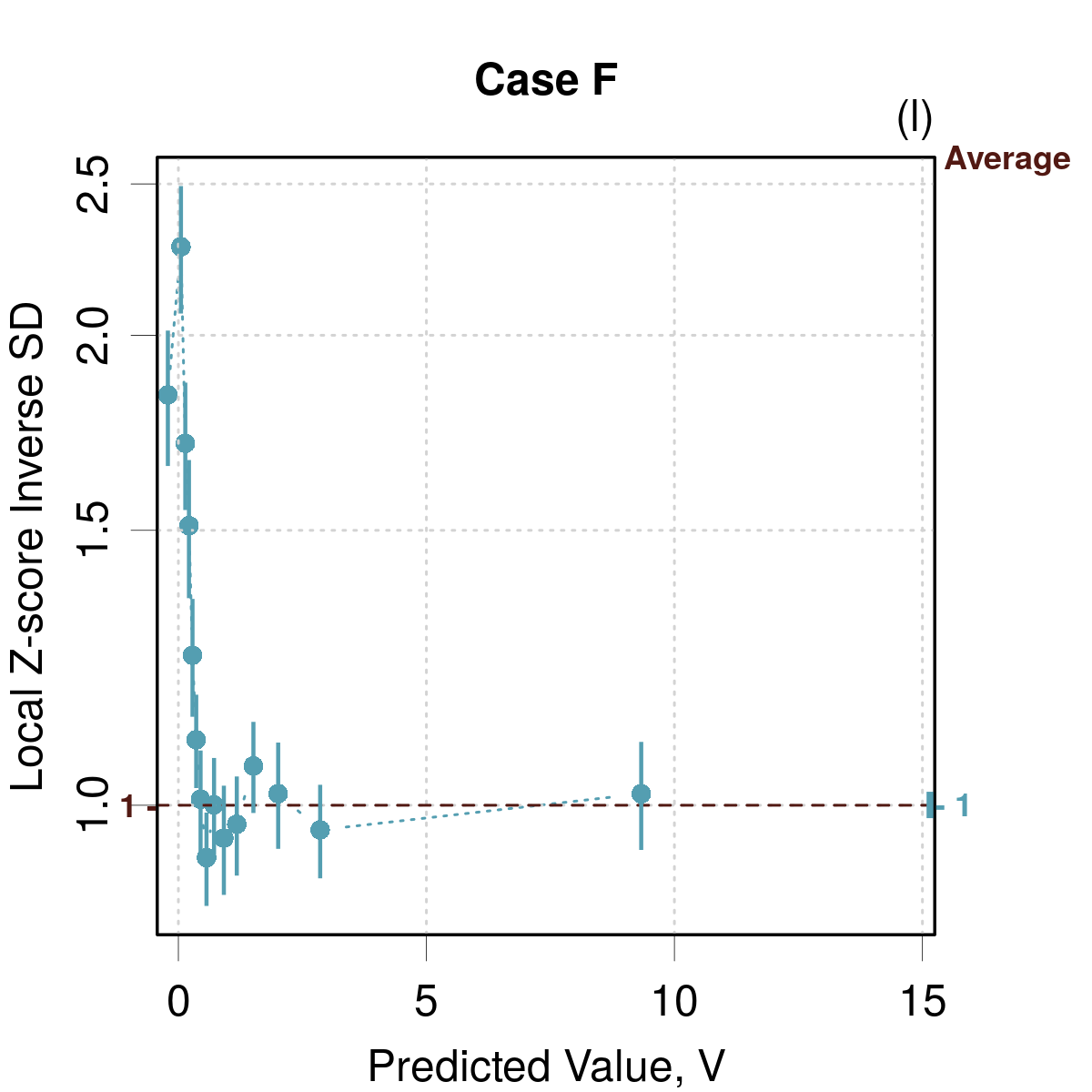}\tabularnewline
\end{tabular}
\par\end{centering}
\caption{\label{fig:validSynth-1}Validation of adaptivity vs predicted values
$V$ for synthetic datasets: (a-f) ''$Z$ vs $V$'' plots; (g-l)
LZISD analysis.}
\end{figure*}

Application of ``$Z$ vs $V$'' plots to the synthetic datasets
is presented in Fig.\ref{fig:validSynth-1}(a-f). Z-score values in
Case A are globally well distributed (horizontal running quantiles
around $\pm2$, but present an anomaly at small $V$ values. This
can be traced back to the correlation of $V$ with $E$ ($E=R-V$),
most visible when $R$ is nearly constant at the bottom of the parabolic
model used to generate the data. This warns us that conditioning on
$V$ might produce some aberrant features, which are not diagnostic
of poor adaptivity. The same artifact can be observed in Cases D,
E and F. For Case E, the excessive dispersion of points, with $|Z|$
values above 5, points to a non-normal distribution, but this should
not be a valid reason to reject adaptivity: a more quantitative analysis
is mandatory. By contrast, the absence of adaptivity is readily visible
for cases B, C, and D : cases B and C present heterogeneous distributions
along $V$, while Case D has $Z$ values mostly contained between
-1 and 1, pointing to an homogeneous overestimation of uncertainties.
Case F is similar to Case A. 

The translation of these observations to the LZISD analysis\textcolor{violet}{{}
}is presented in Fig.\,\ref{fig:validSynth-1}(g-l). The artifact
appears as local deviations of the LZISD values, while the intrinsic
non-adaptivity of Cases B and C is more global. Similar observations
can be made on the conditional calibration curves and LCP analyses
(not shown). 

This shows that using $V$ as a substitute for $X$ to test adaptivity
might lead to ambiguous conclusions when local anomalies are observed.
However, global anomalies are certainly diagnostic of a lack of adaptivity
of the tested UQ metric.

\clearpage{}

\section{Impact of the binning strategy\label{sec:PAL2022---Additional}}

\noindent To implement the LZV analysis, the default binning strategy
is based on intervals with (nearly) equal populations.\citep{Pernot2022a}
This ensures equal testing power in all intervals, but might result
in intervals with very different ranges. This is also often the strategy
chosen for reliability diagrams.\citep{Scalia2020} However, some
authors prefer to use bins with similar widths,\citep{Palmer2022}
a strategy which faces two problems: bins with insufficient population
to derive reliable confidence intervals, and bins containing a large
part of the total sample for uncertainty distributions with large
tails. 

These problems were addressed by Palmer \emph{et al.}\citep{Palmer2022}
by labeling the bins with less than 30 points as unreliable and by
altering the width of bins in the peak area to ensure that the lowest
90\% of the values were spread across at least five bins (assuming
that uncertainty distributions can only have a heavy \emph{upper}
tail). The use of this strategy leads to very regular reliability
diagrams with deviant points mostly in the upper range of uncertainty. 

I present here an alternative binning strategy, considering that for
a positive variable such as uncertainty, a log-transform might be
an efficient way to reduce the skewness of the distribution causing
the accumulation problem in a few bins. Starting from a regular grid
in log space ($n$ bins), the problem of small counts is solved by
merging adjacent bins with insufficient populations, while the problem
of excessively large counts is solved by splitting bins having more
than $M/n$ points ($M$ is the dataset size). Both operations are
iterated until the bins population conforms with the chosen limits
(the lower limit is set to 30). 

Two examples below, taken from Palmer \emph{et al.}\citep{Palmer2022}
(see Sect.\,\ref{subsec:Case-PAL2022}), show that this adaptive
strategy is more efficient to reveal consistency problems. The Diffusion/LR
and Perovskite/GPR\_Bayesian datasets were analyzed by reliability
diagrams and LZISD analysis for increasing number of bins (10, 20
and 40) for the equal-counts bins and adaptive strategies. The reliability
diagrams have also to be compared with those of the original article,
based on 15 equal-width bins.

\begin{figure*}[t]
\noindent \begin{centering}
\begin{tabular}{ccc}
\includegraphics[width=0.33\textwidth]{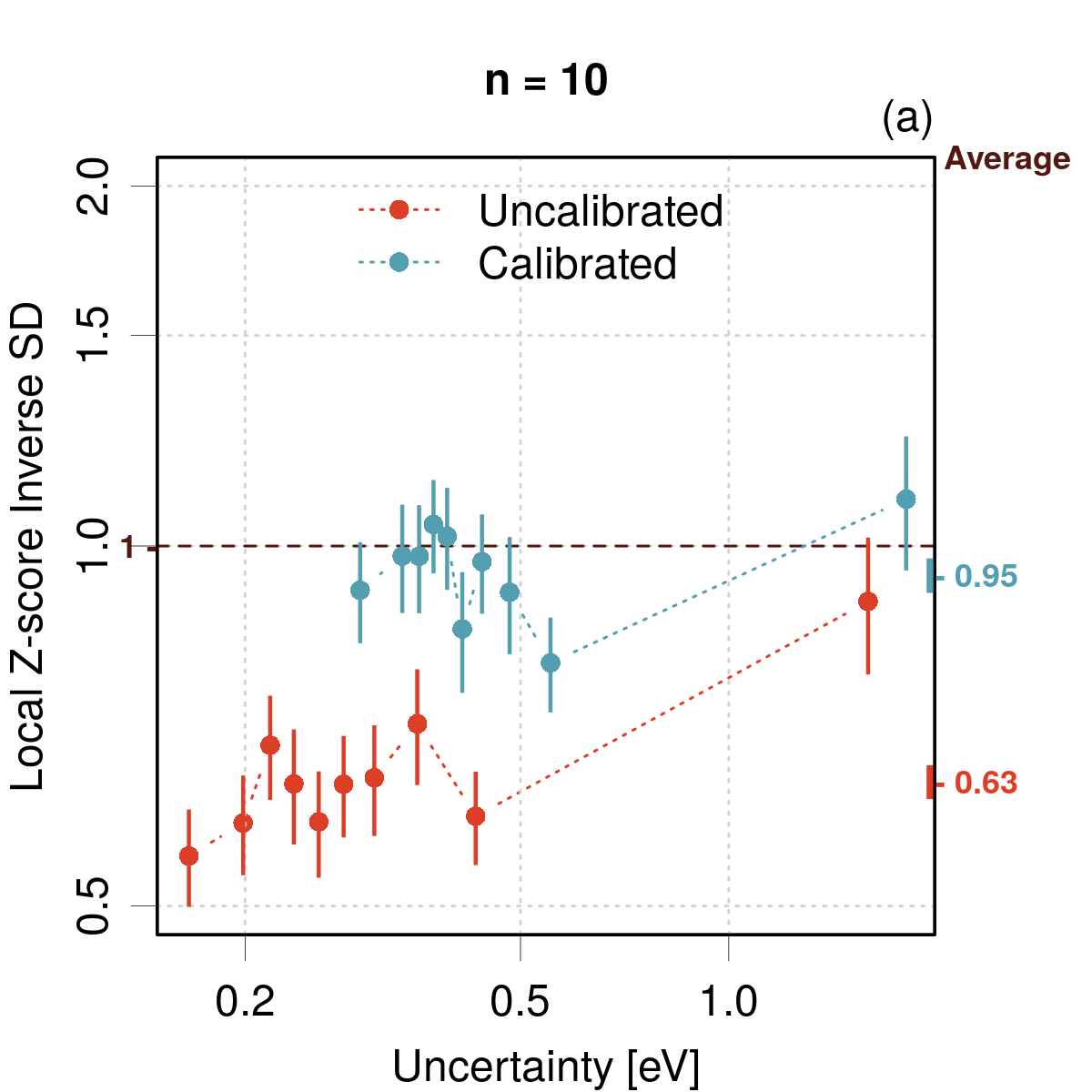} & \includegraphics[width=0.33\textwidth]{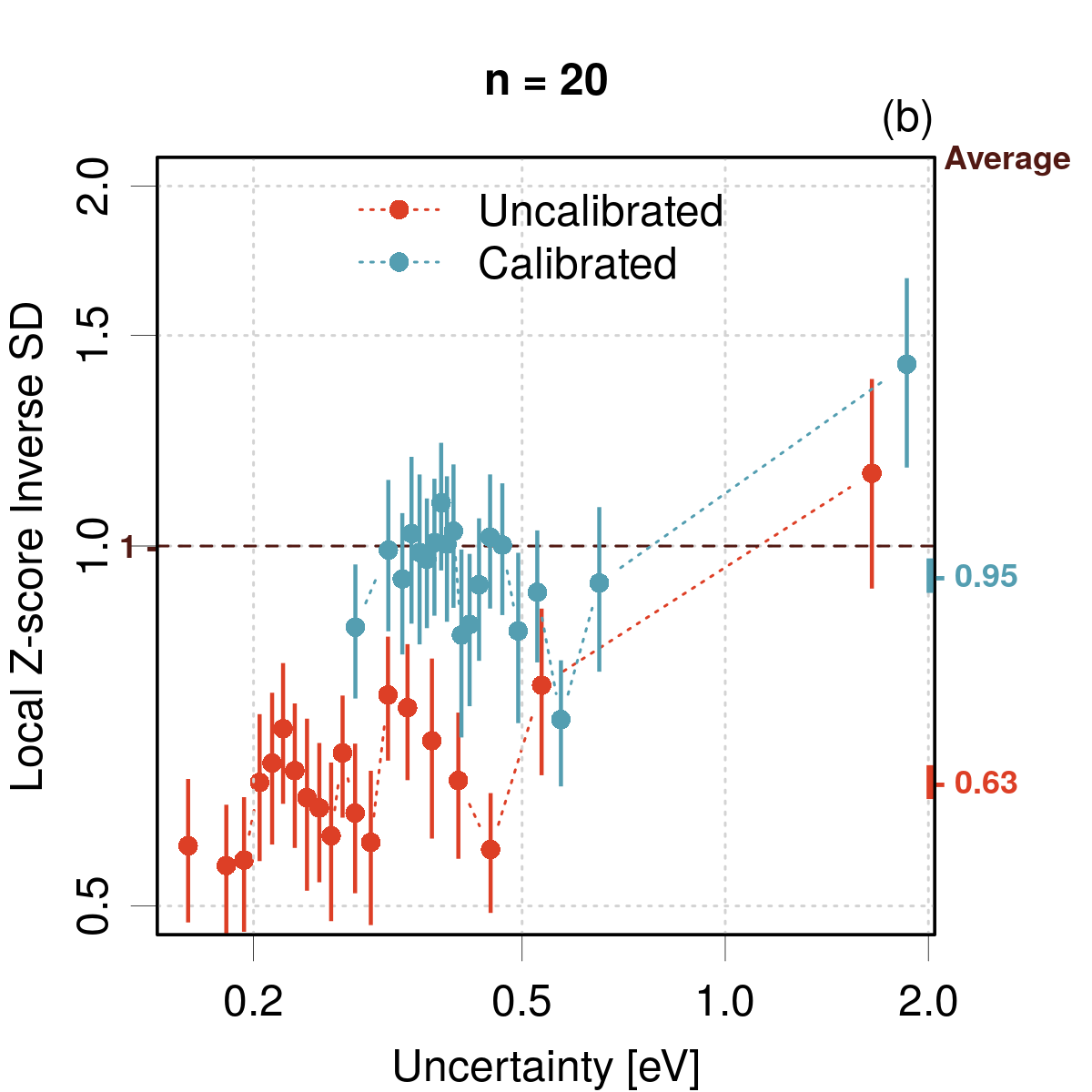} & \includegraphics[width=0.33\textwidth]{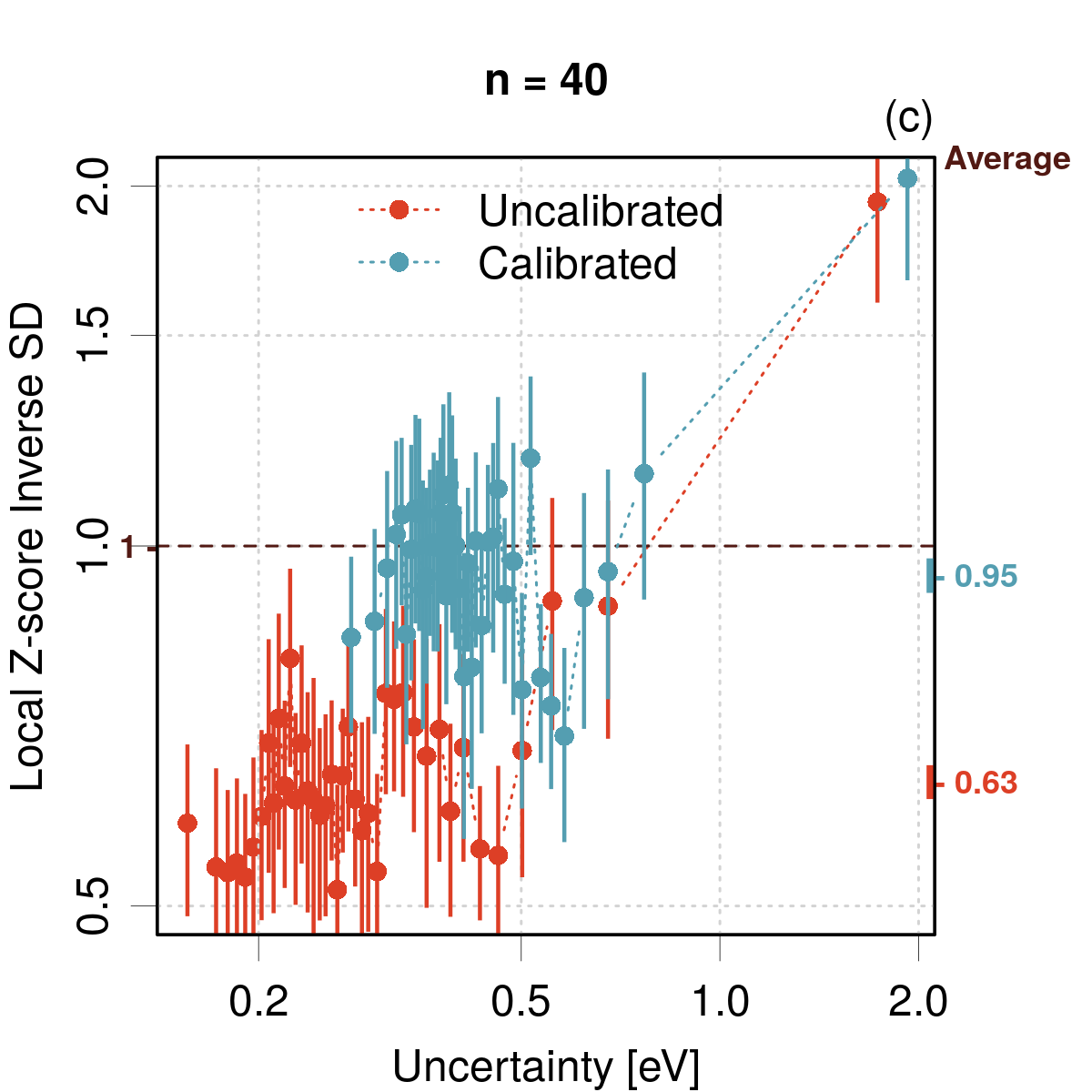}\tabularnewline
\includegraphics[width=0.33\textwidth]{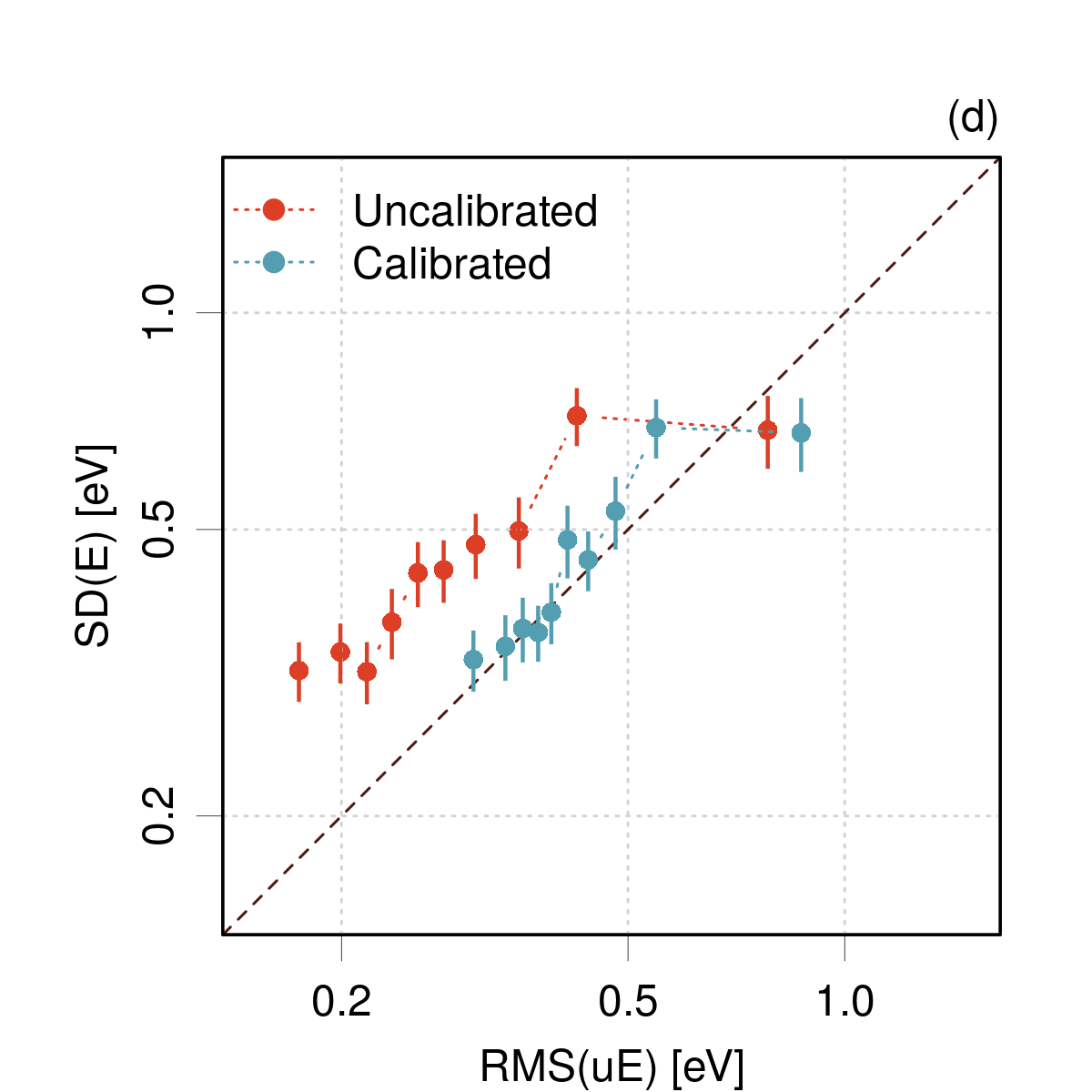} & \includegraphics[width=0.33\textwidth]{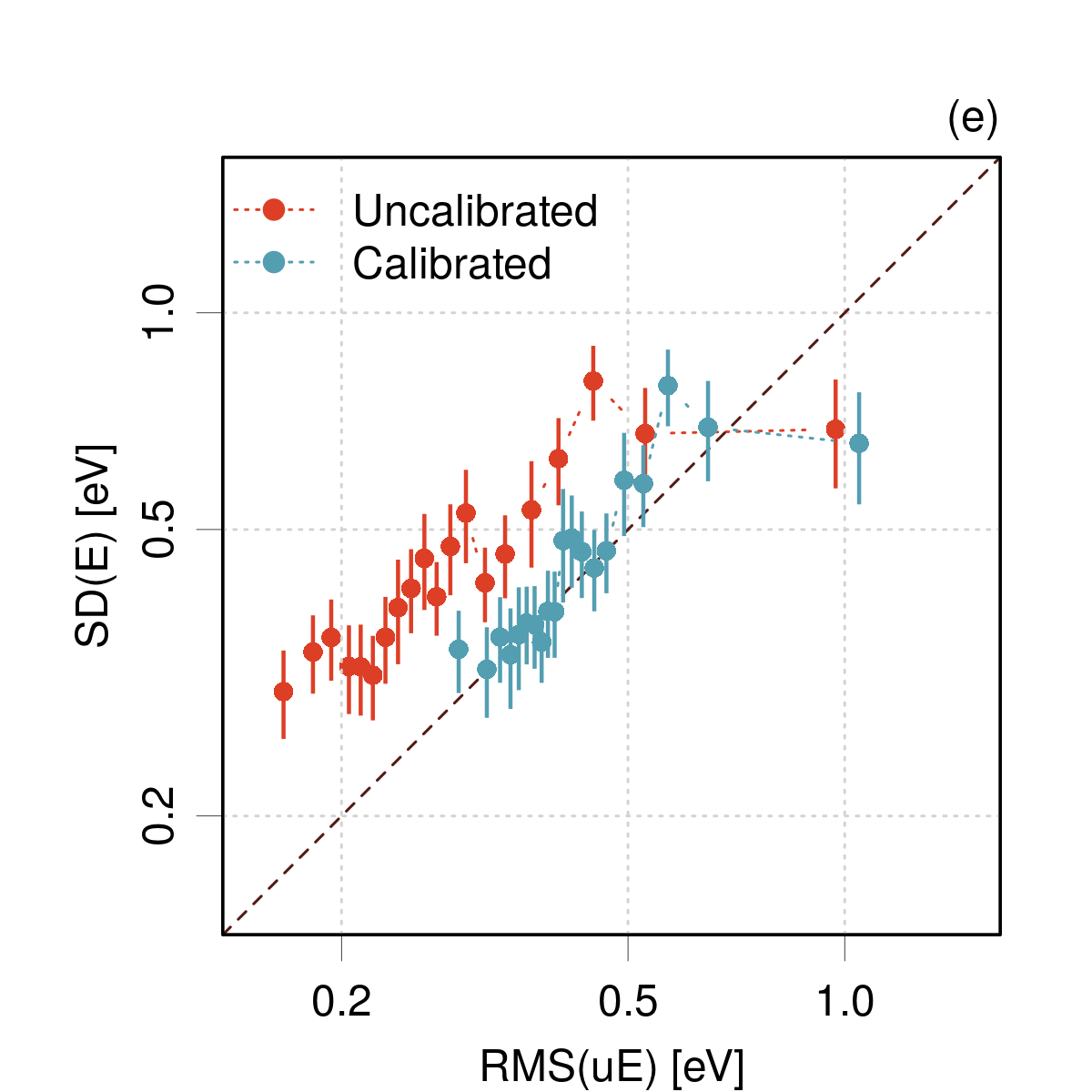} & \includegraphics[width=0.33\textwidth]{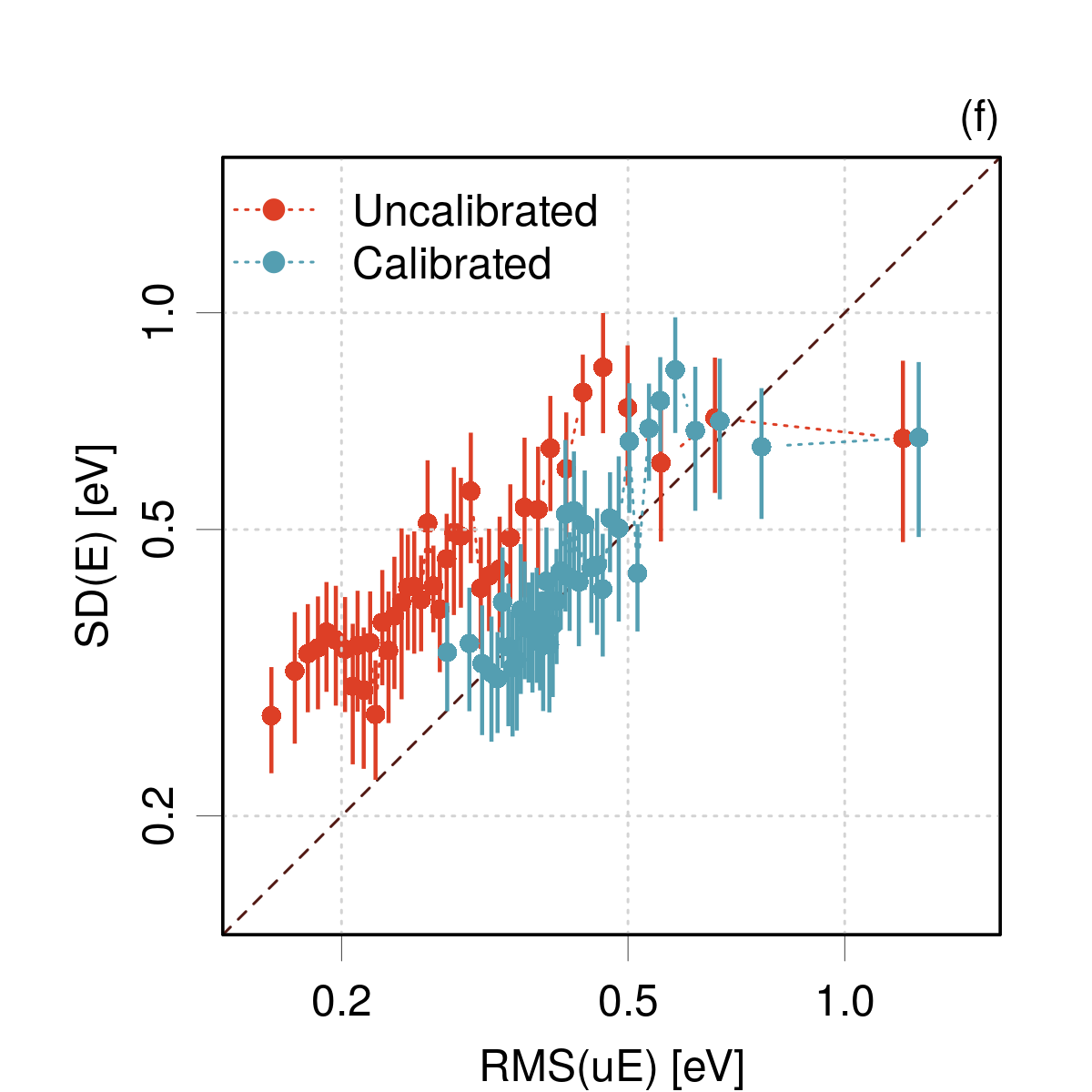}\tabularnewline
\end{tabular}
\par\end{centering}
\noindent \begin{centering}
\begin{tabular}{ccc}
\includegraphics[width=0.33\textwidth]{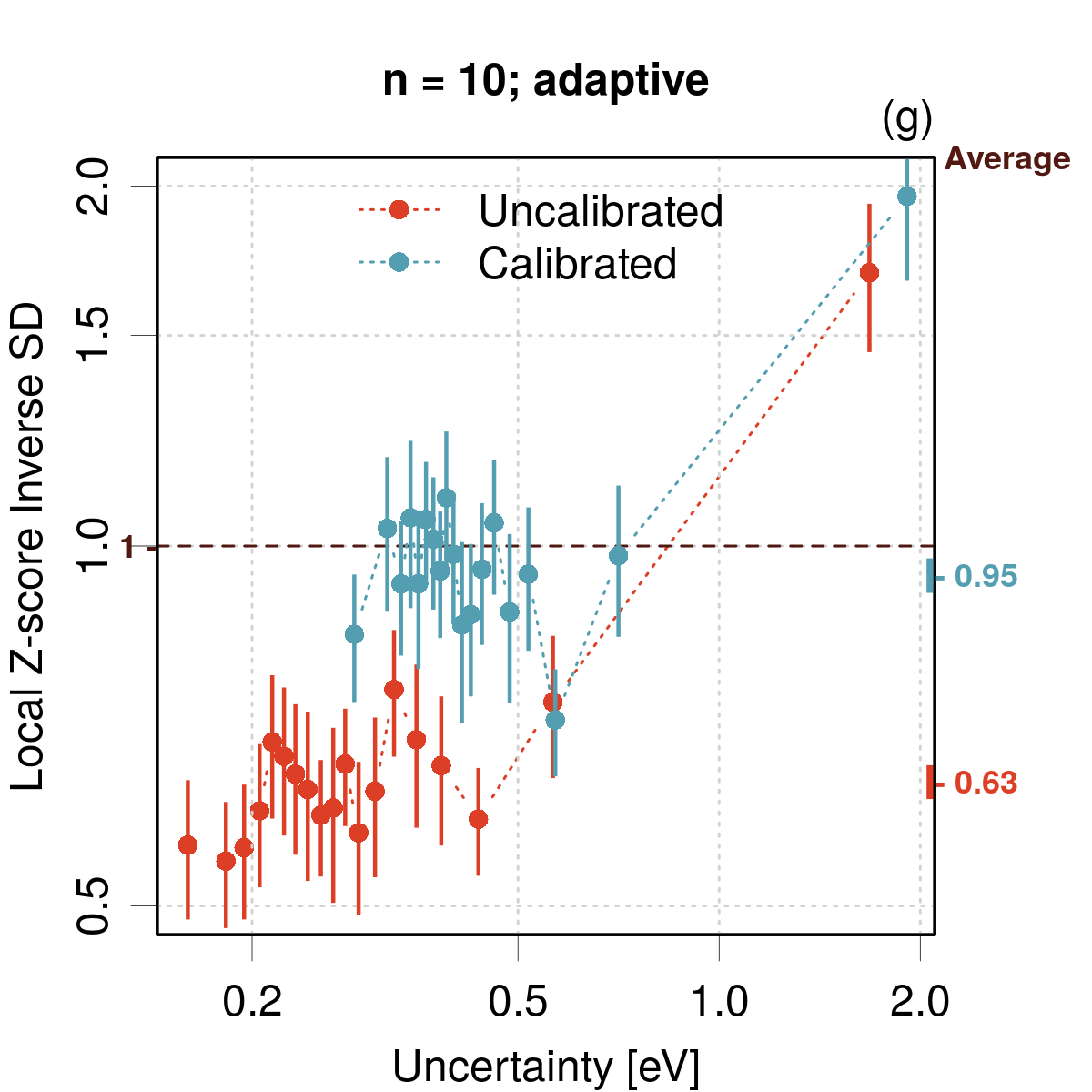} & \includegraphics[width=0.33\textwidth]{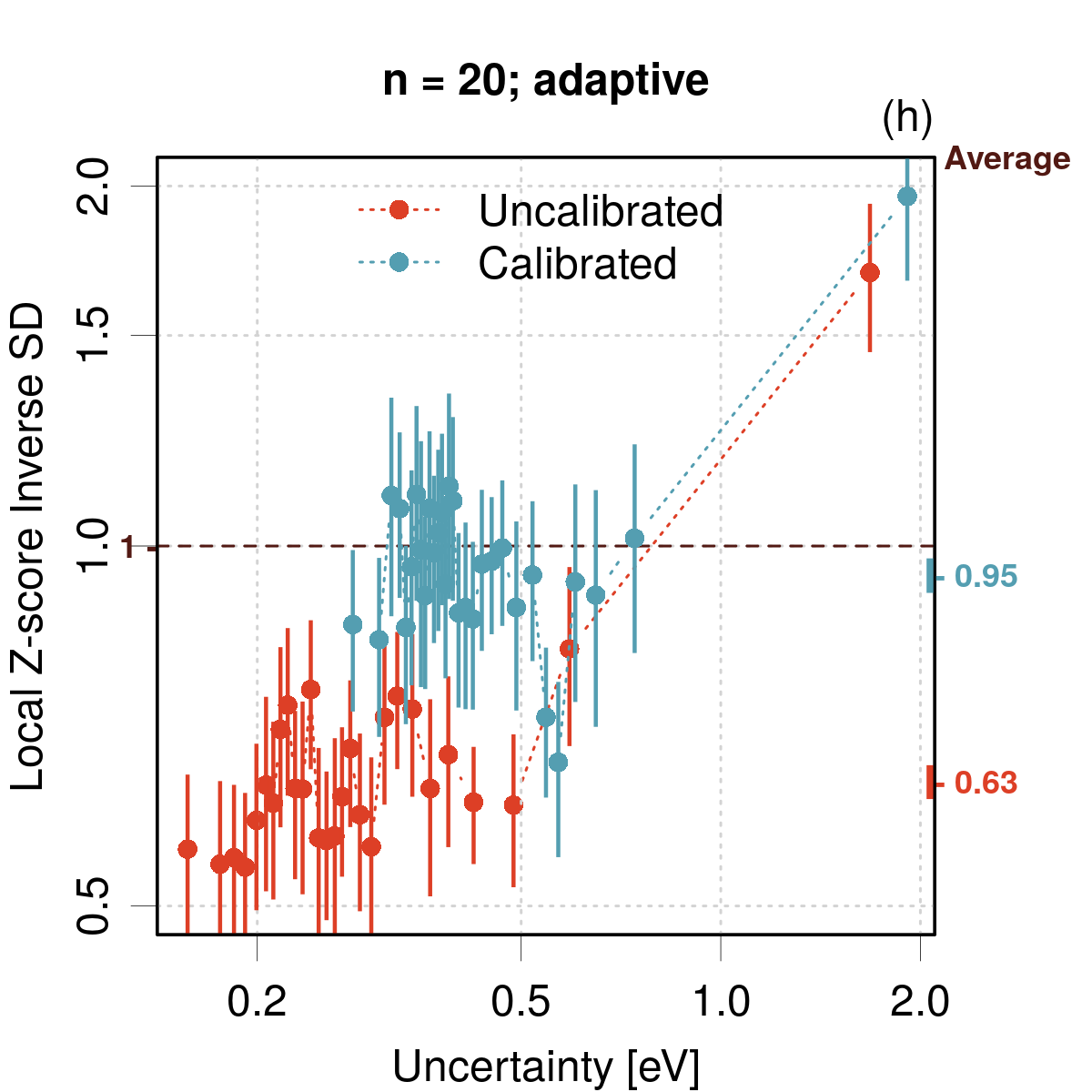} & \includegraphics[width=0.33\textwidth]{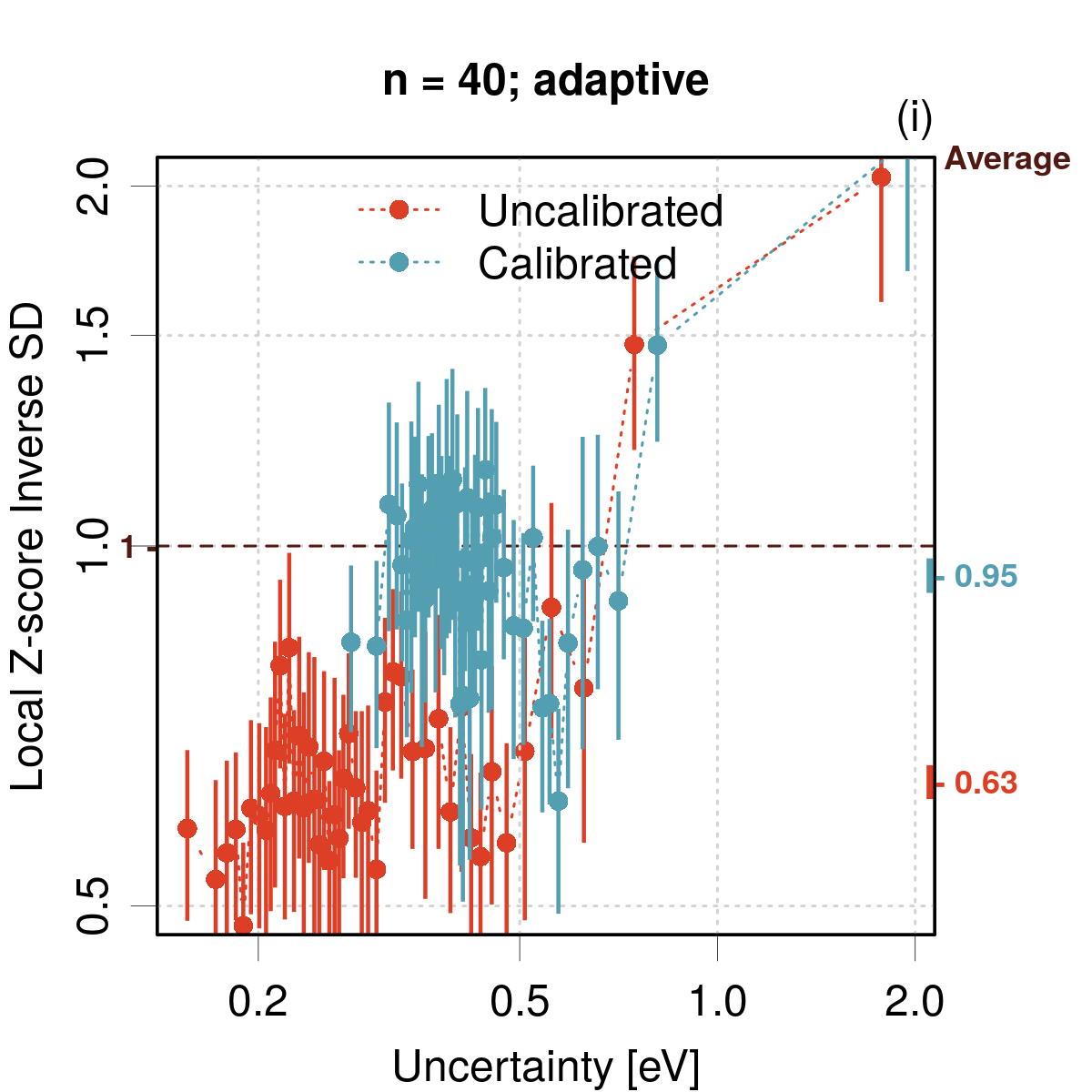}\tabularnewline
\includegraphics[width=0.33\textwidth]{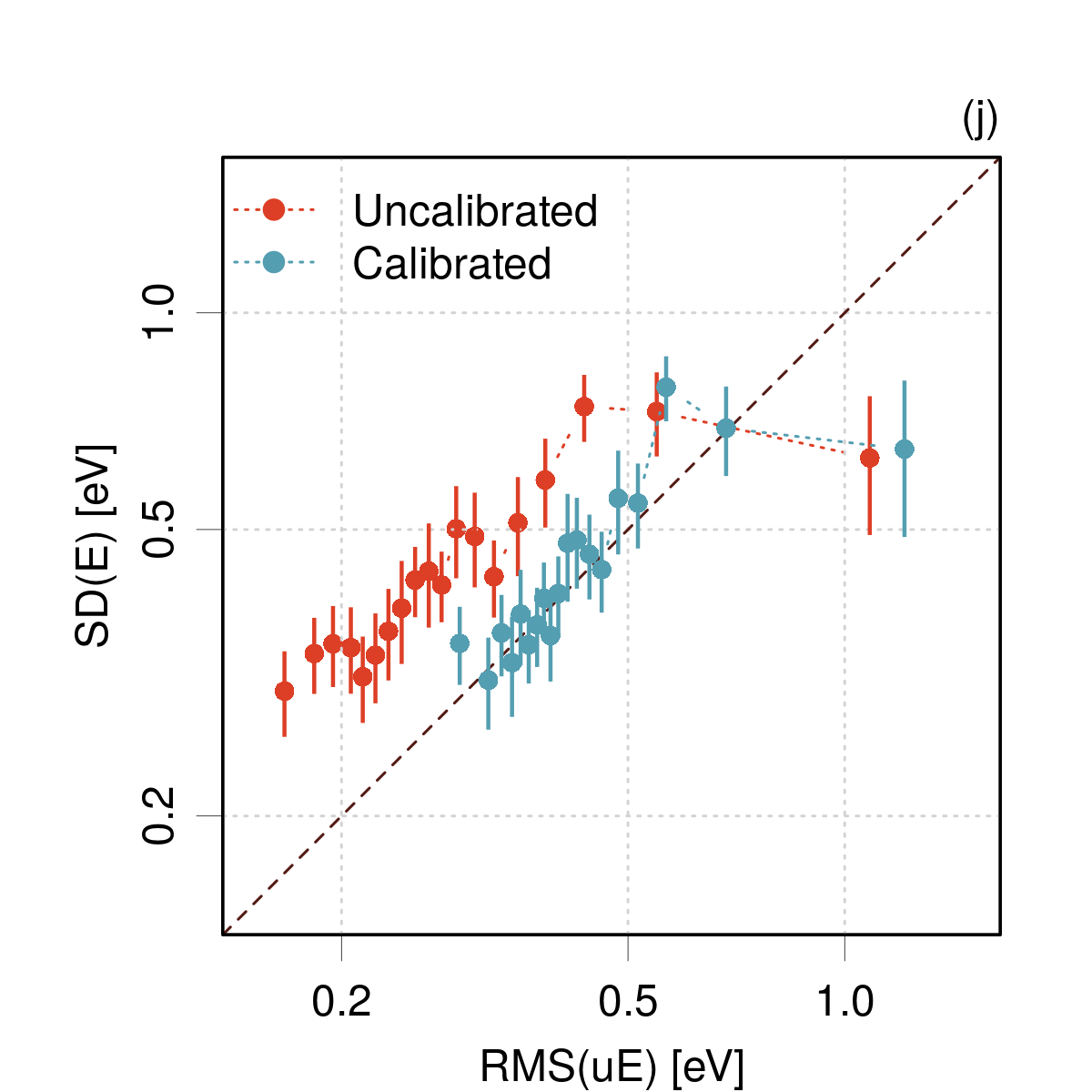} & \includegraphics[width=0.33\textwidth]{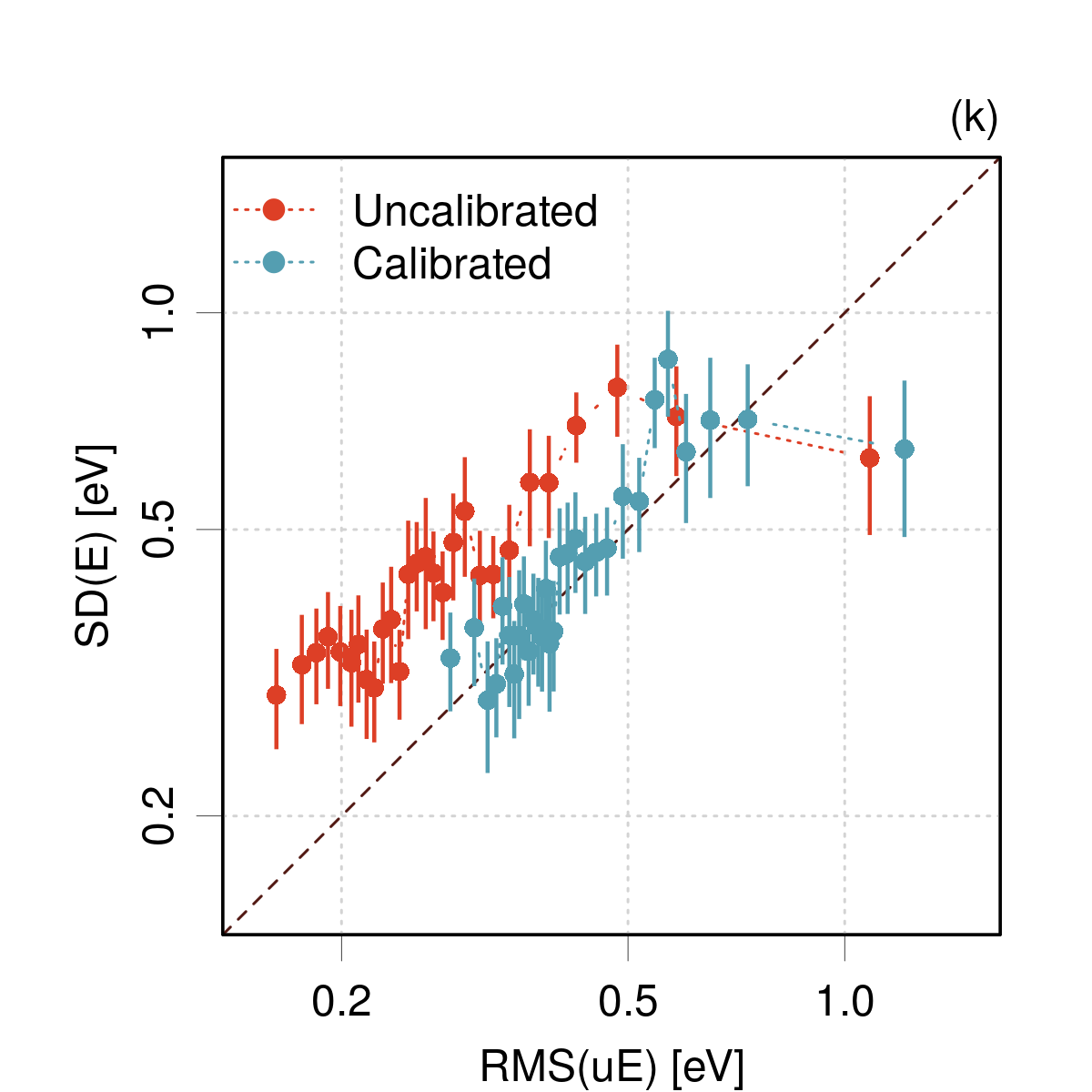} & \includegraphics[width=0.33\textwidth]{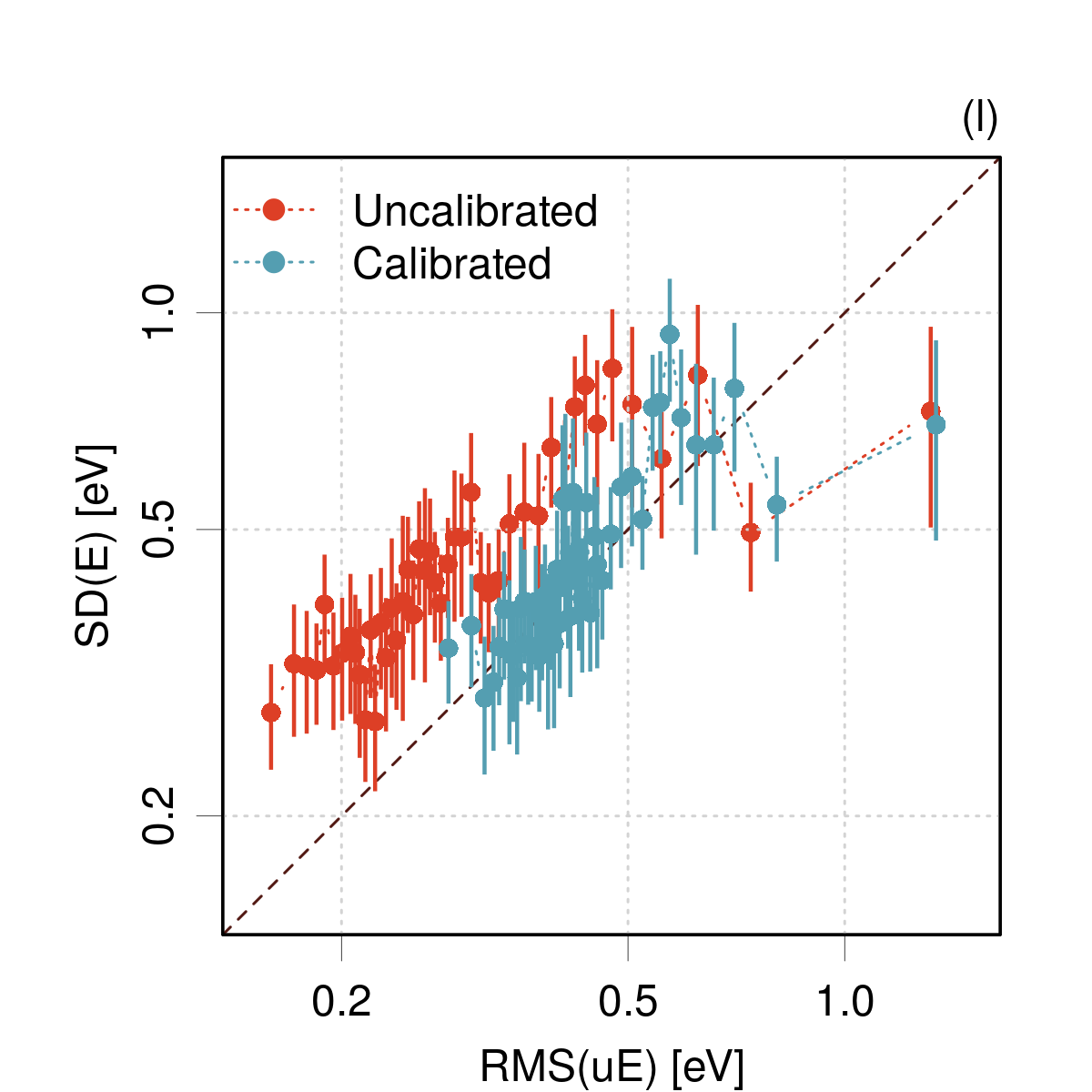}\tabularnewline
\end{tabular}
\par\end{centering}
\noindent \centering{}\caption{\label{fig:Binning-Diffusion-2}Case PAL2022/Diffusion\_LR. Impact
of the number of bins on LZISD analysis and reliability diagrams:
(a-f) bins with equal populations; (g-l) adaptive strategy.}
\end{figure*}

Let us consider first the Diffusion/LR case. For both types of analysis,
the counts-based binning {[}Fig.\,\ref{fig:Binning-Diffusion-2}(a-f){]}
needs at least 20 bins to reveal an overestimation problem for the
largest uncertainties, with a stronger effect for 40 bins. The averaging
effect in large bins ($n=10$) would lend us to believe that calibration
is good in this area, whereas using four times smaller bins reveals
uncertainty in excess by a factor two around 2\,eV. 

By using the new adaptive strategy {[}Fig.\,\ref{fig:Binning-Diffusion-2}(g-l){]},
the problem is apparent for all the specified numbers of bins. In
fact, for a starting point of 10 bins, the merge/split strategy converges
to 20 bins, for 20 bins, one gets 30 and for 40 bins 47. The first
case provides a good binning without the useless details that arise
from the higher bin numbers. 

For this dataset, one clearly has consistency problems with under-
and over-estimation for uncertainties larger than 0.5. These defects
compensate each other and are not detected when using large bins.
This problem is also apparent in the original article {[}Figure\,2(d){]},
albeit for a series of bins with populations below the 30 limit. Aggregating
these bins enables to conclude that the effect is statistically significant.

\begin{figure*}[t]
\noindent \begin{centering}
\begin{tabular}{ccc}
\includegraphics[width=0.33\textwidth]{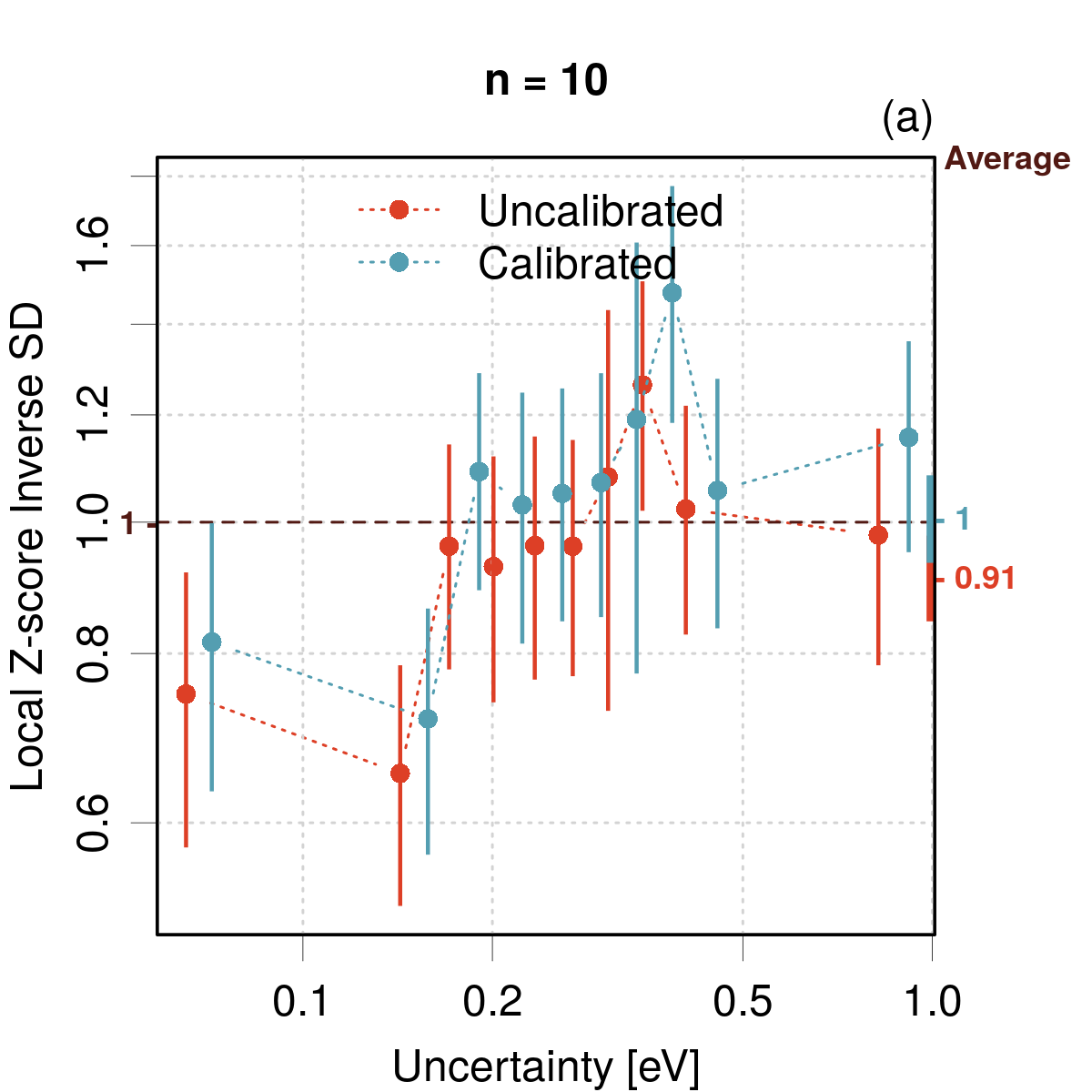} & \includegraphics[width=0.33\textwidth]{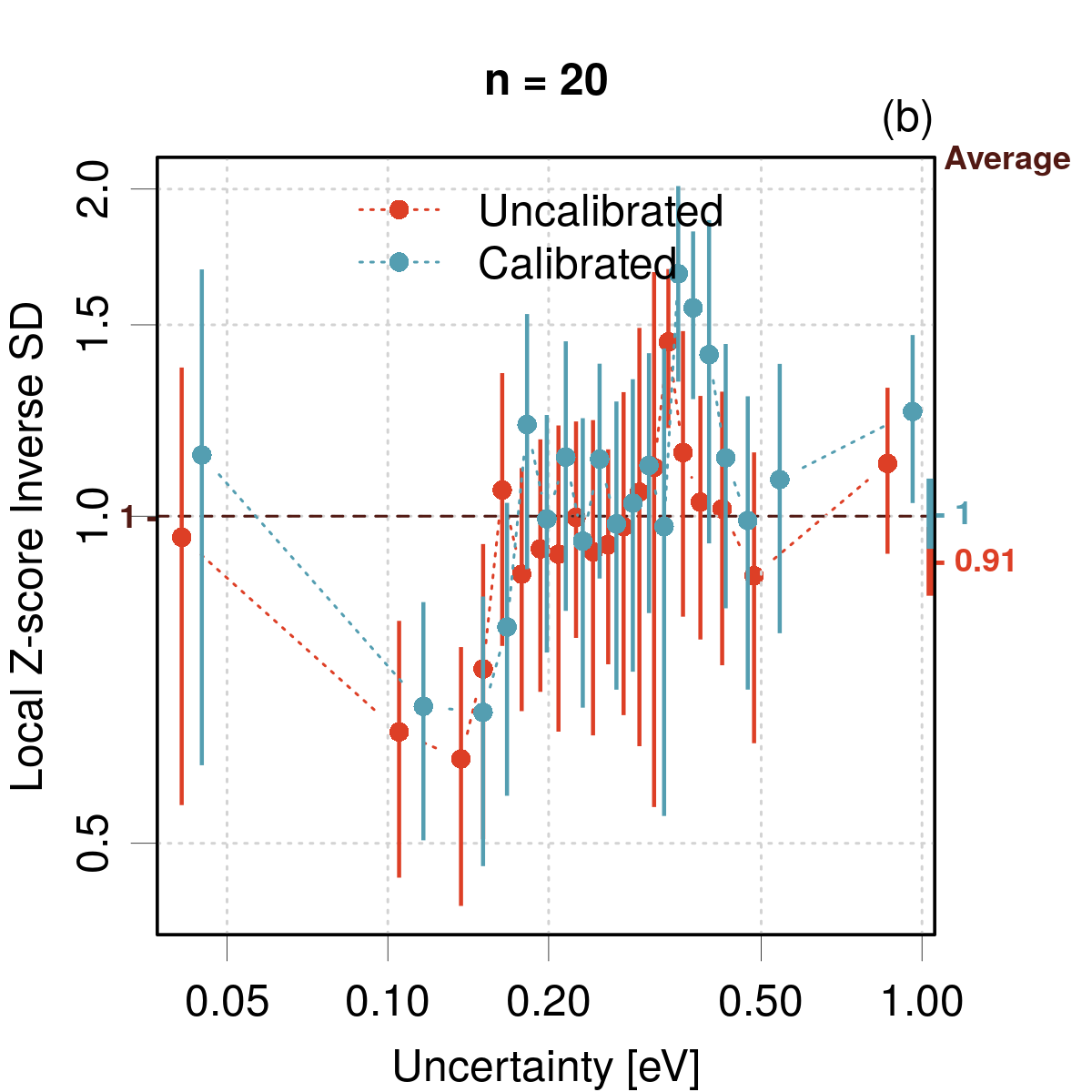} & \includegraphics[width=0.33\textwidth]{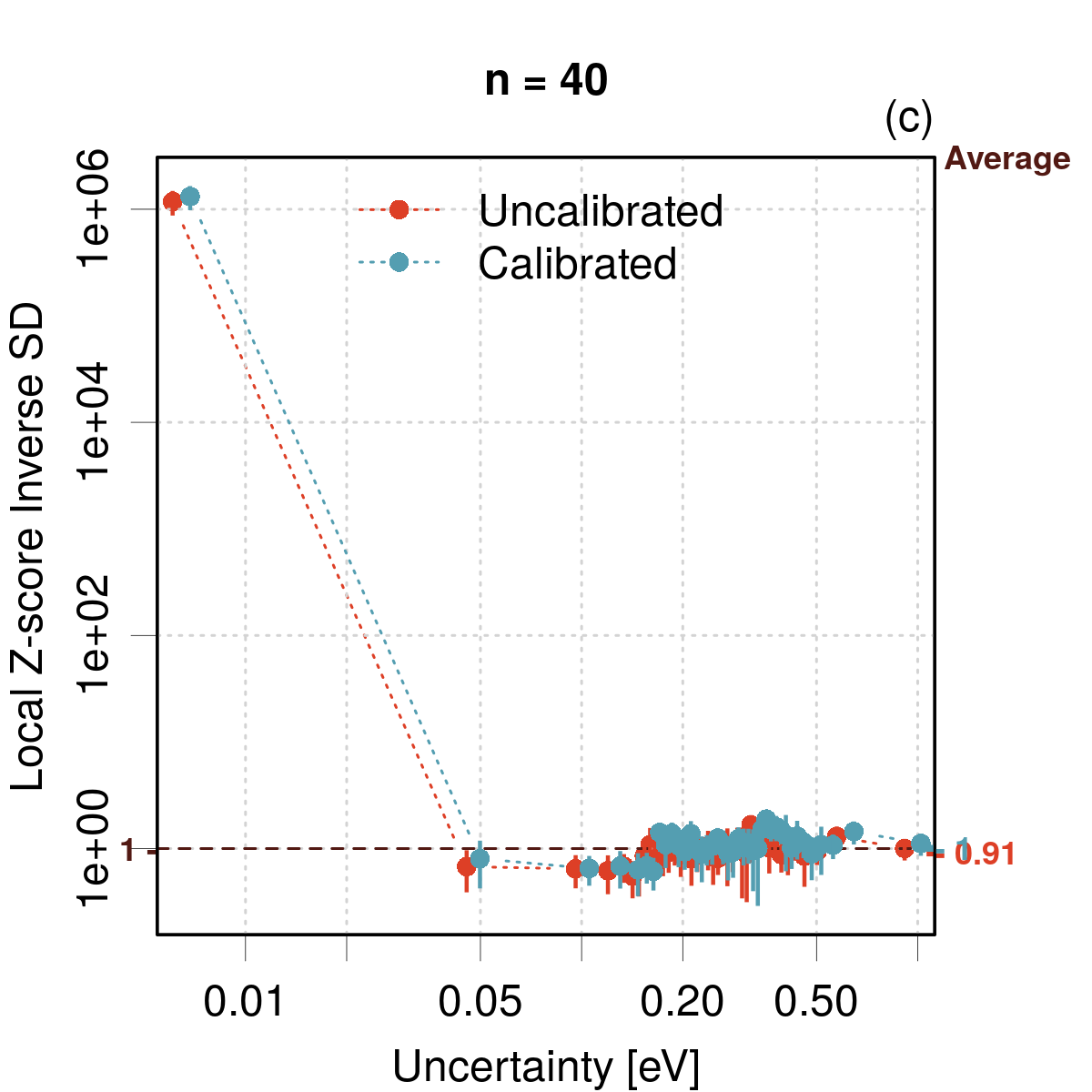}\tabularnewline
\includegraphics[width=0.33\textwidth]{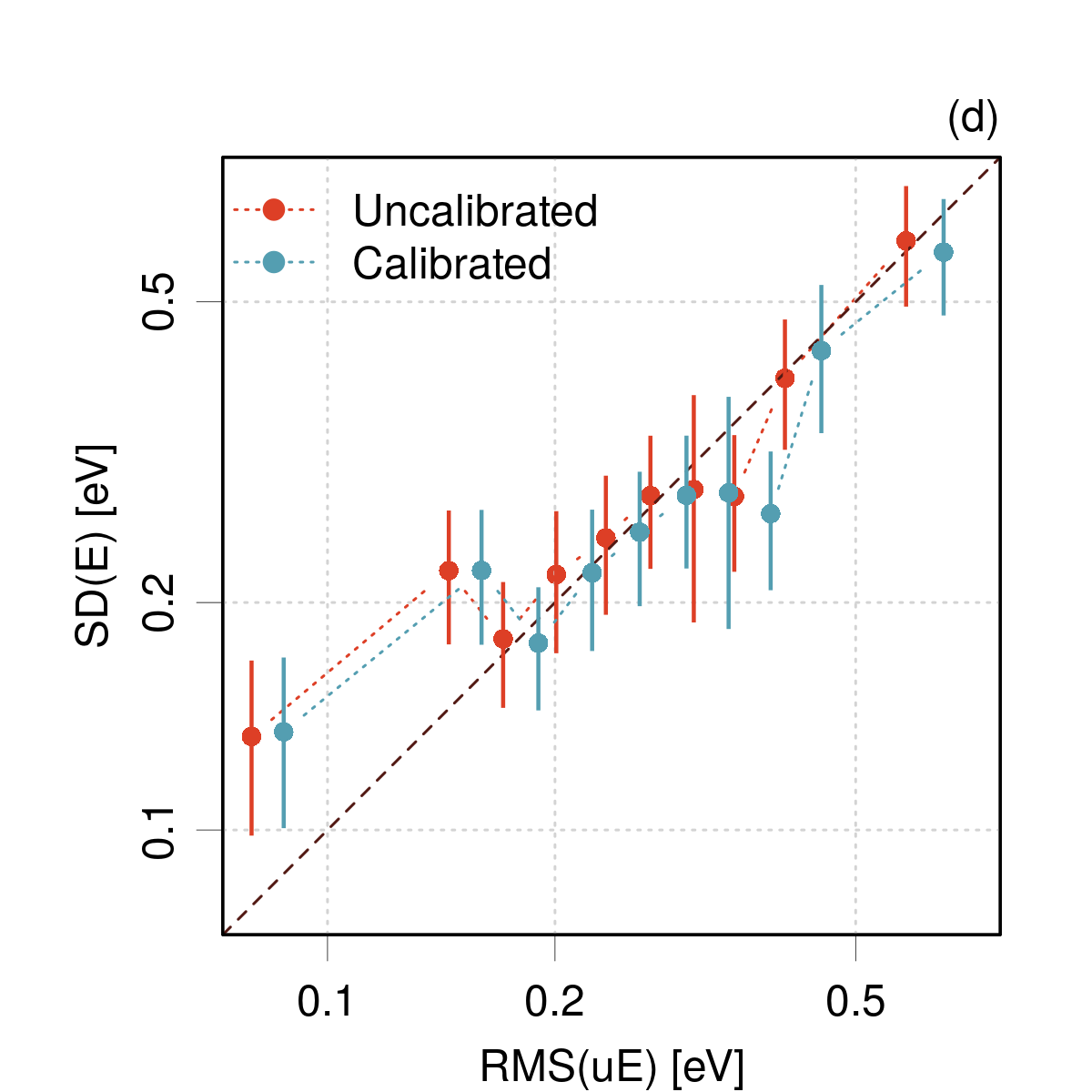} & \includegraphics[width=0.33\textwidth]{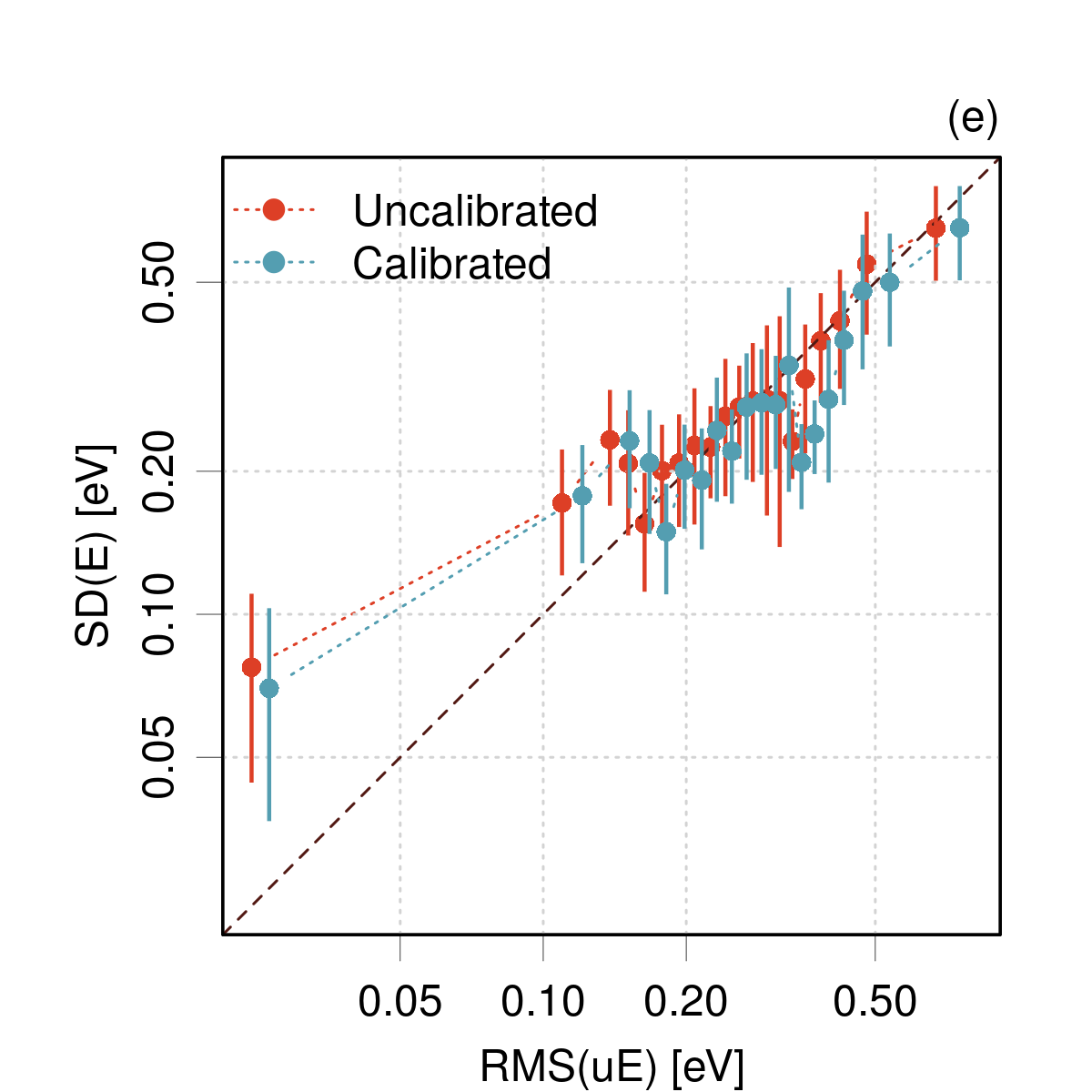} & \includegraphics[width=0.33\textwidth]{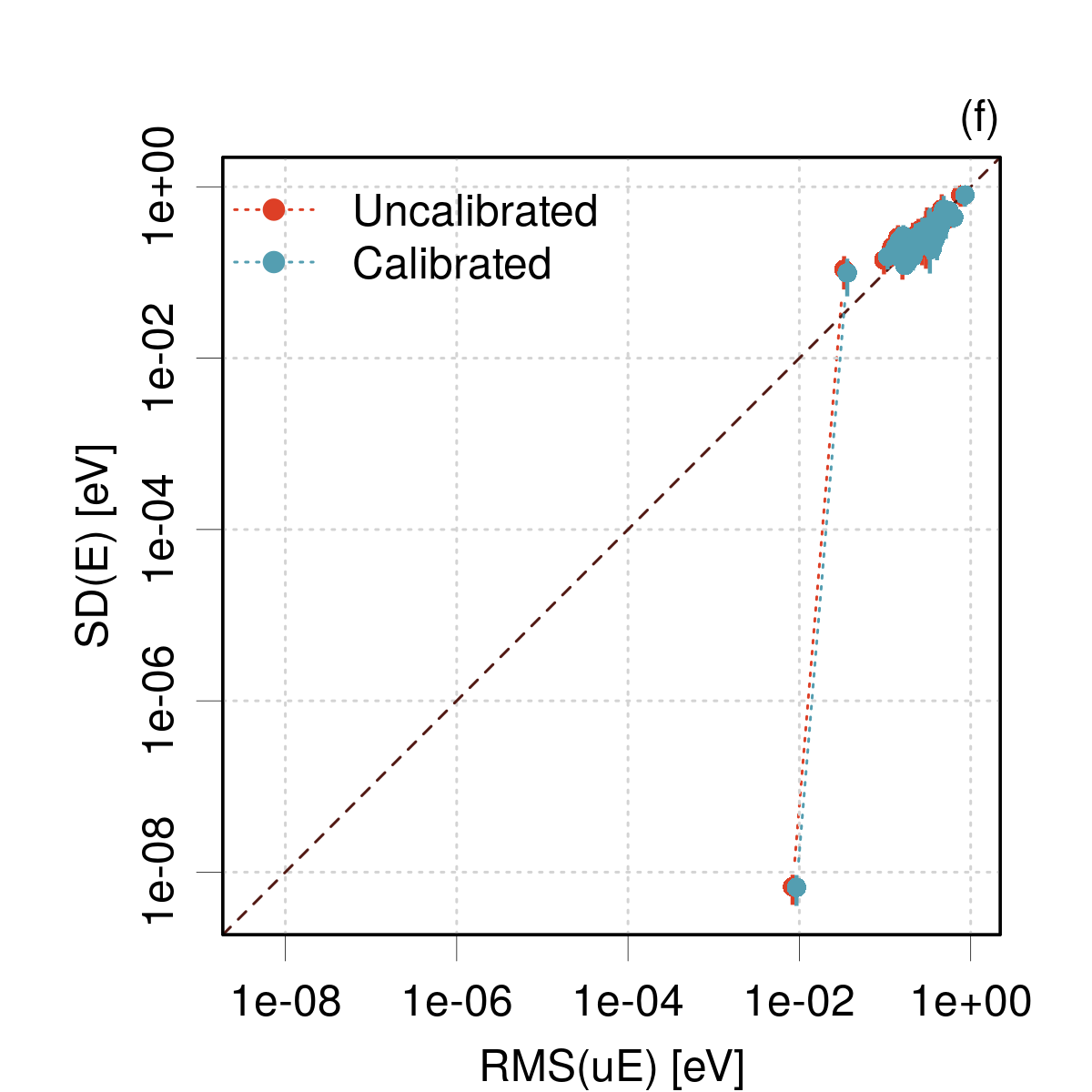}\tabularnewline
\end{tabular}
\par\end{centering}
\noindent \centering{}%
\begin{tabular}{ccc}
\includegraphics[width=0.33\textwidth]{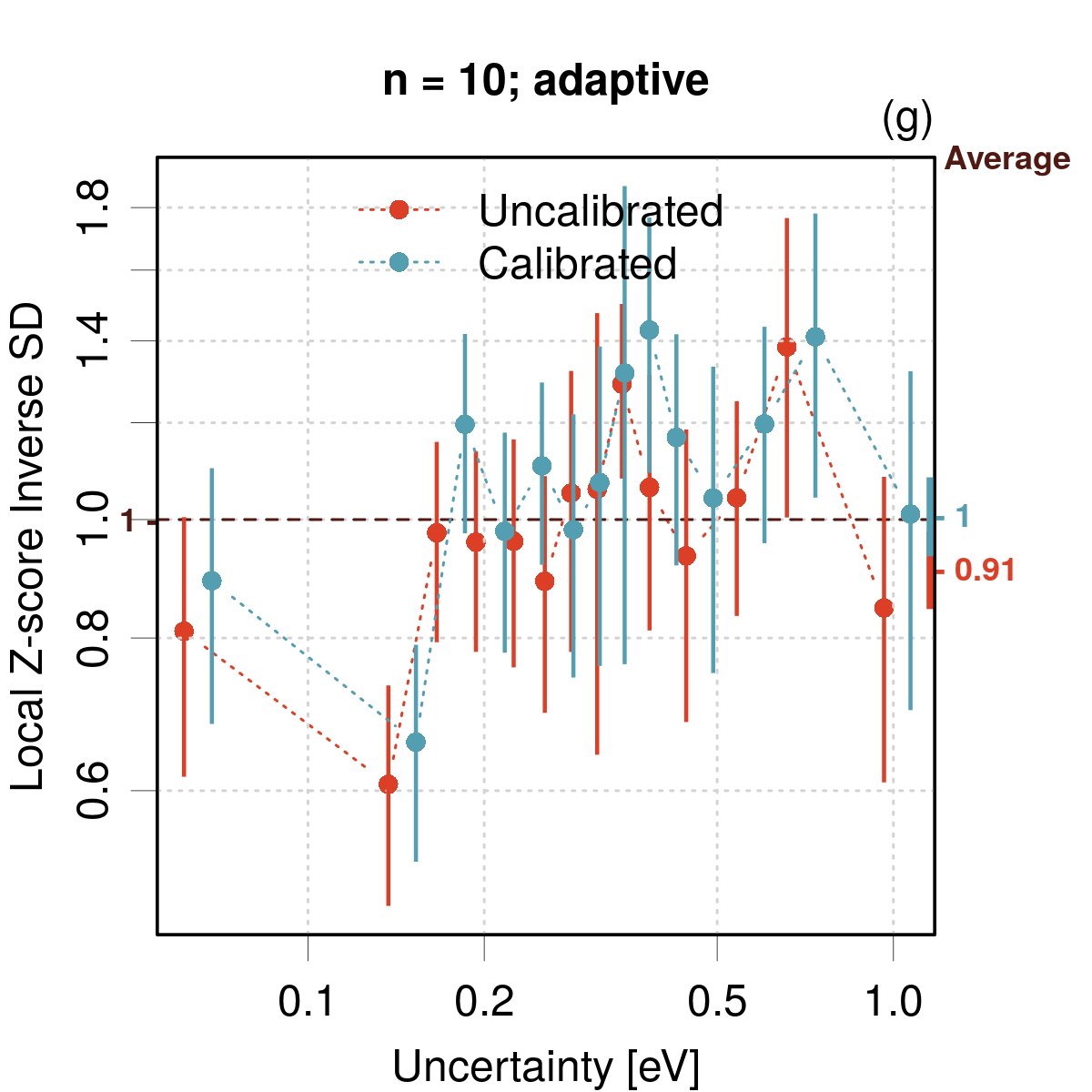} & \includegraphics[width=0.33\textwidth]{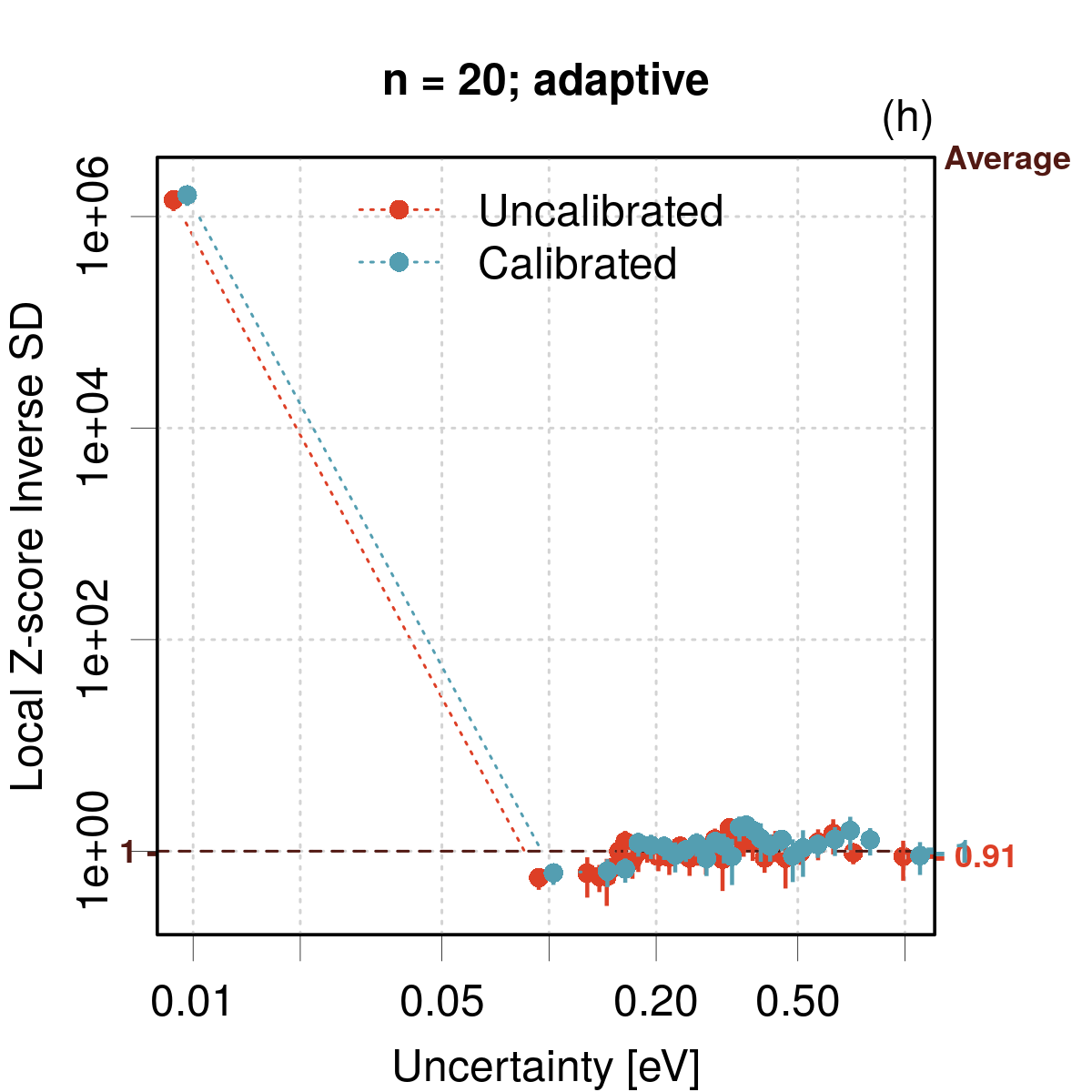} & \includegraphics[width=0.33\textwidth]{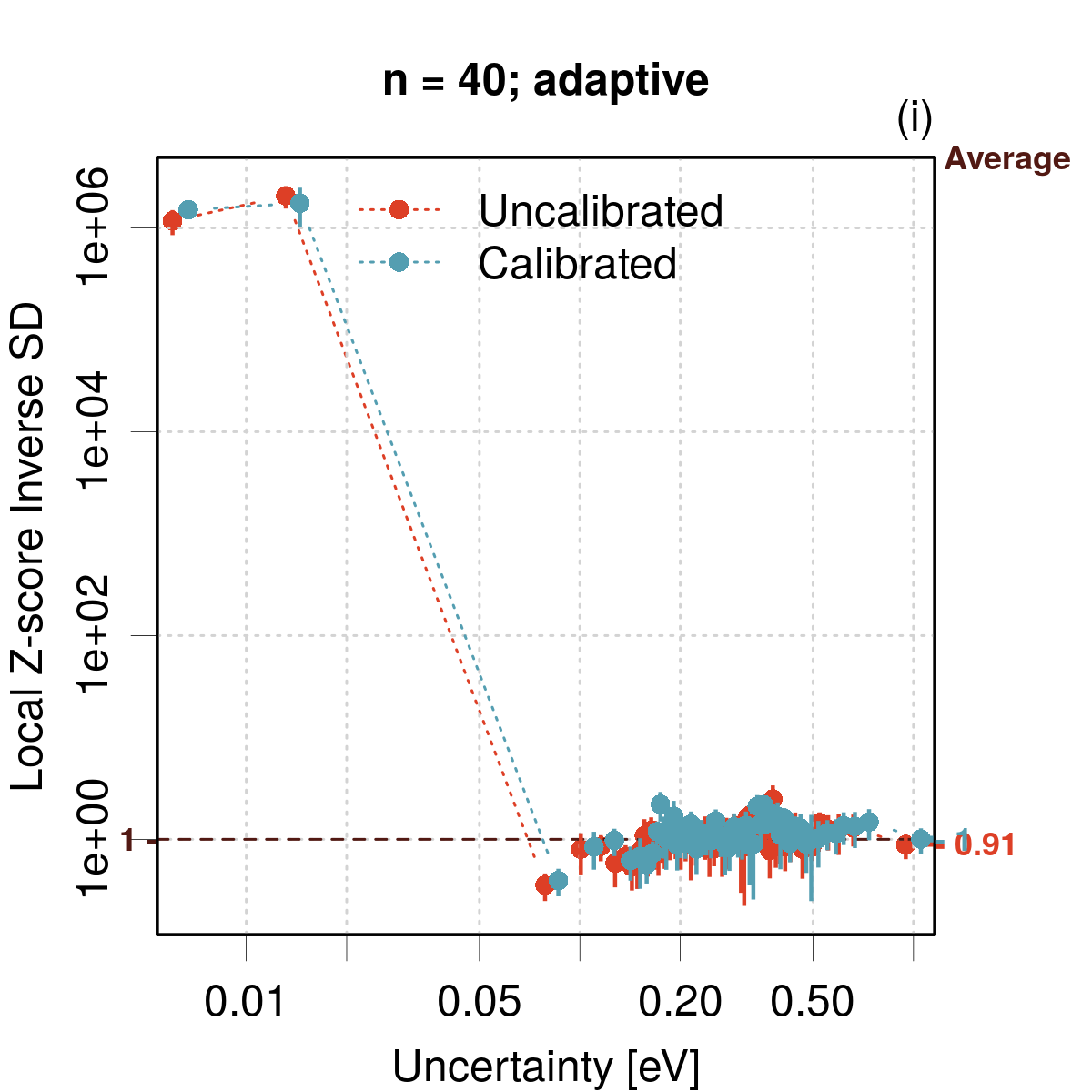}\tabularnewline
\includegraphics[width=0.33\textwidth]{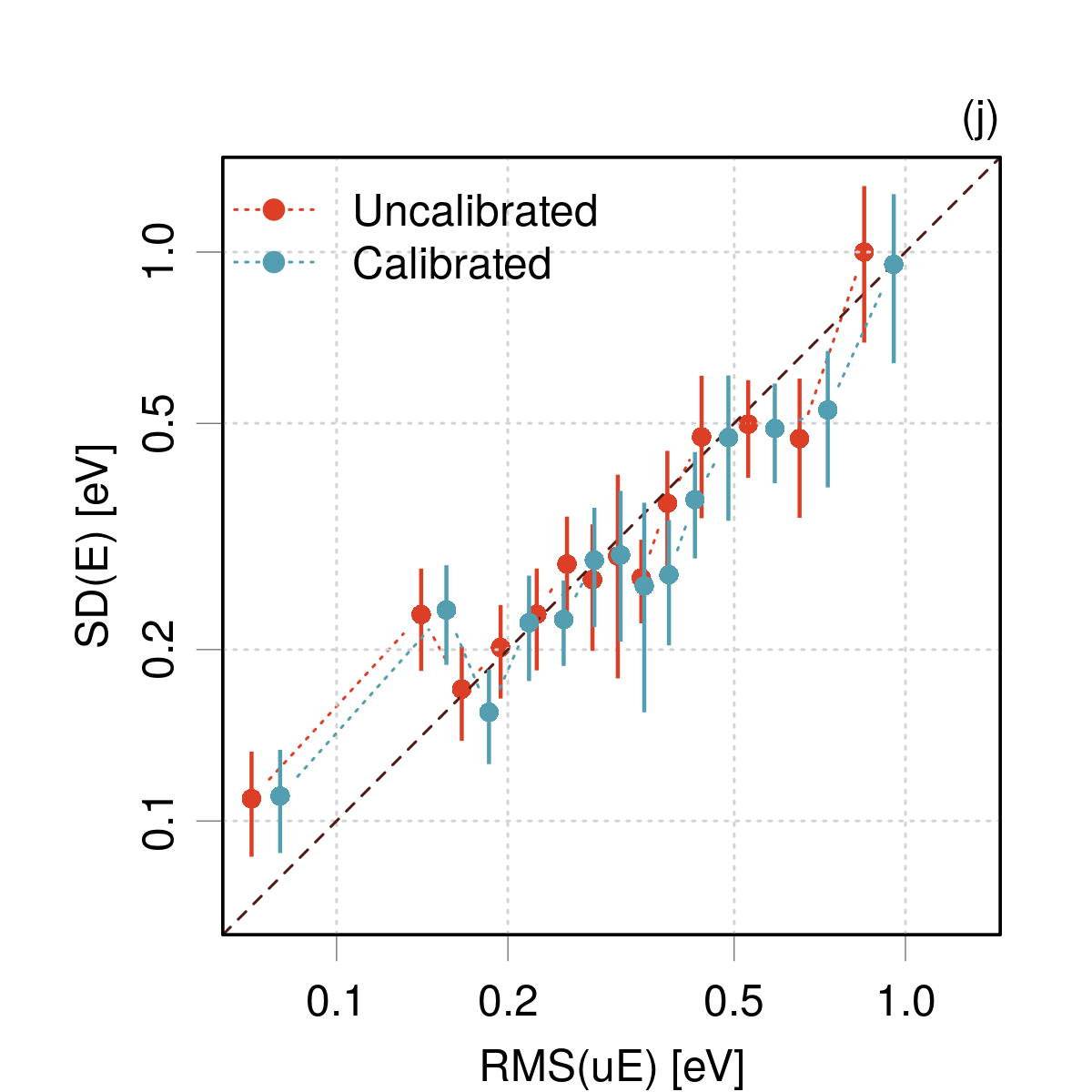} & \includegraphics[width=0.33\textwidth]{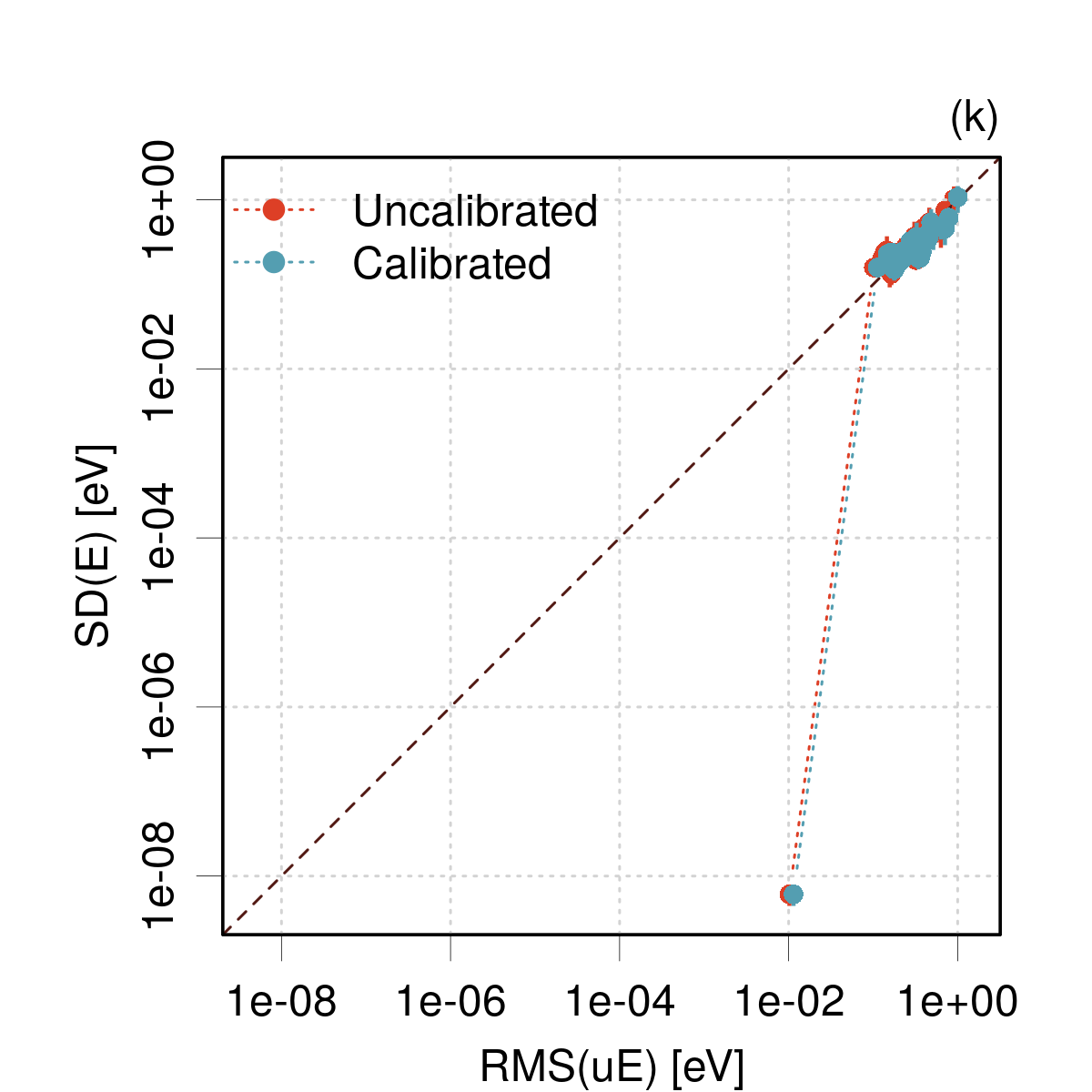} & \includegraphics[width=0.33\textwidth]{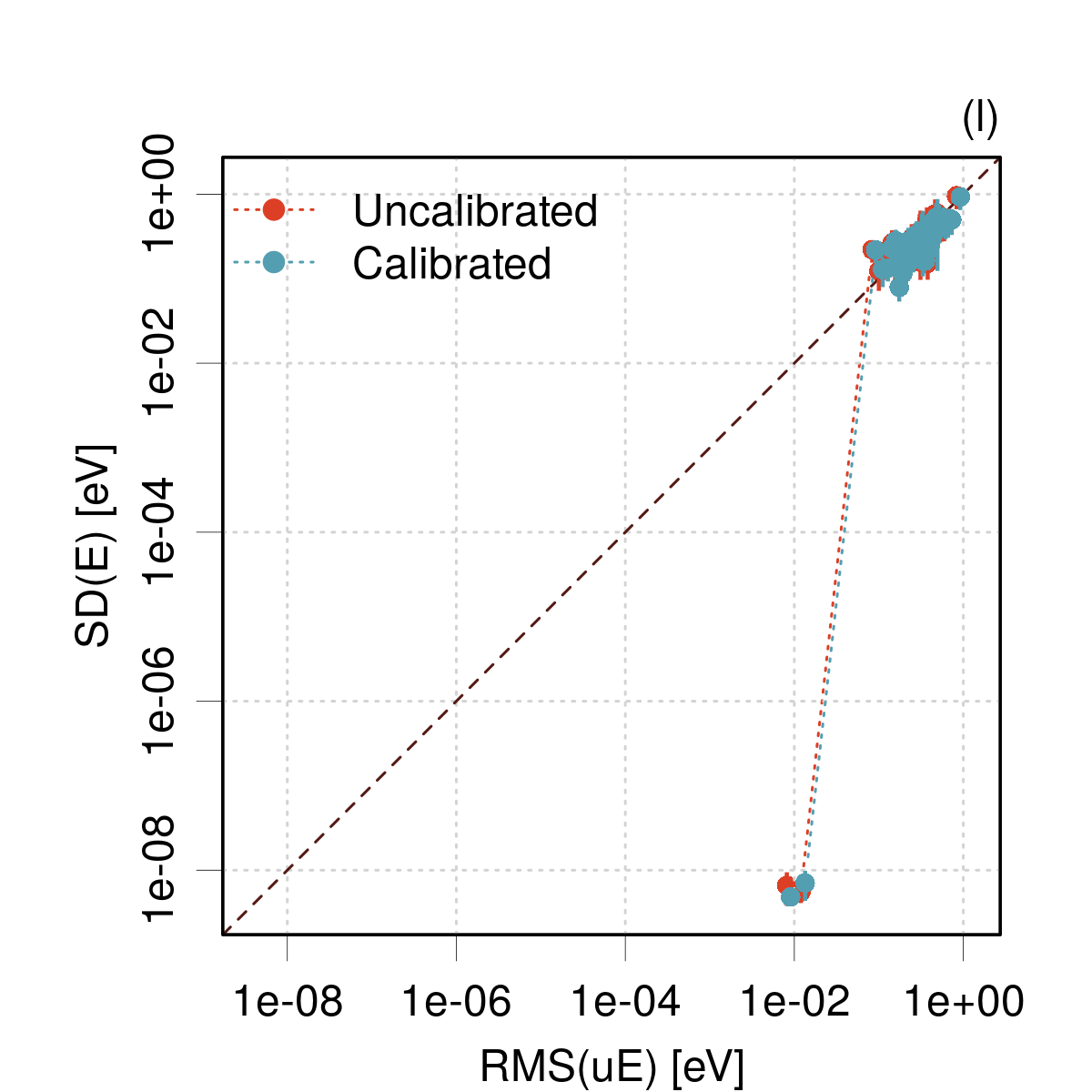}\tabularnewline
\end{tabular}\caption{\label{fig:binning-perovskite-2}Case PAL2022/Perovskite\_GPR\_Bayesian.
Impact of the number of bins on LZISD analysis and reliability diagrams:
(a-f) bins with equal populations; (g-l) adaptive strategy.}
\end{figure*}

The Perovskite/GPR\_Bayesian case is shown in Fig.\,\ref{fig:binning-perovskite-2}.
As in the previous example, the counts-based binning needs more than
20 bins to display a problem at small uncertainties, in both LZV and
RD analyses {[}Fig.\,\ref{fig:binning-perovskite-2}(a-f){]}. The
LZV analysis indicates a strong overestimation of the uncertainties
around 0.01\,eV, the reliability diagram showing that the corresponding
errors have a standard deviation of about 1E-8, which would certainly
deserve a closer inspection. The adaptive analysis needs 20 initial
bins (30 after adaptation) to reveal the problem {[}Fig.\,\ref{fig:binning-perovskite-2}(g-l){]}.
This ``data nugget'' is undetected by the linear equal-width binning
of the original study (see Figure\,38 of the \href{https://static-content.springer.com/esm/art\%3A10.1038\%2Fs41524-022-00794-8/MediaObjects/41524_2022_794_MOESM1_ESM.pdf}{Supplementary Information}
to Palmer \emph{et al.}\citep{Palmer2022}). 

The adaptive binning strategy in logarithmic uncertainty seems to
be able to detect consistency problems more efficiently than binning
in linear uncertainty, either with bins of same counts or same width.
Of course, it does not apply to variables with negative or null values,
where it can be replaced by an adaptive strategy in linear space.
The adaptive binning strategy is used for the LZISD, LCP and reliability
diagrams in the case studies of the present article (Sect.\,\ref{sec:Application})
with a starting point of $n=20$ bins.
\end{document}